\documentclass[structabstract]{aa}  
\usepackage{graphicx}
\usepackage{txfonts}
\usepackage{natbib}
\usepackage{longtable}

\bibpunct[]{(}{)}{;}{a}{}{,}
\begin{document}
   \title{Increased complexity in interstellar chemistry:\\
   Detection and chemical modeling of ethyl formate\\
   and \textit{n}-propyl cyanide in Sgr~B2(N)
   \thanks{Based on observations carried out with the IRAM 30~m telescope. IRAM 
   is supported by INSU/CNRS (France), MPG (Germany) and IGN (Spain).}}


   \author{A. Belloche\inst{1}
          \and
          R.~T. Garrod\inst{2,1}
          \and
          H.~S.~P. M{\"uller}\inst{3,1}
          \and
          K.~M. Menten\inst{1}
          \and
          C. Comito\inst{1}
          \and
          P. Schilke\inst{1}
          }

   \institute{Max-Planck Institut f\"ur Radioastronomie, Auf dem H\"ugel 69,
              53121 Bonn, Germany\\
              \email{[belloche;kmenten;ccomito;schilke]@mpifr-bonn.mpg.de}
         \and
              Department of Astronomy, Cornell University, 
              106 Space Sciences Building, Ithaca, NY 14853, USA\\
              \email{rgarrod@astro.cornell.edu}
         \and
             I. Physikalisches Institut, Universit{\"a}t zu K{\"o}ln,
             Z{\"u}lpicher Str. 77, 50937 K{\"o}ln, Germany\\
             \email{hspm@ph1.uni-koeln.de}
             }

   \date{Received 19 December 2008; accepted 17 February 2009}

 
  \abstract
   {In recent years, organic molecules of increasing complexity have 
   been found toward the prolific Galactic center source Sagittarius~B2.}
   {We wish to explore the degree of complexity that the interstellar 
   chemistry can reach in star-forming regions.}
   {We carried out a complete line survey of the hot cores Sgr~B2(N) and (M) 
   with the IRAM 30~m telescope in the 3~mm range, plus partial surveys at 2 
   and 1.3~mm. We analyzed this spectral survey in the local thermodynamical 
   equilibrium approximation. We modeled the emission of all known molecules 
   simultaneously, which allows us to search for less abundant, more complex 
   molecules. We compared the derived column densities with the predictions of 
   a coupled gas-phase and grain-surface chemical code.}
   {We report the first detection in space of ethyl formate (C$_2$H$_5$OCHO) 
   and \textit{n}-propyl cyanide (C$_3$H$_7$CN) toward Sgr~B2(N). 
   The detection of \textit{n}-propyl cyanide is based on refined spectroscopic 
   parameters derived from combined analyses of available laboratory 
   spectroscopic data.
   For each molecule, we identified spectral features at the predicted 
   frequencies having intensities compatible with a unique rotation 
   temperature. For an assumed source size of 3$\arcsec$, our modeling yields 
   a column density of $5.4 \times 10^{16}$~cm$^{-2}$, a temperature of 100~K, 
   and a linewidth of 7~km~s$^{-1}$ for ethyl formate. \textit{n}-Propyl 
   cyanide is detected with two velocity components having column densities of 
   $1.5 \times 10^{16}$~cm$^{-2}$ and $6.6 \times 10^{15}$~cm$^{-2}$, 
   respectively, for a source size of 3$\arcsec$, a temperature of 150~K, and
   a linewidth of 7~km~s$^{-1}$. The abundances of ethyl formate and 
   \textit{n}-propyl cyanide relative to H$_2$ are estimated to be 
   $3.6 \times 10^{-9}$ and  $1.0 \times 10^{-9}$, respectively.
   We derived column density ratios of \hbox{0.8 / 15 / 1} for the related 
   species \hbox{\textit{t}-HCOOH / CH$_3$OCHO / C$_2$H$_5$OCHO} and 
   \hbox{108 / 80 / 1} for \hbox{CH$_3$CN / C$_2$H$_5$CN / C$_3$H$_7$CN}. 
   Our chemical modeling reproduces these ratios reasonably well. It suggests 
   that the sequential, piecewise construction of ethyl and \textit{n}-propyl 
   cyanide 
   from their constituent functional groups on the grain surfaces is their 
   most likely formation route. Ethyl formate is primarily formed on the 
   grains by adding CH$_3$ to functional-group radicals derived from 
   methyl formate, although ethanol may also be a precursor.}
   {The detection in Sgr~B2(N) of the next stage of complexity in two 
   classes of complex molecule, esters and alkyl cyanides, suggests that 
   greater complexity in other classes of molecule may be present in the 
   interstellar medium.}


   \titlerunning{Detection and chemical modeling of ethyl formate and 
   \textit{n}-propyl cyanide in Sgr~B2(N)}
   \maketitle
%

\section{Introduction}
\label{s:intro}

More than 150 molecules have been discovered in the interstellar medium or in 
circumstellar envelopes over the past four decades
\citep[see, e.g.,][\footnote{\label{fn:cdms1} Visit the Cologne Database for 
Molecular Spectroscopy (CDMS) at http://www.cdms.de for an updated 
list.}]{Mueller05}. 
Among them, ``complex'' organic molecules with up to 13 atoms have been found, 
showing that the interstellar chemistry in some regions is efficient 
enough to achieve a relatively high degree of chemical 
complexity\footnote{These molecules are ``complex'' 
for astronomers, not for biologists!}. In addition, much larger molecules
have been found in meteorites discovered on Earth, including more than 80 
distinct amino acids. The non-terrestrial isotopic ratios of these amino acids, 
as well as their racemic distributions\footnote{A racemic distribution 
means equal amounts of left- and right-handed enantiomers. Enantiomers are 
stereoisomers that are mirror images of each other and non-superposable.},
suggest that they, or at least 
their direct precursors, have an interstellar origin \citep[see, e.g.,][, and 
references therein]{Ehrenfreund01,Bernstein02,Elsila07}. 
Interstellar chemistry is therefore very likely capable of producing
more complex organic molecules than those discovered in the
interstellar medium so far. However, the degree of complexity that may
be reached is still an open question; the partition functions of 
larger molecules are large, making it much more difficult to detect such 
species, even if they are present in reasonably large quantities.

Grain-surface chemistry is 
frequently invoked as the formation mechanism of many complex
species, particularly following recent determinations of some key 
gas-phase reaction rates. Gas-phase production of methyl formate, a molecule 
ubiquitous in hot-core spectra, appears prohibitively slow \citep[][]{Horn04}, 
pointing to an efficient alternative. Additionally, the dissociative
recombination of large organic molecular ions with electrons, which is 
typically the final step in the gas-phase synthesis of complex molecules, 
appears strongly to favor the fragmentation of complex structure
\citep[][]{Geppert06}. 

In the case of hot cores, the granular ice mantles built up during prior phases 
of evolution present a rich source of simple saturated molecules from
which more complex species may form, as has long been realized 
\citep[][]{Millar91}.
However, while the efficiency of complex molecule formation in the gas phase 
is limited (not exclusively) by the need to stabilize the energized complex, 
often resulting in fragmentation, adhesion to a grain surface allows an adduct 
to quickly thermalize.
Thus, molecular radicals derived from 
the ice mantles may combine \textit{in situ} on the grain surfaces to build up 
complex structures efficiently, if dust temperatures are sufficient for the 
reactants to meet by thermal diffusion. The hot-core models of \citet{Garrod06}
and \citet{Garrod08a} have demonstrated the plausibility of such mechanisms in 
reproducing observed abundances of many complex organic species.

The detection of new complex molecules places valuable constraints on
the chemical models. In the context of the model employed, e.g., by 
\citet{Garrod08a}, obtaining abundances of structurally-related molecules 
allows one to isolate the chemical behavior of the functional groups from 
which they are constructed, and to relate these back to more fundamental model
parameters such as photodissociation rates, binding energies, and
initial ice composition. Such an approach then allows further
observational predictions to be made.

One of the current best sources to search for new molecules in the 
interstellar medium is the hot dense core Sagittarius~B2(N) -- hereafter 
\object{Sgr~B2(N)} for short. This source, dubbed
the ``Large Molecule Heimat'' by \citet*{Snyder94}, is extraordinary 
for its rich molecular content: most complex organic molecules such as, e.g., 
acetic acid \citep[CH$_3$COOH,][]{Mehringer97}, glycolaldehyde 
\citep[CH$_2$(OH)CHO,][]{Hollis00}, acetamide 
\citep[CH$_3$CONH$_2$,][]{Hollis06}, and aminoacetonitrile 
\citep[NH$_2$CH$_2$CN,][]{Belloche08a,Belloche08b}, were first discovered
in Sgr~B2(N). This hot core is located in the very massive and extremely active 
region of high-mass star formation Sagittarius B2, at a projected distance of 
$\sim 100$~pc from the Galactic center, whose distance is 
$8.0 \pm 0.5$~kpc from the Sun \citep[][]{Reid93}. 
A second major and somewhat more evolved center of star formation activity, 
\object{Sgr~B2(M)}, is situated in its vicinity ($\sim 2$~pc). 
A more detailed introduction on these two sources and their environment can be
found in, e.g., \citet{Belloche08a}.

Here, we report the detection of warm compact emission from ethyl formate
(C$_2$H$_5$OCHO) and \textit{n}-propyl cyanide (C$_3$H$_7$CN) in Sgr~B2(N) 
with the IRAM 30~m telescope. Section~\ref{s:obs} summarizes the observational 
details. The detections of ethyl formate and \textit{n}-propyl cyanide are 
presented in Sects.~\ref{s:etocho} and \ref{s:prcn}, respectively. 
Implications in terms of interstellar chemistry are discussed in 
Sect.~\ref{s:chemistry} based on a coupled gas-phase and grain-surface 
chemical code. Our conclusions are summarized in Sect.~\ref{s:conclusions}.

\section{Observations and data analysis}
\label{s:obs}

\subsection{Observations}
\label{ss:obs}

We observed the two hot core regions Sgr~B2(N) and Sgr~B2(M) in January 2004, 
September 2004, and January 2005 with the IRAM 30~m telescope on Pico Veleta, 
Spain.
We carried out a complete spectral survey toward both sources in the 3\,mm 
atmospheric window between 80 and 116~GHz. A complete survey was performed in 
parallel in the 1.3\,mm window between 201.8 and 204.6~GHz and between 205.0 
and 217.7~GHz. Additional selected spectra were also obtained in the 2\,mm 
window and between 219 and 268~GHz. The coordinates of the observed positions 
are $\alpha_{\mathrm{J2000}}$=17$^\mathrm{h}$47$^\mathrm{m}$20$\fs$0,
$\delta_{\mathrm{J2000}}$=$-28^\circ$22$\arcmin$19.0$\arcsec$ for Sgr~B2(N) with
a systemic velocity $V_{\mathrm{lsr}}$ = 64 km~s$^{-1}$ and
$\alpha_{\mathrm{J2000}}$=17$^\mathrm{h}$47$^\mathrm{m}$20$\fs$4,
$\delta_{\mathrm{J2000}}$=$-28^\circ$23$\arcmin$07.0$\arcsec$ for  Sgr~B2(M)
with $V_{\mathrm{lsr}}$ = 62 km~s$^{-1}$.
More details about the observational setup and the data 
reduction can be found in \citet{Belloche08a}. 
An rms noise level of 15--20~mK on the $T^\star_{\mathrm{a}}$ scale 
was achieved below 100~GHz, 20--30~mK between 100 and 114.5~GHz, about 50~mK 
between 114.5 and 116~GHz, and 25--60~mK in the 2\,mm 
window. At 1.3~mm, the confusion limit was reached for most of the spectra
obtained toward Sgr~B2(N).

\subsection{Modeling of the spectral survey}
\label{ss:modeling}

The overall goal of our survey was to characterize the molecular content 
of Sgr~B2(N) and (M). It also allows searches for new species once 
lines emitted by known molecules have been identified, including vibrationally 
and torsionally excited states, as well as less abundant isotopologues 
containing, e.g., $^{13}$C, $^{18}$O, $^{17}$O, $^{34}$S, $^{33}$S, or $^{15}$N. 
We detected about 3700 and 950 lines above $3\sigma$ over the whole 3\,mm band 
toward Sgr~B2(N) and (M), respectively. These numbers correspond to an average 
line density of about 100 and 25 features per GHz. Given this high line 
density, the assignment of a line to a given  molecule can be trusted only if 
all lines emitted by this molecule in our frequency coverage are detected with 
the right intensity predicted  by a model (see below) and no predicted line is 
missing in the observed spectrum.

We used the XCLASS software \citep[see][]{Comito05} to model the emission of 
all known molecules in the local thermodynamical equilibrium approximation 
(LTE for short). Each molecule is modeled separately and assumed to be emitted 
by a uniform region. For each molecule, the free parameters are: source size, 
temperature, column density, velocity linewidth, velocity offset with respect 
to the systemic velocity of the source, and a flag indicating if its 
transitions are in emission or in absorption. For some of the molecules, it 
was necessary to include several velocity components to reproduce the observed 
spectra. The velocity components in emission are supposed to be 
non-interacting, i.e. the intensities add up linearly. This approximation is 
valid for two distinct, non-overlapping sources smaller than the beam of the 
telescope, but it is \textit{a priori} less good for, e.g., a source that 
consists of a hot, compact region surrounded by a cold, extended envelope 
or two overlapping sources of spectrally overlapping optically thick emission.
More details about the entire analysis are given in \citet{Belloche08a} and the
detailed results of this modeling will be published in a forthcoming article
describing the complete survey (Belloche et al., in prep.). So far, we have 
identified 49 different molecules, 60 rare isotopologues, and lines arising 
from within 42 vibrationally or torsionally excited states apart from the 
gound state in Sgr~B2(N). This represents about $60\%$ of the lines detected 
above the $3\sigma$ level. In Sgr~B2(M), the corresponding numbers are 42, 53, 
23, and $50\%$, respectively.

\section{Identification of ethyl formate}
\label{s:etocho}

\subsection{Ethyl formate frequencies}
\label{ss:freqetocho}

Ethyl formate, C$_2$H$_5$OCHO, is also known as formic acid ethyl 
ester, or, according to the International Union of Pure and Applied 
Chemistry (IUPAC), as ethyl methanoate. Its rotational spectrum  
was studied in the microwave \citep{Riveros67} and in the millimeter wave 
regions up to 241~GHz \citep{Demaison84}. 
The molecule occurs in two conformers. The heavy atoms C-C-O-C=O form a planar 
zigzag chain in the lowest \textit{anti}-conformer which occasionally is also 
called the \textit{trans}-conformer. 
The two conformers are depicted schematically in \citet{Medvedev09}.
The terminal methyl group is rotated by $\sim 95^{\circ}$ to the left or to the 
right in the \textit{gauche}-conformer. Because of these two options, the 
\textit{gauche}-conformer would be twice as abundant as the 
\textit{anti}-conformer if the energy 
difference between the two were zero. However, the \textit{gauche}-conformer is 
$0.78 \pm 0.25$~kJ\,mol$^{-1}$ or $65 \pm 21$~cm$^{-1}$ or $94 \pm 30$~K 
higher in energy \citep{Riveros67}. 
Therefore, the abundance of the \textit{gauche}-conformer is less than twice 
that of the \textit{anti}-conformer, in particular at lower temperatures.
Since the energy difference has been estimated at room temperature only from 
relative intensities in the ground state spectra and since excited vibrational 
states have not been taken into consideration the error in the energy 
difference may well be larger.

\textit{Anti}-ethyl formate is a strongly prolate molecule 
($A \gg B \approx C$) with electric dipole moments for $a$- and $b$-type 
transitions, $\mu_a$ and $\mu_b$, of 1.85 and 0.70~D, respectively. 
The \textit{gauche}-conformer is more 
asymmetric, $A$ is smaller by approximately one third and $B$ and $C$ are 
larger by about one third. The dipole moment components are $\mu_a = 1.45$, 
$\mu_b = 1.05$, and $\mu_c = 0.25$~D \citep{Riveros67}. Internal rotation 
of the terminal methyl group can be neglected. Tunneling between the two 
\textit{gauche}-conformers has not been observed \citep{Riveros67}.

In the early stages of the current study we received additional ethyl formate 
data from E. Herbst \citep{Medvedev09} based on spectra taken at the Ohio State 
University (OSU) and covering the frequency range 106 -- 378~GHz. The 
predictions used for the current analysis are based on this data set.
An entry for ethyl formate will be available in the catalog section of the 
Cologne Database for Molecular Spectroscopy 
\citep[CDMS\footnote{\label{fn:cdms2} http://www.cdms.de}, 
see][]{Mueller01,Mueller05}. 
The partition function of ethyl formate is $5.690 \times 10^4$ and 
$1.518 \times 10^4$ at 150 and 75~K, respectively.
In the course of the analysis, the two conformers have been treated separately 
on occasion to evaluate if the abundance of either conformer is lower than 
would be expected under LTE conditions.

\subsection{Detection of ethyl formate in Sgr~B2(N)}
\label{ss:detetocho}

\addtocounter{table}{1}
\newcounter{apptable1}
\setcounter{apptable1}{\value{table}}
\onltab{\value{apptable1}}{\clearpage
\begin{table*}
 {\centering
 \caption{
 Transitions of the \textit{anti}-conformer of ethyl formate observed with the IRAM 30 m telescope toward Sgr~B2(N).
The horizontal lines mark discontinuities in the observed frequency coverage.
 Only the transitions associated with a modeled line stronger than 20 mK are listed.
 }
 \label{t:etocho-a}
 \vspace*{0.0ex}
\begin{tabular}{rlrrrrrl}
 \hline\hline
 \multicolumn{1}{c}{$N^a$} & \multicolumn{1}{c}{Transition$^b$} & \multicolumn{1}{c}{Frequency} & \multicolumn{1}{c}{Unc.$^c$} & \multicolumn{1}{c}{$E_\mathrm{l}$$^d$} & \multicolumn{1}{c}{$S\mu^2$} &  \multicolumn{1}{c}{$\sigma^e$} &  \multicolumn{1}{c}{Comments} \\ 
  & & \multicolumn{1}{c}{\scriptsize (MHz)} & \multicolumn{1}{c}{\scriptsize (kHz)} & \multicolumn{1}{c}{\scriptsize (K)} & \multicolumn{1}{c}{\scriptsize (D$^2$)} &  \multicolumn{1}{c}{\scriptsize (mK)} &  \\ 
 \multicolumn{1}{c}{(1)} & \multicolumn{1}{c}{(2)} & \multicolumn{1}{c}{(3)} & \multicolumn{1}{c}{(4)} & \multicolumn{1}{c}{(5)} & \multicolumn{1}{c}{(6)} & \multicolumn{1}{c}{(7)} & \multicolumn{1}{c}{(8)} \\ 
 \hline
   1 & 15$_{ 2,14}$ -- 14$_{ 2,13}$ &   81779.567 &     4 &   30 &        50.5 &   13 & \textbf{Detected}, blend with U-line \\ 
   2 & 15$_{13, 2}$ -- 14$_{13, 1}$$^\star$ &   82297.815 &     4 &  149 &        12.8 &   19 & Blend with C$_2$H$_5$CN, $\varv_{13}$=1/$\varv_{21}$=1 and absorption line of c-HC$^{13}$CCH \\ 
   4 & 15$_{14, 1}$ -- 14$_{14, 0}$$^\star$ &   82297.944 &     4 &  169 &         6.6 &   19 & Blend with C$_2$H$_5$CN, $\varv_{13}$=1/$\varv_{21}$=1 and absorption line of c-HC$^{13}$CCH \\ 
   6 & 15$_{12, 3}$ -- 14$_{12, 2}$$^\star$ &   82298.480 &     4 &  131 &        18.5 &   19 & Blend with C$_2$H$_5$CN, $\varv_{13}$=1/$\varv_{21}$=1 and absorption line of c-HC$^{13}$CCH \\ 
   8 & 15$_{10, 5}$ -- 14$_{10, 4}$$^\star$ &   82303.360 &     4 &  100 &        28.6 &   19 & Blend with HC$_3$N, $\varv_6$=$\varv_7$=1 and absorption line of c-HC$^{13}$CCH  \\
\hline
 \end{tabular}
 }\\[1ex] 
 \end{table*}
\begin{table*}
 {\centering
 \addtocounter{table}{-1}
 \caption{
continued.
 }
 \begin{tabular}{rlrrrrrl}
 \hline\hline
 \multicolumn{1}{c}{$N^a$} & \multicolumn{1}{c}{Transition$^b$} & \multicolumn{1}{c}{Frequency} & \multicolumn{1}{c}{Unc.$^c$} & \multicolumn{1}{c}{$E_\mathrm{l}$$^d$} & \multicolumn{1}{c}{$S\mu^2$} &  \multicolumn{1}{c}{$\sigma^e$} &  \multicolumn{1}{c}{Comments} \\ 
  & & \multicolumn{1}{c}{\scriptsize (MHz)} & \multicolumn{1}{c}{\scriptsize (kHz)} & \multicolumn{1}{c}{\scriptsize (K)} & \multicolumn{1}{c}{\scriptsize (D$^2$)} &  \multicolumn{1}{c}{\scriptsize (mK)} &  \\ 
 \multicolumn{1}{c}{(1)} & \multicolumn{1}{c}{(2)} & \multicolumn{1}{c}{(3)} & \multicolumn{1}{c}{(4)} & \multicolumn{1}{c}{(5)} & \multicolumn{1}{c}{(6)} & \multicolumn{1}{c}{(7)} & \multicolumn{1}{c}{(8)} \\ 
 \hline
  93 & 18$_{ 2,17}$ -- 17$_{ 2,16}$ &   97883.281 &     4 &   43 &        60.9 &   20 & Blend with CH$_3$OCHO, $\varv_{\mathrm{t}}$=1 and NH$_2$CHO \\ 
  94 & 18$_{15, 3}$ -- 17$_{15, 2}$$^\star$ &   98760.931 &     5 &  202 &        18.9 &   18 & \textbf{Group detected}, blend with U-line \\ 
  96 & 18$_{14, 4}$ -- 17$_{14, 3}$$^\star$ &   98761.079 &     5 &  181 &        24.4 &   18 & \textbf{Group detected}, blend with U-line \\ 
  98 & 18$_{16, 2}$ -- 17$_{16, 1}$$^\star$ &   98761.555 &     5 &  224 &        13.0 &   18 & \textbf{Group detected}, blend with U-line \\ 
 100 & 18$_{13, 5}$ -- 17$_{13, 4}$$^\star$ &   98762.218 &     5 &  162 &        29.5 &   18 & \textbf{Group detected}, blend with U-line \\
\hline
 \end{tabular}
 }\\[1ex] 
 \end{table*}
\begin{table*}
 {\centering
 \addtocounter{table}{-1}
 \caption{
continued.
 }
 \begin{tabular}{rlrrrrrl}
 \hline\hline
 \multicolumn{1}{c}{$N^a$} & \multicolumn{1}{c}{Transition$^b$} & \multicolumn{1}{c}{Frequency} & \multicolumn{1}{c}{Unc.$^c$} & \multicolumn{1}{c}{$E_\mathrm{l}$$^d$} & \multicolumn{1}{c}{$S\mu^2$} &  \multicolumn{1}{c}{$\sigma^e$} &  \multicolumn{1}{c}{Comments} \\ 
  & & \multicolumn{1}{c}{\scriptsize (MHz)} & \multicolumn{1}{c}{\scriptsize (kHz)} & \multicolumn{1}{c}{\scriptsize (K)} & \multicolumn{1}{c}{\scriptsize (D$^2$)} &  \multicolumn{1}{c}{\scriptsize (mK)} &  \\ 
 \multicolumn{1}{c}{(1)} & \multicolumn{1}{c}{(2)} & \multicolumn{1}{c}{(3)} & \multicolumn{1}{c}{(4)} & \multicolumn{1}{c}{(5)} & \multicolumn{1}{c}{(6)} & \multicolumn{1}{c}{(7)} & \multicolumn{1}{c}{(8)} \\ 
 \hline
 195 & 20$_{ 5,15}$ -- 19$_{ 5,14}$ &  109976.496 &   100 &   68 &        64.3 &   41 & Blend with U-lines \\ 
 196 & 20$_{ 3,18}$ -- 19$_{ 3,17}$ &  109982.754 &   100 &   57 &        67.0 &   41 & Blend with U-lines \\ 
 197 & 20$_{ 4,17}$ -- 19$_{ 4,16}$ &  110103.554 &   100 &   62 &        65.8 &   24 & Blend with HNCO, $\varv_5$=1 \\ 
 198 & 20$_{ 4,16}$ -- 19$_{ 4,15}$ &  110209.376 &     5 &   62 &        65.8 &   24 & Blend with $^{13}$CO and HC$_3$N, $\varv_5$=1/$\varv_7$=3 \\ 
 199 & 21$_{ 1,21}$ -- 20$_{ 1,20}$ &  110417.565 &     5 &   54 &        71.8 &   24 & Blend with HNCO, $\varv_5$=1 \\
\hline
 \end{tabular}
 }\\[1ex] 
 \end{table*}
\begin{table*}
 {\centering
 \addtocounter{table}{-1}
 \caption{
continued.
 }
 \begin{tabular}{rlrrrrrl}
 \hline\hline
 \multicolumn{1}{c}{$N^a$} & \multicolumn{1}{c}{Transition$^b$} & \multicolumn{1}{c}{Frequency} & \multicolumn{1}{c}{Unc.$^c$} & \multicolumn{1}{c}{$E_\mathrm{l}$$^d$} & \multicolumn{1}{c}{$S\mu^2$} &  \multicolumn{1}{c}{$\sigma^e$} &  \multicolumn{1}{c}{Comments} \\ 
  & & \multicolumn{1}{c}{\scriptsize (MHz)} & \multicolumn{1}{c}{\scriptsize (kHz)} & \multicolumn{1}{c}{\scriptsize (K)} & \multicolumn{1}{c}{\scriptsize (D$^2$)} &  \multicolumn{1}{c}{\scriptsize (mK)} &  \\ 
 \multicolumn{1}{c}{(1)} & \multicolumn{1}{c}{(2)} & \multicolumn{1}{c}{(3)} & \multicolumn{1}{c}{(4)} & \multicolumn{1}{c}{(5)} & \multicolumn{1}{c}{(6)} & \multicolumn{1}{c}{(7)} & \multicolumn{1}{c}{(8)} \\ 
 \hline
 282 & 37$_{28, 9}$ -- 36$_{28, 8}$$^\star$ &  203016.140 &   138 &  738 &        54.2 &  138 & Blend with U-line \\ 
 284 & 37$_{19,18}$ -- 36$_{19,17}$$^\star$ &  203016.766 &   100 &  435 &        93.4 &  138 & Blend with U-line \\ 
 286 & 37$_{29, 8}$ -- 36$_{29, 7}$$^\star$ &  203017.650 &   181 &  779 &        48.9 &  138 & Blend with U-line \\ 
 288 & 37$_{18,19}$ -- 36$_{18,18}$$^\star$ &  203023.193 &     6 &  408 &        96.9 &  138 & Blend with CH$_3$OCH$_3$ \\ 
 290 & 37$_{17,20}$ -- 36$_{17,19}$$^\star$ &  203032.356 &   100 &  383 &       100.1 &  138 & Blend with CH$_3$OCH$_3$ and CH$_3$$^{13}$CN, $\varv_8$=1 \\
\hline
 \end{tabular}
 }\\[1ex] 
 \end{table*}
\begin{table*}
 {\centering
 \addtocounter{table}{-1}
 \caption{
continued.
 }
 \begin{tabular}{rlrrrrrl}
 \hline\hline
 \multicolumn{1}{c}{$N^a$} & \multicolumn{1}{c}{Transition$^b$} & \multicolumn{1}{c}{Frequency} & \multicolumn{1}{c}{Unc.$^c$} & \multicolumn{1}{c}{$E_\mathrm{l}$$^d$} & \multicolumn{1}{c}{$S\mu^2$} &  \multicolumn{1}{c}{$\sigma^e$} &  \multicolumn{1}{c}{Comments} \\ 
  & & \multicolumn{1}{c}{\scriptsize (MHz)} & \multicolumn{1}{c}{\scriptsize (kHz)} & \multicolumn{1}{c}{\scriptsize (K)} & \multicolumn{1}{c}{\scriptsize (D$^2$)} &  \multicolumn{1}{c}{\scriptsize (mK)} &  \\ 
 \multicolumn{1}{c}{(1)} & \multicolumn{1}{c}{(2)} & \multicolumn{1}{c}{(3)} & \multicolumn{1}{c}{(4)} & \multicolumn{1}{c}{(5)} & \multicolumn{1}{c}{(6)} & \multicolumn{1}{c}{(7)} & \multicolumn{1}{c}{(8)} \\ 
 \hline
 375 & 38$_{14,24}$ -- 37$_{14,23}$$^\star$ &  208579.484 &   100 &  326 &       112.6 &  168 & Blend with CH$_3$CH$_3$CO, $\varv_t$=1 and U-line \\ 
 377 & 38$_{13,25}$ -- 37$_{13,24}$$^\star$ &  208612.841 &     4 &  307 &       115.1 &  168 & Strong C$_2$H$_5$CN, $\varv_{13}$=1/$\varv_{21}$=1 \\ 
 379 & 38$_{12,26}$ -- 37$_{12,25}$$^\star$ &  208657.029 &     4 &  289 &       117.3 &  168 & Blend with CH$_3$OCHO and CH$_3$CH$_3$CO \\ 
 381 & 38$_{11,27}$ -- 37$_{11,26}$$^\star$ &  208716.183 &   100 &  272 &       119.4 &  160 & Blend with CH$_3$OCHO \\ 
 383 & 38$_{10,29}$ -- 37$_{10,28}$$^\star$ &  208796.910 &     4 &  257 &       121.3 &  160 & Blend with CH$_2$CH$^{13}$CN and CH$_3$OCHO \\
\hline
 \end{tabular}
 }\\[1ex] 
 \end{table*}
\begin{table*}
 {\centering
 \addtocounter{table}{-1}
 \caption{
continued.
 }
 \begin{tabular}{rlrrrrrl}
 \hline\hline
 \multicolumn{1}{c}{$N^a$} & \multicolumn{1}{c}{Transition$^b$} & \multicolumn{1}{c}{Frequency} & \multicolumn{1}{c}{Unc.$^c$} & \multicolumn{1}{c}{$E_\mathrm{l}$$^d$} & \multicolumn{1}{c}{$S\mu^2$} &  \multicolumn{1}{c}{$\sigma^e$} &  \multicolumn{1}{c}{Comments} \\ 
  & & \multicolumn{1}{c}{\scriptsize (MHz)} & \multicolumn{1}{c}{\scriptsize (kHz)} & \multicolumn{1}{c}{\scriptsize (K)} & \multicolumn{1}{c}{\scriptsize (D$^2$)} &  \multicolumn{1}{c}{\scriptsize (mK)} &  \\ 
 \multicolumn{1}{c}{(1)} & \multicolumn{1}{c}{(2)} & \multicolumn{1}{c}{(3)} & \multicolumn{1}{c}{(4)} & \multicolumn{1}{c}{(5)} & \multicolumn{1}{c}{(6)} & \multicolumn{1}{c}{(7)} & \multicolumn{1}{c}{(8)} \\ 
 \hline
 465 & 39$_{ 8,32}$ -- 38$_{ 8,31}$ &  214608.530 &     4 &  241 &       128.1 &   75 & Blend with CH$_3$$^{13}$CH$_2$CN \\ 
 466 & 39$_{ 8,31}$ -- 38$_{ 8,30}$ &  214609.070 &     4 &  241 &       128.1 &   75 & Blend with CH$_3$$^{13}$CH$_2$CN \\ 
 467 & 39$_{ 7,33}$ -- 38$_{ 7,32}$ &  214870.593 &   100 &  231 &       129.5 &   75 & Blend with $^{13}$CH$_3$CN, $\varv_8$=1  \\ 
 468 & 39$_{ 7,32}$ -- 38$_{ 7,31}$ &  214881.820 &     4 &  231 &       129.5 &   75 & Blend with $^{13}$CH$_2$CHCN \\ 
 469 & 39$_{ 6,34}$ -- 38$_{ 6,33}$ &  215238.445 &   100 &  222 &       130.6 &   74 & Strong C$_2$H$_5$CN, $\varv_{13}$=1/$\varv_{21}$=1 and C$_2$H$_5$CN, $\varv_{20}$=1  \\
\hline
 \end{tabular}
 }\\[1ex] 
 \end{table*}
\begin{table*}
 {\centering
 \addtocounter{table}{-1}
 \caption{
continued.
 }
 \begin{tabular}{rlrrrrrl}
 \hline\hline
 \multicolumn{1}{c}{$N^a$} & \multicolumn{1}{c}{Transition$^b$} & \multicolumn{1}{c}{Frequency} & \multicolumn{1}{c}{Unc.$^c$} & \multicolumn{1}{c}{$E_\mathrm{l}$$^d$} & \multicolumn{1}{c}{$S\mu^2$} &  \multicolumn{1}{c}{$\sigma^e$} &  \multicolumn{1}{c}{Comments} \\ 
  & & \multicolumn{1}{c}{\scriptsize (MHz)} & \multicolumn{1}{c}{\scriptsize (kHz)} & \multicolumn{1}{c}{\scriptsize (K)} & \multicolumn{1}{c}{\scriptsize (D$^2$)} &  \multicolumn{1}{c}{\scriptsize (mK)} &  \\ 
 \multicolumn{1}{c}{(1)} & \multicolumn{1}{c}{(2)} & \multicolumn{1}{c}{(3)} & \multicolumn{1}{c}{(4)} & \multicolumn{1}{c}{(5)} & \multicolumn{1}{c}{(6)} & \multicolumn{1}{c}{(7)} & \multicolumn{1}{c}{(8)} \\ 
 \hline
 544 & 43$_{25,18}$ -- 42$_{25,17}$$^\star$ &  235908.220 &    65 &  687 &        97.6 &  131 & Blend with HCC$^{13}$CN, $\varv_6$=1 \\ 
 546 & 43$_{24,19}$ -- 42$_{24,18}$$^\star$ &  235909.130 &    46 &  652 &       101.6 &  131 & Blend with HCC$^{13}$CN, $\varv_6$=1 \\ 
 548 & 43$_{23,20}$ -- 42$_{23,19}$$^\star$ &  235911.029 &    32 &  618 &       105.3 &  131 & Blend with HCC$^{13}$CN, $\varv_6$=1 \\ 
 550 & 43$_{22,22}$ -- 42$_{22,21}$$^\star$ &  235914.238 &   100 &  586 &       108.9 &  131 & Strong CH$_3$OCHO \\ 
 552 & 43$_{21,22}$ -- 42$_{21,21}$$^\star$ &  235919.208 &    15 &  555 &       112.3 &  131 & Strong CH$_3$OCHO \\
\hline
 \end{tabular}
 }\\[1ex] 
 \end{table*}
\begin{table*}
 {\centering
 \addtocounter{table}{-1}
 \caption{
continued.
 }
 \begin{tabular}{rlrrrrrl}
 \hline\hline
 \multicolumn{1}{c}{$N^a$} & \multicolumn{1}{c}{Transition$^b$} & \multicolumn{1}{c}{Frequency} & \multicolumn{1}{c}{Unc.$^c$} & \multicolumn{1}{c}{$E_\mathrm{l}$$^d$} & \multicolumn{1}{c}{$S\mu^2$} &  \multicolumn{1}{c}{$\sigma^e$} &  \multicolumn{1}{c}{Comments} \\ 
  & & \multicolumn{1}{c}{\scriptsize (MHz)} & \multicolumn{1}{c}{\scriptsize (kHz)} & \multicolumn{1}{c}{\scriptsize (K)} & \multicolumn{1}{c}{\scriptsize (D$^2$)} &  \multicolumn{1}{c}{\scriptsize (mK)} &  \\ 
 \multicolumn{1}{c}{(1)} & \multicolumn{1}{c}{(2)} & \multicolumn{1}{c}{(3)} & \multicolumn{1}{c}{(4)} & \multicolumn{1}{c}{(5)} & \multicolumn{1}{c}{(6)} & \multicolumn{1}{c}{(7)} & \multicolumn{1}{c}{(8)} \\ 
 \hline
 632 & 48$_{ 2,47}$ -- 47$_{ 2,46}$ &  254081.225 &     6 &  293 &       163.9 &   32 & Strong CH$_3$CH$_3$CO and C$_2$H$_5$CN \\ 
 633 & 86$_{ 6,80}$ -- 86$_{ 5,81}$ &  254082.060 &    64 & 1013 &        20.4 &   32 & Strong CH$_3$CH$_3$CO and C$_2$H$_5$CN \\ 
 634 & 48$_{ 1,47}$ -- 47$_{ 1,46}$ &  254100.087 &     6 &  293 &       164.0 &   32 & Blend with SiS \\ 
 635 & 32$_{ 9,23}$ -- 32$_{ 8,24}$$^\star$ &  254170.380 &    13 &  185 &         7.7 &   32 & Blend with $^{13}$CH$_3$CH$_2$CN \\ 
 637 & 48$_{ 2,47}$ -- 47$_{ 1,46}$ &  254171.401 &     6 &  293 &        18.1 &   32 & Blend with $^{13}$CH$_3$CH$_2$CN \\
\hline
 \end{tabular}
 }\\[1ex] 
 Notes:
 $^a$ Numbering of the observed transitions associated with a modeled line stronger than 20 mK.
 $^b$ Transitions marked with a $^\star$ are double with a frequency difference less than 0.1 MHz. The quantum numbers of the second one are not shown.
 $^c$ Frequency uncertainty.
 $^d$ Lower energy level in temperature units ($E_\mathrm{l}/k_\mathrm{B}$).
 $^e$ Calculated rms noise level in $T_{\mathrm{mb}}$ scale.
 \end{table*}

}

\addtocounter{table}{1}
\newcounter{apptable2}
\setcounter{apptable2}{\value{table}}
\onltab{\value{apptable2}}{\clearpage
\begin{table*}
 {\centering
 \caption{
 Transitions of the \textit{gauche}-conformer of ethyl formate observed with the IRAM 30 m telescope toward Sgr~B2(N).
The horizontal lines mark discontinuities in the observed frequency coverage.
 Only the transitions associated with a modeled line stronger than 20 mK are listed.
 }
 \label{t:etocho-g}
 \vspace*{0.0ex}
\begin{tabular}{rlrrrrrl}
 \hline\hline
 \multicolumn{1}{c}{$N^a$} & \multicolumn{1}{c}{Transition$^b$} & \multicolumn{1}{c}{Frequency} & \multicolumn{1}{c}{Unc.$^c$} & \multicolumn{1}{c}{$E_\mathrm{l}$$^d$} & \multicolumn{1}{c}{$S\mu^2$} &  \multicolumn{1}{c}{$\sigma^e$} &  \multicolumn{1}{c}{Comments} \\ 
  & & \multicolumn{1}{c}{\scriptsize (MHz)} & \multicolumn{1}{c}{\scriptsize (kHz)} & \multicolumn{1}{c}{\scriptsize (K)} & \multicolumn{1}{c}{\scriptsize (D$^2$)} &  \multicolumn{1}{c}{\scriptsize (mK)} &  \\ 
 \multicolumn{1}{c}{(1)} & \multicolumn{1}{c}{(2)} & \multicolumn{1}{c}{(3)} & \multicolumn{1}{c}{(4)} & \multicolumn{1}{c}{(5)} & \multicolumn{1}{c}{(6)} & \multicolumn{1}{c}{(7)} & \multicolumn{1}{c}{(8)} \\ 
 \hline
   1 & 14$_{ 7, 8}$ -- 13$_{ 7, 7}$ &   99252.267 &     9 &  140 &        43.9 &   19 & Blend with a(CH$_2$OH)$_2$ and C$_2$H$_5$CN \\ 
   2 & 14$_{ 7, 7}$ -- 13$_{ 7, 6}$ &   99252.460 &     9 &  140 &        43.9 &   19 & Blend with a(CH$_2$OH)$_2$ and C$_2$H$_5$CN \\ 
   3 & 16$_{ 1,16}$ -- 15$_{ 1,15}$ &  104834.473 &    10 &  132 &        66.3 &   25 & Blend with C$_2$H$_5$CN, $\varv_{13}$=1/$\varv_{21}$=1 and U-line \\ 
   4 & 16$_{ 0,16}$ -- 15$_{ 0,15}$ &  104848.839 &    10 &  132 &        66.3 &   25 & Blend with C$_2$H$_3$CN, $\varv_{11}$=1/$\varv_{15}$=1 \\ 
   5 & 15$_{ 9, 7}$ -- 14$_{ 9, 6}$$^\star$ &  106216.238 &     9 &  154 &        40.2 &   25 & Blend with C$_2$H$_3$CN, $\varv_{11}$=1/$\varv_{15}$=1 \\
\hline
 \end{tabular}
 }\\[1ex] 
 \end{table*}
\begin{table*}
 {\centering
 \addtocounter{table}{-1}
 \caption{
continued.
 }
 \begin{tabular}{rlrrrrrl}
 \hline\hline
 \multicolumn{1}{c}{$N^a$} & \multicolumn{1}{c}{Transition$^b$} & \multicolumn{1}{c}{Frequency} & \multicolumn{1}{c}{Unc.$^c$} & \multicolumn{1}{c}{$E_\mathrm{l}$$^d$} & \multicolumn{1}{c}{$S\mu^2$} &  \multicolumn{1}{c}{$\sigma^e$} &  \multicolumn{1}{c}{Comments} \\ 
  & & \multicolumn{1}{c}{\scriptsize (MHz)} & \multicolumn{1}{c}{\scriptsize (kHz)} & \multicolumn{1}{c}{\scriptsize (K)} & \multicolumn{1}{c}{\scriptsize (D$^2$)} &  \multicolumn{1}{c}{\scriptsize (mK)} &  \\ 
 \multicolumn{1}{c}{(1)} & \multicolumn{1}{c}{(2)} & \multicolumn{1}{c}{(3)} & \multicolumn{1}{c}{(4)} & \multicolumn{1}{c}{(5)} & \multicolumn{1}{c}{(6)} & \multicolumn{1}{c}{(7)} & \multicolumn{1}{c}{(8)} \\ 
 \hline
\hline
  72 & 59$_{17,43}$ -- 59$_{16,44}$ &  205092.813 &    17 &  774 &        72.3 &  100 & Strong C$_2$H$_5$CN, CH$_3$OCH$_3$, and CH$_3$$^{13}$CH$_2$CN \\ 
  73 & 28$_{ 5,23}$ -- 27$_{ 5,22}$ &  205093.048 &     8 &  232 &       113.5 &  100 & Strong C$_2$H$_5$CN, CH$_3$OCH$_3$, and CH$_3$$^{13}$CH$_2$CN \\ 
  74 & 31$_{ 1,30}$ -- 30$_{ 2,29}$ &  205095.436 &    11 &  245 &        57.4 &  100 & Strong CH$_3$$^{13}$CH$_2$CN and CH$_3$OCH$_3$ \\ 
  75 & 31$_{ 2,30}$ -- 30$_{ 2,29}$ &  205095.823 &    11 &  245 &       127.9 &  100 & Strong CH$_3$$^{13}$CH$_2$CN and CH$_3$OCH$_3$ \\ 
  76 & 31$_{ 1,30}$ -- 30$_{ 1,29}$ &  205096.091 &    11 &  245 &       127.9 &  100 & Strong CH$_3$$^{13}$CH$_2$CN and CH$_3$OCH$_3$ \\
\hline
 \end{tabular}
 }\\[1ex] 
 \end{table*}
\begin{table*}
 {\centering
 \addtocounter{table}{-1}
 \caption{
continued.
 }
 \begin{tabular}{rlrrrrrl}
 \hline\hline
 \multicolumn{1}{c}{$N^a$} & \multicolumn{1}{c}{Transition$^b$} & \multicolumn{1}{c}{Frequency} & \multicolumn{1}{c}{Unc.$^c$} & \multicolumn{1}{c}{$E_\mathrm{l}$$^d$} & \multicolumn{1}{c}{$S\mu^2$} &  \multicolumn{1}{c}{$\sigma^e$} &  \multicolumn{1}{c}{Comments} \\ 
  & & \multicolumn{1}{c}{\scriptsize (MHz)} & \multicolumn{1}{c}{\scriptsize (kHz)} & \multicolumn{1}{c}{\scriptsize (K)} & \multicolumn{1}{c}{\scriptsize (D$^2$)} &  \multicolumn{1}{c}{\scriptsize (mK)} &  \\ 
 \multicolumn{1}{c}{(1)} & \multicolumn{1}{c}{(2)} & \multicolumn{1}{c}{(3)} & \multicolumn{1}{c}{(4)} & \multicolumn{1}{c}{(5)} & \multicolumn{1}{c}{(6)} & \multicolumn{1}{c}{(7)} & \multicolumn{1}{c}{(8)} \\ 
 \hline
 158 & 18$_{17, 1}$ -- 18$_{16, 2}$$^\star$ &  211857.177 &    16 &  231 &         4.4 &   47 & Blend with C$_2$H$_3$CN, $\varv_{15}$=1 and U-line \\ 
 160 & 29$_{17,12}$ -- 29$_{16,13}$$^\star$ &  211859.692 &    10 &  320 &        24.2 &   47 & Blend with C$_2$H$_3$CN, $\varv_{15}$=1 and U-line \\ 
 162 & 29$_{17,12}$ -- 29$_{16,14}$$^\star$ &  211859.692 &    10 &  320 &         1.2 &   47 & Blend with C$_2$H$_3$CN, $\varv_{15}$=1 and U-line \\ 
 164 & 28$_{17,11}$ -- 28$_{16,12}$$^\star$ &  211880.904 &    10 &  311 &        22.6 &   47 & Blend with CH$_2$CH$^{13}$CN \\ 
 166 & 28$_{17,11}$ -- 28$_{16,13}$$^\star$ &  211880.904 &    10 &  311 &         1.1 &   47 & Blend with CH$_2$CH$^{13}$CN \\
\hline
 \end{tabular}
 }\\[1ex] 
 \end{table*}
\begin{table*}
 {\centering
 \addtocounter{table}{-1}
 \caption{
continued.
 }
 \begin{tabular}{rlrrrrrl}
 \hline\hline
 \multicolumn{1}{c}{$N^a$} & \multicolumn{1}{c}{Transition$^b$} & \multicolumn{1}{c}{Frequency} & \multicolumn{1}{c}{Unc.$^c$} & \multicolumn{1}{c}{$E_\mathrm{l}$$^d$} & \multicolumn{1}{c}{$S\mu^2$} &  \multicolumn{1}{c}{$\sigma^e$} &  \multicolumn{1}{c}{Comments} \\ 
  & & \multicolumn{1}{c}{\scriptsize (MHz)} & \multicolumn{1}{c}{\scriptsize (kHz)} & \multicolumn{1}{c}{\scriptsize (K)} & \multicolumn{1}{c}{\scriptsize (D$^2$)} &  \multicolumn{1}{c}{\scriptsize (mK)} &  \\ 
 \multicolumn{1}{c}{(1)} & \multicolumn{1}{c}{(2)} & \multicolumn{1}{c}{(3)} & \multicolumn{1}{c}{(4)} & \multicolumn{1}{c}{(5)} & \multicolumn{1}{c}{(6)} & \multicolumn{1}{c}{(7)} & \multicolumn{1}{c}{(8)} \\ 
 \hline
 257 & 32$_{ 2,30}$ -- 31$_{ 3,29}$ &  215600.637 &     8 &  261 &        51.8 &   74 & Strong C$_2$H$_5$CN \\ 
 258 & 32$_{ 3,30}$ -- 31$_{ 3,29}$ &  215608.230 &   100 &  261 &       131.0 &   74 & Strong C$_2$H$_5$CN \\ 
 259 & 32$_{ 2,30}$ -- 31$_{ 2,29}$ &  215612.952 &   100 &  261 &       131.0 &   74 & Strong C$_2$H$_5$CN \\ 
 260 & 32$_{ 3,30}$ -- 31$_{ 2,29}$ &  215620.426 &     8 &  261 &        51.8 &   74 & Strong C$_2$H$_5$CN \\ 
 261 & 36$_{ 9,27}$ -- 36$_{ 6,30}$ &  215623.049 &    44 &  336 &         2.1 &   74 & Strong C$_2$H$_5$CN \\
\hline
 \end{tabular}
 }\\[1ex] 
 \end{table*}
\begin{table*}
 {\centering
 \addtocounter{table}{-1}
 \caption{
continued.
 }
 \begin{tabular}{rlrrrrrl}
 \hline\hline
 \multicolumn{1}{c}{$N^a$} & \multicolumn{1}{c}{Transition$^b$} & \multicolumn{1}{c}{Frequency} & \multicolumn{1}{c}{Unc.$^c$} & \multicolumn{1}{c}{$E_\mathrm{l}$$^d$} & \multicolumn{1}{c}{$S\mu^2$} &  \multicolumn{1}{c}{$\sigma^e$} &  \multicolumn{1}{c}{Comments} \\ 
  & & \multicolumn{1}{c}{\scriptsize (MHz)} & \multicolumn{1}{c}{\scriptsize (kHz)} & \multicolumn{1}{c}{\scriptsize (K)} & \multicolumn{1}{c}{\scriptsize (D$^2$)} &  \multicolumn{1}{c}{\scriptsize (mK)} &  \\ 
 \multicolumn{1}{c}{(1)} & \multicolumn{1}{c}{(2)} & \multicolumn{1}{c}{(3)} & \multicolumn{1}{c}{(4)} & \multicolumn{1}{c}{(5)} & \multicolumn{1}{c}{(6)} & \multicolumn{1}{c}{(7)} & \multicolumn{1}{c}{(8)} \\ 
 \hline
\hline
 352 & 32$_{11,22}$ -- 31$_{11,21}$ &  227412.121 &     7 &  300 &       118.1 &   85 & Strong HC$_3$N \\ 
 353 & 32$_{11,21}$ -- 31$_{11,20}$ &  227412.786 &     7 &  300 &       118.1 &   85 & Strong HC$_3$N \\ 
 354 & 31$_{ 6,25}$ -- 30$_{ 6,24}$ &  227600.701 &    10 &  265 &       125.4 &   85 & Strong C$_2$H$_5$OH and NH$_2$CHO \\ 
 355 & 32$_{10,23}$ -- 31$_{10,22}$ &  227826.118 &     8 &  293 &       120.8 &   85 & Blend with CH$_3$OH \\ 
 356 & 32$_{10,22}$ -- 31$_{10,21}$ &  227836.414 &     8 &  293 &       120.8 &   85 & Blend with CH$_3$CH$_3$CO \\
\hline
 \end{tabular}
 }\\[1ex] 
 \end{table*}
\begin{table*}
 {\centering
 \addtocounter{table}{-1}
 \caption{
continued.
 }
 \begin{tabular}{rlrrrrrl}
 \hline\hline
 \multicolumn{1}{c}{$N^a$} & \multicolumn{1}{c}{Transition$^b$} & \multicolumn{1}{c}{Frequency} & \multicolumn{1}{c}{Unc.$^c$} & \multicolumn{1}{c}{$E_\mathrm{l}$$^d$} & \multicolumn{1}{c}{$S\mu^2$} &  \multicolumn{1}{c}{$\sigma^e$} &  \multicolumn{1}{c}{Comments} \\ 
  & & \multicolumn{1}{c}{\scriptsize (MHz)} & \multicolumn{1}{c}{\scriptsize (kHz)} & \multicolumn{1}{c}{\scriptsize (K)} & \multicolumn{1}{c}{\scriptsize (D$^2$)} &  \multicolumn{1}{c}{\scriptsize (mK)} &  \\ 
 \multicolumn{1}{c}{(1)} & \multicolumn{1}{c}{(2)} & \multicolumn{1}{c}{(3)} & \multicolumn{1}{c}{(4)} & \multicolumn{1}{c}{(5)} & \multicolumn{1}{c}{(6)} & \multicolumn{1}{c}{(7)} & \multicolumn{1}{c}{(8)} \\ 
 \hline
\hline
 436 & 36$_{14,23}$ -- 35$_{14,22}$$^\star$ &  255232.834 &     8 &  368 &       127.8 &  217 & Strong CH$_3$OH \\ 
 438 & 17$_{11, 6}$ -- 16$_{10, 7}$$^\star$ &  255442.928 &    14 &  171 &        23.8 &  217 & Strong CH$_2$CH$^{13}$CN \\ 
 440 & 17$_{11, 6}$ -- 16$_{10, 6}$$^\star$ &  255442.928 &    14 &  171 &         1.4 &  217 & Strong CH$_2$CH$^{13}$CN \\ 
 442 & 36$_{13,24}$ -- 35$_{13,23}$$^\star$ &  255458.533 &     8 &  360 &       131.0 &  217 & Strong C$_2$H$_5$CN, $\varv_{13}$=1/$\varv_{21}$=1 and CH$_2$$^{13}$CHCN \\ 
 444 & 37$_{ 4,33}$ -- 36$_{ 5,32}$ &  255646.981 &     8 &  327 &        46.2 &  217 & Strong H$^{13}$CCCN and NH$_2$CHO, $\varv_{12}$=1 \\
\hline
 \end{tabular}
 }\\[1ex] 
 Notes:
 $^a$ Numbering of the observed transitions associated with a modeled line stronger than 20 mK.
 $^b$ Transitions marked with a $^\star$ are double with a frequency difference less than 0.1 MHz. The quantum numbers of the second one are not shown.
 $^c$ Frequency uncertainty.
 $^d$ Lower energy level in temperature units ($E_\mathrm{l}/k_\mathrm{B}$).
 $^e$ Calculated rms noise level in $T_{\mathrm{mb}}$ scale.
 \end{table*}

}

\begin{table*}
 {\centering
 \caption{
 Transitions of the \textit{anti}-conformer of ethyl formate detected toward Sgr~B2(N) with the IRAM 30 m telescope.
}
 \label{t:detectetocho-a}
 \vspace*{-1.0ex}
 \begin{tabular}{rlrrrrrcrrrrl}
 \hline\hline
 \multicolumn{1}{c}{$N^a$} & \multicolumn{1}{c}{\hspace*{-1ex} Transition} & \multicolumn{1}{c}{\hspace*{-2ex} Frequency} & \multicolumn{1}{c}{\hspace*{-3ex} Unc.$^b$} & \multicolumn{1}{c}{\hspace*{-1ex} $E_\mathrm{l}$$^c$} & \multicolumn{1}{c}{\hspace*{-1ex} $S\mu^2$} & \multicolumn{1}{c}{\hspace*{-2ex} $\sigma^d$} & \multicolumn{1}{c}{$F^e$} & \multicolumn{1}{c}{$\tau^f$} & \multicolumn{1}{c}{$I_{\mathrm{obs}}$$^g$} & \multicolumn{1}{c}{\hspace*{-1ex} $I_{\mathrm{mod}}$$^g$} & \multicolumn{1}{c}{\hspace*{-1ex} $I_{\mathrm{all}}$$^g$} & \multicolumn{1}{c}{Comments} \\ 
  & & \multicolumn{1}{c}{\hspace*{-2ex} \scriptsize (MHz)} & \multicolumn{1}{c}{\hspace*{-3ex} \scriptsize (kHz)} & \multicolumn{1}{c}{\hspace*{-1ex} \scriptsize (K)} & \multicolumn{1}{c}{\hspace*{-1ex} \scriptsize (D$^2$)} & \multicolumn{1}{c}{\hspace*{-2ex} \scriptsize (mK)} & & & \multicolumn{1}{c}{\scriptsize (K~km~s$^{-1}$)} & \multicolumn{2}{c}{\hspace*{-1ex} \scriptsize (K~km~s$^{-1}$)} & \\ 
 \multicolumn{1}{c}{(1)} & \multicolumn{1}{c}{\hspace*{-1ex} (2)} & \multicolumn{1}{c}{\hspace*{-2ex} (3)} & \multicolumn{1}{c}{\hspace*{-3ex} (4)} & \multicolumn{1}{c}{\hspace*{-1ex} (5)} & \multicolumn{1}{c}{\hspace*{-1ex} (6)} & \multicolumn{1}{c}{\hspace*{-2ex} (7)} & \multicolumn{1}{c}{(8)} & \multicolumn{1}{c}{(9)} & \multicolumn{1}{c}{(10)} & \multicolumn{1}{c}{(11)} & \multicolumn{1}{c}{\hspace*{-1ex} (12)} & \multicolumn{1}{c}{\hspace*{-1ex} (13)} \\ 
 \hline
   1 & \hspace*{-1ex} 15$_{ 2,14}$ -- 14$_{ 2,13}$ & \hspace*{-2ex}   81779.567 & \hspace*{-3ex}    4 & \hspace*{-1ex}   30 & \hspace*{-1ex}          51 & \hspace*{-2ex}    13 &    1 & 0.05 &        0.52(06) & \hspace*{-2ex}        0.32 & \hspace*{-2ex}        0.44 &  blend with U-line \\ 
  16 & \hspace*{-1ex} 15$_{ 6,10}$ -- 14$_{ 6, 9}$ & \hspace*{-2ex}   82351.854 & \hspace*{-3ex}    4 & \hspace*{-1ex}   54 & \hspace*{-1ex}          43 & \hspace*{-2ex}    19 &    2 & 0.07 &        1.02(08) & \hspace*{-2ex}        0.44 & \hspace*{-2ex}        0.55 &  partial blend with C$_2$H$_5$CN, \\ 
 & & & & & & & & & & & &  $\varv_{13}$=1/$\varv_{21}$=1 and U-line \\ 
  17 & \hspace*{-1ex} 15$_{ 6, 9}$ -- 14$_{ 6, 8}$ & \hspace*{-2ex}   82351.858 & \hspace*{-3ex}    4 & \hspace*{-1ex}   54 & \hspace*{-1ex}          43 & \hspace*{-2ex}    19 &    2 & -- & -- & \hspace*{-2ex} -- & \hspace*{-2ex} -- & -- \\ 
  26 & \hspace*{-1ex} 15$_{ 2,13}$ -- 14$_{ 2,12}$ & \hspace*{-2ex}   84081.357 & \hspace*{-3ex}    4 & \hspace*{-1ex}   31 & \hspace*{-1ex}          51 & \hspace*{-2ex}    19 &    3 & 0.05 &        0.53(08) & \hspace*{-2ex}        0.35 & \hspace*{-2ex}        0.42 &  partial blend with CH$_3$CH$_3$CO, $\varv_t$=1 \\ 
  28 & \hspace*{-1ex} 16$_{ 0,16}$ -- 15$_{ 0,15}$ & \hspace*{-2ex}   85065.106 & \hspace*{-3ex}    4 & \hspace*{-1ex}   31 & \hspace*{-1ex}          55 & \hspace*{-2ex}    22 &    4 & 0.06 &        0.73(10) & \hspace*{-2ex}        0.39 & \hspace*{-2ex}        0.57 &  partial blend with c-C$_2$H$_4$O \\ 
  44 & \hspace*{-1ex} 16$_{ 8, 8}$ -- 15$_{ 8, 7}$ & \hspace*{-2ex}   87810.372 & \hspace*{-3ex}    4 & \hspace*{-1ex}   78 & \hspace*{-1ex}          41 & \hspace*{-2ex}    17 &    5 & 0.05 &        1.28(07) & \hspace*{-2ex}        0.40 & \hspace*{-2ex}        1.22 &  partial blend with CH$_2$(OH)CHO \\ 
 & & & & & & & & & & & &  and C$_2$H$_5$CN \\ 
  45 & \hspace*{-1ex} 16$_{ 8, 9}$ -- 15$_{ 8, 8}$ & \hspace*{-2ex}   87810.372 & \hspace*{-3ex}    4 & \hspace*{-1ex}   78 & \hspace*{-1ex}          41 & \hspace*{-2ex}    17 &    5 & -- & -- & \hspace*{-2ex} -- & \hspace*{-2ex} -- & -- \\ 
  46 & \hspace*{-1ex} 16$_{ 7, 9}$ -- 15$_{ 7, 8}$ & \hspace*{-2ex}   87826.665 & \hspace*{-3ex}    4 & \hspace*{-1ex}   67 & \hspace*{-1ex}          44 & \hspace*{-2ex}    17 &    6 & 0.06 &        0.99(07) & \hspace*{-2ex}        0.48 & \hspace*{-2ex}        0.86 &  partial blend with HNCO, $\varv_5$=1 \\ 
  47 & \hspace*{-1ex} 16$_{ 7,10}$ -- 15$_{ 7, 9}$ & \hspace*{-2ex}   87826.665 & \hspace*{-3ex}    4 & \hspace*{-1ex}   67 & \hspace*{-1ex}          44 & \hspace*{-2ex}    17 &    6 & -- & -- & \hspace*{-2ex} -- & \hspace*{-2ex} -- & -- \\ 
  53 & \hspace*{-1ex} 16$_{ 3,14}$ -- 15$_{ 3,13}$ & \hspace*{-2ex}   87993.944 & \hspace*{-3ex}    4 & \hspace*{-1ex}   38 & \hspace*{-1ex}          53 & \hspace*{-2ex}    19 &    7 & 0.05 &        1.01(08) & \hspace*{-2ex}        0.39 & \hspace*{-2ex}        0.64 &  partial blend with CH$_3$CH$_3$CO \\ 
 & & & & & & & & & & & &  and U-line \\ 
  54 & \hspace*{-1ex} 16$_{ 4,12}$ -- 15$_{ 4,11}$ & \hspace*{-2ex}   88001.562 & \hspace*{-3ex}    4 & \hspace*{-1ex}   43 & \hspace*{-1ex}          51 & \hspace*{-2ex}    19 &    8 & 0.05 &        0.66(08) & \hspace*{-2ex}        0.35 & \hspace*{-2ex}        0.62 &  blend with C$_2$H$_5$CN, $\varv_{13}$=1/$\varv_{21}$=1 \\ 
  73 & \hspace*{-1ex} 17$_{10, 7}$ -- 16$_{10, 6}$ & \hspace*{-2ex}   93284.077 & \hspace*{-3ex}    4 & \hspace*{-1ex}  108 & \hspace*{-1ex}          38 & \hspace*{-2ex}    22 &    9 & 0.04 &        0.07(09) & \hspace*{-2ex}        0.33 & \hspace*{-2ex}        0.35 &  uncertain baseline \\ 
  74 & \hspace*{-1ex} 17$_{10, 8}$ -- 16$_{10, 7}$ & \hspace*{-2ex}   93284.077 & \hspace*{-3ex}    4 & \hspace*{-1ex}  108 & \hspace*{-1ex}          38 & \hspace*{-2ex}    22 &    9 & -- & -- & \hspace*{-2ex} -- & \hspace*{-2ex} -- & -- \\ 
  75 & \hspace*{-1ex} 17$_{ 9, 8}$ -- 16$_{ 9, 7}$ & \hspace*{-2ex}   93292.297 & \hspace*{-3ex}    4 & \hspace*{-1ex}   94 & \hspace*{-1ex}          42 & \hspace*{-2ex}    22 &   10 & 0.05 &        0.56(09) & \hspace*{-2ex}        0.42 & \hspace*{-2ex}        0.50 &  uncertain baseline \\ 
  76 & \hspace*{-1ex} 17$_{ 9, 9}$ -- 16$_{ 9, 8}$ & \hspace*{-2ex}   93292.297 & \hspace*{-3ex}    4 & \hspace*{-1ex}   94 & \hspace*{-1ex}          42 & \hspace*{-2ex}    22 &   10 & -- & -- & \hspace*{-2ex} -- & \hspace*{-2ex} -- & -- \\ 
  77 & \hspace*{-1ex} 17$_{ 8, 9}$ -- 16$_{ 8, 8}$ & \hspace*{-2ex}   93304.955 & \hspace*{-3ex}    4 & \hspace*{-1ex}   82 & \hspace*{-1ex}          45 & \hspace*{-2ex}    22 &   11 & 0.06 &        0.96(09) & \hspace*{-2ex}        0.51 & \hspace*{-2ex}        0.67 &  partial blend with U-line \\ 
  78 & \hspace*{-1ex} 17$_{ 8,10}$ -- 16$_{ 8, 9}$ & \hspace*{-2ex}   93304.955 & \hspace*{-3ex}    4 & \hspace*{-1ex}   82 & \hspace*{-1ex}          45 & \hspace*{-2ex}    22 &   11 & -- & -- & \hspace*{-2ex} -- & \hspace*{-2ex} -- & -- \\ 
  79 & \hspace*{-1ex} 17$_{ 7,11}$ -- 16$_{ 7,10}$ & \hspace*{-2ex}   93324.728 & \hspace*{-3ex}    4 & \hspace*{-1ex}   71 & \hspace*{-1ex}          48 & \hspace*{-2ex}    22 &   12 & 0.07 &        1.32(09) & \hspace*{-2ex}        0.60 & \hspace*{-2ex}        0.74 &  partial blend with U-lines \\ 
  80 & \hspace*{-1ex} 17$_{ 7,10}$ -- 16$_{ 7, 9}$ & \hspace*{-2ex}   93324.728 & \hspace*{-3ex}    4 & \hspace*{-1ex}   71 & \hspace*{-1ex}          48 & \hspace*{-2ex}    22 &   12 & -- & -- & \hspace*{-2ex} -- & \hspace*{-2ex} -- & -- \\ 
  81 & \hspace*{-1ex} 17$_{ 6,12}$ -- 16$_{ 6,11}$ & \hspace*{-2ex}   93356.821 & \hspace*{-3ex}    4 & \hspace*{-1ex}   62 & \hspace*{-1ex}          51 & \hspace*{-2ex}    22 &   13 & 0.08 &        1.68(10) & \hspace*{-2ex}        0.70 & \hspace*{-2ex}        1.24 &  blend with CH$_3$CH$_3$CO and U-line? \\ 
  82 & \hspace*{-1ex} 17$_{ 6,11}$ -- 16$_{ 6,10}$ & \hspace*{-2ex}   93356.838 & \hspace*{-3ex}    4 & \hspace*{-1ex}   62 & \hspace*{-1ex}          51 & \hspace*{-2ex}    22 &   13 & -- & -- & \hspace*{-2ex} -- & \hspace*{-2ex} -- & -- \\ 
  83 & \hspace*{-1ex} 17$_{ 5,13}$ -- 16$_{ 5,12}$ & \hspace*{-2ex}   93412.160 & \hspace*{-3ex}    4 & \hspace*{-1ex}   54 & \hspace*{-1ex}          53 & \hspace*{-2ex}    22 &   14 & 0.08 &        1.12(10) & \hspace*{-2ex}        0.79 & \hspace*{-2ex}        0.85 &  partial blend with U-line? \\ 
  84 & \hspace*{-1ex} 17$_{ 5,12}$ -- 16$_{ 5,11}$ & \hspace*{-2ex}   93413.168 & \hspace*{-3ex}    4 & \hspace*{-1ex}   54 & \hspace*{-1ex}          53 & \hspace*{-2ex}    22 &   14 & -- & -- & \hspace*{-2ex} -- & \hspace*{-2ex} -- & -- \\ 
  86 & \hspace*{-1ex} 17$_{ 4,14}$ -- 16$_{ 4,13}$ & \hspace*{-2ex}   93504.972 & \hspace*{-3ex}    4 & \hspace*{-1ex}   47 & \hspace*{-1ex}          55 & \hspace*{-2ex}    24 &   15 & 0.05 &        1.47(10) & \hspace*{-2ex}        0.44 & \hspace*{-2ex}        0.87 &  blend with CH$_3$$^{18}$OH \\ 
  87 & \hspace*{-1ex} 17$_{ 4,13}$ -- 16$_{ 4,12}$ & \hspace*{-2ex}   93539.303 & \hspace*{-3ex}    4 & \hspace*{-1ex}   47 & \hspace*{-1ex}          55 & \hspace*{-2ex}    24 &   16 & 0.05 &        0.84(10) & \hspace*{-2ex}        0.44 & \hspace*{-2ex}        0.45 & no blend \\ 
  94 & \hspace*{-1ex} 18$_{15, 3}$ -- 17$_{15, 2}$ & \hspace*{-2ex}   98760.931 & \hspace*{-3ex}    5 & \hspace*{-1ex}  202 & \hspace*{-1ex}          19 & \hspace*{-2ex}    18 &   17 & 0.04 &        0.87(09) & \hspace*{-2ex}        0.68 & \hspace*{-2ex}        0.78 &  blend with U-line \\ 
  95 & \hspace*{-1ex} 18$_{15, 4}$ -- 17$_{15, 3}$ & \hspace*{-2ex}   98760.931 & \hspace*{-3ex}    5 & \hspace*{-1ex}  202 & \hspace*{-1ex}          19 & \hspace*{-2ex}    18 &   17 & -- & -- & \hspace*{-2ex} -- & \hspace*{-2ex} -- & -- \\ 
  96 & \hspace*{-1ex} 18$_{14, 4}$ -- 17$_{14, 3}$ & \hspace*{-2ex}   98761.079 & \hspace*{-3ex}    5 & \hspace*{-1ex}  181 & \hspace*{-1ex}          24 & \hspace*{-2ex}    18 &   17 & -- & -- & \hspace*{-2ex} -- & \hspace*{-2ex} -- & -- \\ 
  97 & \hspace*{-1ex} 18$_{14, 5}$ -- 17$_{14, 4}$ & \hspace*{-2ex}   98761.079 & \hspace*{-3ex}    5 & \hspace*{-1ex}  181 & \hspace*{-1ex}          24 & \hspace*{-2ex}    18 &   17 & -- & -- & \hspace*{-2ex} -- & \hspace*{-2ex} -- & -- \\ 
  98 & \hspace*{-1ex} 18$_{16, 2}$ -- 17$_{16, 1}$ & \hspace*{-2ex}   98761.555 & \hspace*{-3ex}    5 & \hspace*{-1ex}  224 & \hspace*{-1ex}          13 & \hspace*{-2ex}    18 &   17 & -- & -- & \hspace*{-2ex} -- & \hspace*{-2ex} -- & -- \\ 
  99 & \hspace*{-1ex} 18$_{16, 3}$ -- 17$_{16, 2}$ & \hspace*{-2ex}   98761.555 & \hspace*{-3ex}    5 & \hspace*{-1ex}  224 & \hspace*{-1ex}          13 & \hspace*{-2ex}    18 &   17 & -- & -- & \hspace*{-2ex} -- & \hspace*{-2ex} -- & -- \\ 
 100 & \hspace*{-1ex} 18$_{13, 5}$ -- 17$_{13, 4}$ & \hspace*{-2ex}   98762.218 & \hspace*{-3ex}    5 & \hspace*{-1ex}  162 & \hspace*{-1ex}          30 & \hspace*{-2ex}    18 &   17 & -- & -- & \hspace*{-2ex} -- & \hspace*{-2ex} -- & -- \\ 
 101 & \hspace*{-1ex} 18$_{13, 6}$ -- 17$_{13, 5}$ & \hspace*{-2ex}   98762.218 & \hspace*{-3ex}    5 & \hspace*{-1ex}  162 & \hspace*{-1ex}          30 & \hspace*{-2ex}    18 &   17 & -- & -- & \hspace*{-2ex} -- & \hspace*{-2ex} -- & -- \\ 
 102 & \hspace*{-1ex} 18$_{17, 1}$ -- 17$_{17, 0}$ & \hspace*{-2ex}   98762.790 & \hspace*{-3ex}    5 & \hspace*{-1ex}  248 & \hspace*{-1ex}           7 & \hspace*{-2ex}    18 &   17 & -- & -- & \hspace*{-2ex} -- & \hspace*{-2ex} -- & -- \\ 
 103 & \hspace*{-1ex} 18$_{17, 2}$ -- 17$_{17, 1}$ & \hspace*{-2ex}   98762.790 & \hspace*{-3ex}    5 & \hspace*{-1ex}  248 & \hspace*{-1ex}           7 & \hspace*{-2ex}    18 &   17 & -- & -- & \hspace*{-2ex} -- & \hspace*{-2ex} -- & -- \\ 
 104 & \hspace*{-1ex} 18$_{12, 6}$ -- 17$_{12, 5}$ & \hspace*{-2ex}   98764.650 & \hspace*{-3ex}    5 & \hspace*{-1ex}  144 & \hspace*{-1ex}          34 & \hspace*{-2ex}    18 &   17 & -- & -- & \hspace*{-2ex} -- & \hspace*{-2ex} -- & -- \\ 
 105 & \hspace*{-1ex} 18$_{12, 7}$ -- 17$_{12, 6}$ & \hspace*{-2ex}   98764.650 & \hspace*{-3ex}    5 & \hspace*{-1ex}  144 & \hspace*{-1ex}          34 & \hspace*{-2ex}    18 &   17 & -- & -- & \hspace*{-2ex} -- & \hspace*{-2ex} -- & -- \\ 
 118 & \hspace*{-1ex} 18$_{ 5,14}$ -- 17$_{ 5,13}$ & \hspace*{-2ex}   98928.453 & \hspace*{-3ex}    4 & \hspace*{-1ex}   58 & \hspace*{-1ex}          57 & \hspace*{-2ex}    18 &   18 & 0.07 &        2.37(09) & \hspace*{-2ex}        0.96 & \hspace*{-2ex}        2.23 &  blend with C$_2$H$_5$CN, $\varv_{13}$=1/$\varv_{21}$=1 \\ 
 119 & \hspace*{-1ex} 18$_{ 5,13}$ -- 17$_{ 5,12}$ & \hspace*{-2ex}   98930.153 & \hspace*{-3ex}    4 & \hspace*{-1ex}   58 & \hspace*{-1ex}          57 & \hspace*{-2ex}    18 &   18 & -- & -- & \hspace*{-2ex} -- & \hspace*{-2ex} -- & -- \\ 
 160 & \hspace*{-1ex} 20$_{ 1,20}$ -- 19$_{ 1,19}$ & \hspace*{-2ex}  105234.713 & \hspace*{-3ex}    5 & \hspace*{-1ex}   49 & \hspace*{-1ex}          68 & \hspace*{-2ex}    28 &   19 & 0.07 &        1.66(11) & \hspace*{-2ex}        0.76 & \hspace*{-2ex}        1.27 &  blend with C$_2$H$_5$OH and CH$_3$OCHO \\ 
 161 & \hspace*{-1ex} 19$_{ 1,18}$ -- 18$_{ 1,17}$ & \hspace*{-2ex}  105272.047 & \hspace*{-3ex}    5 & \hspace*{-1ex}   47 & \hspace*{-1ex}          65 & \hspace*{-2ex}    28 &   20 & 0.07 &        2.07(11) & \hspace*{-2ex}        0.73 & \hspace*{-2ex}        0.81 &  blend with U-lines \\ 
 162 & \hspace*{-1ex} 19$_{ 3,16}$ -- 18$_{ 3,15}$ & \hspace*{-2ex}  105447.141 & \hspace*{-3ex}    5 & \hspace*{-1ex}   52 & \hspace*{-1ex}          64 & \hspace*{-2ex}    37 &   21 & 0.06 &        1.07(15) & \hspace*{-2ex}        0.69 & \hspace*{-2ex}        1.07 & no blend \\ 
 165 & \hspace*{-1ex} 20$_{ 2,19}$ -- 19$_{ 2,18}$ & \hspace*{-2ex}  108552.378 & \hspace*{-3ex}  100 & \hspace*{-1ex}   53 & \hspace*{-1ex}          68 & \hspace*{-2ex}    20 &   22 & 0.07 &        2.05(08) & \hspace*{-2ex}        0.79 & \hspace*{-2ex}        1.24 &  partial blend with C$_2$H$_3$CN, $\varv_{15}$=1 \\ 
 & & & & & & & & & & & &  and U-lines \\ 
 233 & \hspace*{-1ex} 21$_{ 6,16}$ -- 20$_{ 6,15}$ & \hspace*{-2ex}  115395.797 & \hspace*{-3ex}    5 & \hspace*{-1ex}   81 & \hspace*{-1ex}          66 & \hspace*{-2ex}    60 &   23 & 0.11 &        2.94(23) & \hspace*{-2ex}        1.37 & \hspace*{-2ex}        1.37 &  partial blend with U-line \\ 
 234 & \hspace*{-1ex} 21$_{ 6,15}$ -- 20$_{ 6,14}$ & \hspace*{-2ex}  115395.986 & \hspace*{-3ex}    5 & \hspace*{-1ex}   81 & \hspace*{-1ex}          66 & \hspace*{-2ex}    60 &   23 & -- & -- & \hspace*{-2ex} -- & \hspace*{-2ex} -- & -- \\ 
 238 & \hspace*{-1ex} 22$_{ 1,22}$ -- 21$_{ 1,21}$ & \hspace*{-2ex}  115595.764 & \hspace*{-3ex}    5 & \hspace*{-1ex}   59 & \hspace*{-1ex}          75 & \hspace*{-2ex}    79 &   24 & 0.08 &       -0.05(30) & \hspace*{-2ex}        0.99 & \hspace*{-2ex}        2.22 &  blend with CH$_3$CH$_3$CO, $\varv_t$=1,  \\ 
 & & & & & & & & & & & & uncertain baseline \\ 
 \hline
 \end{tabular}
 }\\[1ex] 
 Notes:
 $^a$ Numbering of the observed transitions associated with a modeled line stronger than 20 mK (see Table~\ref{t:etocho-a}).
 $^b$ Frequency uncertainty.
 $^c$ Lower energy level in temperature units ($E_\mathrm{l}/k_\mathrm{B}$).
 $^d$ Calculated rms noise level in $T_{\mathrm{mb}}$ scale.
 $^e$ Numbering of the observed features.
 $^f$ Peak opacity of the modeled feature.
 $^g$ Integrated intensity in $T_{\mathrm{mb}}$ scale for the observed spectrum (Col. 10), the ethyl formate model (Col. 11), and the model including all molecules (Col. 12). The uncertainty in Col. 10 is given in parentheses in units of the last digit.
 \end{table*}

For us to claim a reliable detection of a new molecule, it is
essential that many lines of this molecule be detected
in our spectral survey \textit{and} 
that all the other expected lines, as predicted by our LTE model, either be
blended with lines of other species or be below our detection limit 
\citep[see][]{Belloche08a}. Therefore, in the following, we inspect all 
transitions of ethyl formate in our frequency range. 
We list in Tables~\ref{t:etocho-a} 
and \ref{t:etocho-g} (\textit{online material}) only the transitions that our 
LTE modeling predicts to be stronger than 20 mK in the main-beam 
brightness temperature 
scale. 711 transitions of the \textit{anti}-conformer and 478 transitions of 
the \textit{gauche}-conformer are above this threshold that is conservative 
since it is below 
1.5 times the rms noise level of the \textit{best} part of our survey (and 
even below the rms noise level of \textit{most} parts of our survey). To save 
some space, when two transitions have a 
frequency difference smaller than 0.1 MHz that cannot be resolved, we list 
only the first one. We number the transitions in Col.~1 and give their quantum 
numbers in Col.~2. The frequencies, the frequency uncertainties, the energies 
of the lower levels in temperature units, and the $S\mu^2$ 
values are listed in Col.~3, 4, 5, and 6, respectively. Since the spectra are 
in most cases close to the line confusion limit and it is difficult to measure 
the noise level, we give in Col.~7 the rms sensitivity computed from the 
system temperature and the integration time: 
$\sigma = \frac{F_{\mathrm{eff}}}{B_{\mathrm{eff}}} \times 
\frac{2\,T_{\mathrm{sys}}}{\sqrt{\delta f \, t}}$, with $F_{\mathrm{eff}}$ and 
$B_{\mathrm{eff}}$ the forward and beam efficiencies, $T_{\mathrm{sys}}$ the 
system temperature, $\delta f$ the spectral resolution, and $t$ the total 
integration time (on-source plus off-source).

\addtocounter{figure}{1}
\newcounter{appfig1}
\setcounter{appfig1}{\value{figure}}
\onlfig{\value{appfig1}}{\clearpage
\begin{figure*}
 
\centerline{\resizebox{0.85\hsize}{!}{\includegraphics[angle=270]{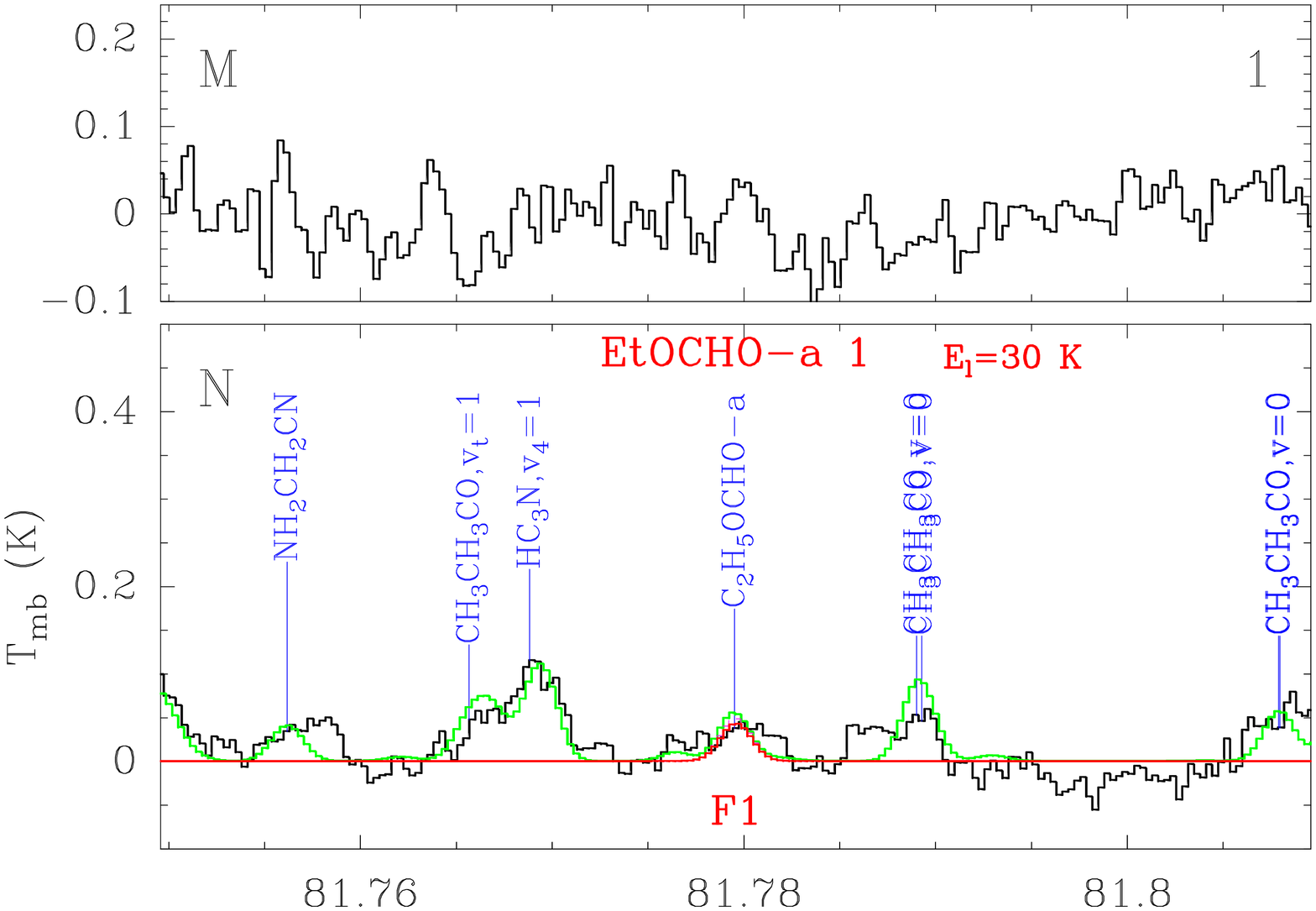}\includegraphics[angle=270]{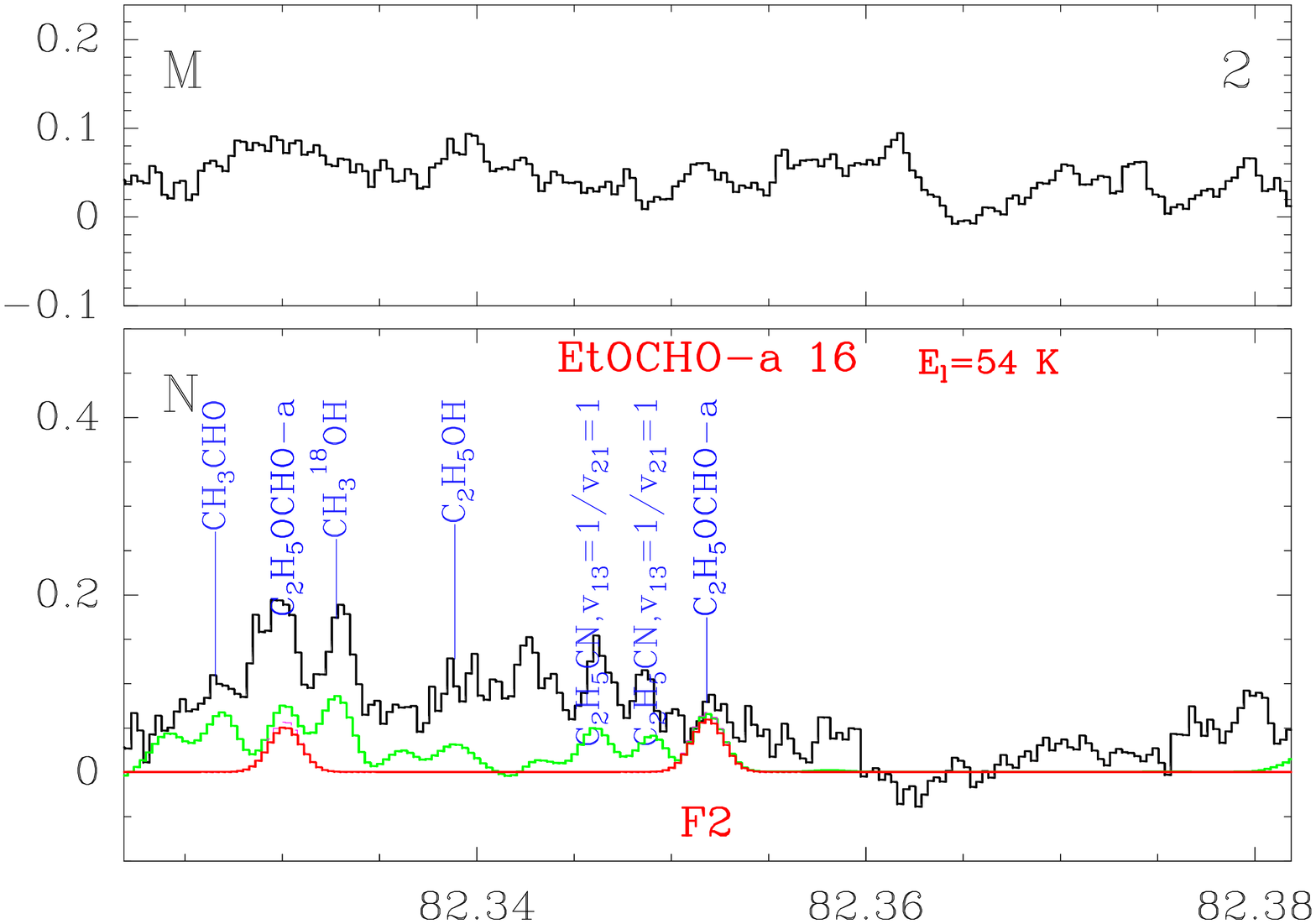}}}
\vspace*{-0.4ex}
\centerline{\resizebox{0.85\hsize}{!}{\includegraphics[angle=270]{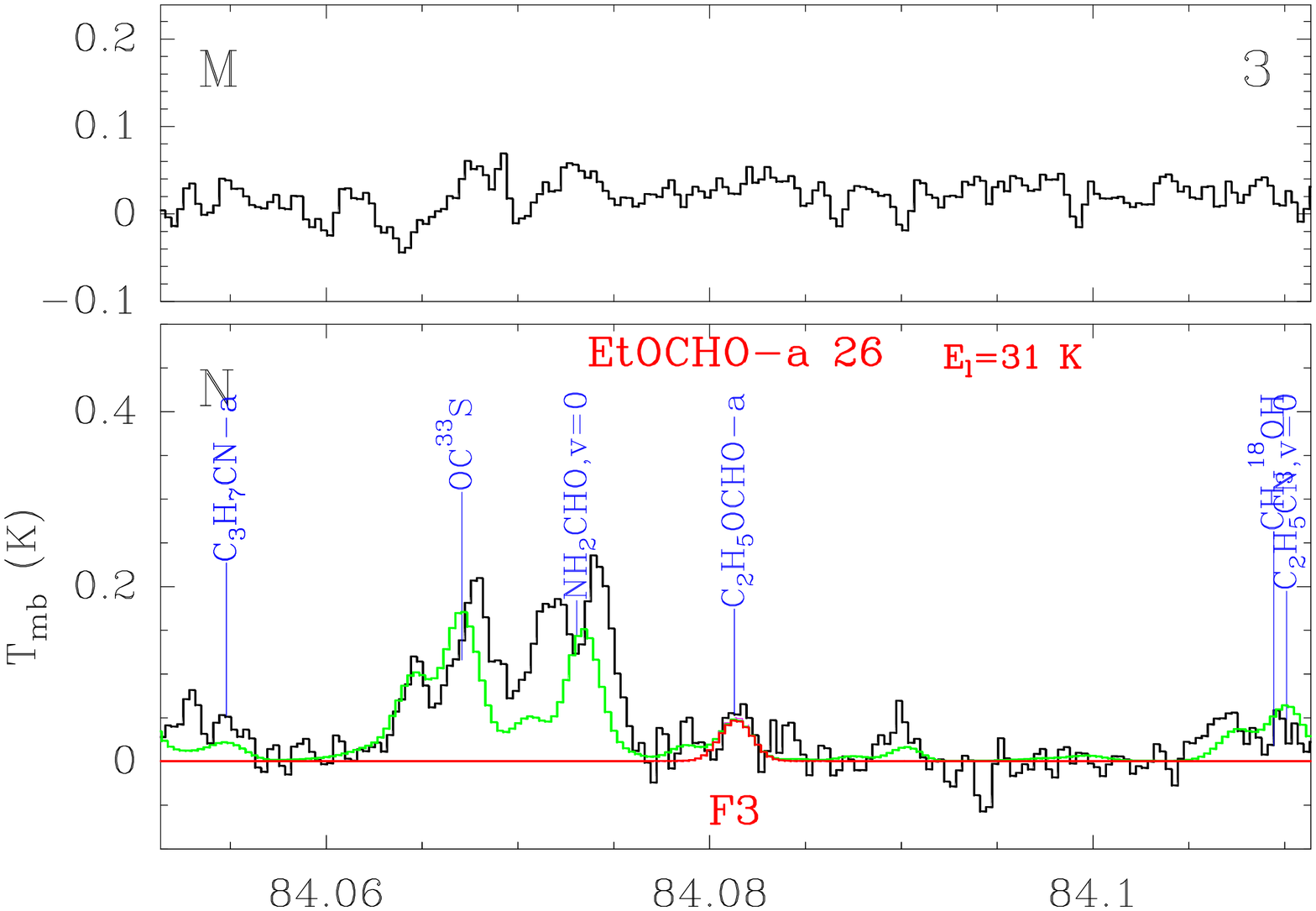}\includegraphics[angle=270]{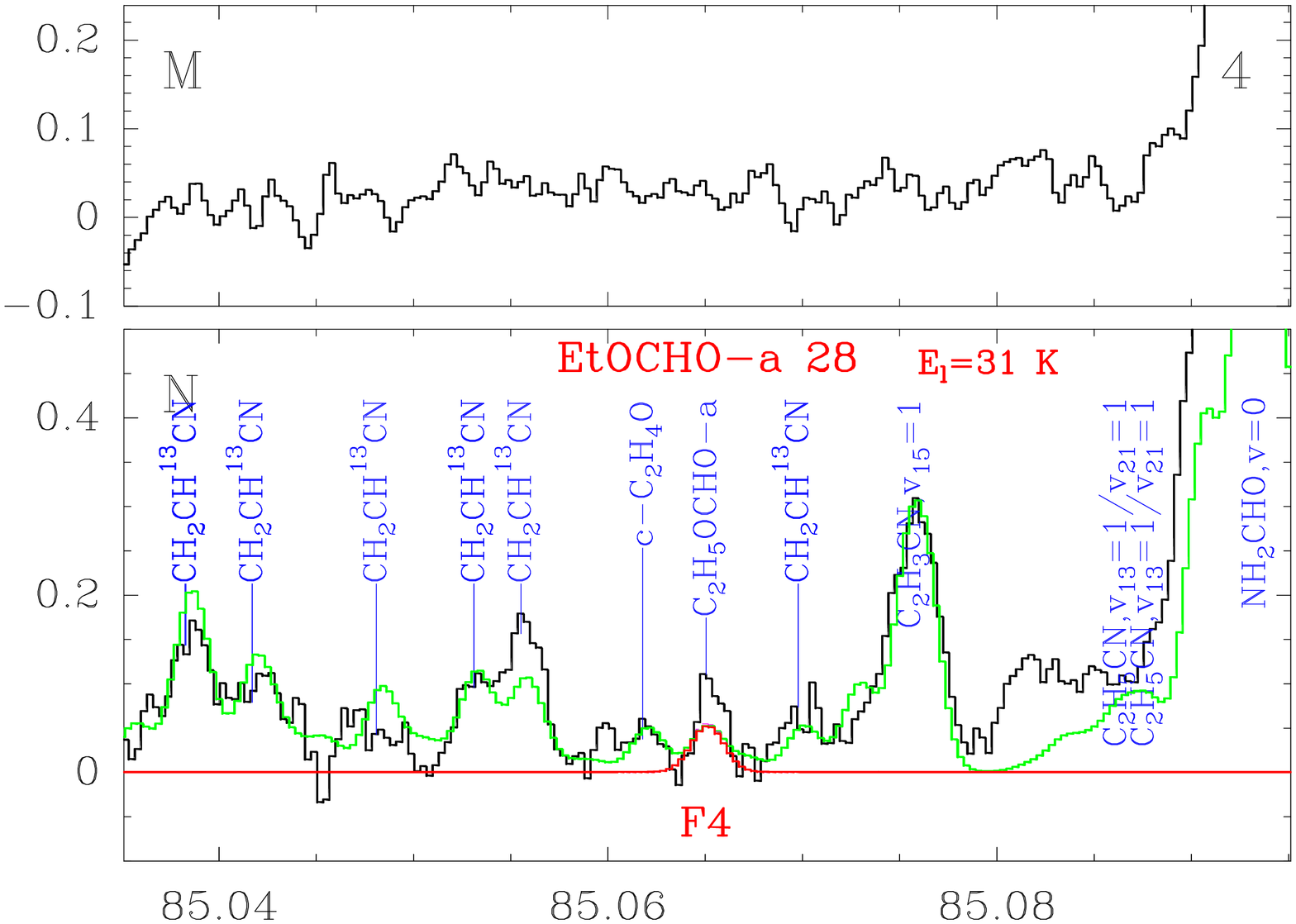}}}
\vspace*{-0.4ex}
\centerline{\resizebox{0.85\hsize}{!}{\includegraphics[angle=270]{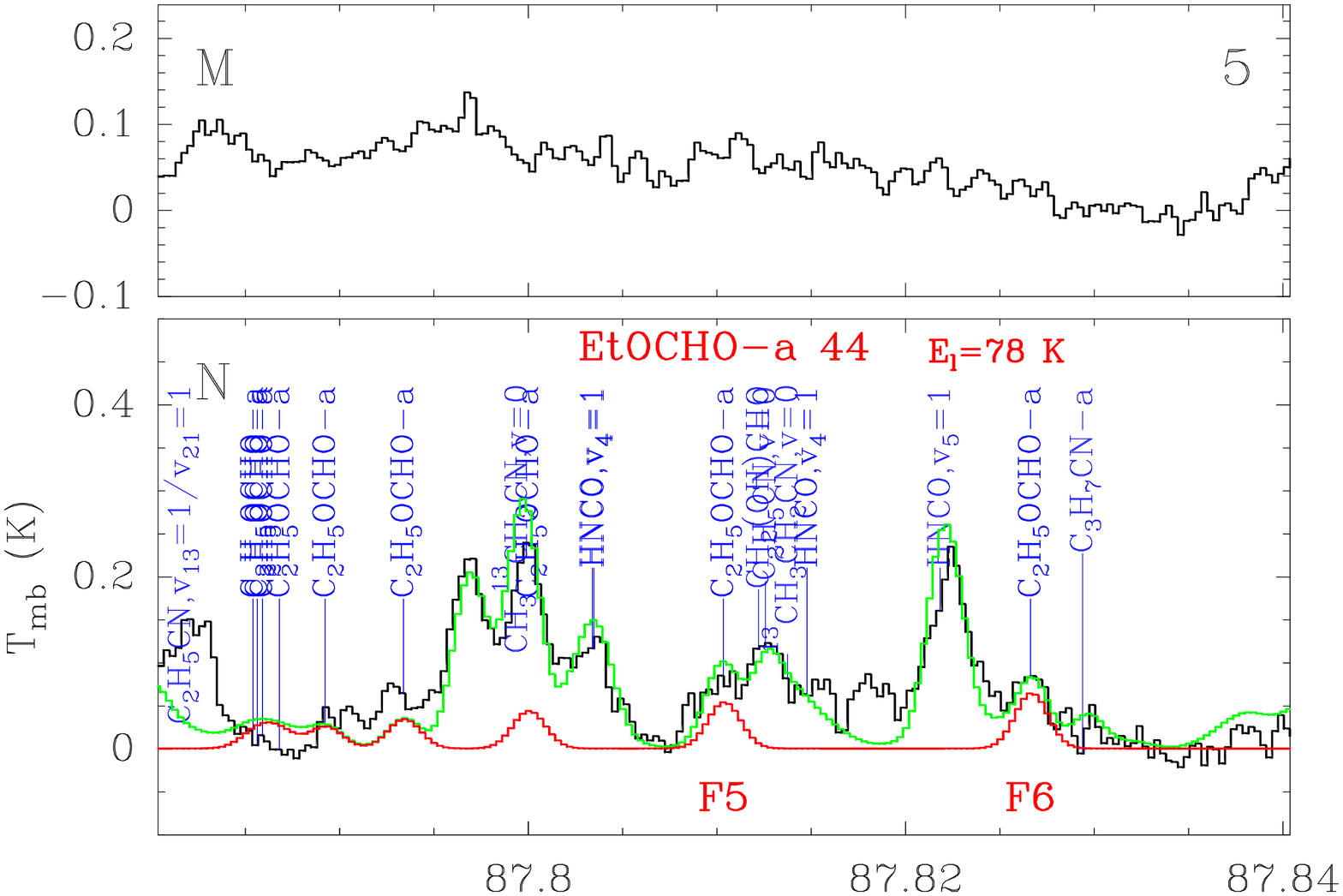}\includegraphics[angle=270]{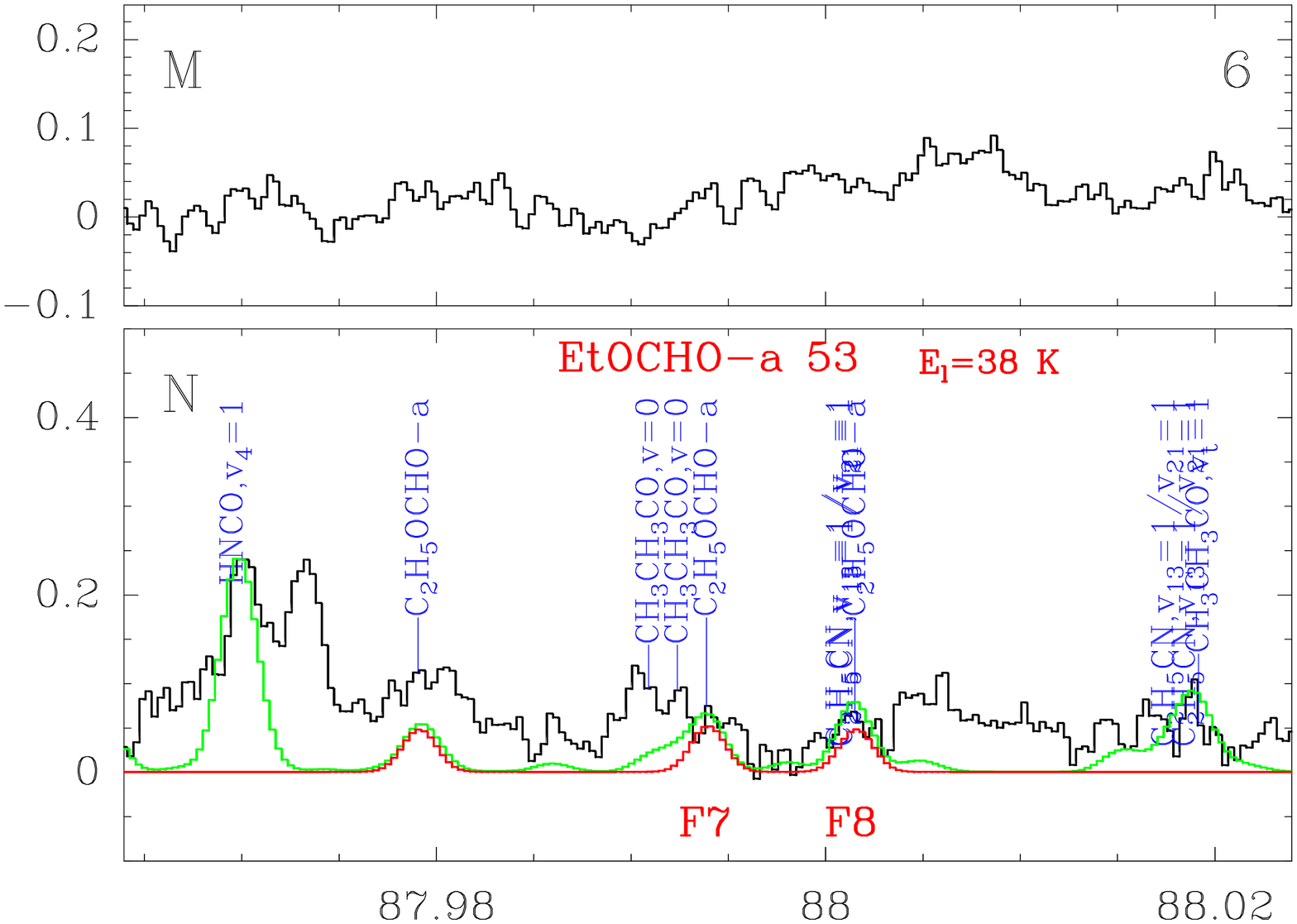}}}
\vspace*{-0.4ex}
\centerline{\resizebox{0.85\hsize}{!}{\includegraphics[angle=270]{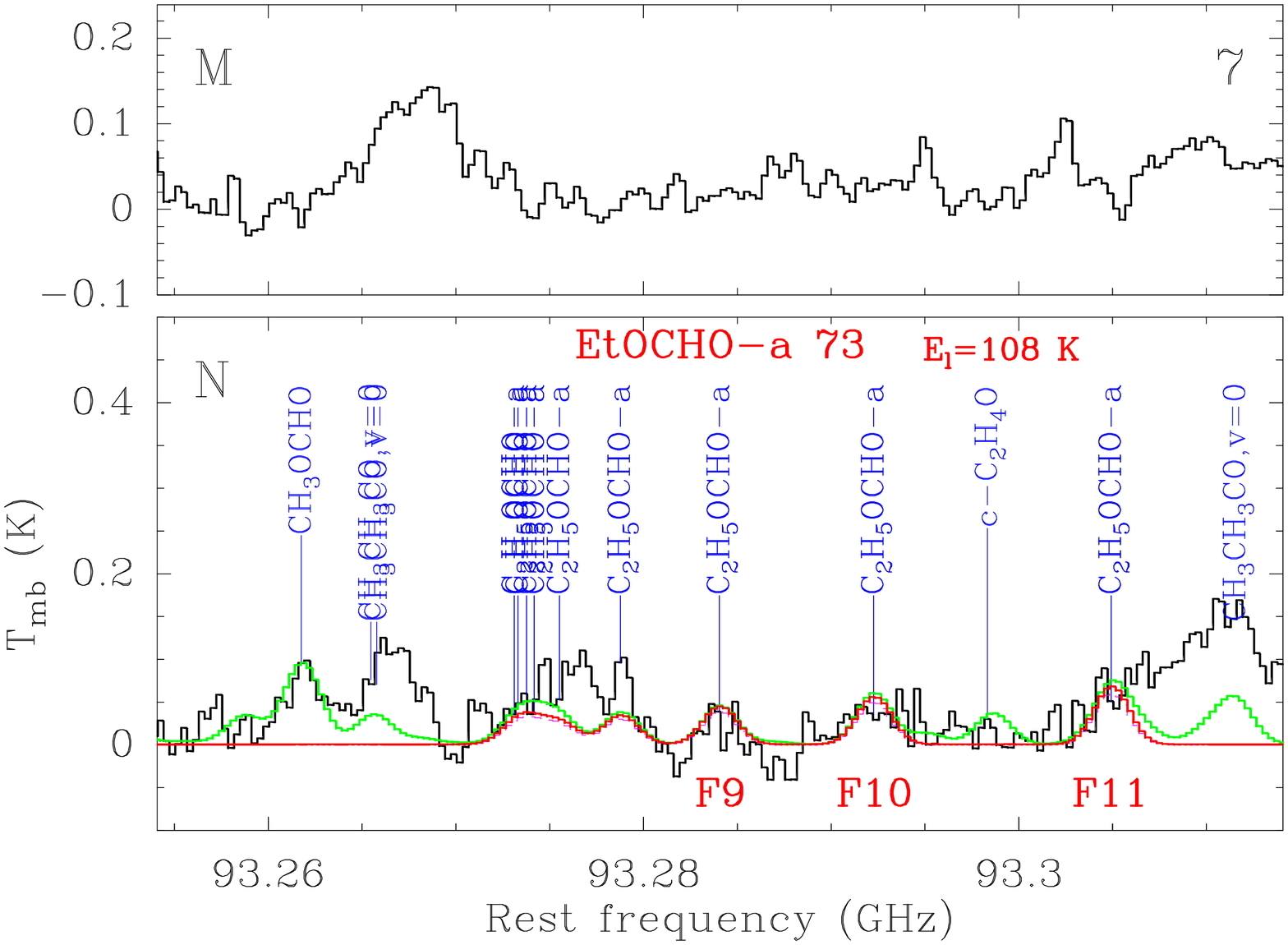}\includegraphics[angle=270]{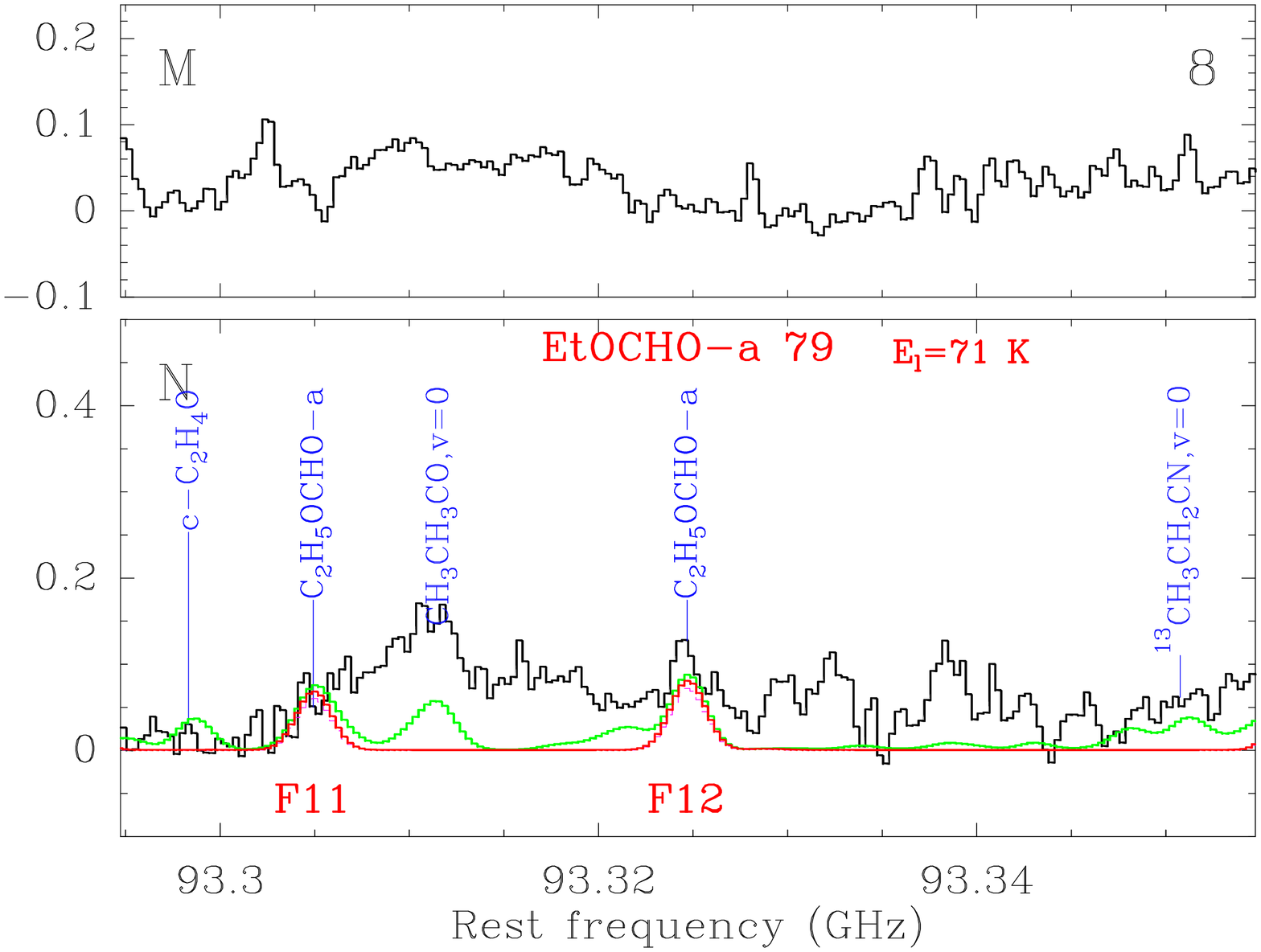}}}
\vspace*{-0.4ex}
\caption{
Transitions of the \textit{anti}-conformer of ethyl formate (EtOCHO-a) detected with the IRAM 30 m telescope.
Each panel consists of two plots and is labeled in black in the upper right corner.
The lower plot shows in black the spectrum obtained toward Sgr~B2(N) in main-beam brightness temperature scale (K), while the upper plot shows the spectrum toward Sgr~B2(M). The rest frequency axis is labeled in GHz. The systemic velocities assumed for Sgr~B2(N) and (M) are 64 and 62 km~s$^{-1}$, respectively.
The lines identified in the Sgr~B2(N) spectrum are labeled in blue. The top red label indicates the EtOCHO-a transition centered in each plot (numbered like in Col.~1 of Table~\ref{t:detectetocho-a}), along with the energy of its lower level in K ($E_l/k_{\mathrm{B}}$).
The other EtOCHO-a lines are labeled in blue only.
The bottom red label is the feature number (see Col.~8 of Table~\ref{t:detectetocho-a}).
The green spectrum shows our LTE model containing all identified molecules, including EtOCHO-a.
The LTE synthetic spectrum of EtOCHO-a alone is overlaid in red, and its opacity in dashed violet.
All observed lines which have no counterpart in the green spectrum are still unidentified in Sgr~B2(N).
}
\label{f:detectetocho-a}
\end{figure*}
\begin{figure*}

\centerline{\resizebox{0.85\hsize}{!}{\includegraphics[angle=270]{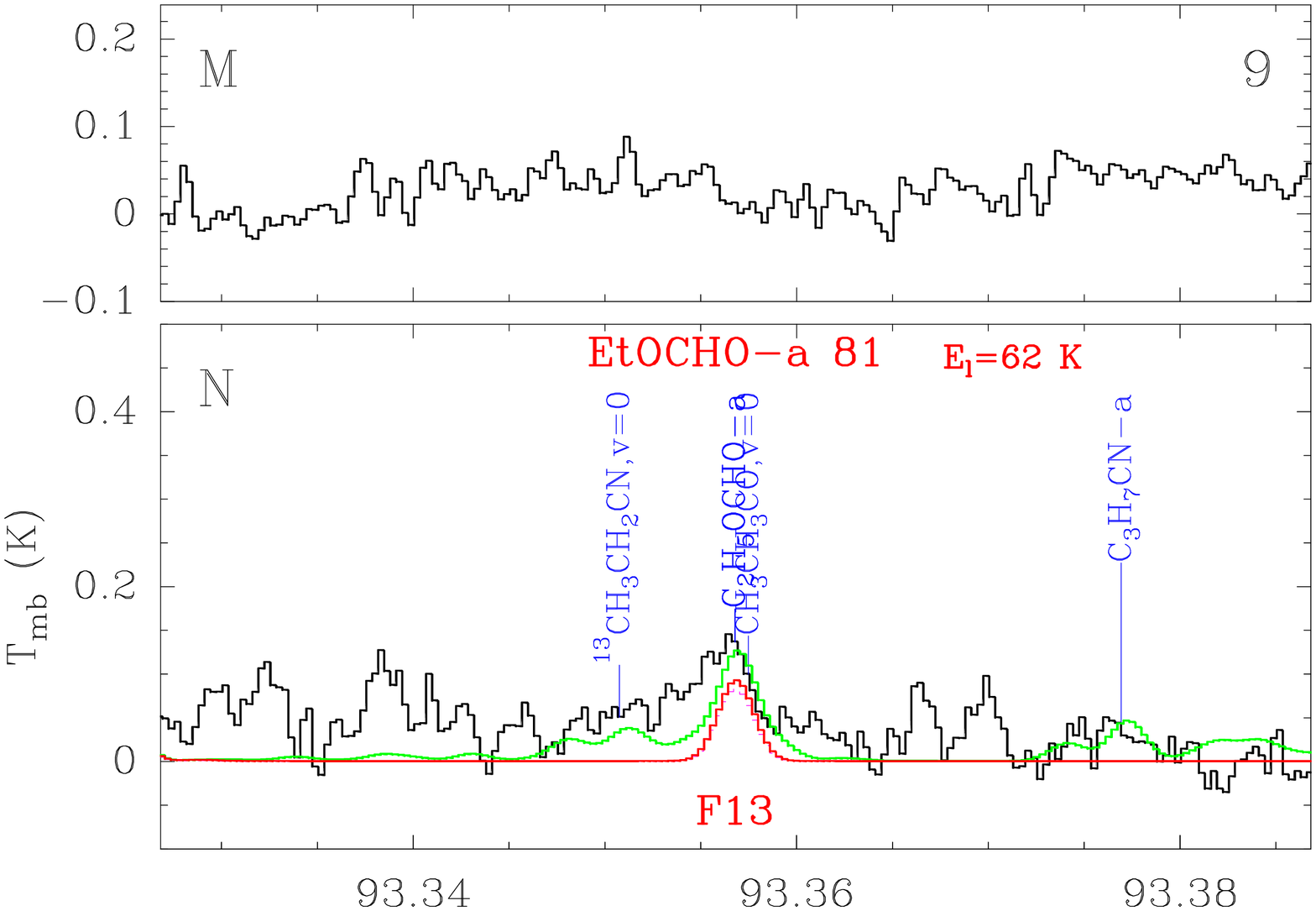}\includegraphics[angle=270]{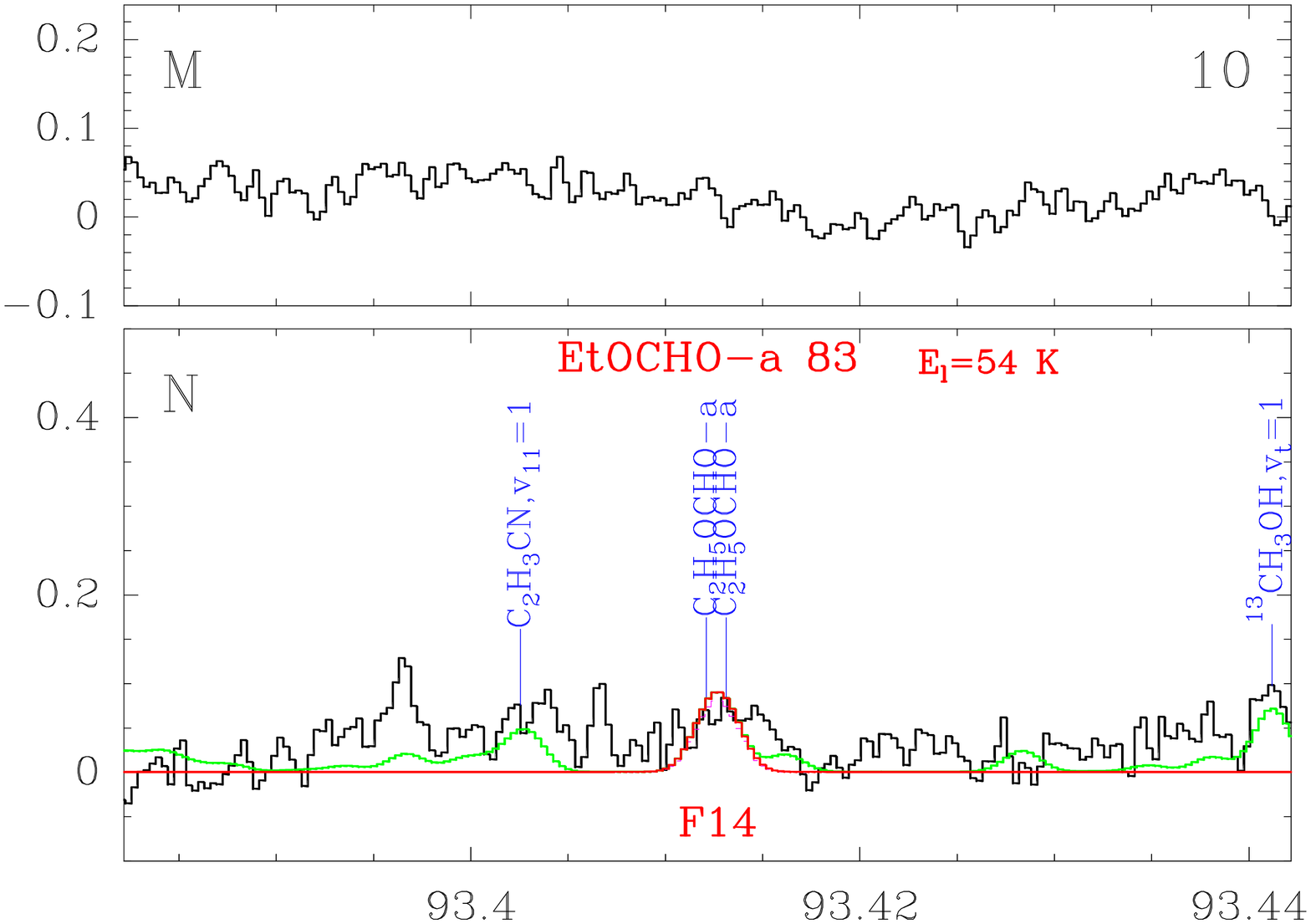}}}
\vspace*{-0.4ex}
\centerline{\resizebox{0.85\hsize}{!}{\includegraphics[angle=270]{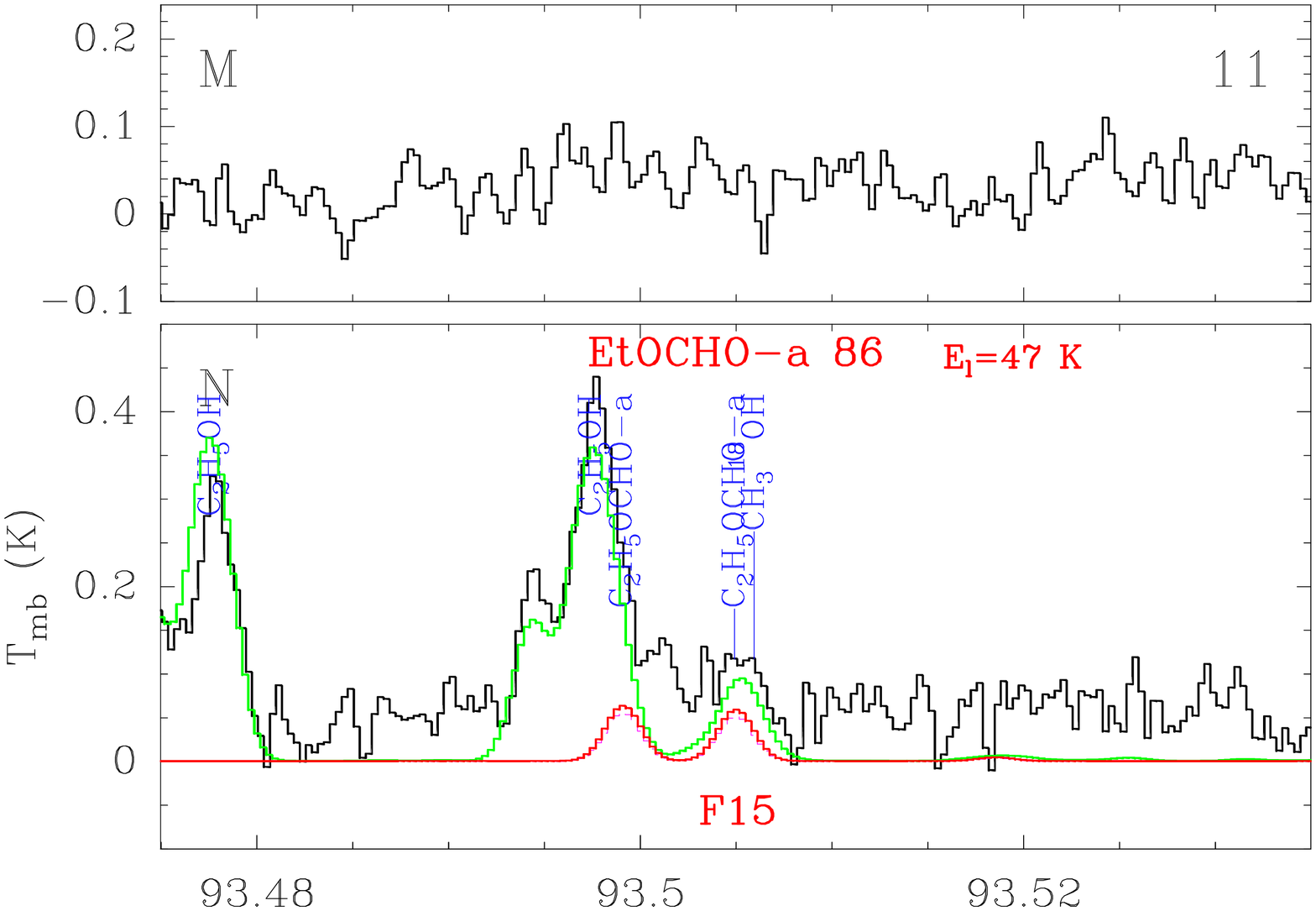}\includegraphics[angle=270]{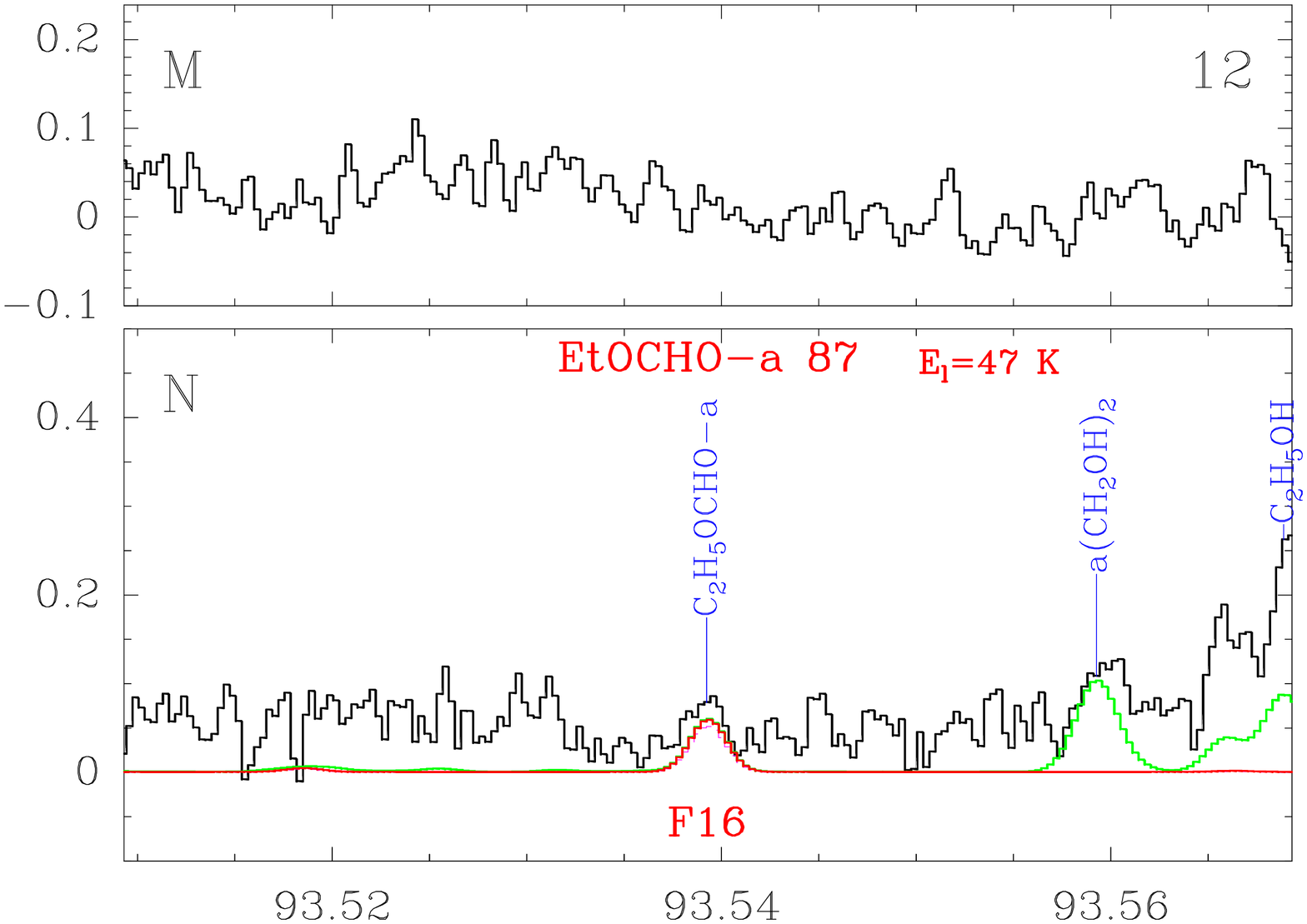}}}
\vspace*{-0.4ex}
\centerline{\resizebox{0.85\hsize}{!}{\includegraphics[angle=270]{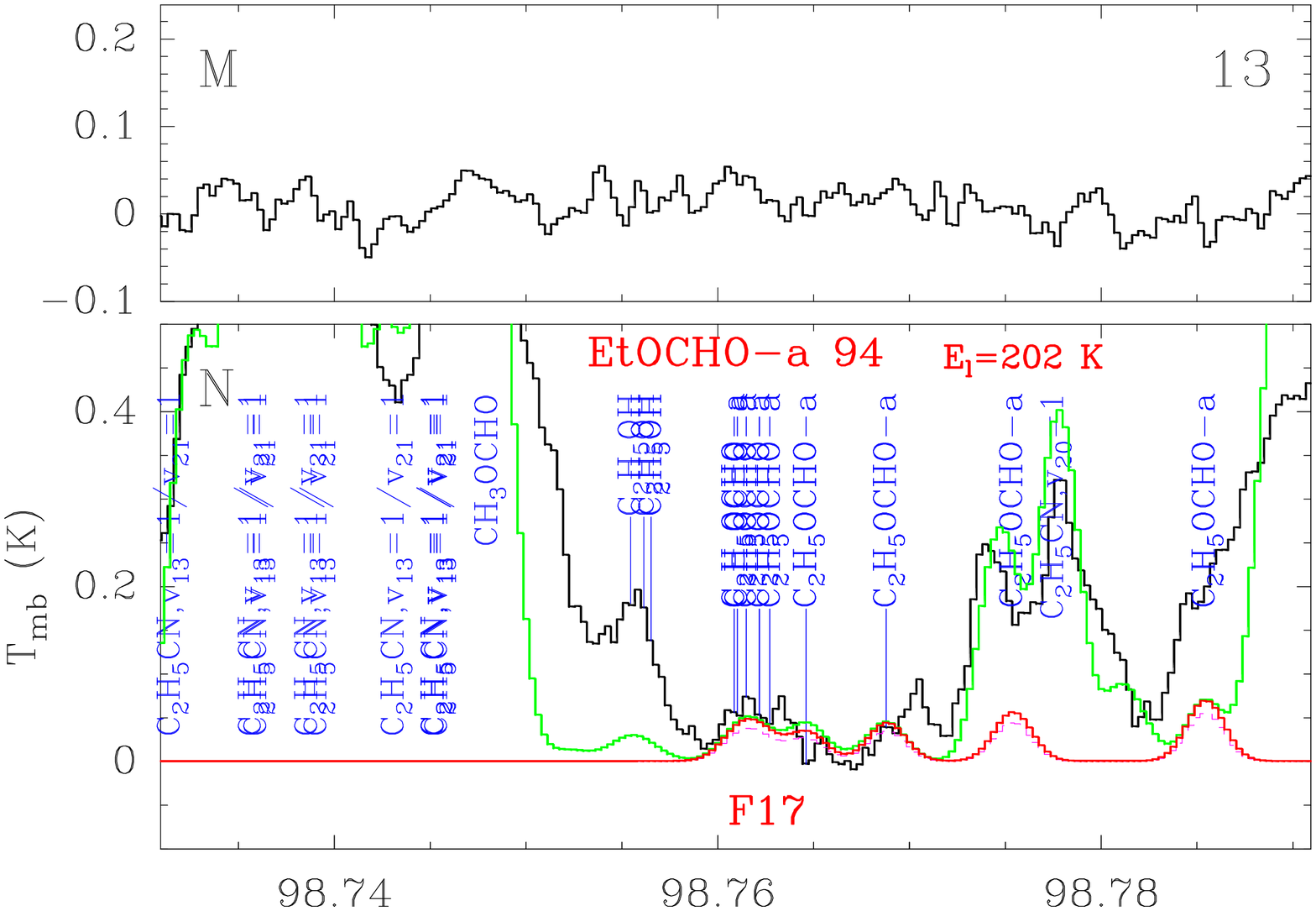}\includegraphics[angle=270]{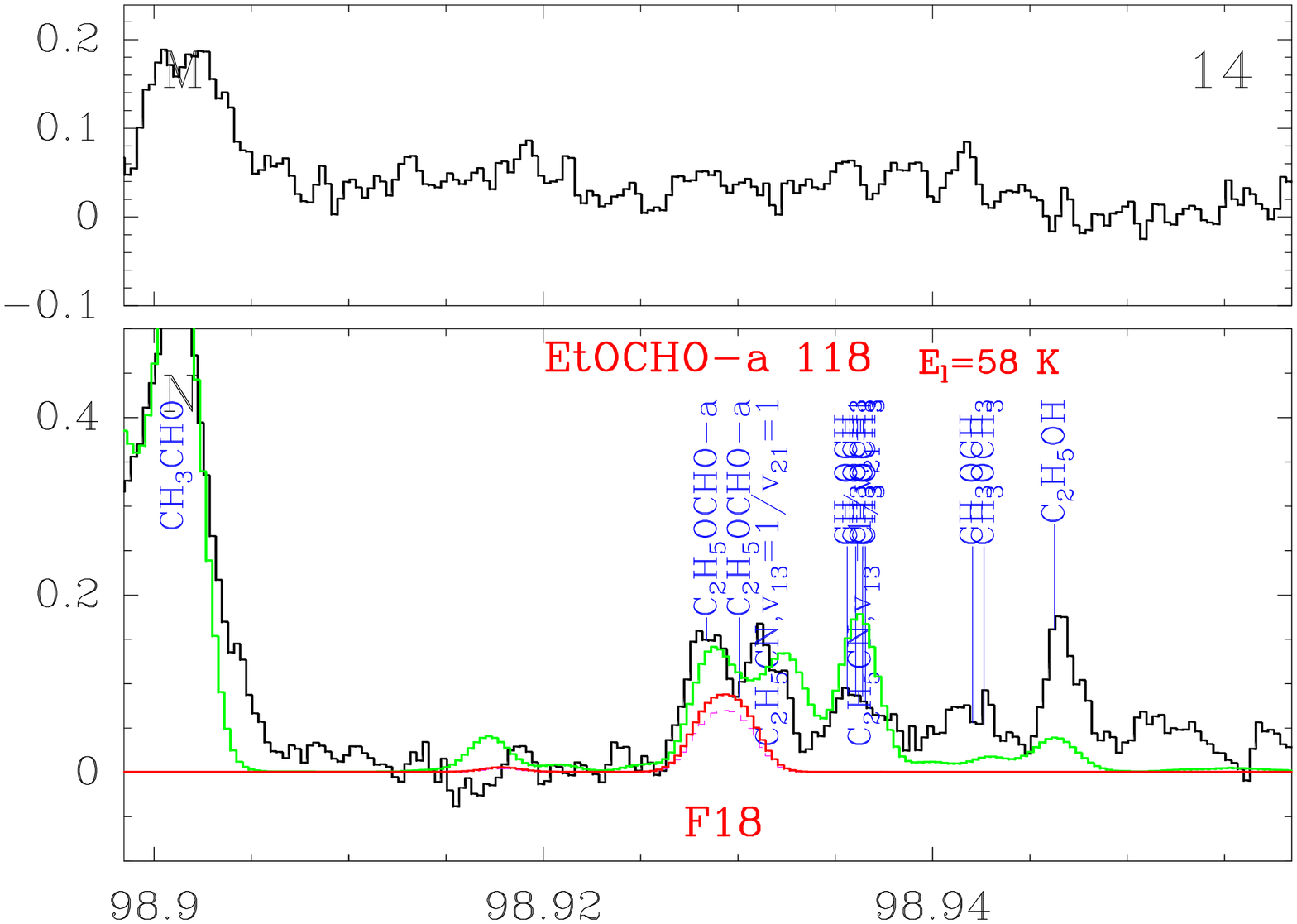}}}
\vspace*{-0.4ex}
\centerline{\resizebox{0.85\hsize}{!}{\includegraphics[angle=270]{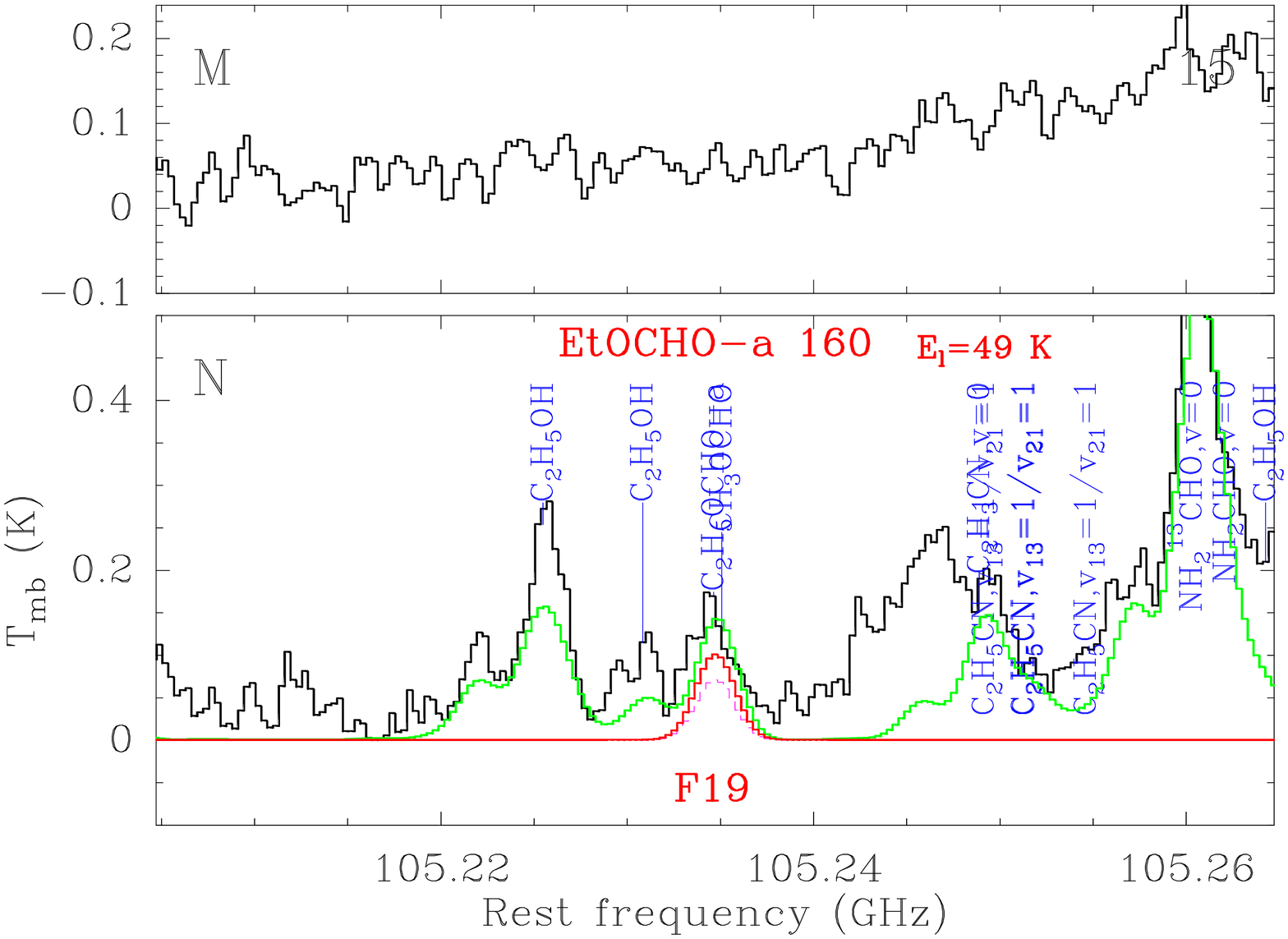}\includegraphics[angle=270]{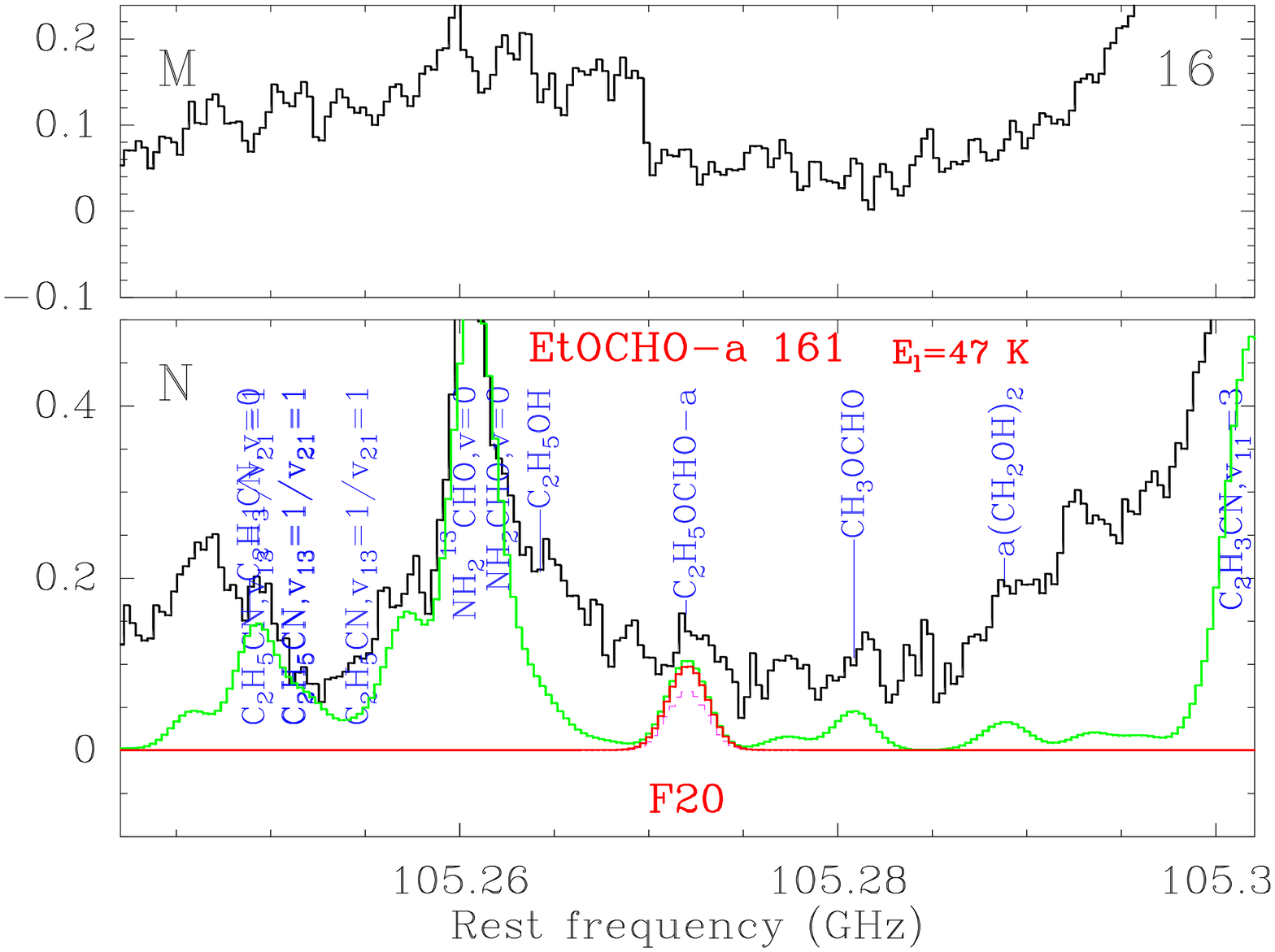}}}
\vspace*{-0.4ex}
\addtocounter{figure}{-1}
\caption{
(continued)
}
\label{f:detectetocho-a}
\end{figure*}
\begin{figure*}

\centerline{\resizebox{0.85\hsize}{!}{\includegraphics[angle=270]{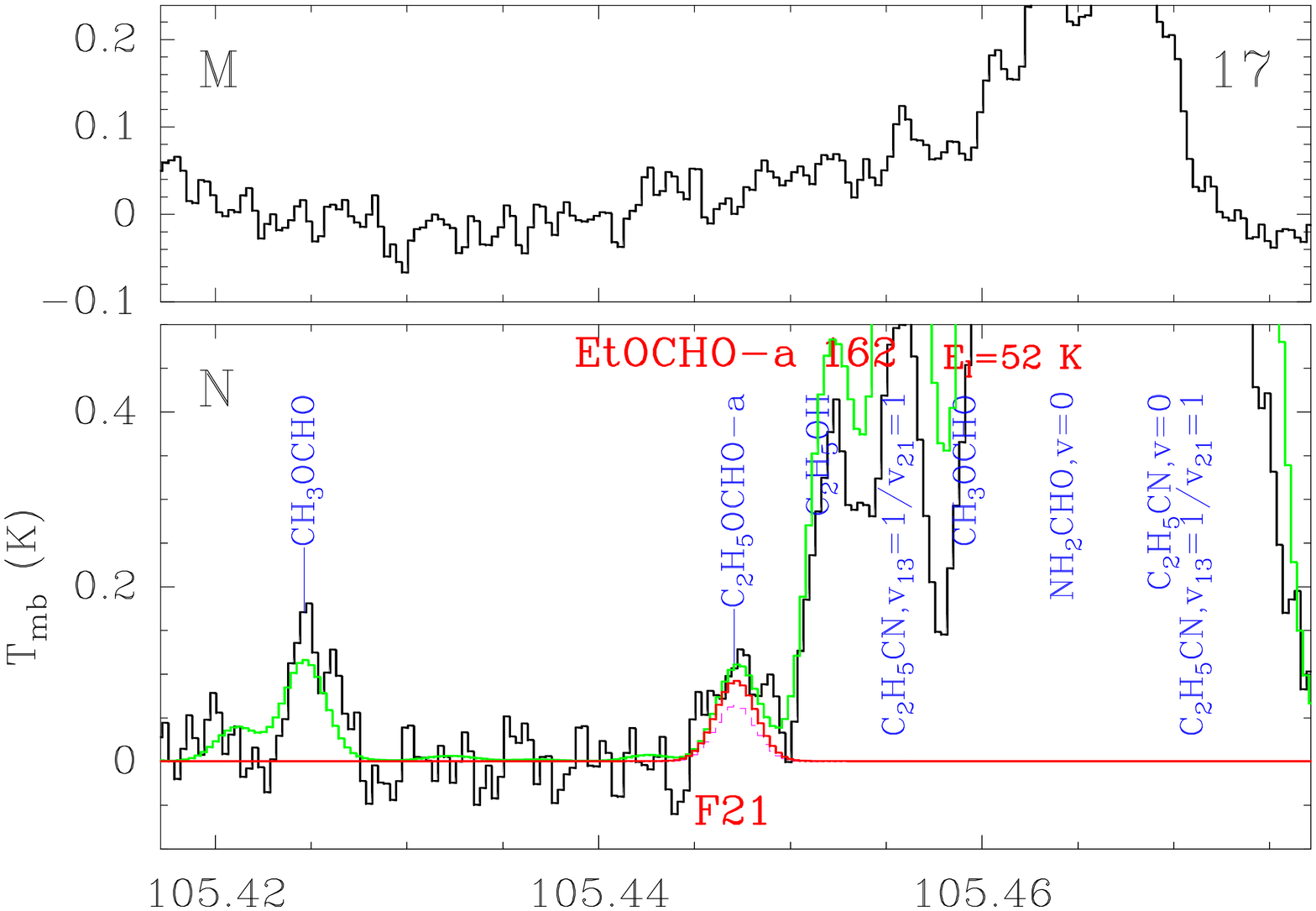}\includegraphics[angle=270]{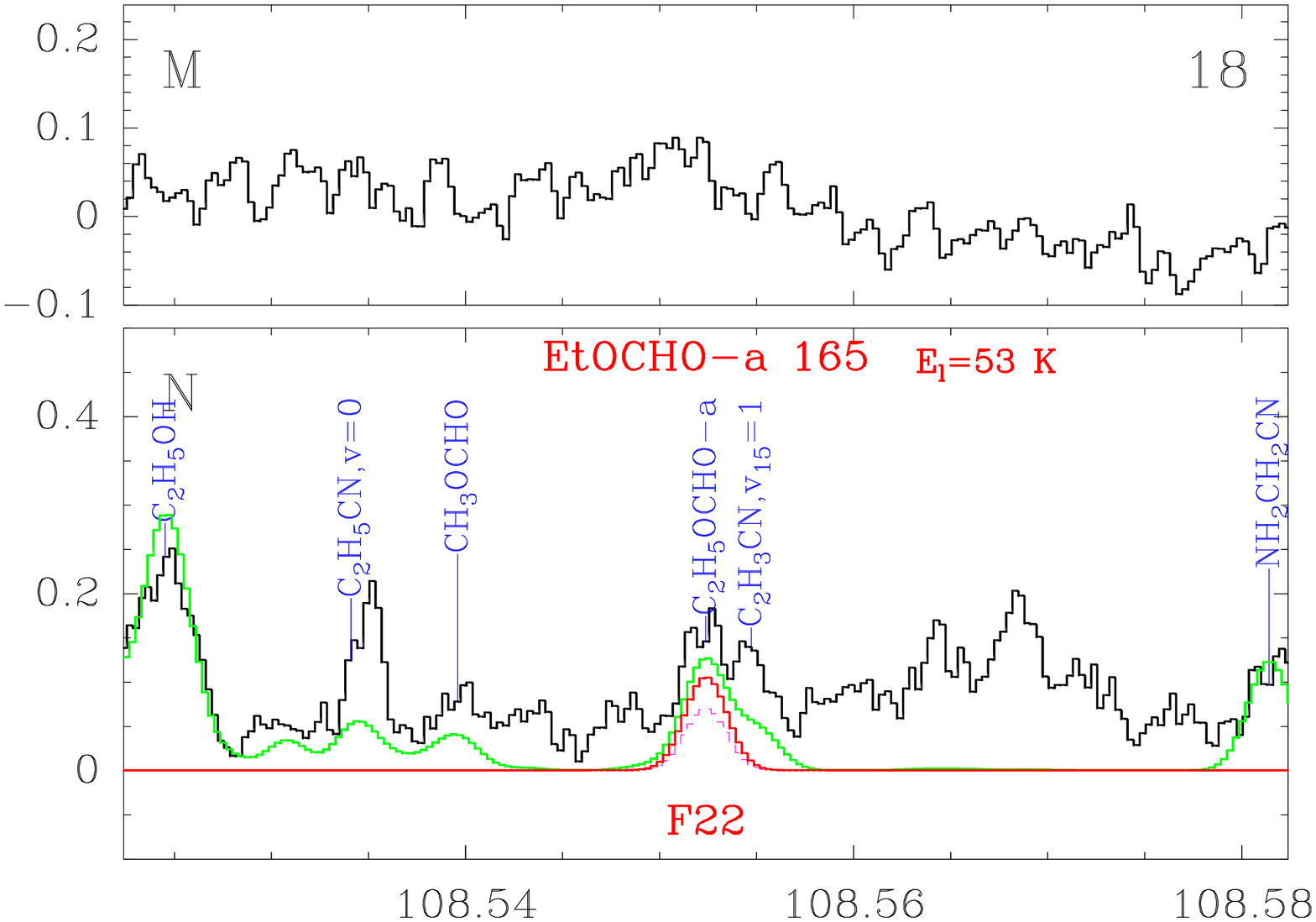}}}
\vspace*{-0.4ex}
\centerline{\resizebox{0.85\hsize}{!}{\includegraphics[angle=270]{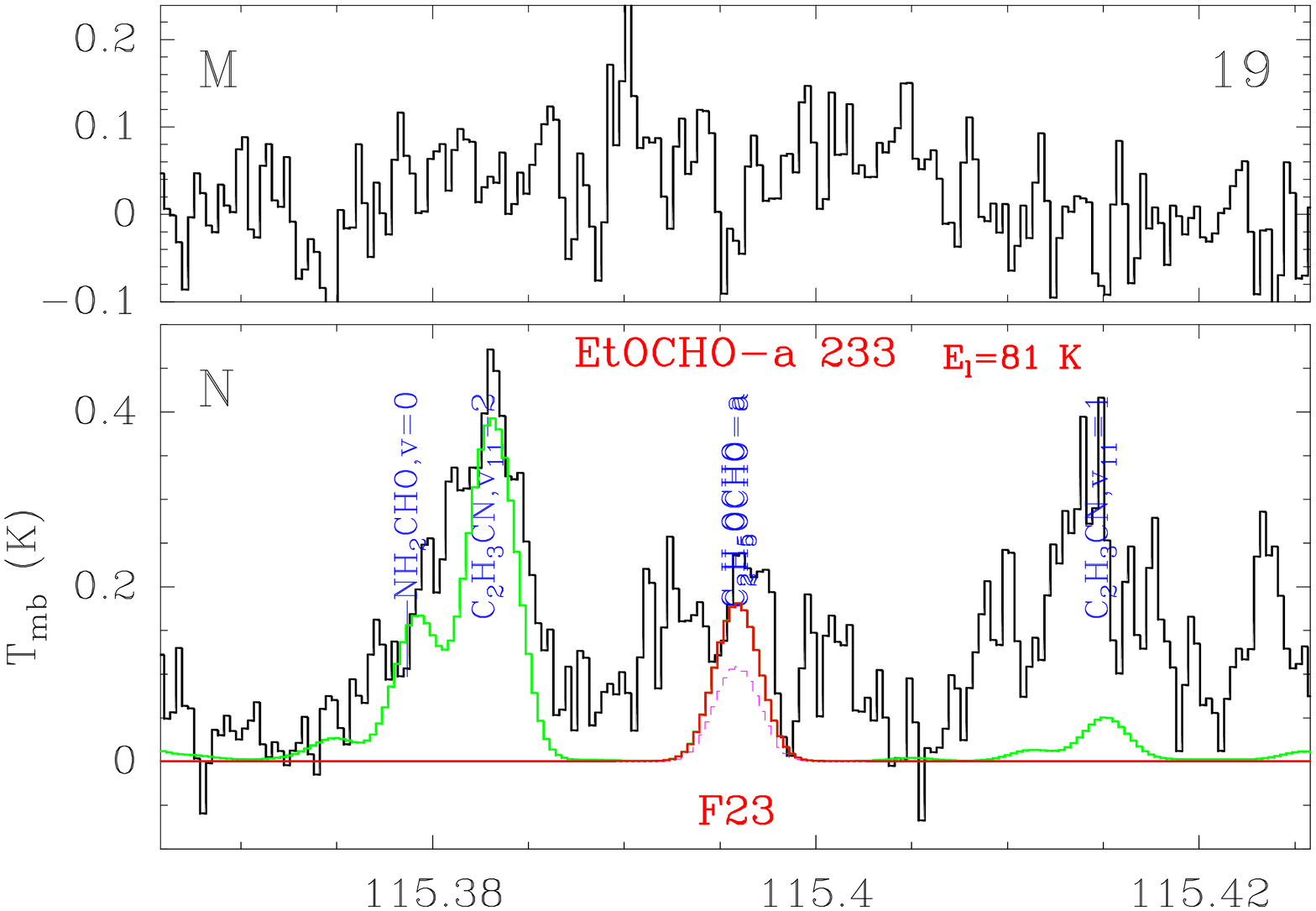}\includegraphics[angle=270]{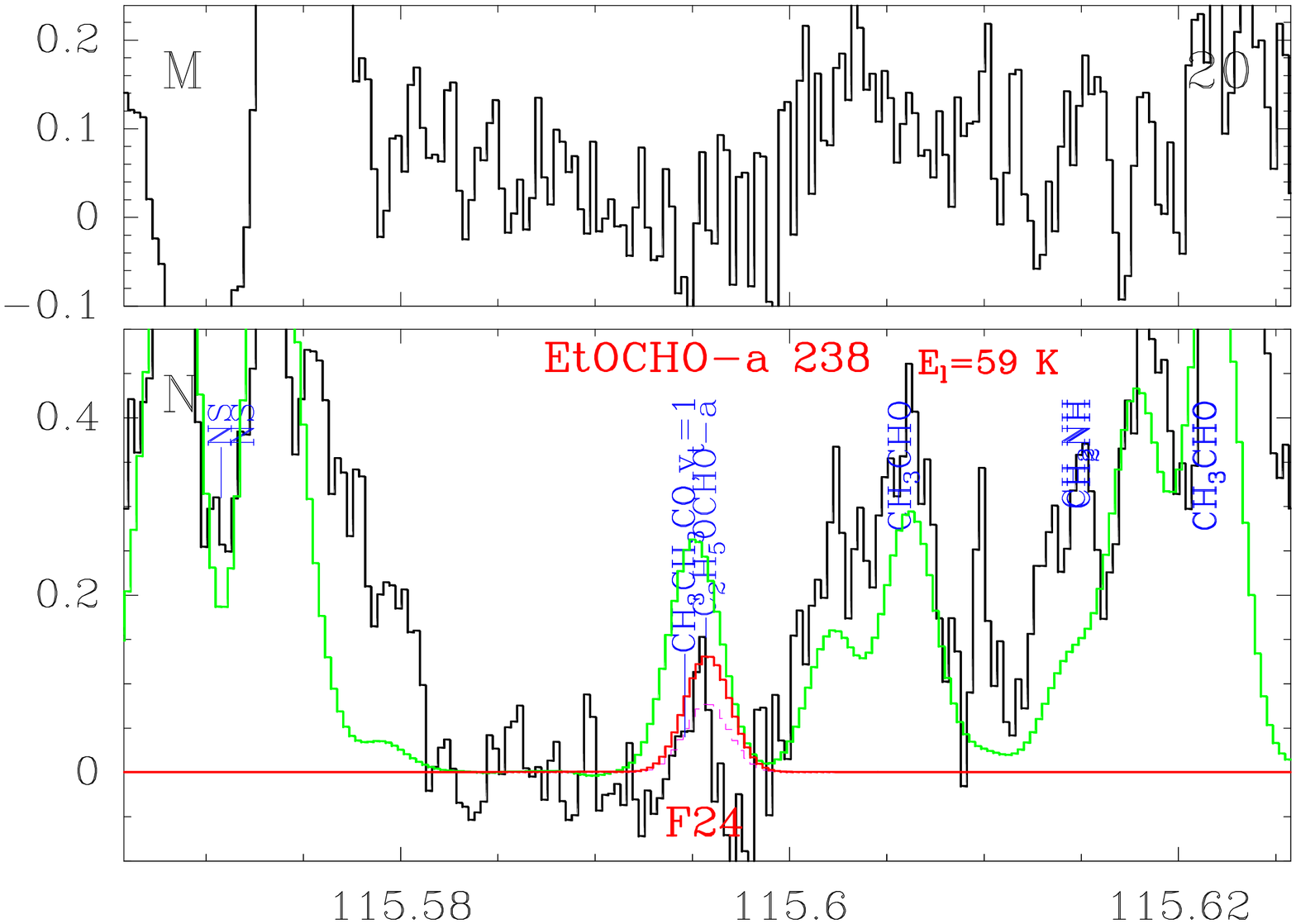}}}
\vspace*{-0.4ex}
\addtocounter{figure}{-1}
\caption{
(continued)
}
\label{f:detectetocho-a}
\end{figure*}

}


We list in Col.~8 of Tables~\ref{t:etocho-a} and \ref{t:etocho-g} comments 
about the blends affecting the transitions of the \textit{anti}- and 
\textit{gauche}-conformers 
of ethyl formate. As can be seen in these tables, most of the ethyl formate 
lines covered by our survey of Sgr~B2(N) are heavily blended with lines of 
other molecules and therefore cannot be identified in this source based on our
single-dish data. Only 46 of the 711 transitions of the 
\textit{anti}-conformer are 
relatively free of contamination from other molecules, known or still 
unidentified according to our modeling. They are marked ``Detected'' or 
``Group detected'' in Col.~8 of Table~\ref{t:etocho-a}, and are listed with 
more information in Table~\ref{t:detectetocho-a}. We stress that all 
transitions of sufficient strength  predicted in the frequency range of our
spectral survey are either detected or blended, i.e. no predicted transition 
is missing in the observed spectrum. The 46 detected transitions correspond to 
24 observed features that are shown in Fig.~\ref{f:detectetocho-a} 
(\textit{online material}) and labeled in Col.~8 of 
Table~\ref{t:detectetocho-a}. For reference, we show the spectrum observed 
toward Sgr~B2(M) in these figures also. We identified the ethyl formate lines 
and the blends affecting them with the LTE model of this molecule and the LTE 
model including all molecules (see Sect.~\ref{ss:modeling}). The parameters 
of our best-fit LTE model of ethyl formate are listed in 
Table~\ref{t:etochomodel}, and the model is overlaid in red on the spectrum 
observed toward Sgr~B2(N) in Fig.~\ref{f:detectetocho-a}. The best-fit LTE 
model including all molecules is shown in green in the same figures.

For the frequency range corresponding to each detected ethyl formate feature, 
we list in Table~\ref{t:detectetocho-a} the integrated intensities of the 
observed spectrum (Col.~10), of the best-fit model of ethyl formate (Col.~11), 
and of the best-fit model including all molecules (Col.~12). In these columns, 
the dash symbol indicates transitions belonging to the same feature. Columns 1 
to 7 of Table~\ref{t:detectetocho-a} are the same as in 
Table~\ref{t:etocho-a}. The $1\sigma$ uncertainty given for the integrated 
intensity in Col.~10 was computed using the estimated noise level of Col.~7. 

\begin{figure*}
\centerline{\resizebox{1.0\hsize}{!}{\includegraphics[angle=270]{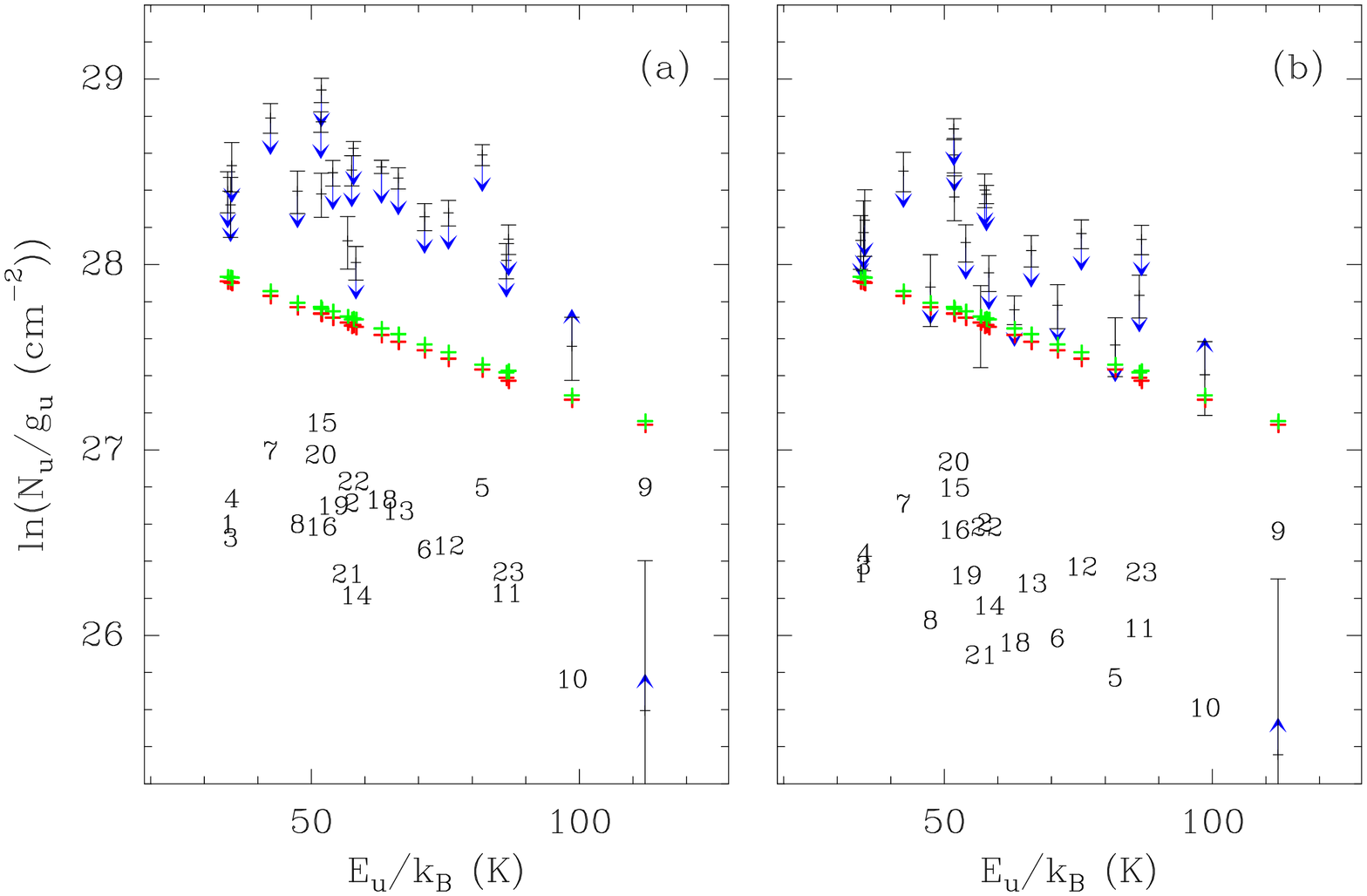}}}
\caption{\textbf{a)} Population diagram of the \textit{anti}-conformer of 
ethyl formate in Sgr~B2(N). The red points 
were computed in the optically thin approximation using the integrated 
intensities of our best-fit \textit{model} of ethyl formate, while the 
green points were corrected for the opacity. The black points were computed in 
the optically thin approximation using the integrated intensities of the 
spectrum \textit{observed} with the IRAM 30~m telescope. 
The error bars are $1 \sigma$ uncertainties on $N_u/g_u$. Blue arrows pointing
downwards mark the transitions blended with transitions from other molecules, 
while blue arrows pointing upwards indicate that the baseline removed in the 
observed spectrum is uncertain. The arrow length is arbitrary. The feature 
labels are shown in black shifted by -1.8 along the Y-axis for clarity, 
except for feature 9 for which it is shifted by +1.2. The measurement 
corresponding to feature 24 (at $E_u/k_B$ = 65~K) is not shown since the 
integrated intensity measured toward Sgr~B2(N) is negative, most likely because 
the level of the baseline was overestimated. Feature 17 is a blend of several 
transitions with different energy levels and was therefore also omitted. 
\textbf{b)} Same as \textbf{a)} but with the expected contribution from the 
contaminating molecules removed from the integrated intensities of the 
observed spectrum.}
\label{f:popdiagetocho}
\end{figure*}

The measurements of the \textit{anti}-conformer of ethyl formate are plotted 
in the form of a population diagram in Fig.~\ref{f:popdiagetocho}a, which 
plots upper level column density divided by statistical weight, $N_u/g_u$, 
versus the upper level energy in Kelvins \citep[see][]{Goldsmith99}.
The data are shown in black and our best-fit model of ethyl formate in 
red. Out of 12 features encompassing several transitions, 
one contains transitions with different energy levels and was ignored in the 
population diagram (feature 17). We used equation A5 of \citet{Snyder05} to
compute the ordinate values:
\begin{equation}
\ln \left( \frac{N_u}{g_u} \right) = 
\ln \left( \frac{1.67 W_T \times 10^{14}}{S \mu^2 B \nu} \right) = 
- \frac{E_u}{T_{rot}} + \ln \left( \frac{N_T}{Z} \right),
\end{equation}
where $W_T$ is the integrated intensity in K~km~s$^{-1}$ in main-beam 
brightness temperature scale, $S \mu^2$ the line strength times the dipole
moment squared 
in D$^2$, $B$ the beam filling factor, $\nu$ the frequency in GHz, $T_{rot}$ 
the rotation temperature in K, $N_T$ the molecular column density in 
cm$^{-2}$, and Z the partition function.
This equation assumes optically thin emission. To 
estimate by how much line opacities affect this diagram, we applied the 
opacity correction factor $C_\tau = \frac{\tau}{1-e^{-\tau}}$ 
\citep[see][]{Goldsmith99,Snyder05} to the modeled intensities, using the 
opacities from our radiative transfer calculations (Col.~9 of 
Table~\ref{t:detectetocho-a}); the result is shown in green in 
Fig.~\ref{f:popdiagetocho}a. The population diagram derived from the modeled 
spectrum is slightly shifted upwards but its shape, in particular its slope 
(the inverse of which \textit{approximately} determines the rotation 
temperature), is not significantly changed, since $ln\, C_{\tau}$ does not 
vary much (from 0.019 to 0.053). The populations derived from the 
\textit{observed} spectrum in the optically thin approximation are therefore 
not significantly affected by the optical depth of the ethyl formate
transitions\footnote{Note that our modeled spectrum is anyway calculated with 
the full LTE radiative transfer that takes into account the optical depth 
effects (see Sect.~\ref{ss:modeling}).}. 
The scatter of the black crosses in Fig.~\ref{f:popdiagetocho}a is  therefore
dominated by the blends with other molecules and uncertainties in the baseline 
removal (indicated by the downwards and upwards blue arrows, respectively). 

The population diagram derived from the modeled spectrum in 
Fig.~\ref{f:popdiagetocho}a is systematically below the measurements. Since 
most of the detected features of the \textit{anti}-conformer of ethyl formate 
are partially blended with lines from other molecules (see Col.~13 of 
Table~\ref{t:detectetocho-a}), we can use our model including all identified 
molecules (shown in green in Fig.~\ref{f:detectetocho-a}) to remove the 
expected contribution from the contaminating molecules. Instead of computing 
$N_u/g_u$ with the integrated intensities $I_{\mathrm{obs}}$ listed in Col.~10 
of Table~\ref{t:detectetocho-a}, we can use the value 
$I_{\mathrm{obs}} - (I_{\mathrm{all}}-I_{\mathrm{mod}})$ derived from Col.~10, 
11, and 12. The corrected population diagram is shown in 
Fig.~\ref{f:popdiagetocho}b. The predicted (red) and measured (black) 
points are much closer to each other. A close inspection of 
Fig.~\ref{f:detectetocho-a} shows however that the wings of most detected 
features of ethyl formate are still contaminated by U-lines, which explains 
why the measured points are still above the predicted ones in the population 
diagram (our fitting method with XCLASS is mainly focused on the peak 
intensity, not on the integrated intensity). The only exception is feature 9 
for which the level of the baseline was obviously overestimated (see panel 7 of 
Fig.~\ref{f:detectetocho-a}). 

\begin{table}
 {\centering
 \caption{
 Parameters of our best-fit LTE model of ethyl formate.
}
 \label{t:etochomodel}
 \vspace*{0.0ex}
 \begin{tabular}{ccccc}
 \hline\hline
 \multicolumn{1}{c}{Size$^{a}$} & \multicolumn{1}{c}{$T_{\mathrm{rot}}$$^{b}$} & \multicolumn{1}{c}{$N^{c}$} & \multicolumn{1}{c}{$\Delta V^{d}$} & \multicolumn{1}{c}{$V_{\mathrm{off}}$$^{e}$} \\ 
 \multicolumn{1}{c}{\scriptsize ($''$)} & \multicolumn{1}{c}{\scriptsize (K)} & \multicolumn{1}{c}{\scriptsize (cm$^{-2}$)} & \multicolumn{1}{c}{\scriptsize (km~s$^{-1}$)} & \multicolumn{1}{c}{\scriptsize (km~s$^{-1}$)} \\ 
 \multicolumn{1}{c}{(1)} & \multicolumn{1}{c}{(2)} & \multicolumn{1}{c}{(3)} & \multicolumn{1}{c}{(4)} & \multicolumn{1}{c}{(5)} \\ 
 \hline
3.0 &  100 & $ 5.40 \times 10^{16}$ & 7.0 & 0.0 \\  \hline
 \end{tabular}
 }\\[1ex] 
 Notes:
 $^a$ Source diameter (\textit{FWHM}).
 $^b$ Temperature.
 $^c$ Column density.
 $^d$ Linewidth (\textit{FWHM}).
 $^e$ Velocity offset with respect to the systemic velocity of Sgr~B2(N) V$_{\mathrm{lsr}} = 64$ km~s$^{-1}$.
 \end{table}

Given the remaining uncertainties due to the contamination from U-lines, it is 
difficult to derive the temperature with high accuracy. However, feature 17,
which can unfortunately not be shown in the population diagram since it is a 
blend of several transitions with different energy levels (from 149 to 253~K), 
is significantly detected in panel 13 of Fig.~\ref{f:detectetocho-a}. This is 
a strong indication that the temperature cannot be much lower than 100~K. 
Overall, we estimate the resulting uncertainty on the derived column density 
to be on the order of 25$\%$. Finally, since all detected transitions are 
optically 
thin and the region emitting in ethyl formate is most likely compact given its 
high temperature, column density and source size are degenerate. We fixed the 
source size to 3$\arcsec$. This is approximately the size of the region 
emitting in the chemically related molecule methyl formate (CH$_3$OCHO) that 
we measured with the IRAM Plateau de Bure interferometer 
\citep[see Table~5 of][]{Belloche08b}. 

From this analysis, we conclude that our best-fit model for 
the \textit{anti}-conformer of ethyl formate is fully consistent with our 30~m 
data of Sgr~B2(N). This detection of ethyl formate is, to our knowledge, the 
first one in space\footnote{\citet{Jones07} tentatively identified three lines 
detected with the Australia Telescope Compact Array at $\sim$86.2738,
$\sim$86.9784, and $\sim$86.9787~GHz as two transitions of the 
\textit{anti}-conformer of ethyl formate, the second one with two velocity 
components.
However, our model predicts a peak temperature of the ethyl formate 
transition at 86.977087~GHz on the order of 2~mK whereas the two lines detected 
with the 30~m telescope close to this frequency have peak temperatures of 0.38 
and 0.65~K, respectively! We identified these two lines with two velocity 
components of a transition of the vibrationally excited 
$\varv_{13}$=1/$\varv_{21}$=1 state of ethyl cyanide, and our modeled spectrum 
matches the observed lines very well. The tentative identification of 
\citet{Jones07} at this frequency is therefore not confirmed. On the other 
hand, the line detected at $\sim$86.2738~GHz in our survey is still
unidentified. The frequency of the $45_{5,41}$--$44_{6,38}$ transition of 
ethyl formate mentioned by \citet{Jones07} comes from the JPL catalog 
\citep[][, see http://spec.jpl.nasa.gov/]{Pickett98}. Our catalog contains a 
significantly different frequency for this transition 
($86256.5339 \pm 0.0114$~MHz 
instead of $86273.7945 \pm 0.2103$~MHz), and our model anyway predicts a very
low peak temperature on the order of 0.3 mK for this transition. Our catalog
contains two other overlapping transitions closer to 86.2738 GHz 
($35_{10,26}$--$36_{9,27}$ and $35_{10,25}$--$36_{9,28}$ at 86.2703101 and 
86.2703225~GHz, respectively). However, our model predicts a very low peak 
temperature of 0.6~mK for these transitions as well. Therefore, 
this tentative identification of \citet{Jones07} is not confirmed either.}.

No feature of the \textit{gauche}-conformer of ethyl formate is clearly 
detected in our spectral survey of Sgr~B2(N). Only one feature at 213.6~GHz is 
possibly detected, but the baseline in this frequency range is very uncertain 
and the feature is blended with a transition of H$^{13}$CCCN
(see Table \ref{t:etocho-g}). If we consider this feature as a detection, then 
it implies a column density a factor 2 smaller than for the 
\textit{anti}-conformer. This may suggest that the distribution of ethyl 
formate molecules in the two conformers is not in thermodynamical equilibrium. 
However, we first have to evaluate the uncertainty on the ratio of the 
\textit{anti}- and \textit{gauche}-conformer 
populations coming from the uncertainty on the energy difference between the
two conformers ($\Delta E = 65 \pm 21$~cm$^{-1}$, see 
Sect.~\ref{ss:freqetocho}). With $\Delta E = 0$, the ratio would be $1/2$. For 
the preferred energy difference of 65~cm$^{-1}$, we have a ratio of about 
$0.56/0.44$ at 100~K. If we assume an energy difference of 86~cm$^{-1}$ this 
ratio would change to $0.62/0.38$, i.e. a variation of $\sim 30\%$. This is 
not enough to compensate for the factor 2 mentioned above, but can have a 
significant contribution. In addition, a model of the emission spectrum of the 
\textit{gauche}-conformer with the same parameters as for the 
\textit{anti}-conformer is not 
excluded because of the large uncertainty on the baseline at 213.6~GHz. 
Therefore, given the large densities characterizing the hot core in Sgr~B2(N) 
\citep[see, e.g.,][]{Belloche08a,Belloche08b}, it seems unlikely that the 
population in the 
\textit{gauche}-conformer is subthermal compared to the \textit{anti}-conformer.

\subsection{Upper limit in Sgr~B2(M)}
\label{ss:b2metocho}

We do not detect ethyl formate in our spectral survey toward Sgr~B2(M). Using 
the same source size, linewidth, and temperature as for Srg~B2(N) (see 
Table~\ref{t:etochomodel}), we find $\sim 3 \sigma$ column density upper 
limits of $2.0 \times 10^{16}$~cm$^{-2}$ and $4.0 \times 10^{16}$~cm$^{-2}$ in 
the LTE approximation for the \textit{anti}- and \textit{gauche}-conformers, 
respectively. The 
column density of ethyl formate is thus at least a factor $\sim 3$ lower 
toward Sgr~B2(M) than toward Sgr~B2(N). This is not surprising since, e.g., 
\citet{Nummelin00} found that hot-core-type molecules are more abundant in 
Sgr~B2(N) by factors 3--8 as compared to Sgr~B2(M).

\subsection{Comparison to related species}
\label{ss:competocho}

We easily detect the already known molecules formic acid in the $trans$ form
($t$-HCOOH or $t$-HOCHO), methyl formate (CH$_3$OCHO), ethanol (C$_2$H$_5$OH), 
and dimethyl ether (CH$_3$OCH$_3$) in our survey toward 
Sgr~B2(N) \citep[see also, e.g.,][]{Nummelin00,Liu01}. The parameters of 
our current best fit models of these molecules are listed in 
Table~\ref{t:xochomodel}. All species have two 
velocity components that correspond to the two hot cores embedded in Sgr~B2(N) 
\citep[see, e.g.,][ for a discussion about these two sources]{Belloche08a}. 
Ethyl formate may have a second velocity component too, but our survey is not
sensitive enough to detect it with a significant signal-to-noise ratio. Using 
the same parameters as for the first velocity component but a velocity shift of 
10~km~s$^{-1}$, we estimate a $3\sigma$ upper limit of 
$\sim 2.4 \times 10^{16}$~cm$^{-2}$ for the column density of a second velocity 
component of ethyl formate.

The lines of formic acid are optically thin in our model, so the size of the 
emitting region 
cannot be measured with our single-dish data. It was here fixed to 5$\arcsec$, 
assuming that a more extended region would have a lower temperature. 
\citet{Nummelin00} derived a temperature of $74^{+82}_{-30}$~K and a 
beam-averaged column density of 
$\sim 4.2^{+2.0}_{-1.0} \times 10^{14}$~cm$^{-2}$ 
in the LTE approximation with the SEST telescope (\textit{HPBW} 
$\sim 23\arcsec$ at 1.3~mm). They used a linewidth of 13~km~s$^{-1}$ ($FWHM$), 
which more or less corresponds to the combination of the two velocity 
components we identified. Their column density translates 
into a column density of $\sim 9.3^{+4.4}_{-2.1} \times 10^{15}$~cm$^{-2}$ for 
a source size of 5$\arcsec$, i.e. about a factor 2 smaller than the sum of the 
column densities of both velocity components in Table~\ref{t:xochomodel}.
At least two reasons may explain this discrepancy. First of all, as we noticed 
in our own partial survey at 1.3~mm, the level of the baseline in this 
wavelength range is very uncertain for Sgr~B2(N) because of the line 
confusion and it may easily be overestimated. Second, at these high 
frequencies in Sgr~B2(N), the dust is partially optically thick and should 
partially absorb the line emission\footnote{\citet{Lis93} measured a peak flux 
of 20~Jy/$4.5\arcsec\times3.7\arcsec$-beam at 227~GHz toward Sgr~B2(N), i.e. 
28~K in temperature unit. For a temperature of $\sim 100$~K, this yields a 
dust optical depth of $\sim 0.34$. On larger scales ($\sim 10\arcsec$), 
\citet{Gordon93} estimated that the dust opacity toward Sgr~B2(N) reaches a 
value of 1 at 850~$\mu$m, which implies an opacity of $\sim$0.43--0.53 at 
1.3~mm. As a result, if not taken into account, these significant opacities 
imply an underestimate of the line intensities by a factor $\sim$ 1.4--1.7.}. 
We estimate that the combination of these two effects can lead to 
underestimating the true
line intensities by about a factor 2 or 3. In addition, assuming a temperature 
of 200~K, \citet{Liu01} measured a beam-averaged column density of 
$1.1 \pm 0.3 \times 10^{16}$~cm$^{-2}$ with the BIMA interferometer at 
86--90~GHz (\textit{HPBW} $\sim 14\arcsec \times 4\arcsec$). This translates 
into a column density of $\sim 6.3 \pm 1.5 \times 10^{15}$~cm$^{-2}$
for a source size of 5$\arcsec$ and a temperature of 70~K. The interferometric 
detection of \citet{Liu01} is somewhat uncertain but suggests that about half 
of the 30~m flux may be emitted by an extended region filtered out by the 
interferometer. The formic acid column density of the compact sources listed 
in Table~\ref{t:xochomodel} may therefore be overestimated by up to a factor 2.

\begin{table}
 {\centering
 \caption{
 Parameters of our best-fit LTE models of formic acid, methyl formate, ethanol, and dimethyl ether.
}
 \label{t:xochomodel}
 \vspace*{0.0ex}
 \begin{tabular}{lccccc}
 \hline\hline
 \multicolumn{1}{c}{Molecule$^{a}$} & \multicolumn{1}{c}{Size$^{b}$} & \multicolumn{1}{c}{$T_{\mathrm{rot}}$$^{c}$} & \multicolumn{1}{c}{$N^{d}$} & \multicolumn{1}{c}{$\Delta V^{e}$} & \multicolumn{1}{c}{$V_{\mathrm{off}}$$^{f}$} \\ 
  & \multicolumn{1}{c}{\scriptsize ($''$)} & \multicolumn{1}{c}{\scriptsize (K)} & \multicolumn{1}{c}{\scriptsize (cm$^{-2}$)} & \multicolumn{1}{c}{\scriptsize (km~s$^{-1}$)} & \multicolumn{1}{c}{\scriptsize (km~s$^{-1}$)} \\ 
 \multicolumn{1}{c}{(1)} & \multicolumn{1}{c}{(2)} & \multicolumn{1}{c}{(3)} & \multicolumn{1}{c}{(4)} & \multicolumn{1}{c}{(5)} & \multicolumn{1}{c}{(6)} \\ 
 \hline
\noalign{\smallskip} $t$-HCOOH & 5.0 &   70 & $ 1.50 \times 10^{16}$ & 8.0 & -1.0 \\  & 5.0 &   70 & $ 7.50 \times 10^{15}$ & 8.0 & 9.0 \\ \noalign{\smallskip} CH$_3$OCHO & 4.0 &   80 & $ 4.50 \times 10^{17}$ & 7.2 & 0.0 \\  & 4.0 &   80 & $ 1.50 \times 10^{17}$ & 7.2 & 10.0 \\ \noalign{\smallskip} C$_2$H$_5$OH & 3.0 &  100 & $ 8.40 \times 10^{17}$ & 8.0 & 0.0 \\  & 3.0 &  100 & $ 3.00 \times 10^{17}$ & 7.0 & 10.0 \\ \noalign{\smallskip} CH$_3$OCH$_3$ & 2.5 &  130 & $ 2.30 \times 10^{18}$ & 6.0 & 0.0 \\  & 2.5 &  130 & $ 1.10 \times 10^{18}$ & 6.0 & 10.0 \\  \hline
 \end{tabular}
 }\\[1ex] 
 Notes:
 $^a$ We used the CDMS entry for \textit{t}-HCOOH (version 1), and the JPL entries for CH$_3$OCHO (ver. 1), C$_2$H$_5$OH (ver. 4), and CH$_3$OCH$_3$ (ver. 1). See references to the laboratory data therein.
 $^b$ Source diameter (\textit{FWHM}).
 $^c$ Temperature.
 $^d$ Column density.
 $^e$ Linewidth (\textit{FWHM}).
 $^f$ Velocity offset with respect to the systemic velocity of Sgr~B2(N) V$_{\mathrm{lsr}} = 64$ km~s$^{-1}$.
 \end{table}

The lines of methyl formate have opacities of up to about 1 in our model of 
the 3~mm spectrum, which puts only weak constraints on the source size that we 
fixed to 4$\arcsec$. Assuming a temperature of 200~K, 
\citet{Nummelin00} derived a beam-averaged column density of 
$\sim 5.6^{+0.3}_{-0.1} \times 10^{15}$~cm$^{-2}$ with the SEST telescope for 
the $a$-type lines and, assuming a temperature of 500~K,
$\sim 4.0^{+0.3}_{-0.4} \times 10^{16}$~cm$^{-2}$ for the $b$-type lines.
For a temperature of 80~K and a source size of 4$\arcsec$, the column density
of the $a$-type lines translates into a column density of 
$\sim 1.2 \times 10^{17}$~cm$^{-2}$, which is about a factor 5 smaller than the 
one we derived here for the sum of the two velocity components. Again, the
uncertainty on the level of the baseline and the partial dust absorption at 
1.3~mm may explain part of this discrepancy. In addition, we note that our 
model at 3~mm reproduces quite well both the $a$- and $b$-type lines with the
same temperature and column density (see Appendix~\ref{a:meocho}, 
\textit{online material}), while \citet{Nummelin00} found an order
of magnitude difference between the column densities of the two types. We 
believe that this discrepancy results from the fact that they did not properly 
take into account the line blending, which is large in Sgr~B2(N) and should 
affect the (weak) $b$-type lines the most, and that they underestimated the 
line opacities of the (strong) $a$-type lines that our model predicts to be on 
the order of 1--3 in the 1.3~mm range.
Using the BIMA interferometer at 90.15~GHz with a beam size of 
$14\arcsec \times 4\arcsec$, \citet{Liu01} found a beam-averaged column 
density of $1.1 \times 10^{17}$~cm$^{-2}$ for an assumed temperature of 200~K 
in the optically thin approximation. This translates into a column density of 
$1.5 \times 10^{17}$~cm$^{-2}$ for a source size of 4$\arcsec$ and a 
temperature of 80~K. However, our model predicts an opacity of $\sim 0.6$ for
this transition, which implies a higher column density of 
$2.0 \times 10^{17}$~cm$^{-2}$. This is still about a factor 2 times lower than 
our estimate and suggests that, like in the case of formic acid, half of the 
single-dish flux may actually come from a region more extended than the size of 
our model and may be filtered out by the interferometer. This conclusion is 
further supported by the flux ratio of 1.7 between the 12~m telescope 
($HPBW = 71\arcsec$) and BIMA ($HPBW = 25.2\arcsec \times 6.3\arcsec$) 
measurements of \citet{Friedel04} at 86--90~GHz, and the flux ratio of 2.3 we
found between the measurements done with the 30~m telescope and the Plateau de 
Bure interferometer at 82.2~GHz \citep[see Table~5 of][]{Belloche08b}. As a 
result, the methyl formate column density of the compact sources listed in 
Table~\ref{t:xochomodel} may be overestimated by up to a factor 2.

Most lines of ethanol are optically thin at 3~mm ($\tau < 0.7$), except for 
three lines that are marginally optically thick ($\tau \sim 1 - 1.2$). As a
result, the source size is not well contrained and we fixed it to 3$\arcsec$. 
\citet{Nummelin00} derived a beam-averaged column density of 
$4.2 \pm 0.2 \times 10^{15}$~cm$^{-2}$ for a temperature of 73$^{+5}_{-4}$~K 
with the SEST telescope. This translates into a column
density of $2.5 \times 10^{17}$~cm$^{-2}$ for a source size of 3$\arcsec$, which 
is significantly lower than our measurement. However, \citet{Nummelin00} used
an earlier version of the JPL entry for ethanol that turned out to be 
inaccurate (J. Pearson, \textit{private communication}). With this older 
version, we determined column densities of $2.8 \times 10^{17}$ and 
$8.9 \times 10^{16}$~cm$^{-2}$ for both velocity components, which was 
consistent with the result of \citet{Nummelin00}. The column densities given 
in Table~\ref{t:xochomodel} were obtained with the latest JPL entry for 
ethanol \citep*[][]{Pearson08}. The high-energy lines 
($E_{\mathrm{l}}/k_{\mathrm{B}} \sim 40-80$~K) detected by \citet{Friedel04} with
the NRAO 12~m telescope and the BIMA interferometer have the same fluxes with
both instruments, implying that they are emitted by a compact region. Only the 
$4_{1,4}-3_{0.3}$ line with $E_{\mathrm{l}}/k_{\mathrm{B}} = 5.0$~K has an 
interferometric flux significantly lower than the single-dish flux. Our LTE 
model is also too weak for this transition compared to the spectrum obtained 
with the 30~m telescope. However, it fits well the low-energy transitions 
at 84.595868~GHz ($E_{\mathrm{l}}/k_{\mathrm{B}} = 9.4$~K) and 112.807174~GHz 
($E_{\mathrm{l}}/k_{\mathrm{B}} = 2.1$~K) detected in our survey. Therefore,
it is unclear whether the BIMA missing flux of the $4_{1,4}-3_{0.3}$ transition 
suggests an additional cold, extended component, or this line is heavily 
blended with a transition of another molecule.

Our model of dimethyl ether predicts line opacities up to 2. The size of the
emitting region is thus reasonably well constrained for this molecule. 
\citet{Nummelin00} derived a beam-averaged 
column density of $7.9^{+0.8}_{-0.7} \times 10^{15}$~cm$^{-2}$ for a temperature 
of 197$^{+31}_{-22}$~K with the SEST telescope. This translates into a column
density of $6.8 \times 10^{17}$~cm$^{-2}$ for a source size of 2.5$\arcsec$,
which is a factor 4 lower than derived here. The discrepancy most likely comes 
from the beam filling factor of unity assumed by \citet{Nummelin00} that leads 
to underestimating the line opacities. Our LTE model indeed predicts 
line optical depths up to 9 in the 1.3~mm window.

After rescaling to the same size of 3$\arcsec$, the relative column densities
of the three related molecules \hbox{$t$-HCOOH / CH$_3$OCHO / C$_2$H$_5$OCHO} 
are about \hbox{0.8 / 15 / 1} for the first velocity component, and 
\hbox{0.9 / 11 / 1} for the second velocity component using the upper limit 
found for ethyl formate. 
We discuss these ratios and the implications for the 
interstellar chemistry in Sect.~\ref{s:chemistry}.

\section{Identification of  \textit{n}-propyl cyanide}
\label{s:prcn}

\subsection{\textit{n}-Propyl cyanide frequencies}
\label{ss:freqprcn}

\textit{n}-Butanenitrile, C$_3$H$_7$CN, is more commonly known as 
\textit{n}-propyl cyanide or \textit{n}-butyronitrile. Its rotational spectrum 
has been investigated in the microwave \citep{Hirota62,Demaison82,Vormann88} 
and in the millimeter wave regions up to 284~GHz \citep{Wlodarczak88}. The 
\textit{n} indicates the \textit{normal} isomer with the carbon atoms forming 
a chain, in contrast to the \textit{iso} isomer which has a branched 
structure. This isomer has been studied to a lesser extent. 
However, its rotational spectrum is currently under investigation in Cologne. 

\textit{n}-Propyl cyanide exists in two conformers, \textit{anti} and 
\textit{gauche}, just as does ethyl formate. Again, the 
\textit{anti}-conformer is the lower energy form, is strongly prolate, and has 
a large $a$-dipole moment component of 3.60~D and a still sizable $b$-dipole 
moment component of 0.98~D. The \textit{gauche}-conformer is 
$1.1 \pm 0.3$~kJ\,mol$^{-1}$ or $92 \pm 25$~cm$^{-1}$ or $132 \pm 36$~K
higher in energy, more 
asymmetric, and has $\mu_a = 3.27$ and $\mu_b = 2.14$~D \citep{Wlodarczak88}. 
The energy difference has been estimated at room temperature and at 233~K 
from relative intensities in the ground state spectra. Since excited 
vibrational states have not been taken into consideration the error in the 
energy difference may well be slightly larger than mentioned above.
The residuals quoted in the most recent study \citep{Vormann88} for their 
measurements are frequently much larger than the suggested uncertainties of 
about 5~kHz suggesting an insufficient set of spectroscopic parameters 
was used. Moreover, only newly determined rotational and centrifugal distortion 
parameters were given for the \textit{gauche}-conformer. Therefore, new sets of 
rotational and centrifugal distortion parameters were determined for both 
conformers in the present study. 

In the initial fits transition frequencies were taken from all four studies 
\citep[][]{Hirota62,Demaison82,Wlodarczak88,Vormann88}. Two $b$-type 
transitions 
from \citet{Wlodarczak88} were omitted from the fits as suggested in the 
erratum to this paper \citep{Wlodarczak91}. On the other hand, transition 
frequencies not given in \citet{Vormann88}, but deposited at the library 
of the University of Kiel were obtained from there and included in the fits.
Uncertainties of 200, 10, 50, and 5~kHz were assigned to the transitions 
from \citet{Hirota62}, \citet{Demaison82}, \citet{Wlodarczak88}, and 
\citet{Vormann88}, respectively. \citet{Demaison82} and \citet{Vormann88} 
resolved in part internal rotation of the 
methyl group or quadrupole splitting of the $^{14}$N nucleus in their 
laboratory measurements. The methyl internal rotation is unlikely to be 
resolved in astronomical observations. The quadrupole splitting may be 
resolvable for some low energy transitions, but these will be generally too 
weak. Therefore, only the unsplit frequencies were used from these two studies. 
In the unlikely event of detecting \textit{n}-propyl cyanide in cold sources, 
quadrupole parameters published in \citet{Vormann88} would be adequate. 

There were comparatively few transitions reported in \citet{Hirota62}, and 
their uncertainties were fairly large. Trial fits with these transitions 
omitted from the fits caused essentially no change in the values and in the 
uncertainties of the spectroscopic parameters. Therefore, these transitions 
were omitted from the final fits. 
Two transitions, $36_{1,36} - 35_{0,35}$ of the \textit{anti}-conformer and 
$31_{5,27} - 30_{5,26}$ of the \textit{gauche}-conformer, had residuals 
between observed and calculated frequencies larger than four times the 
experimental uncertainties. Therefore, these transitions were omitted from 
the data sets. The final line list for the \textit{anti}-conformer contained 
4, 93, and 50 different transition frequencies from \citet{Demaison82},
\citet{Wlodarczak88}, and \citet{Vormann88}, respectively. The total 
number of transitions 
is larger by 62 because of unresolved asymmetry splitting. 
The corresponding numbers of different transition frequencies for the 
\textit{gauche}-conformer are 4, 119, and 46. Unresolved asymmetry splitting 
causes the total number of transitions to be larger by 46. The final line lists
for both conformers are given in Tables~\ref{t:lines-anti-PrCN} and 
\ref{t:lines-gauche-PrCN} (\textit{online material}).

The asymmetry parameter $\kappa = (2B - A - C)/(A - C)$ is $-0.9893$ for 
\textit{anti}-\textit{n}-propyl cyanide, rather close to the symmetric prolate 
limit of --1. In such cases it is advisable to avoid using Watson's 
$A$-reduction and use the $S$-reduction instead. In the case of the 
\textit{gauche}-conformer one finds $\kappa = -0.8471$. In this case both 
reductions may be used. In the present work the $S$-reduction was used 
throughout for consistency reason. The sextic distortion parameter $H_K$ 
of the \textit{anti}-conformer was initially estimated to be smaller than 
$D_K$ by the same factor that that parameter is smaller than $A$. 
This is certainly only a crude estimate. Trial fits with $H_K$ released 
suggested its value to be slightly larger than this estimate. 
But since the uncertainty was more than a third of its value and since 
the difference was smaller than the uncertainty, $H_K$ was finally fixed 
to the estimated value. The final spectroscopic parameters are given 
in Table~\ref{PrCN_parameters}. Overall, the transition frequencies 
have been reproduced within experimental uncertainties as the dimensionless
rms errors are 0.75 and 0.66 for the \textit{anti} and 
\textit{gauche}-conformer, 
respectively. The values for the individual data sets do not differ 
very much from these values. Moreover, this is reasonably close to 1.0 and 
suggests the ascribed uncertainties are quite appropriate.

\addtocounter{table}{1}
\newcounter{aprcn}
\setcounter{aprcn}{\value{table}}

\addtocounter{table}{1}
\newcounter{gprcn}
\setcounter{gprcn}{\value{table}}


\begin{table}
\caption{Spectroscopic parameters$^a$ (MHz) of \textit{n}-propyl cyanide.}
\label{PrCN_parameters}
\smallskip
\begin{tabular}{lr@{}lr@{}l}
\hline \hline
Parameter & \multicolumn{2}{c}{\textit{anti}} & \multicolumn{2}{c}{\textit{gauche}}  \\
\hline
$A$                        & 23\,668&.319\,31\,(143)  &  10\,060&.416\,49\,(108)  \\
$B$                        &  2\,268&.146\,892\,(147) &   3\,267&.662\,408\,(301) \\
$C$                        &  2\,152&.963\,946\,(168) &   2\,705&.459\,572\,(290) \\
$D_K \times 10^3$          &     240&.653\,(29)       &       60&.235\,(6)        \\
$D_{JK} \times 10^3$       &   $-$10&.826\,31\,(92)   &    $-$18&.264\,70\,(117)  \\
$D_J \times 10^6$          &     398&.674\,(69)       &   3\,195&.064\,(207)      \\
$d_1 \times 10^6$          &   $-$46&.637\,(42)       &$-$1\,037&.470\,(55)       \\
$d_2 \times 10^6$          &    $-$0&.590\,1\,(59)    &    $-$77&.186\,4\,(183)   \\
$H_K \times 10^6$          &       2&.5               &        1&.806\,(18)       \\
$H_{KJ} \times 10^9$       &     372&.4\,(24)         &   $-$517&.3\,(35)         \\
$H_{JK} \times 10^9$       &   $-$20&.67\,(20)        &        9&.92\,(68)        \\
$H_J \times 10^9$          &       0&.353\,(11)       &        4&.486\,(56)       \\
$h_1 \times 10^9$          &       0&.117\,(14)       &        2&.505\,(29)       \\
$h_2 \times 10^{12}$       &        &$-$              &      524&.6\,(135)        \\
$h_3 \times 10^{12}$       &        &$-$              &      111&.3\,(31)         \\
$L_{KKJ} \times 10^{12}$   &        &$-$              &       30&.6\,(31)         \\
$L_{JK} \times 10^{12}$    &        &$-$              &     $-$4&.11\,(78)        \\
\hline
\end{tabular}\\[2pt]
$^a$ Watson's $S$-reduction was used in the representation $I^r$. Numbers in 
parentheses are one standard deviation in units of the least significant 
figures. Parameter values with no uncertainties given were estimated. 
A long dash indicates parameters that are determinable in theory, but could
not be determined with significance here.\\
\end{table}


The \textit{gauche}-conformer is considerably more asymmetric than the 
\textit{anti}-conformer. Therefore, it is probably not surprising that 
the distortion parameters describing the asymmetry splitting (the 
off-diagonal $d_i$ and the $h_i$) are not only larger in magnitude for the 
former conformer, but also more of these terms are required in the fits. 
In addition, two octic centrifugal distortion parameters $L$ were needed 
in the fit of the \textit{gauche}-conformer resulting in an overall 
much larger parameter set and thus a much slower converging Hamiltonian 
compared with the \textit{anti}-conformer. 
A similar situation occured in the recent investigation of ethyl formate 
\citep{Medvedev09} where also a much larger set of spectroscopic parameters 
was needed to fit the data of the \textit{gauche}-conformer compared to the 
\textit{anti}-conformer. 

The predictions used for the current analysis will be made available in the 
CDMS \citep[][, see footnote \ref{fn:cdms2}]{Mueller01,Mueller05}. 
The partition 
function of \textit{n}-propyl cyanide is $5.608 \times 10^4$ at 150~K.
In the course of the analysis, the two conformers again have been treated 
separately on occasion to evaluate if the abundance of either conformer is 
lower than would be expected under LTE conditions.

\subsection{Detection of \textit{n}-propyl cyanide in Sgr~B2(N)}
\label{ss:detprcn}

\addtocounter{table}{1}
\newcounter{apptable3}
\setcounter{apptable3}{\value{table}}
\onltab{\value{apptable3}}{\clearpage
\begin{table*}
 {\centering
 \caption{
 Transitions of the \textit{anti}-conformer of \textit{n}-propyl cyanide observed with the IRAM 30 m telescope toward Sgr~B2(N).
The horizontal lines mark discontinuities in the observed frequency coverage.
 Only the transitions associated with a modeled line stronger than 20 mK are listed.
 }
 \label{t:prcn-a}
 \vspace*{0.0ex}
\begin{tabular}{rlrrrrrl}
 \hline\hline
 \multicolumn{1}{c}{$N^a$} & \multicolumn{1}{c}{Transition$^b$} & \multicolumn{1}{c}{Frequency} & \multicolumn{1}{c}{Unc.$^c$} & \multicolumn{1}{c}{$E_\mathrm{l}$$^d$} & \multicolumn{1}{c}{$S\mu^2$} &  \multicolumn{1}{c}{$\sigma^e$} &  \multicolumn{1}{c}{Comments} \\ 
  & & \multicolumn{1}{c}{\scriptsize (MHz)} & \multicolumn{1}{c}{\scriptsize (kHz)} & \multicolumn{1}{c}{\scriptsize (K)} & \multicolumn{1}{c}{\scriptsize (D$^2$)} &  \multicolumn{1}{c}{\scriptsize (mK)} &  \\ 
 \multicolumn{1}{c}{(1)} & \multicolumn{1}{c}{(2)} & \multicolumn{1}{c}{(3)} & \multicolumn{1}{c}{(4)} & \multicolumn{1}{c}{(5)} & \multicolumn{1}{c}{(6)} & \multicolumn{1}{c}{(7)} & \multicolumn{1}{c}{(8)} \\ 
 \hline
   1 & 18$_{ 1,17}$ -- 17$_{ 1,16}$ &   80486.371 &     5 &   34 &       232.1 &   24 & Blend with C$_2$H$_5$CN, $\varv_{13}$=1/$\varv_{21}$=1 and U-line \\ 
   2 & 19$_{ 1,19}$ -- 18$_{ 1,18}$ &   82778.264 &     4 &   37 &       245.1 &   17 & Blend with HC$_3$N, $\varv_5$=1/$\varv_7$=3 and C$_2$H$_5$CN \\ 
   3 & 19$_{ 0,19}$ -- 18$_{ 0,18}$ &   83487.658 &     4 &   36 &       245.7 &   17 & Blend with C$_2$H$_3$CN and U-line \\ 
   4 & 19$_{ 2,18}$ -- 18$_{ 2,17}$ &   83906.011 &     5 &   40 &       243.1 &   16 & Weak, blend with U-lines \\ 
   5 & 19$_{ 6,14}$ -- 18$_{ 6,13}$$^\star$ &   84021.555 &     4 &   73 &       221.3 &   19 & \textbf{Group detected}, partial blend with C$_2$H$_5$CN \\
\hline
 \end{tabular}
 }\\[1ex] 
 \end{table*}
\begin{table*}
 {\centering
 \addtocounter{table}{-1}
 \caption{
continued.
 }
 \begin{tabular}{rlrrrrrl}
 \hline\hline
 \multicolumn{1}{c}{$N^a$} & \multicolumn{1}{c}{Transition$^b$} & \multicolumn{1}{c}{Frequency} & \multicolumn{1}{c}{Unc.$^c$} & \multicolumn{1}{c}{$E_\mathrm{l}$$^d$} & \multicolumn{1}{c}{$S\mu^2$} &  \multicolumn{1}{c}{$\sigma^e$} &  \multicolumn{1}{c}{Comments} \\ 
  & & \multicolumn{1}{c}{\scriptsize (MHz)} & \multicolumn{1}{c}{\scriptsize (kHz)} & \multicolumn{1}{c}{\scriptsize (K)} & \multicolumn{1}{c}{\scriptsize (D$^2$)} &  \multicolumn{1}{c}{\scriptsize (mK)} &  \\ 
 \multicolumn{1}{c}{(1)} & \multicolumn{1}{c}{(2)} & \multicolumn{1}{c}{(3)} & \multicolumn{1}{c}{(4)} & \multicolumn{1}{c}{(5)} & \multicolumn{1}{c}{(6)} & \multicolumn{1}{c}{(7)} & \multicolumn{1}{c}{(8)} \\ 
 \hline
  81 & 22$_{ 2,21}$ -- 21$_{ 2,20}$ &   97115.607 &     5 &   53 &       282.3 &   21 & Blend with U-line \\ 
  82 & 22$_{ 7,15}$ -- 21$_{ 7,14}$$^\star$ &   97289.589 &     5 &   99 &       255.8 &   21 & Blend with C$_2$H$_3$CN and OCS \\ 
  84 & 22$_{ 6,17}$ -- 21$_{ 6,16}$$^\star$ &   97289.742 &     5 &   86 &       263.5 &   21 & Blend with C$_2$H$_3$CN and OCS \\ 
  86 & 22$_{ 8,14}$ -- 21$_{ 8,13}$$^\star$ &   97292.679 &     5 &  115 &       247.0 &   21 & Blend with OCS \\ 
  88 & 22$_{ 5,18}$ -- 21$_{ 5,17}$$^\star$ &   97295.253 &     5 &   75 &       269.9 &   21 & Strong OCS \\
\hline
 \end{tabular}
 }\\[1ex] 
 \end{table*}
\begin{table*}
 {\centering
 \addtocounter{table}{-1}
 \caption{
continued.
 }
 \begin{tabular}{rlrrrrrl}
 \hline\hline
 \multicolumn{1}{c}{$N^a$} & \multicolumn{1}{c}{Transition$^b$} & \multicolumn{1}{c}{Frequency} & \multicolumn{1}{c}{Unc.$^c$} & \multicolumn{1}{c}{$E_\mathrm{l}$$^d$} & \multicolumn{1}{c}{$S\mu^2$} &  \multicolumn{1}{c}{$\sigma^e$} &  \multicolumn{1}{c}{Comments} \\ 
  & & \multicolumn{1}{c}{\scriptsize (MHz)} & \multicolumn{1}{c}{\scriptsize (kHz)} & \multicolumn{1}{c}{\scriptsize (K)} & \multicolumn{1}{c}{\scriptsize (D$^2$)} &  \multicolumn{1}{c}{\scriptsize (mK)} &  \\ 
 \multicolumn{1}{c}{(1)} & \multicolumn{1}{c}{(2)} & \multicolumn{1}{c}{(3)} & \multicolumn{1}{c}{(4)} & \multicolumn{1}{c}{(5)} & \multicolumn{1}{c}{(6)} & \multicolumn{1}{c}{(7)} & \multicolumn{1}{c}{(8)} \\ 
 \hline
 173 & 25$_{ 7,18}$ -- 24$_{ 7,17}$$^\star$ &  110556.359 &     5 &  114 &       298.1 &   32 & Blend with U-lines and CH$_3$OCHO \\ 
 175 & 25$_{ 8,17}$ -- 24$_{ 8,16}$$^\star$ &  110558.521 &     5 &  130 &       290.3 &   32 & Blend with U-lines and CH$_3$OCHO \\ 
 177 & 25$_{ 6,20}$ -- 24$_{ 6,19}$$^\star$ &  110558.636 &     5 &  101 &       304.8 &   32 & Blend with U-lines and CH$_3$OCHO \\ 
 179 & 25$_{ 9,16}$ -- 24$_{ 9,15}$$^\star$ &  110563.690 &     5 &  147 &       281.5 &   32 & Blend with U-line \\ 
 181 & 25$_{ 5,21}$ -- 24$_{ 5,20}$$^\star$ &  110568.450 &     5 &   89 &       310.5 &   32 & Blend with CH$_3$OCHO, $\varv_{\mathrm{t}}$=1 \\
\hline
 \end{tabular}
 }\\[1ex] 
 \end{table*}
\begin{table*}
 {\centering
 \addtocounter{table}{-1}
 \caption{
continued.
 }
 \begin{tabular}{rlrrrrrl}
 \hline\hline
 \multicolumn{1}{c}{$N^a$} & \multicolumn{1}{c}{Transition$^b$} & \multicolumn{1}{c}{Frequency} & \multicolumn{1}{c}{Unc.$^c$} & \multicolumn{1}{c}{$E_\mathrm{l}$$^d$} & \multicolumn{1}{c}{$S\mu^2$} &  \multicolumn{1}{c}{$\sigma^e$} &  \multicolumn{1}{c}{Comments} \\ 
  & & \multicolumn{1}{c}{\scriptsize (MHz)} & \multicolumn{1}{c}{\scriptsize (kHz)} & \multicolumn{1}{c}{\scriptsize (K)} & \multicolumn{1}{c}{\scriptsize (D$^2$)} &  \multicolumn{1}{c}{\scriptsize (mK)} &  \\ 
 \multicolumn{1}{c}{(1)} & \multicolumn{1}{c}{(2)} & \multicolumn{1}{c}{(3)} & \multicolumn{1}{c}{(4)} & \multicolumn{1}{c}{(5)} & \multicolumn{1}{c}{(6)} & \multicolumn{1}{c}{(7)} & \multicolumn{1}{c}{(8)} \\ 
 \hline
 275 & 39$_{11,28}$ -- 38$_{11,27}$$^\star$ &  172471.720 &     7 &  282 &       464.4 &   44 & Strong CH$_3$OCHO, $\varv_{\mathrm{t}}$=1 and C$_2$H$_3$CN \\ 
 277 & 39$_{12,27}$ -- 38$_{12,26}$$^\star$ &  172485.116 &     8 &  305 &       456.8 &   44 & Blend with C$_2$H$_5$CN \\ 
 279 & 39$_{ 6,34}$ -- 38$_{ 6,33}$$^\star$ &  172492.223 &     5 &  194 &       492.7 &   44 & Blend with C$_2$H$_5$CN \\ 
 281 & 39$_{13,26}$ -- 38$_{13,25}$$^\star$ &  172501.606 &     8 &  331 &       448.5 &   44 & Blend with C$_2$H$_5$CN, $\varv_{13}$=1/$\varv_{21}$=1 \\ 
 283 & 52$_{ 2,50}$ -- 51$_{ 3,49}$ &  172516.550 &    77 &  291 &        14.0 &   44 & Blend with C$_2$H$_5$CN, $\varv_{13}$=1/$\varv_{21}$=1 and CH$_3$OCHO, $\varv_{\mathrm{t}}$=1 \\
\hline
 \end{tabular}
 }\\[1ex] 
 \end{table*}
\begin{table*}
 {\centering
 \addtocounter{table}{-1}
 \caption{
continued.
 }
 \begin{tabular}{rlrrrrrl}
 \hline\hline
 \multicolumn{1}{c}{$N^a$} & \multicolumn{1}{c}{Transition$^b$} & \multicolumn{1}{c}{Frequency} & \multicolumn{1}{c}{Unc.$^c$} & \multicolumn{1}{c}{$E_\mathrm{l}$$^d$} & \multicolumn{1}{c}{$S\mu^2$} &  \multicolumn{1}{c}{$\sigma^e$} &  \multicolumn{1}{c}{Comments} \\ 
  & & \multicolumn{1}{c}{\scriptsize (MHz)} & \multicolumn{1}{c}{\scriptsize (kHz)} & \multicolumn{1}{c}{\scriptsize (K)} & \multicolumn{1}{c}{\scriptsize (D$^2$)} &  \multicolumn{1}{c}{\scriptsize (mK)} &  \\ 
 \multicolumn{1}{c}{(1)} & \multicolumn{1}{c}{(2)} & \multicolumn{1}{c}{(3)} & \multicolumn{1}{c}{(4)} & \multicolumn{1}{c}{(5)} & \multicolumn{1}{c}{(6)} & \multicolumn{1}{c}{(7)} & \multicolumn{1}{c}{(8)} \\ 
 \hline
 377 & 15$_{ 7, 8}$ -- 16$_{ 6,11}$$^\star$ &  207955.703 &    25 &   66 &         1.4 &  173 & Blend with C$_2$H$_5$CN and CH$_2$$^{13}$CHCN \\ 
 379 & 47$_{ 5,43}$ -- 46$_{ 5,42}$ &  207983.470 &     6 &  255 &       601.2 &  173 & Blend with C$_2$H$_5$OH and CH$_2$$^{13}$CHCN \\ 
 380 & 47$_{18,29}$ -- 46$_{18,28}$$^\star$ &  207985.767 &    10 &  562 &       518.9 &  173 & Blend with C$_2$H$_5$OH and CH$_2$$^{13}$CHCN \\ 
 382 & 47$_{ 5,42}$ -- 46$_{ 5,41}$ &  207996.582 &     6 &  255 &       601.2 &  173 & Blend with C$_2$H$_5$CN, C$_2$H$_5$OH, and CH$_2$CH$^{13}$CN \\ 
 383 & 47$_{19,28}$ -- 46$_{19,27}$$^\star$ &  208021.478 &    11 &  600 &       508.7 &  173 & Blend with CH$_2$$^{13}$CHCN \\
\hline
 \end{tabular}
 }\\[1ex] 
 \end{table*}
\begin{table*}
 {\centering
 \addtocounter{table}{-1}
 \caption{
continued.
 }
 \begin{tabular}{rlrrrrrl}
 \hline\hline
 \multicolumn{1}{c}{$N^a$} & \multicolumn{1}{c}{Transition$^b$} & \multicolumn{1}{c}{Frequency} & \multicolumn{1}{c}{Unc.$^c$} & \multicolumn{1}{c}{$E_\mathrm{l}$$^d$} & \multicolumn{1}{c}{$S\mu^2$} &  \multicolumn{1}{c}{$\sigma^e$} &  \multicolumn{1}{c}{Comments} \\ 
  & & \multicolumn{1}{c}{\scriptsize (MHz)} & \multicolumn{1}{c}{\scriptsize (kHz)} & \multicolumn{1}{c}{\scriptsize (K)} & \multicolumn{1}{c}{\scriptsize (D$^2$)} &  \multicolumn{1}{c}{\scriptsize (mK)} &  \\ 
 \multicolumn{1}{c}{(1)} & \multicolumn{1}{c}{(2)} & \multicolumn{1}{c}{(3)} & \multicolumn{1}{c}{(4)} & \multicolumn{1}{c}{(5)} & \multicolumn{1}{c}{(6)} & \multicolumn{1}{c}{(7)} & \multicolumn{1}{c}{(8)} \\ 
 \hline
 473 & 49$_{17,32}$ -- 48$_{17,31}$$^\star$ &  216789.070 &    50 &  546 &       557.7 &   50 & Blend with CH$_3$CH$_3$CO, $\varv_t$=1 \\ 
 475 & 13$_{ 7, 6}$ -- 14$_{ 6, 9}$$^\star$ &  216795.931 &    26 &   59 &         1.0 &   50 & Blend with CH$_3$CH$_3$CO, $\varv_t$=1 \\ 
 477 & 49$_{18,31}$ -- 48$_{18,30}$$^\star$ &  216823.235 &    10 &  582 &       548.3 &   50 & Strong CH$_3$OCHO \\ 
 479 & 49$_{ 5,45}$ -- 48$_{ 5,44}$ &  216847.000 &     7 &  275 &       627.3 &   50 & Strong C$_2$H$_5$CN, $\varv_{13}$=1/$\varv_{21}$=1 \\ 
 480 & 49$_{19,30}$ -- 48$_{19,29}$$^\star$ &  216860.211 &    11 &  620 &       538.7 &   50 & Blend with C$_2$H$_5$CN, $\varv_{13}$=1/$\varv_{21}$=1 and U-line \\
\hline
 \end{tabular}
 }\\[1ex] 
 \end{table*}
\begin{table*}
 {\centering
 \addtocounter{table}{-1}
 \caption{
continued.
 }
 \begin{tabular}{rlrrrrrl}
 \hline\hline
 \multicolumn{1}{c}{$N^a$} & \multicolumn{1}{c}{Transition$^b$} & \multicolumn{1}{c}{Frequency} & \multicolumn{1}{c}{Unc.$^c$} & \multicolumn{1}{c}{$E_\mathrm{l}$$^d$} & \multicolumn{1}{c}{$S\mu^2$} &  \multicolumn{1}{c}{$\sigma^e$} &  \multicolumn{1}{c}{Comments} \\ 
  & & \multicolumn{1}{c}{\scriptsize (MHz)} & \multicolumn{1}{c}{\scriptsize (kHz)} & \multicolumn{1}{c}{\scriptsize (K)} & \multicolumn{1}{c}{\scriptsize (D$^2$)} &  \multicolumn{1}{c}{\scriptsize (mK)} &  \\ 
 \multicolumn{1}{c}{(1)} & \multicolumn{1}{c}{(2)} & \multicolumn{1}{c}{(3)} & \multicolumn{1}{c}{(4)} & \multicolumn{1}{c}{(5)} & \multicolumn{1}{c}{(6)} & \multicolumn{1}{c}{(7)} & \multicolumn{1}{c}{(8)} \\ 
 \hline
 553 & 19$_{ 6,13}$ -- 19$_{ 5,14}$$^\star$ &  235882.182 &    13 &   66 &         8.8 &  131 & Strong $^{13}$CH$_3$OH, CH$_3$OCHO, and HC$^{13}$CCN, $\varv_6$=1 \\ 
 555 & 10$_{ 6, 4}$ -- 10$_{ 5, 5}$$^\star$ &  235883.423 &    15 &   37 &         3.7 &  131 & Strong $^{13}$CH$_3$OH, CH$_3$OCHO, and HC$^{13}$CCN, $\varv_6$=1 \\ 
 557 & 18$_{ 6,12}$ -- 18$_{ 5,13}$$^\star$ &  235884.587 &    13 &   62 &         8.2 &  131 & Strong $^{13}$CH$_3$OH, CH$_3$OCHO, and HC$^{13}$CCN, $\varv_6$=1 \\ 
 559 & 11$_{ 6, 5}$ -- 11$_{ 5, 6}$$^\star$ &  235884.702 &    15 &   40 &         4.3 &  131 & Strong $^{13}$CH$_3$OH, CH$_3$OCHO, and HC$^{13}$CCN, $\varv_6$=1 \\ 
 561 & 12$_{ 6, 6}$ -- 12$_{ 5, 7}$$^\star$ &  235885.818 &    15 &   42 &         4.9 &  131 & Strong $^{13}$CH$_3$OH, CH$_3$OCHO, and HC$^{13}$CCN, $\varv_6$=1 \\
\hline
 \end{tabular}
 }\\[1ex] 
 Notes:
 $^a$ Numbering of the observed transitions associated with a modeled line stronger than 20 mK.
 $^b$ Transitions marked with a $^\star$ are double with a frequency difference less than 0.1 MHz. The quantum numbers of the second one are not shown.
 $^c$ Frequency uncertainty.
 $^d$ Lower energy level in temperature units ($E_\mathrm{l}/k_\mathrm{B}$).
 $^e$ Calculated rms noise level in $T_{\mathrm{mb}}$ scale.
 \end{table*}

}

\addtocounter{table}{1}
\newcounter{apptable4}
\setcounter{apptable4}{\value{table}}
\onltab{\value{apptable4}}{\clearpage
\begin{table*}
 {\centering
 \caption{
 Transitions of the \textit{gauche}-conformer of \textit{n}-propyl cyanide observed with the IRAM 30 m telescope toward Sgr~B2(N).
The horizontal lines mark discontinuities in the observed frequency coverage.
 Only the transitions associated with a modeled line stronger than 20 mK are listed.
 }
 \label{t:prcn-g}
 \vspace*{0.0ex}
\begin{tabular}{rlrrrrrl}
 \hline\hline
 \multicolumn{1}{c}{$N^a$} & \multicolumn{1}{c}{Transition$^b$} & \multicolumn{1}{c}{Frequency} & \multicolumn{1}{c}{Unc.$^c$} & \multicolumn{1}{c}{$E_\mathrm{l}$$^d$} & \multicolumn{1}{c}{$S\mu^2$} &  \multicolumn{1}{c}{$\sigma^e$} &  \multicolumn{1}{c}{Comments} \\ 
  & & \multicolumn{1}{c}{\scriptsize (MHz)} & \multicolumn{1}{c}{\scriptsize (kHz)} & \multicolumn{1}{c}{\scriptsize (K)} & \multicolumn{1}{c}{\scriptsize (D$^2$)} &  \multicolumn{1}{c}{\scriptsize (mK)} &  \\ 
 \multicolumn{1}{c}{(1)} & \multicolumn{1}{c}{(2)} & \multicolumn{1}{c}{(3)} & \multicolumn{1}{c}{(4)} & \multicolumn{1}{c}{(5)} & \multicolumn{1}{c}{(6)} & \multicolumn{1}{c}{(7)} & \multicolumn{1}{c}{(8)} \\ 
 \hline
   1 & 17$_{ 9, 9}$ -- 16$_{ 9, 8}$$^\star$ &  101965.911 &     6 &  199 &       262.1 &   34 & Blend with U-line \\ 
   3 & 17$_{ 8,10}$ -- 16$_{ 8, 9}$$^\star$ &  102048.097 &     6 &  193 &       283.5 &   34 & Blend with NH$_2$CHO, $\varv_{12}$=1 \\ 
   5 & 29$_{ 2,27}$ -- 29$_{ 1,28}$ &  102048.362 &     6 &  253 &        66.0 &   34 & Blend with NH$_2$CHO, $\varv_{12}$=1 \\ 
   6 & 34$_{ 5,30}$ -- 34$_{ 3,31}$ &  102175.037 &     5 &  308 &        17.0 &   30 & Blend with C$_2$H$_5$CN, $\varv_{13}$=1/$\varv_{21}$=1, $^{13}$CH$_2$CHCN, and  \\ 
 & & & & & & & CH$_3$OCHO, $\varv_{\mathrm{t}}$=1 \\
\hline
 \end{tabular}
 }\\[1ex] 
 \end{table*}
\begin{table*}
 {\centering
 \addtocounter{table}{-1}
 \caption{
continued.
 }
 \begin{tabular}{rlrrrrrl}
 \hline\hline
 \multicolumn{1}{c}{$N^a$} & \multicolumn{1}{c}{Transition$^b$} & \multicolumn{1}{c}{Frequency} & \multicolumn{1}{c}{Unc.$^c$} & \multicolumn{1}{c}{$E_\mathrm{l}$$^d$} & \multicolumn{1}{c}{$S\mu^2$} &  \multicolumn{1}{c}{$\sigma^e$} &  \multicolumn{1}{c}{Comments} \\ 
  & & \multicolumn{1}{c}{\scriptsize (MHz)} & \multicolumn{1}{c}{\scriptsize (kHz)} & \multicolumn{1}{c}{\scriptsize (K)} & \multicolumn{1}{c}{\scriptsize (D$^2$)} &  \multicolumn{1}{c}{\scriptsize (mK)} &  \\ 
 \multicolumn{1}{c}{(1)} & \multicolumn{1}{c}{(2)} & \multicolumn{1}{c}{(3)} & \multicolumn{1}{c}{(4)} & \multicolumn{1}{c}{(5)} & \multicolumn{1}{c}{(6)} & \multicolumn{1}{c}{(7)} & \multicolumn{1}{c}{(8)} \\ 
 \hline
  76 & 34$_{ 5,30}$ -- 33$_{ 5,29}$ &  202090.888 &     6 &  303 &       707.1 &  138 & Strong CH$_2$CO \\ 
  77 & 56$_{ 3,53}$ -- 56$_{ 3,54}$$^\star$ &  202474.090 &   224 &  572 &        15.7 &  108 & Blend with U-line and H$_2$CS \\ 
  79 & 56$_{ 4,53}$ -- 56$_{ 3,54}$$^\star$ &  202474.109 &   224 &  572 &       100.1 &  108 & Blend with U-line and H$_2$CS \\ 
  81 & 41$_{15,26}$ -- 41$_{14,27}$$^\star$ &  202477.934 &    26 &  447 &       185.2 &  108 & Blend with U-line and H$_2$CS \\ 
  83 & 40$_{15,25}$ -- 40$_{14,26}$$^\star$ &  202689.659 &    25 &  435 &       178.9 &  108 & Blend with C$_2$H$_3$CN, $\varv_{15}$=1 and NH$_2$CHO \\
\hline
 \end{tabular}
 }\\[1ex] 
 \end{table*}
\begin{table*}
 {\centering
 \addtocounter{table}{-1}
 \caption{
continued.
 }
 \begin{tabular}{rlrrrrrl}
 \hline\hline
 \multicolumn{1}{c}{$N^a$} & \multicolumn{1}{c}{Transition$^b$} & \multicolumn{1}{c}{Frequency} & \multicolumn{1}{c}{Unc.$^c$} & \multicolumn{1}{c}{$E_\mathrm{l}$$^d$} & \multicolumn{1}{c}{$S\mu^2$} &  \multicolumn{1}{c}{$\sigma^e$} &  \multicolumn{1}{c}{Comments} \\ 
  & & \multicolumn{1}{c}{\scriptsize (MHz)} & \multicolumn{1}{c}{\scriptsize (kHz)} & \multicolumn{1}{c}{\scriptsize (K)} & \multicolumn{1}{c}{\scriptsize (D$^2$)} &  \multicolumn{1}{c}{\scriptsize (mK)} &  \\ 
 \multicolumn{1}{c}{(1)} & \multicolumn{1}{c}{(2)} & \multicolumn{1}{c}{(3)} & \multicolumn{1}{c}{(4)} & \multicolumn{1}{c}{(5)} & \multicolumn{1}{c}{(6)} & \multicolumn{1}{c}{(7)} & \multicolumn{1}{c}{(8)} \\ 
 \hline
 191 & 34$_{12,23}$ -- 33$_{12,22}$$^\star$ &  204453.939 &     8 &  343 &       637.5 &  316 & Blend with CH$_3$CH$_3$CO and U-line \\ 
\hline
 193 & 34$_{10,25}$ -- 33$_{10,24}$ &  205107.534 &     6 &  328 &       665.3 &  100 & Strong CH$_3$OCH$_3$, C$_2$H$_5$$^{13}$CN, and CH$_3$$^{13}$CH$_2$CN \\ 
 194 & 34$_{10,24}$ -- 33$_{10,23}$ &  205113.060 &     6 &  328 &       665.2 &  100 & Strong CH$_3$OCH$_3$, C$_2$H$_5$$^{13}$CN, and CH$_3$$^{13}$CH$_2$CN \\ 
 195 & 34$_{ 6,29}$ -- 33$_{ 6,28}$ &  205176.196 &     6 &  307 &       703.2 &  100 & Blend with CH$_3$CHO and CH$_3$$^{13}$CH$_2$CN \\ 
 196 & 35$_{ 4,31}$ -- 34$_{ 5,30}$ &  205501.651 &     7 &  313 &       152.2 &  100 & Strong CH$_3$OCHO and c-C$_2$H$_4$O \\
\hline
 \end{tabular}
 }\\[1ex] 
 \end{table*}
\begin{table*}
 {\centering
 \addtocounter{table}{-1}
 \caption{
continued.
 }
 \begin{tabular}{rlrrrrrl}
 \hline\hline
 \multicolumn{1}{c}{$N^a$} & \multicolumn{1}{c}{Transition$^b$} & \multicolumn{1}{c}{Frequency} & \multicolumn{1}{c}{Unc.$^c$} & \multicolumn{1}{c}{$E_\mathrm{l}$$^d$} & \multicolumn{1}{c}{$S\mu^2$} &  \multicolumn{1}{c}{$\sigma^e$} &  \multicolumn{1}{c}{Comments} \\ 
  & & \multicolumn{1}{c}{\scriptsize (MHz)} & \multicolumn{1}{c}{\scriptsize (kHz)} & \multicolumn{1}{c}{\scriptsize (K)} & \multicolumn{1}{c}{\scriptsize (D$^2$)} &  \multicolumn{1}{c}{\scriptsize (mK)} &  \\ 
 \multicolumn{1}{c}{(1)} & \multicolumn{1}{c}{(2)} & \multicolumn{1}{c}{(3)} & \multicolumn{1}{c}{(4)} & \multicolumn{1}{c}{(5)} & \multicolumn{1}{c}{(6)} & \multicolumn{1}{c}{(7)} & \multicolumn{1}{c}{(8)} \\ 
 \hline
 281 & 35$_{ 9,26}$ -- 34$_{ 9,25}$ &  211938.010 &    50 &  332 &       700.1 &   36 & Strong NH$_2$$^{13}$CHO and NH$_2$$^{13}$CHO, $\varv_{12}$=1 \\ 
 282 & 13$_{10, 3}$ -- 12$_{ 9, 4}$$^\star$ &  211955.981 &    12 &  182 &        87.6 &   36 & Strong NH$_2$$^{13}$CHO, $\varv_{12}$=1 and NH$_2$$^{13}$CHO \\ 
 284 & 34$_{ 6,28}$ -- 33$_{ 6,27}$ &  212120.060 &    50 &  309 &       708.0 &   36 & Strong CH$_3$CHO and NH$_2$$^{13}$CHO \\ 
 285 & 35$_{ 8,28}$ -- 34$_{ 8,27}$ &  212458.290 &    50 &  326 &       710.4 &   99 & Blend with C$_2$H$_3$CN and C$_2$H$_5$OH \\ 
 286 & 35$_{ 7,29}$ -- 34$_{ 7,28}$ &  212483.750 &    50 &  321 &       719.0 &   99 & Blend with CH$_3$OCHO, a-C$_3$H$_7$CN, and $^{13}$CH$_3$OH, $\varv_t$=1 \\
\hline
 \end{tabular}
 }\\[1ex] 
 \end{table*}
\begin{table*}
 {\centering
 \addtocounter{table}{-1}
 \caption{
continued.
 }
 \begin{tabular}{rlrrrrrl}
 \hline\hline
 \multicolumn{1}{c}{$N^a$} & \multicolumn{1}{c}{Transition$^b$} & \multicolumn{1}{c}{Frequency} & \multicolumn{1}{c}{Unc.$^c$} & \multicolumn{1}{c}{$E_\mathrm{l}$$^d$} & \multicolumn{1}{c}{$S\mu^2$} &  \multicolumn{1}{c}{$\sigma^e$} &  \multicolumn{1}{c}{Comments} \\ 
  & & \multicolumn{1}{c}{\scriptsize (MHz)} & \multicolumn{1}{c}{\scriptsize (kHz)} & \multicolumn{1}{c}{\scriptsize (K)} & \multicolumn{1}{c}{\scriptsize (D$^2$)} &  \multicolumn{1}{c}{\scriptsize (mK)} &  \\ 
 \multicolumn{1}{c}{(1)} & \multicolumn{1}{c}{(2)} & \multicolumn{1}{c}{(3)} & \multicolumn{1}{c}{(4)} & \multicolumn{1}{c}{(5)} & \multicolumn{1}{c}{(6)} & \multicolumn{1}{c}{(7)} & \multicolumn{1}{c}{(8)} \\ 
 \hline
\hline
 375 & 39$_{ 8,31}$ -- 39$_{ 5,34}$ &  219371.585 &    37 &  370 &        11.6 &   92 & Blend with C$_2$H$_3$CN, $\varv_{11}$=1/$\varv_{15}$=1 \\ 
 376 & 19$_{ 8,12}$ -- 18$_{ 7,11}$ &  219376.897 &    12 &  198 &        77.1 &   92 & Blend with C$_2$H$_3$CN, $\varv_{11}$=1/$\varv_{15}$=1 and C$_2$H$_5$OH \\ 
 377 & 19$_{ 8,11}$ -- 18$_{ 7,12}$ &  219380.063 &    12 &  198 &        77.1 &   92 & Blend with C$_2$H$_3$CN, $\varv_{11}$=1/$\varv_{15}$=1 and C$_2$H$_5$OH \\ 
 378 & 38$_{ 3,35}$ -- 37$_{ 4,34}$ &  219549.670 &    50 &  338 &       223.9 &   92 & Strong C$^{18}$O \\ 
 379 & 36$_{ 5,31}$ -- 35$_{ 5,30}$ &  219565.844 &     7 &  327 &       748.9 &   92 & Strong C$^{18}$O \\
\hline
 \end{tabular}
 }\\[1ex] 
 \end{table*}
\begin{table*}
 {\centering
 \addtocounter{table}{-1}
 \caption{
continued.
 }
 \begin{tabular}{rlrrrrrl}
 \hline\hline
 \multicolumn{1}{c}{$N^a$} & \multicolumn{1}{c}{Transition$^b$} & \multicolumn{1}{c}{Frequency} & \multicolumn{1}{c}{Unc.$^c$} & \multicolumn{1}{c}{$E_\mathrm{l}$$^d$} & \multicolumn{1}{c}{$S\mu^2$} &  \multicolumn{1}{c}{$\sigma^e$} &  \multicolumn{1}{c}{Comments} \\ 
  & & \multicolumn{1}{c}{\scriptsize (MHz)} & \multicolumn{1}{c}{\scriptsize (kHz)} & \multicolumn{1}{c}{\scriptsize (K)} & \multicolumn{1}{c}{\scriptsize (D$^2$)} &  \multicolumn{1}{c}{\scriptsize (mK)} &  \\ 
 \multicolumn{1}{c}{(1)} & \multicolumn{1}{c}{(2)} & \multicolumn{1}{c}{(3)} & \multicolumn{1}{c}{(4)} & \multicolumn{1}{c}{(5)} & \multicolumn{1}{c}{(6)} & \multicolumn{1}{c}{(7)} & \multicolumn{1}{c}{(8)} \\ 
 \hline
 467 & 40$_{ 5,35}$ -- 39$_{ 6,34}$ &  235342.332 &    14 &  370 &       156.8 &  131 & Blend with U-line and CH$_3$CH$_3$CO \\ 
 468 & 62$_{ 2,60}$ -- 62$_{ 1,61}$$^\star$ &  235610.226 &   562 &  656 &        67.0 &  131 & Blend with CH$_3$CH$_3$CO, $^{13}$CH$_3$OH, $\varv_t$=1, and  \\ 
 & & & & & & & C$_2$H$_5$CN, $\varv_{13}$=1/$\varv_{21}$=1 \\ 
 470 & 62$_{ 2,60}$ -- 62$_{ 2,61}$$^\star$ &  235610.226 &   562 &  656 &        10.9 &  131 & Blend with CH$_3$CH$_3$CO, $^{13}$CH$_3$OH, $\varv_t$=1, and  \\ 
 & & & & & & & C$_2$H$_5$CN, $\varv_{13}$=1/$\varv_{21}$=1 \\
\hline
 \end{tabular}
 }\\[1ex] 
 \end{table*}
\begin{table*}
 {\centering
 \addtocounter{table}{-1}
 \caption{
continued.
 }
 \begin{tabular}{rlrrrrrl}
 \hline\hline
 \multicolumn{1}{c}{$N^a$} & \multicolumn{1}{c}{Transition$^b$} & \multicolumn{1}{c}{Frequency} & \multicolumn{1}{c}{Unc.$^c$} & \multicolumn{1}{c}{$E_\mathrm{l}$$^d$} & \multicolumn{1}{c}{$S\mu^2$} &  \multicolumn{1}{c}{$\sigma^e$} &  \multicolumn{1}{c}{Comments} \\ 
  & & \multicolumn{1}{c}{\scriptsize (MHz)} & \multicolumn{1}{c}{\scriptsize (kHz)} & \multicolumn{1}{c}{\scriptsize (K)} & \multicolumn{1}{c}{\scriptsize (D$^2$)} &  \multicolumn{1}{c}{\scriptsize (mK)} &  \\ 
 \multicolumn{1}{c}{(1)} & \multicolumn{1}{c}{(2)} & \multicolumn{1}{c}{(3)} & \multicolumn{1}{c}{(4)} & \multicolumn{1}{c}{(5)} & \multicolumn{1}{c}{(6)} & \multicolumn{1}{c}{(7)} & \multicolumn{1}{c}{(8)} \\ 
 \hline
 556 & 41$_{30,11}$ -- 40$_{30,10}$$^\star$ &  245972.316 &    19 &  672 &       408.0 &   53 & Blend with C$_2$H$_5$CN, $\varv_{13}$=1/$\varv_{21}$=1 and $\varv$=0, and U-line \\ 
 558 & 38$_{18,20}$ -- 38$_{17,21}$$^\star$ &  245972.593 &    78 &  443 &       150.7 &   53 & Blend with C$_2$H$_5$CN, $\varv_{13}$=1/$\varv_{21}$=1 and $\varv$=0, and U-line \\ 
 560 & 41$_{31,10}$ -- 40$_{31, 9}$$^\star$ &  246034.256 &    22 &  692 &       376.2 &   53 & Blend with C$_2$H$_5$CN, $\varv_{13}$=1/$\varv_{21}$=1 and CH$_3$OCHO \\ 
 562 & 37$_{18,19}$ -- 37$_{17,20}$$^\star$ &  246036.674 &    78 &  432 &       144.4 &   53 & Blend with C$_2$H$_5$CN, $\varv_{13}$=1/$\varv_{21}$=1 and CH$_3$OCHO \\ 
\hline
 564 & 27$_{ 7,21}$ -- 26$_{ 6,20}$ &  247058.498 &    12 &  246 &        70.8 &   68 & Strong CH$_3$OCHO \\
\hline
 \end{tabular}
 }\\[1ex] 
 \end{table*}
\begin{table*}
 {\centering
 \addtocounter{table}{-1}
 \caption{
continued.
 }
 \begin{tabular}{rlrrrrrl}
 \hline\hline
 \multicolumn{1}{c}{$N^a$} & \multicolumn{1}{c}{Transition$^b$} & \multicolumn{1}{c}{Frequency} & \multicolumn{1}{c}{Unc.$^c$} & \multicolumn{1}{c}{$E_\mathrm{l}$$^d$} & \multicolumn{1}{c}{$S\mu^2$} &  \multicolumn{1}{c}{$\sigma^e$} &  \multicolumn{1}{c}{Comments} \\ 
  & & \multicolumn{1}{c}{\scriptsize (MHz)} & \multicolumn{1}{c}{\scriptsize (kHz)} & \multicolumn{1}{c}{\scriptsize (K)} & \multicolumn{1}{c}{\scriptsize (D$^2$)} &  \multicolumn{1}{c}{\scriptsize (mK)} &  \\ 
 \multicolumn{1}{c}{(1)} & \multicolumn{1}{c}{(2)} & \multicolumn{1}{c}{(3)} & \multicolumn{1}{c}{(4)} & \multicolumn{1}{c}{(5)} & \multicolumn{1}{c}{(6)} & \multicolumn{1}{c}{(7)} & \multicolumn{1}{c}{(8)} \\ 
 \hline
 657 & 28$_{19, 9}$ -- 28$_{18,10}$$^\star$ &  260327.281 &   114 &  359 &        79.6 &  413 & Strong CH$_3$OCH$_3$ and C$_2$H$_5$CN \\ 
 659 & 27$_{19, 8}$ -- 27$_{18, 9}$$^\star$ &  260327.999 &   114 &  351 &        72.5 &  413 & Strong CH$_3$OCH$_3$ and C$_2$H$_5$CN \\ 
\hline
 661 & 43$_{ 9,34}$ -- 42$_{ 9,33}$ &  263500.082 &    18 &  422 &       881.1 &   74 & Blend with C$_2$H$_5$$^{13}$CN and CH$_3$OCH$_3$ \\ 
 662 & 37$_{ 7,31}$ -- 36$_{ 6,30}$ &  263656.862 &    14 &  340 &        85.0 &  108 & Strong CH$_3$OCH$_3$ \\ 
 663 & 44$_{23,21}$ -- 43$_{23,20}$$^\star$ &  263662.413 &    12 &  583 &       684.9 &  108 & Strong CH$_3$OCH$_3$, HNCO, and HC$^{13}$CCN, $\varv_7$=1 \\
\hline
 \end{tabular}
 }\\[1ex] 
 Notes:
 $^a$ Numbering of the observed transitions associated with a modeled line stronger than 20 mK.
 $^b$ Transitions marked with a $^\star$ are double with a frequency difference less than 0.1 MHz. The quantum numbers of the second one are not shown.
 $^c$ Frequency uncertainty.
 $^d$ Lower energy level in temperature units ($E_\mathrm{l}/k_\mathrm{B}$).
 $^e$ Calculated rms noise level in $T_{\mathrm{mb}}$ scale.
 \end{table*}

}

\begin{table*}
 {\centering
 \caption{
 Transitions of the \textit{anti}-conformer of \textit{n}-propyl cyanide detected toward Sgr~B2(N) with the IRAM 30 m telescope.
}
 \label{t:detectprcn-a}
 \vspace*{-1.0ex}
 \begin{tabular}{rlrrrrrcrrrrl}
 \hline\hline
 \multicolumn{1}{c}{$N^a$} & \multicolumn{1}{c}{\hspace*{-1ex} Transition} & \multicolumn{1}{c}{\hspace*{-2ex} Frequency} & \multicolumn{1}{c}{\hspace*{-3ex} Unc.$^b$} & \multicolumn{1}{c}{\hspace*{-1ex} $E_\mathrm{l}$$^c$} & \multicolumn{1}{c}{\hspace*{-1ex} $S\mu^2$} & \multicolumn{1}{c}{\hspace*{-2ex} $\sigma^d$} & \multicolumn{1}{c}{$F^e$} & \multicolumn{1}{c}{$\tau^f$} & \multicolumn{1}{c}{$I_{\mathrm{obs}}$$^g$} & \multicolumn{1}{c}{\hspace*{-1ex} $I_{\mathrm{mod}}$$^g$} & \multicolumn{1}{c}{\hspace*{-1ex} $I_{\mathrm{all}}$$^g$} & \multicolumn{1}{c}{Comments} \\ 
  & & \multicolumn{1}{c}{\hspace*{-2ex} \scriptsize (MHz)} & \multicolumn{1}{c}{\hspace*{-3ex} \scriptsize (kHz)} & \multicolumn{1}{c}{\hspace*{-1ex} \scriptsize (K)} & \multicolumn{1}{c}{\hspace*{-1ex} \scriptsize (D$^2$)} & \multicolumn{1}{c}{\hspace*{-2ex} \scriptsize (mK)} & & & \multicolumn{1}{c}{\scriptsize (K~km~s$^{-1}$)} & \multicolumn{2}{c}{\hspace*{-1ex} \scriptsize (K~km~s$^{-1}$)} & \\ 
 \multicolumn{1}{c}{(1)} & \multicolumn{1}{c}{\hspace*{-1ex} (2)} & \multicolumn{1}{c}{\hspace*{-2ex} (3)} & \multicolumn{1}{c}{\hspace*{-3ex} (4)} & \multicolumn{1}{c}{\hspace*{-1ex} (5)} & \multicolumn{1}{c}{\hspace*{-1ex} (6)} & \multicolumn{1}{c}{\hspace*{-2ex} (7)} & \multicolumn{1}{c}{(8)} & \multicolumn{1}{c}{(9)} & \multicolumn{1}{c}{(10)} & \multicolumn{1}{c}{(11)} & \multicolumn{1}{c}{\hspace*{-1ex} (12)} & \multicolumn{1}{c}{\hspace*{-1ex} (13)} \\ 
 \hline
   5 & \hspace*{-1ex} 19$_{ 6,14}$ -- 18$_{ 6,13}$ & \hspace*{-2ex}   84021.555 & \hspace*{-3ex}    4 & \hspace*{-1ex}   73 & \hspace*{-1ex}         221 & \hspace*{-2ex}    19 &    1 & 0.06 &        3.27(13) & \hspace*{-2ex}        1.90 & \hspace*{-2ex}        2.68 &  partial blend with C$_2$H$_5$CN \\ 
   6 & \hspace*{-1ex} 19$_{ 6,13}$ -- 18$_{ 6,12}$ & \hspace*{-2ex}   84021.556 & \hspace*{-3ex}    4 & \hspace*{-1ex}   73 & \hspace*{-1ex}         221 & \hspace*{-2ex}    19 &    1 & -- & -- & \hspace*{-2ex} -- & \hspace*{-2ex} -- & -- \\ 
   7 & \hspace*{-1ex} 19$_{ 7,12}$ -- 18$_{ 7,11}$ & \hspace*{-2ex}   84022.819 & \hspace*{-3ex}    5 & \hspace*{-1ex}   87 & \hspace*{-1ex}         212 & \hspace*{-2ex}    19 &    1 & -- & -- & \hspace*{-2ex} -- & \hspace*{-2ex} -- & -- \\ 
   8 & \hspace*{-1ex} 19$_{ 7,13}$ -- 18$_{ 7,12}$ & \hspace*{-2ex}   84022.819 & \hspace*{-3ex}    5 & \hspace*{-1ex}   87 & \hspace*{-1ex}         212 & \hspace*{-2ex}    19 &    1 & -- & -- & \hspace*{-2ex} -- & \hspace*{-2ex} -- & -- \\ 
   9 & \hspace*{-1ex} 19$_{ 5,15}$ -- 18$_{ 5,14}$ & \hspace*{-2ex}   84023.956 & \hspace*{-3ex}    4 & \hspace*{-1ex}   62 & \hspace*{-1ex}         229 & \hspace*{-2ex}    19 &    1 & -- & -- & \hspace*{-2ex} -- & \hspace*{-2ex} -- & -- \\ 
  10 & \hspace*{-1ex} 19$_{ 5,14}$ -- 18$_{ 5,13}$ & \hspace*{-2ex}   84023.960 & \hspace*{-3ex}    4 & \hspace*{-1ex}   62 & \hspace*{-1ex}         229 & \hspace*{-2ex}    19 &    1 & -- & -- & \hspace*{-2ex} -- & \hspace*{-2ex} -- & -- \\ 
  11 & \hspace*{-1ex} 19$_{ 8,11}$ -- 18$_{ 8,10}$ & \hspace*{-2ex}   84026.382 & \hspace*{-3ex}    5 & \hspace*{-1ex}  102 & \hspace*{-1ex}         202 & \hspace*{-2ex}    19 &    1 & -- & -- & \hspace*{-2ex} -- & \hspace*{-2ex} -- & -- \\ 
  12 & \hspace*{-1ex} 19$_{ 8,12}$ -- 18$_{ 8,11}$ & \hspace*{-2ex}   84026.382 & \hspace*{-3ex}    5 & \hspace*{-1ex}  102 & \hspace*{-1ex}         202 & \hspace*{-2ex}    19 &    1 & -- & -- & \hspace*{-2ex} -- & \hspace*{-2ex} -- & -- \\ 
  32 & \hspace*{-1ex} 20$_{ 6,14}$ -- 19$_{ 6,13}$ & \hspace*{-2ex}   88444.212 & \hspace*{-3ex}    5 & \hspace*{-1ex}   77 & \hspace*{-1ex}         235 & \hspace*{-2ex}    17 &    2 & 0.06 &        4.30(11) & \hspace*{-2ex}        2.31 & \hspace*{-2ex}        3.28 &  partial blend with CH$_3$CH$_3$CO, \\ 
 & & & & & & & & & & & &  C$_2$H$_5$OH, and U-line? \\ 
  33 & \hspace*{-1ex} 20$_{ 6,15}$ -- 19$_{ 6,14}$ & \hspace*{-2ex}   88444.212 & \hspace*{-3ex}    5 & \hspace*{-1ex}   77 & \hspace*{-1ex}         235 & \hspace*{-2ex}    17 &    2 & -- & -- & \hspace*{-2ex} -- & \hspace*{-2ex} -- & -- \\ 
  34 & \hspace*{-1ex} 20$_{ 7,13}$ -- 19$_{ 7,12}$ & \hspace*{-2ex}   88445.075 & \hspace*{-3ex}    5 & \hspace*{-1ex}   91 & \hspace*{-1ex}         227 & \hspace*{-2ex}    17 &    2 & -- & -- & \hspace*{-2ex} -- & \hspace*{-2ex} -- & -- \\ 
  35 & \hspace*{-1ex} 20$_{ 7,14}$ -- 19$_{ 7,13}$ & \hspace*{-2ex}   88445.075 & \hspace*{-3ex}    5 & \hspace*{-1ex}   91 & \hspace*{-1ex}         227 & \hspace*{-2ex}    17 &    2 & -- & -- & \hspace*{-2ex} -- & \hspace*{-2ex} -- & -- \\ 
  36 & \hspace*{-1ex} 20$_{ 5,16}$ -- 19$_{ 5,15}$ & \hspace*{-2ex}   88447.526 & \hspace*{-3ex}    5 & \hspace*{-1ex}   66 & \hspace*{-1ex}         243 & \hspace*{-2ex}    17 &    2 & -- & -- & \hspace*{-2ex} -- & \hspace*{-2ex} -- & -- \\ 
  37 & \hspace*{-1ex} 20$_{ 5,15}$ -- 19$_{ 5,14}$ & \hspace*{-2ex}   88447.532 & \hspace*{-3ex}    5 & \hspace*{-1ex}   66 & \hspace*{-1ex}         243 & \hspace*{-2ex}    17 &    2 & -- & -- & \hspace*{-2ex} -- & \hspace*{-2ex} -- & -- \\ 
  38 & \hspace*{-1ex} 20$_{ 8,12}$ -- 19$_{ 8,11}$ & \hspace*{-2ex}   88448.528 & \hspace*{-3ex}    5 & \hspace*{-1ex}  106 & \hspace*{-1ex}         217 & \hspace*{-2ex}    17 &    2 & -- & -- & \hspace*{-2ex} -- & \hspace*{-2ex} -- & -- \\ 
  39 & \hspace*{-1ex} 20$_{ 8,13}$ -- 19$_{ 8,12}$ & \hspace*{-2ex}   88448.528 & \hspace*{-3ex}    5 & \hspace*{-1ex}  106 & \hspace*{-1ex}         217 & \hspace*{-2ex}    17 &    2 & -- & -- & \hspace*{-2ex} -- & \hspace*{-2ex} -- & -- \\ 
  40 & \hspace*{-1ex} 20$_{ 9,11}$ -- 19$_{ 9,10}$ & \hspace*{-2ex}   88453.836 & \hspace*{-3ex}    5 & \hspace*{-1ex}  124 & \hspace*{-1ex}         206 & \hspace*{-2ex}    17 &    3 & 0.05 &        1.90(11) & \hspace*{-2ex}        1.40 & \hspace*{-2ex}        2.42 &  partial blend with C$_2$H$_5$$^{13}$CN, \\ 
 & & & & & & & & & & & &  $^{13}$CH$_3$CH$_2$CN, and CH$_2$(OH)CHO \\ 
  41 & \hspace*{-1ex} 20$_{ 9,12}$ -- 19$_{ 9,11}$ & \hspace*{-2ex}   88453.836 & \hspace*{-3ex}    5 & \hspace*{-1ex}  124 & \hspace*{-1ex}         206 & \hspace*{-2ex}    17 &    3 & -- & -- & \hspace*{-2ex} -- & \hspace*{-2ex} -- & -- \\ 
  42 & \hspace*{-1ex} 20$_{ 4,17}$ -- 19$_{ 4,16}$ & \hspace*{-2ex}   88458.794 & \hspace*{-3ex}    5 & \hspace*{-1ex}   57 & \hspace*{-1ex}         248 & \hspace*{-2ex}    17 &    3 & -- & -- & \hspace*{-2ex} -- & \hspace*{-2ex} -- & -- \\ 
  43 & \hspace*{-1ex} 20$_{ 4,16}$ -- 19$_{ 4,15}$ & \hspace*{-2ex}   88459.387 & \hspace*{-3ex}    5 & \hspace*{-1ex}   57 & \hspace*{-1ex}         248 & \hspace*{-2ex}    17 &    3 & -- & -- & \hspace*{-2ex} -- & \hspace*{-2ex} -- & -- \\ 
  44 & \hspace*{-1ex} 20$_{10,10}$ -- 19$_{10, 9}$ & \hspace*{-2ex}   88460.625 & \hspace*{-3ex}    5 & \hspace*{-1ex}  143 & \hspace*{-1ex}         194 & \hspace*{-2ex}    17 &    3 & -- & -- & \hspace*{-2ex} -- & \hspace*{-2ex} -- & -- \\ 
  45 & \hspace*{-1ex} 20$_{10,11}$ -- 19$_{10,10}$ & \hspace*{-2ex}   88460.625 & \hspace*{-3ex}    5 & \hspace*{-1ex}  143 & \hspace*{-1ex}         194 & \hspace*{-2ex}    17 &    3 & -- & -- & \hspace*{-2ex} -- & \hspace*{-2ex} -- & -- \\ 
  53 & \hspace*{-1ex} 20$_{ 1,19}$ -- 19$_{ 1,18}$ & \hspace*{-2ex}   89391.938 & \hspace*{-3ex}    5 & \hspace*{-1ex}   42 & \hspace*{-1ex}         258 & \hspace*{-2ex}    16 &    4 & 0.03 &        0.65(08) & \hspace*{-2ex}        0.44 & \hspace*{-2ex}        0.62 &  blend with CH$_3$OCHO, $\varv_t$=1 \\ 
  55 & \hspace*{-1ex} 21$_{ 0,21}$ -- 20$_{ 0,20}$ & \hspace*{-2ex}   92164.912 & \hspace*{-3ex}    5 & \hspace*{-1ex}   44 & \hspace*{-1ex}         272 & \hspace*{-2ex}    27 &    5 & 0.03 &        1.15(14) & \hspace*{-2ex}        0.49 & \hspace*{-2ex}        0.66 &  blend with CH$_3$$^{13}$CN, $\varv_8$=1 and \\ 
 & & & & & & & & & & & &  U-line? \\ 
  57 & \hspace*{-1ex} 21$_{ 6,16}$ -- 20$_{ 6,15}$ & \hspace*{-2ex}   92866.939 & \hspace*{-3ex}    5 & \hspace*{-1ex}   82 & \hspace*{-1ex}         250 & \hspace*{-2ex}    28 &    6 & 0.07 &        1.54(21) & \hspace*{-2ex}        3.17 & \hspace*{-2ex}        3.21 &  uncertain baseline \\ 
  58 & \hspace*{-1ex} 21$_{ 6,15}$ -- 20$_{ 6,14}$ & \hspace*{-2ex}   92866.939 & \hspace*{-3ex}    5 & \hspace*{-1ex}   82 & \hspace*{-1ex}         250 & \hspace*{-2ex}    28 &    6 & -- & -- & \hspace*{-2ex} -- & \hspace*{-2ex} -- & -- \\ 
  59 & \hspace*{-1ex} 21$_{ 7,14}$ -- 20$_{ 7,13}$ & \hspace*{-2ex}   92867.332 & \hspace*{-3ex}    5 & \hspace*{-1ex}   95 & \hspace*{-1ex}         242 & \hspace*{-2ex}    28 &    6 & -- & -- & \hspace*{-2ex} -- & \hspace*{-2ex} -- & -- \\ 
  60 & \hspace*{-1ex} 21$_{ 7,15}$ -- 20$_{ 7,14}$ & \hspace*{-2ex}   92867.332 & \hspace*{-3ex}    5 & \hspace*{-1ex}   95 & \hspace*{-1ex}         242 & \hspace*{-2ex}    28 &    6 & -- & -- & \hspace*{-2ex} -- & \hspace*{-2ex} -- & -- \\ 
  61 & \hspace*{-1ex} 21$_{ 8,13}$ -- 20$_{ 8,12}$ & \hspace*{-2ex}   92870.627 & \hspace*{-3ex}    5 & \hspace*{-1ex}  110 & \hspace*{-1ex}         232 & \hspace*{-2ex}    28 &    6 & -- & -- & \hspace*{-2ex} -- & \hspace*{-2ex} -- & -- \\ 
  62 & \hspace*{-1ex} 21$_{ 8,14}$ -- 20$_{ 8,13}$ & \hspace*{-2ex}   92870.627 & \hspace*{-3ex}    5 & \hspace*{-1ex}  110 & \hspace*{-1ex}         232 & \hspace*{-2ex}    28 &    6 & -- & -- & \hspace*{-2ex} -- & \hspace*{-2ex} -- & -- \\ 
  63 & \hspace*{-1ex} 21$_{ 5,17}$ -- 20$_{ 5,16}$ & \hspace*{-2ex}   92871.289 & \hspace*{-3ex}    5 & \hspace*{-1ex}   70 & \hspace*{-1ex}         256 & \hspace*{-2ex}    28 &    6 & -- & -- & \hspace*{-2ex} -- & \hspace*{-2ex} -- & -- \\ 
  64 & \hspace*{-1ex} 21$_{ 5,16}$ -- 20$_{ 5,15}$ & \hspace*{-2ex}   92871.298 & \hspace*{-3ex}    5 & \hspace*{-1ex}   70 & \hspace*{-1ex}         256 & \hspace*{-2ex}    28 &    6 & -- & -- & \hspace*{-2ex} -- & \hspace*{-2ex} -- & -- \\ 
  65 & \hspace*{-1ex} 21$_{ 9,12}$ -- 20$_{ 9,11}$ & \hspace*{-2ex}   92875.977 & \hspace*{-3ex}    5 & \hspace*{-1ex}  128 & \hspace*{-1ex}         222 & \hspace*{-2ex}    28 &    6 & -- & -- & \hspace*{-2ex} -- & \hspace*{-2ex} -- & -- \\ 
  66 & \hspace*{-1ex} 21$_{ 9,13}$ -- 20$_{ 9,12}$ & \hspace*{-2ex}   92875.977 & \hspace*{-3ex}    5 & \hspace*{-1ex}  128 & \hspace*{-1ex}         222 & \hspace*{-2ex}    28 &    6 & -- & -- & \hspace*{-2ex} -- & \hspace*{-2ex} -- & -- \\ 
  77 & \hspace*{-1ex} 21$_{ 2,19}$ -- 20$_{ 2,18}$ & \hspace*{-2ex}   93376.934 & \hspace*{-3ex}    5 & \hspace*{-1ex}   49 & \hspace*{-1ex}         269 & \hspace*{-2ex}    22 &    7 & 0.03 &        0.52(11) & \hspace*{-2ex}        0.49 & \hspace*{-2ex}        0.51 & no blend \\ 
 104 & \hspace*{-1ex} 22$_{ 2,20}$ -- 21$_{ 2,19}$ & \hspace*{-2ex}   97867.394 & \hspace*{-3ex}    5 & \hspace*{-1ex}   53 & \hspace*{-1ex}         282 & \hspace*{-2ex}    20 &    8 & 0.03 &        0.61(10) & \hspace*{-2ex}        0.58 & \hspace*{-2ex}        0.63 &  uncertain baseline \\ 
 108 & \hspace*{-1ex} 10$_{ 4, 6}$ -- 11$_{ 3, 9}$ & \hspace*{-2ex}  101512.348 & \hspace*{-3ex}    5 & \hspace*{-1ex}   23 & \hspace*{-1ex}           1 & \hspace*{-2ex}    16 &    9 & 0.03 &        0.67(08) & \hspace*{-2ex}        0.66 & \hspace*{-2ex}        0.96 &  small blend with CH$_3$CH$_3$CO, $\varv_t$=1 \\ 
 109 & \hspace*{-1ex} 23$_{ 2,22}$ -- 22$_{ 2,21}$ & \hspace*{-2ex}  101515.201 & \hspace*{-3ex}    5 & \hspace*{-1ex}   58 & \hspace*{-1ex}         295 & \hspace*{-2ex}    16 &    9 & -- & -- & \hspace*{-2ex} -- & \hspace*{-2ex} -- & -- \\ 
 110 & \hspace*{-1ex} 23$_{ 7,16}$ -- 22$_{ 7,15}$ & \hspace*{-2ex}  101711.846 & \hspace*{-3ex}    5 & \hspace*{-1ex}  104 & \hspace*{-1ex}         270 & \hspace*{-2ex}    16 &   10 & 0.08 &        5.74(11) & \hspace*{-2ex}        4.35 & \hspace*{-2ex}        6.38 &  partial blend with $^{13}$CH$_2$CHCN, \\ 
 & & & & & & & & & & & &  uncertain baseline \\ 
 111 & \hspace*{-1ex} 23$_{ 7,17}$ -- 22$_{ 7,16}$ & \hspace*{-2ex}  101711.846 & \hspace*{-3ex}    5 & \hspace*{-1ex}  104 & \hspace*{-1ex}         270 & \hspace*{-2ex}    16 &   10 & -- & -- & \hspace*{-2ex} -- & \hspace*{-2ex} -- & -- \\ 
 112 & \hspace*{-1ex} 23$_{ 6,18}$ -- 22$_{ 6,17}$ & \hspace*{-2ex}  101712.624 & \hspace*{-3ex}    5 & \hspace*{-1ex}   91 & \hspace*{-1ex}         277 & \hspace*{-2ex}    16 &   10 & -- & -- & \hspace*{-2ex} -- & \hspace*{-2ex} -- & -- \\ 
 113 & \hspace*{-1ex} 23$_{ 6,17}$ -- 22$_{ 6,16}$ & \hspace*{-2ex}  101712.624 & \hspace*{-3ex}    5 & \hspace*{-1ex}   91 & \hspace*{-1ex}         277 & \hspace*{-2ex}    16 &   10 & -- & -- & \hspace*{-2ex} -- & \hspace*{-2ex} -- & -- \\ 
 114 & \hspace*{-1ex} 23$_{ 8,15}$ -- 22$_{ 8,14}$ & \hspace*{-2ex}  101714.680 & \hspace*{-3ex}    5 & \hspace*{-1ex}  120 & \hspace*{-1ex}         262 & \hspace*{-2ex}    16 &   10 & -- & -- & \hspace*{-2ex} -- & \hspace*{-2ex} -- & -- \\ 
 115 & \hspace*{-1ex} 23$_{ 8,16}$ -- 22$_{ 8,15}$ & \hspace*{-2ex}  101714.680 & \hspace*{-3ex}    5 & \hspace*{-1ex}  120 & \hspace*{-1ex}         262 & \hspace*{-2ex}    16 &   10 & -- & -- & \hspace*{-2ex} -- & \hspace*{-2ex} -- & -- \\ 
 116 & \hspace*{-1ex} 23$_{ 5,19}$ -- 22$_{ 5,18}$ & \hspace*{-2ex}  101719.429 & \hspace*{-3ex}    5 & \hspace*{-1ex}   79 & \hspace*{-1ex}         283 & \hspace*{-2ex}    16 &   10 & -- & -- & \hspace*{-2ex} -- & \hspace*{-2ex} -- & -- \\ 
 117 & \hspace*{-1ex} 23$_{ 5,18}$ -- 22$_{ 5,17}$ & \hspace*{-2ex}  101719.450 & \hspace*{-3ex}    5 & \hspace*{-1ex}   79 & \hspace*{-1ex}         283 & \hspace*{-2ex}    16 &   10 & -- & -- & \hspace*{-2ex} -- & \hspace*{-2ex} -- & -- \\ 
 118 & \hspace*{-1ex} 23$_{ 9,14}$ -- 22$_{ 9,13}$ & \hspace*{-2ex}  101720.011 & \hspace*{-3ex}    5 & \hspace*{-1ex}  137 & \hspace*{-1ex}         252 & \hspace*{-2ex}    16 &   10 & -- & -- & \hspace*{-2ex} -- & \hspace*{-2ex} -- & -- \\ 
 119 & \hspace*{-1ex} 23$_{ 9,15}$ -- 22$_{ 9,14}$ & \hspace*{-2ex}  101720.011 & \hspace*{-3ex}    5 & \hspace*{-1ex}  137 & \hspace*{-1ex}         252 & \hspace*{-2ex}    16 &   10 & -- & -- & \hspace*{-2ex} -- & \hspace*{-2ex} -- & -- \\ 
 161 & \hspace*{-1ex} 24$_{ 3,22}$ -- 23$_{ 3,21}$ & \hspace*{-2ex}  106188.197 & \hspace*{-3ex}    5 & \hspace*{-1ex}   68 & \hspace*{-1ex}         306 & \hspace*{-2ex}    25 &   11 & 0.03 &        0.97(11) & \hspace*{-2ex}        0.76 & \hspace*{-2ex}        0.82 &  noisy \\ 
 205 & \hspace*{-1ex} 25$_{ 1,24}$ -- 24$_{ 1,23}$ & \hspace*{-2ex}  111593.662 & \hspace*{-3ex}    5 & \hspace*{-1ex}   65 & \hspace*{-1ex}         323 & \hspace*{-2ex}    29 &   12 & 0.03 &        1.48(14) & \hspace*{-2ex}        0.90 & \hspace*{-2ex}        1.08 & no blend \\ 
 \hline
 \end{tabular}
 }\\[0ex] 
 Notes:
 $^a$ Numbering of the observed transitions associated with a modeled line stronger than 20 mK (see Table~\ref{t:prcn-a}).
 $^b$ Frequency uncertainty.
 $^c$ Lower energy level in temperature units ($E_\mathrm{l}/k_\mathrm{B}$).
 $^d$ Calculated rms noise level in $T_{\mathrm{mb}}$ scale.
 $^e$ Numbering of the observed features.
 $^f$ Peak opacity of the modeled feature.
 $^g$ Integrated intensity in $T_{\mathrm{mb}}$ scale for the observed spectrum (Col. 10), the \textit{n}-propyl cyanide model (Col. 11), and the model including all molecules (Col. 12). The uncertainty in Col. 10 is given in parentheses in units of the last digit.
 \end{table*}

To identify \textit{n}-propyl cyanide, we used the same method as for ethyl 
formate (see Sect.~\ref{ss:detetocho}). 
In our spectral survey, 636 transitions of the \textit{anti}-conformer and 
706 transitions of the \textit{gauche}-conformer are predicted above the 
threshold of 20~mK defined in Sect.~\ref{ss:detetocho}. They are listed in 
Tables~\ref{t:prcn-a} and \ref{t:prcn-g} (\textit{online material}), 
respectively, which are presented 
in the same way as Tables~\ref{t:etocho-a} and \ref{t:etocho-g}. Again, as can 
be seen in these tables, most of the \textit{n}-propyl cyanide lines covered 
by our survey of Sgr~B2(N) are heavily blended with lines of other molecules 
and therefore cannot be identified in this source. Only 50 of the 636 
transitions of the \textit{anti}-conformer are relatively free of 
contamination from other molecules, known or still unidentified according to
our modeling. They are marked ``Detected'' or ``Group detected'' in Col.~8 of 
Table~\ref{t:prcn-a}, and are listed with more information in 
Table~\ref{t:detectprcn-a}. We stress that all transitions of sufficient 
strength  predicted in the frequency range of our spectral survey are either 
detected or blended, i.e. no predicted transition is missing in the observed 
spectrum. The 50 detected transitions correspond to 12 observed features that 
are shown in Fig.~\ref{f:detectprcn-a} (\textit{online material}) and labeled 
in Col.~8 of Table~\ref{t:detectprcn-a}. For reference, we show the spectrum 
observed toward Sgr~B2(M) in these figures also. We identified the 
\textit{n}-propyl cyanide lines and the blends affecting them with the LTE 
model of this molecule and the LTE model including all molecules (see 
Sect.~\ref{ss:modeling}). The parameters of our best-fit LTE model of 
\textit{n}-propyl cyanide are listed in Table~\ref{t:prcnmodel}, and the model 
is overlaid in red on the spectrum observed toward Sgr~B2(N) in 
Fig.~\ref{f:detectprcn-a}. The best-fit LTE model including all molecules is 
shown in green in the same figures.

\addtocounter{figure}{1}
\newcounter{appfig3}
\setcounter{appfig3}{\value{figure}}
\onlfig{\value{appfig3}}{\clearpage
\begin{figure*}
 
\centerline{\resizebox{0.85\hsize}{!}{\includegraphics[angle=270]{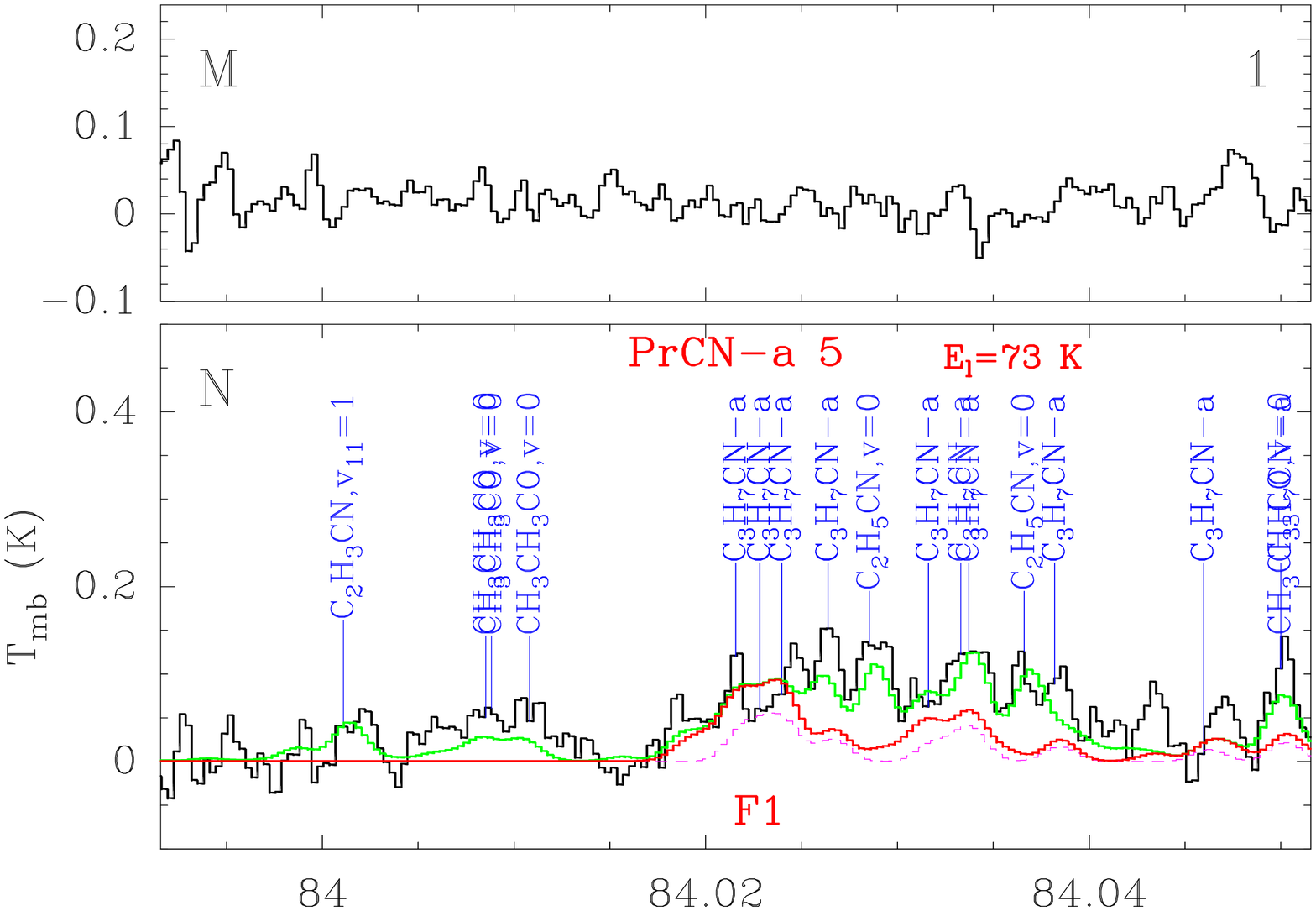}\includegraphics[angle=270]{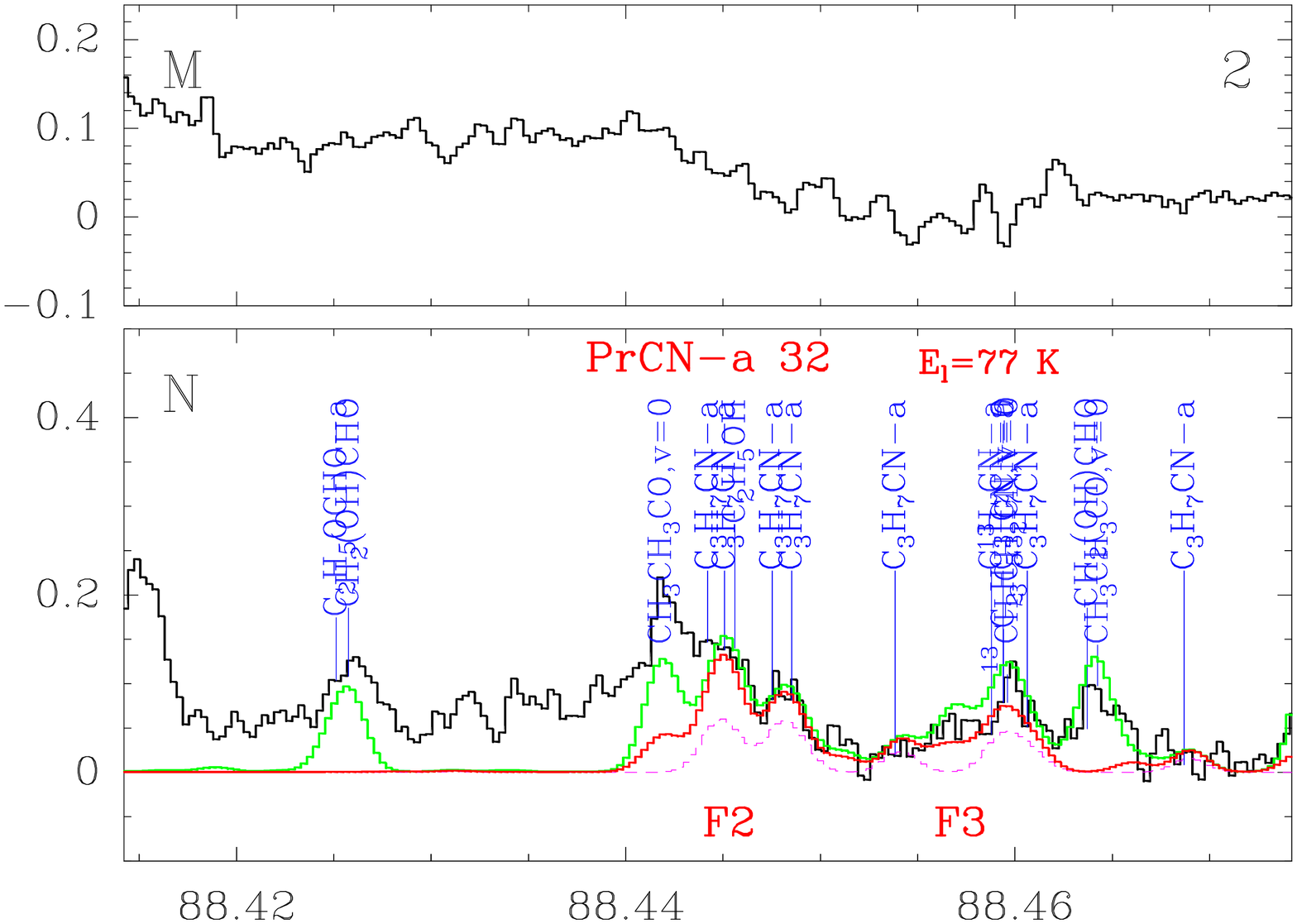}}}
\vspace*{-0.4ex}
\centerline{\resizebox{0.85\hsize}{!}{\includegraphics[angle=270]{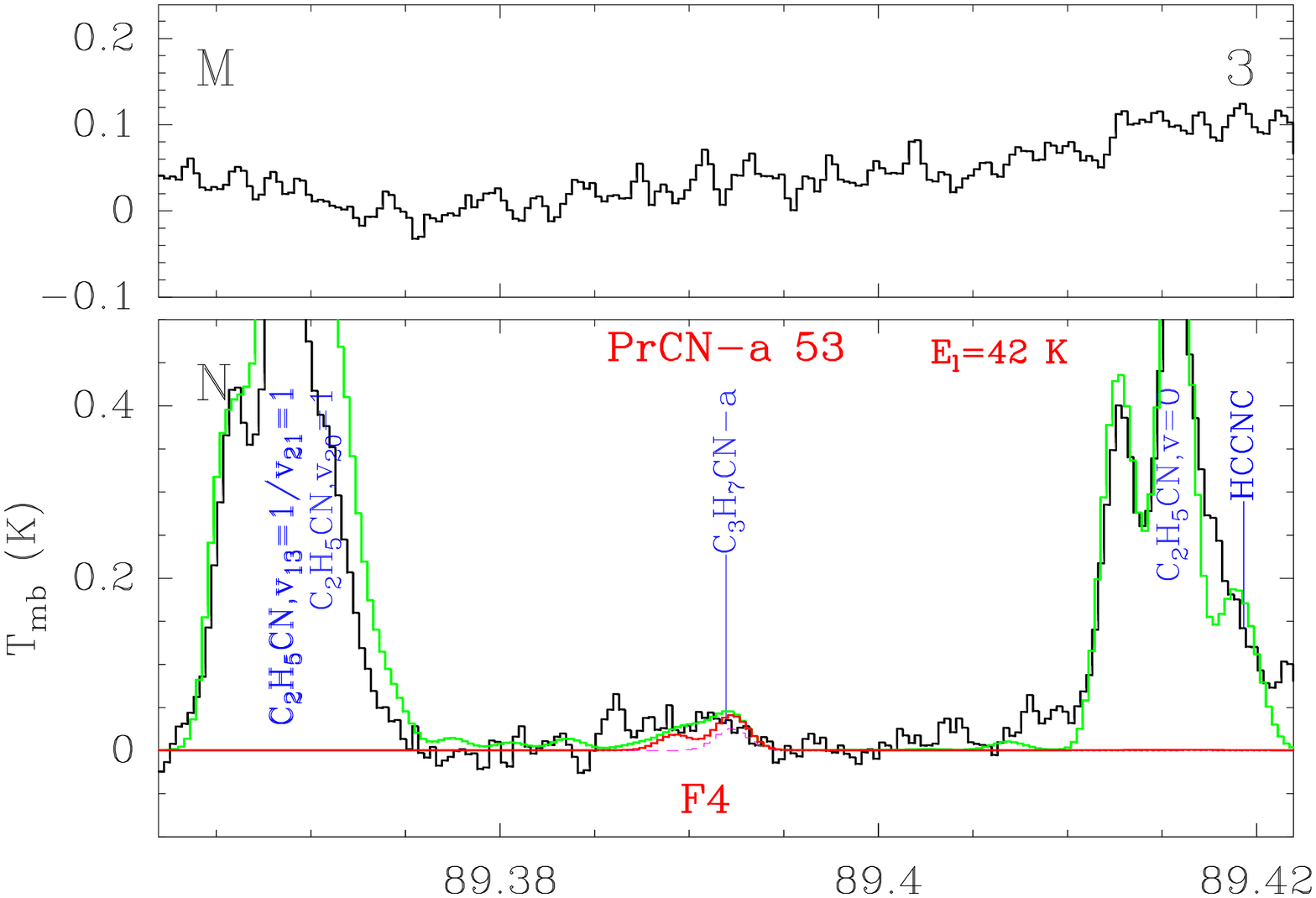}\includegraphics[angle=270]{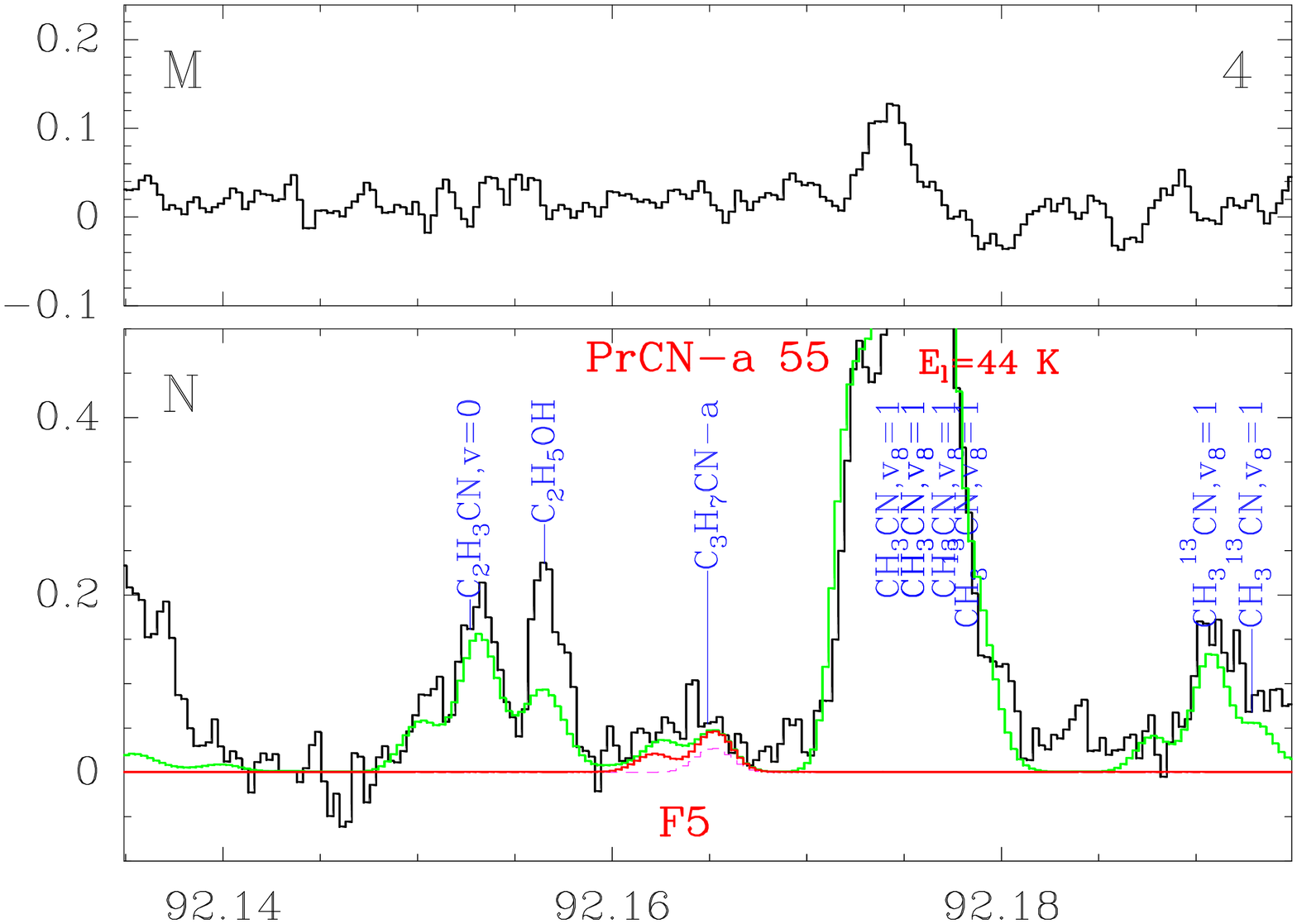}}}
\vspace*{-0.4ex}
\centerline{\resizebox{0.85\hsize}{!}{\includegraphics[angle=270]{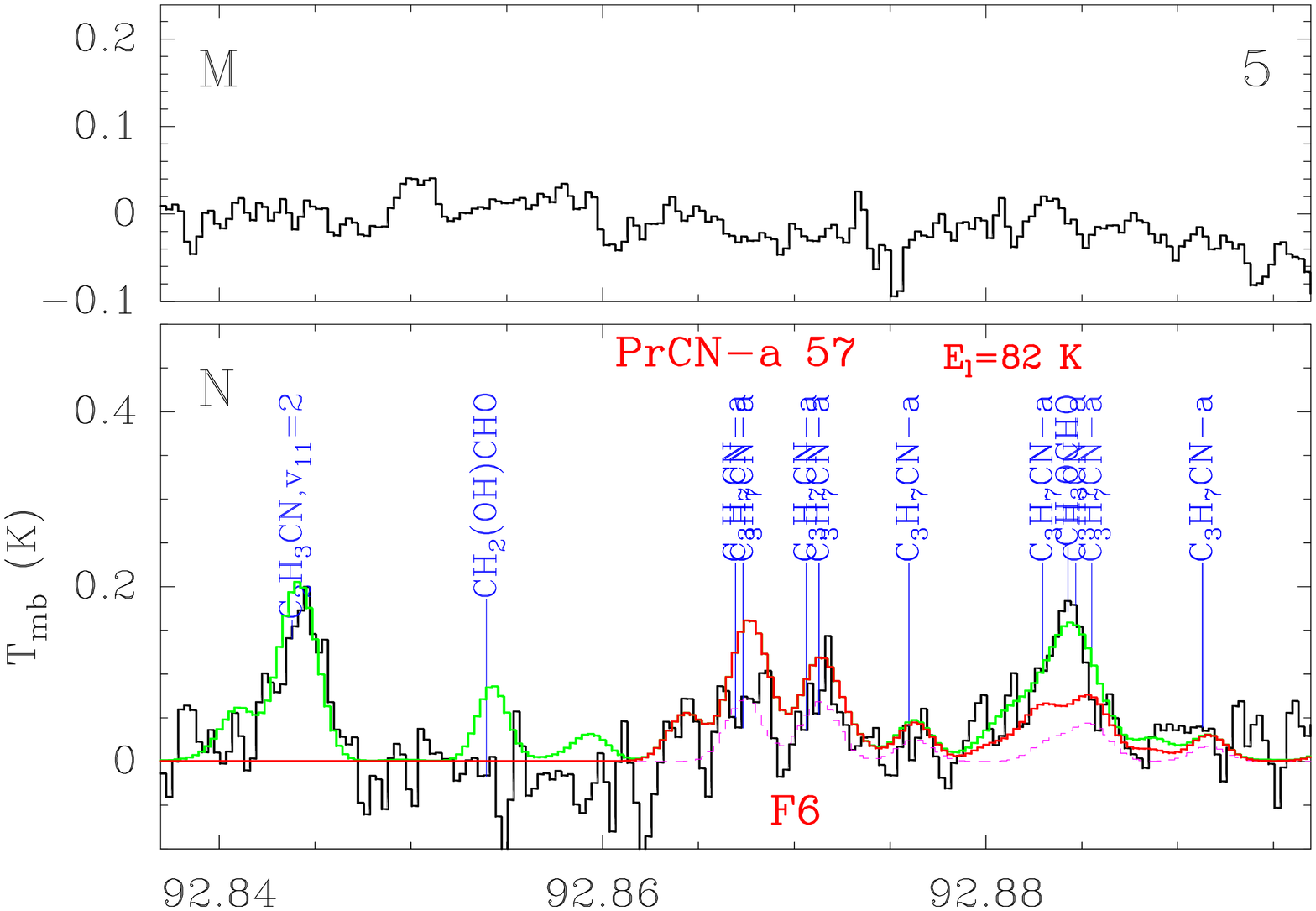}\includegraphics[angle=270]{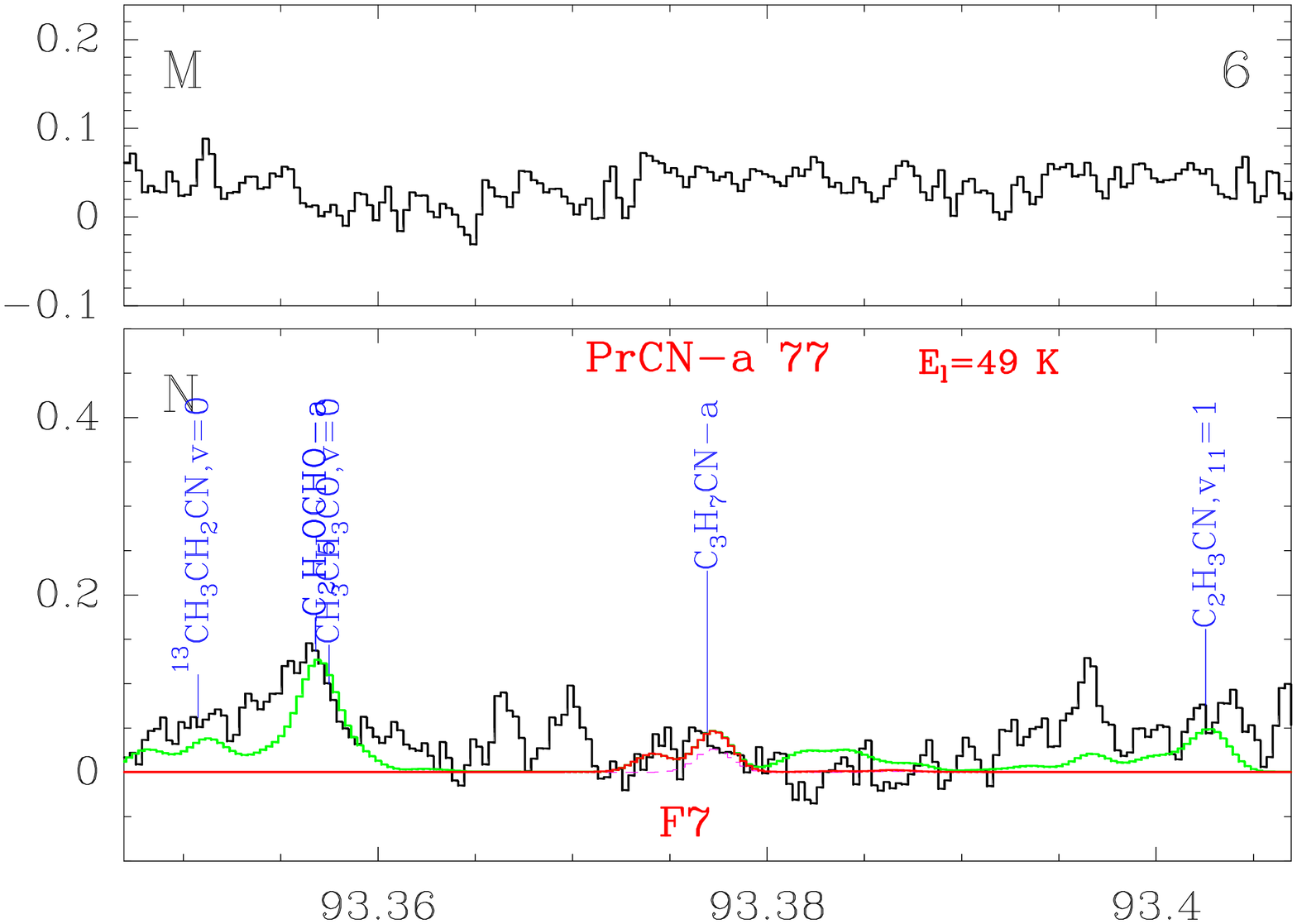}}}
\vspace*{-0.4ex}
\centerline{\resizebox{0.85\hsize}{!}{\includegraphics[angle=270]{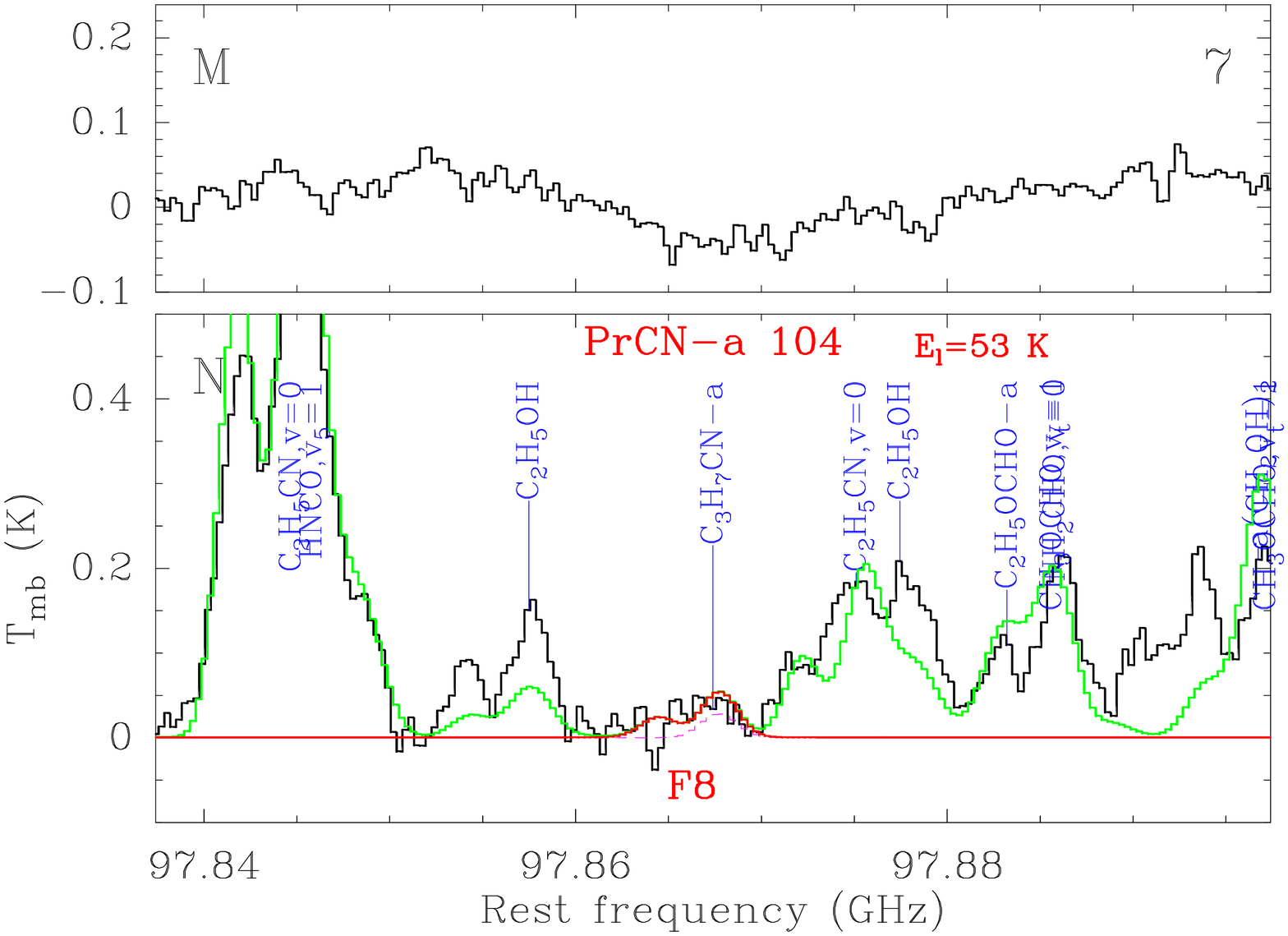}\includegraphics[angle=270]{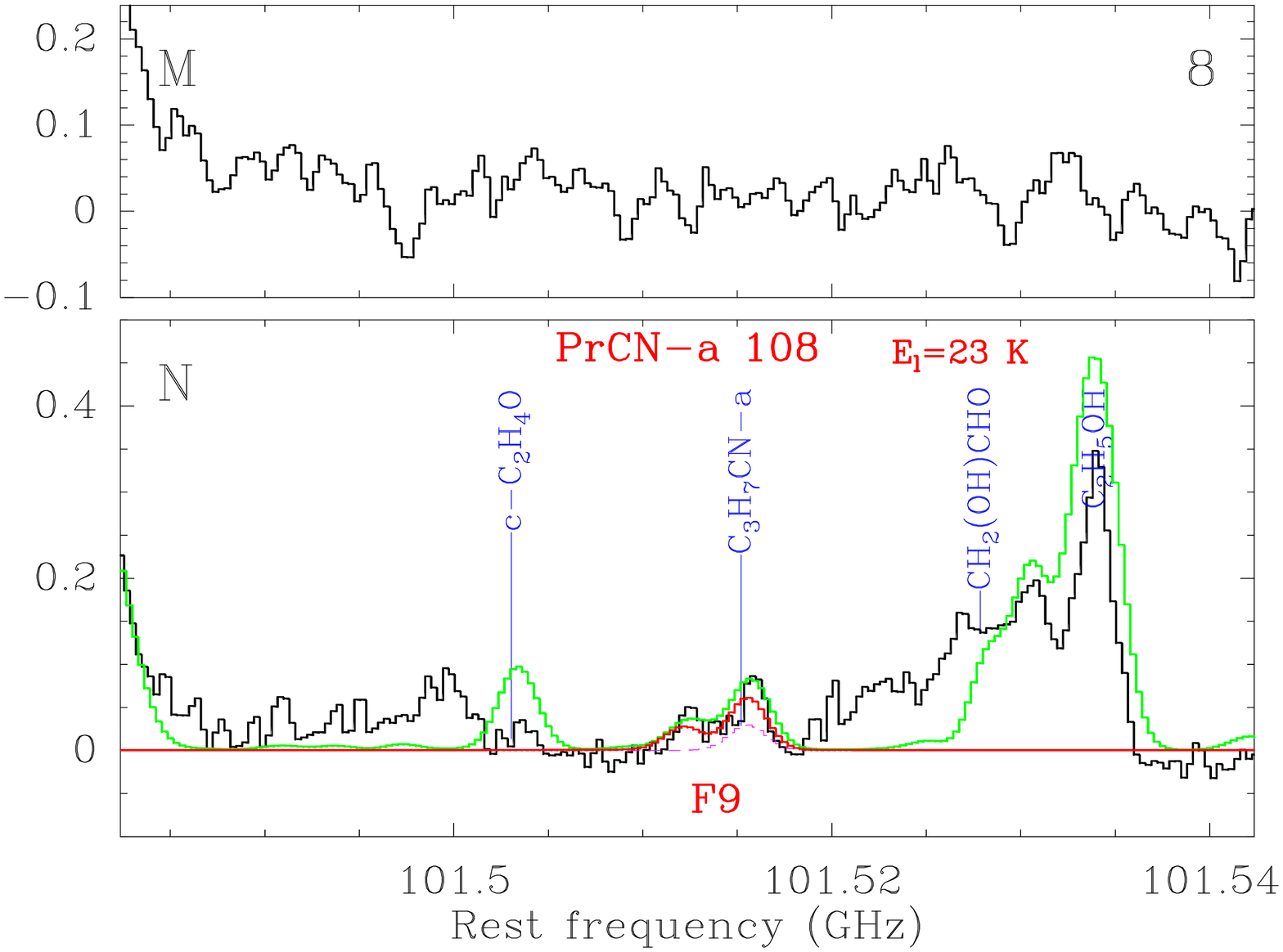}}}
\vspace*{-0.4ex}
\caption{
Transitions of the \textit{anti}-conformer of \textit{n}-propyl cyanide (PrCN-a) detected with the IRAM 30 m telescope.
Each panel consists of two plots and is labeled in black in the upper right corner.
The lower plot shows in black the spectrum obtained toward Sgr~B2(N) in main-beam brightness temperature scale (K), while the upper plot shows the spectrum toward Sgr~B2(M). The rest frequency axis is labeled in GHz. The systemic velocities assumed for Sgr~B2(N) and (M) are 64 and 62 km~s$^{-1}$, respectively.
The lines identified in the Sgr~B2(N) spectrum are labeled in blue. The top red label indicates the PrCN-a transition centered in each plot (numbered like in Col.~1 of Table~\ref{t:detectprcn-a}), along with the energy of its lower level in K ($E_l/k_{\mathrm{B}}$).
The other PrCN-a lines are labeled in blue only.
The bottom red label is the feature number (see Col.~8 of Table~\ref{t:detectprcn-a}).
The green spectrum shows our LTE model containing all identified molecules, including PrCN-a.
The LTE synthetic spectrum of PrCN-a alone is overlaid in red, and its opacity in dashed violet.
All observed lines which have no counterpart in the green spectrum are still unidentified in Sgr~B2(N).
}
\label{f:detectprcn-a}
\end{figure*}
\begin{figure*}

\centerline{\resizebox{0.85\hsize}{!}{\includegraphics[angle=270]{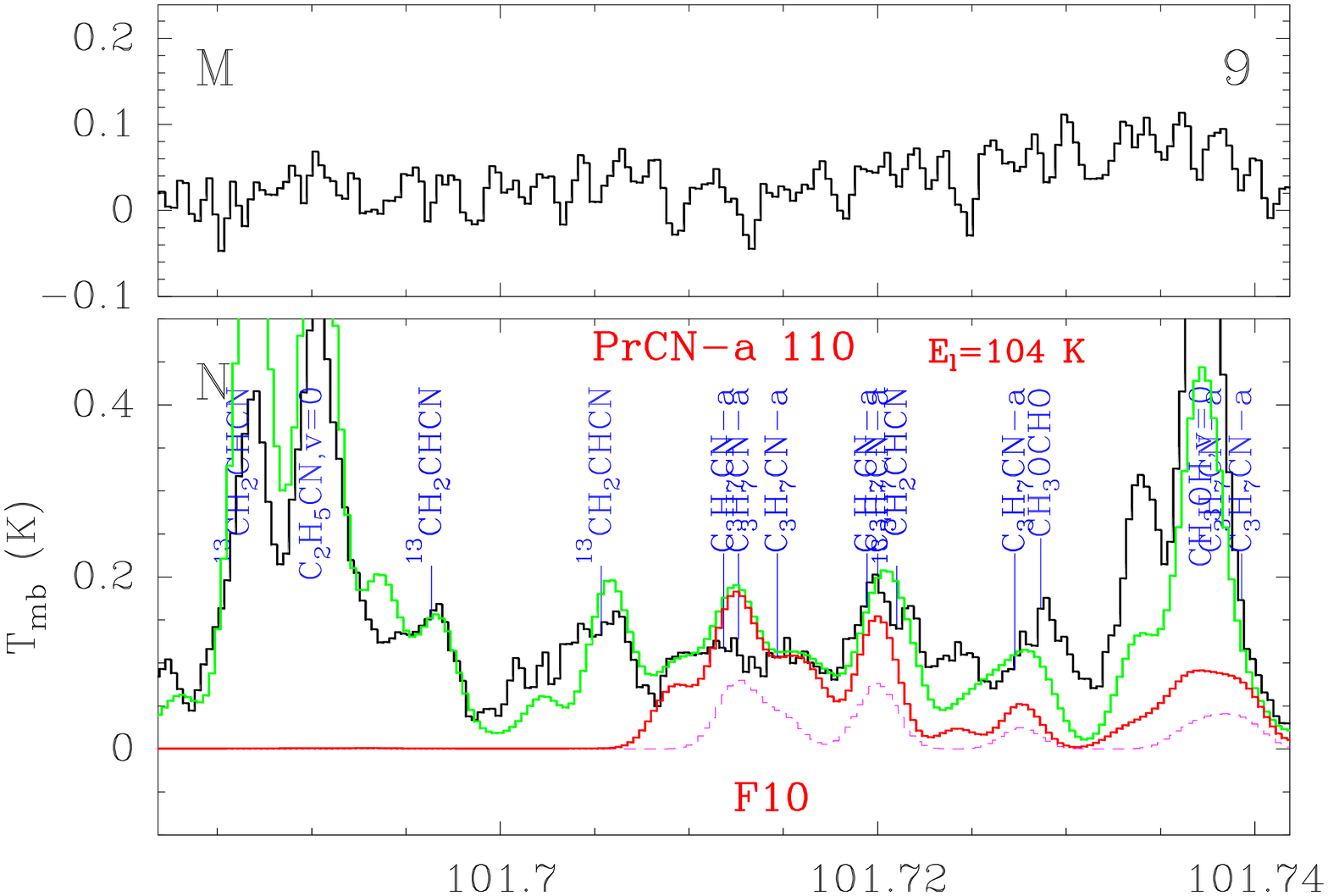}\includegraphics[angle=270]{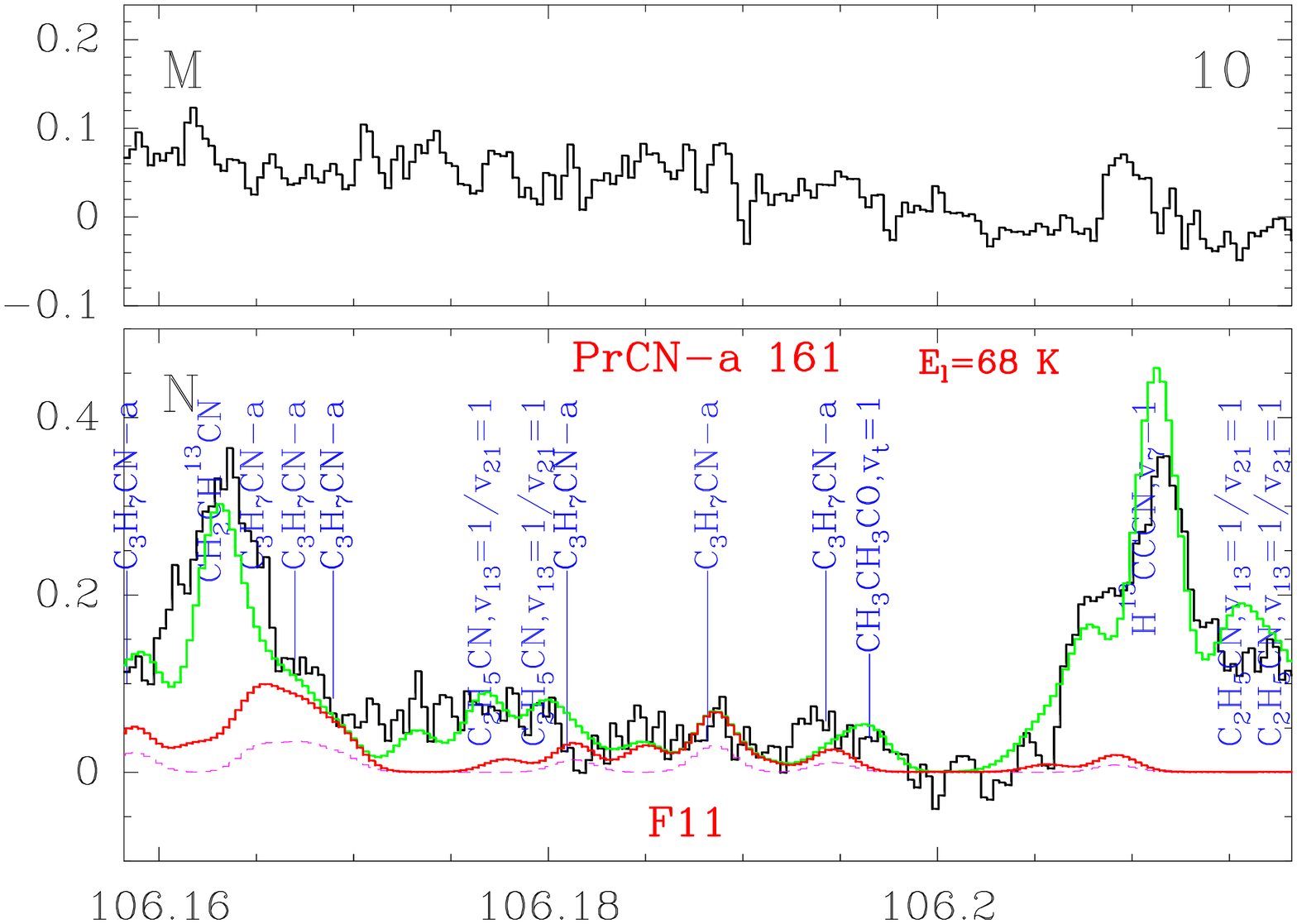}}}
\vspace*{-0.4ex}
\centerline{\resizebox{0.425\hsize}{!}{\includegraphics[angle=270]{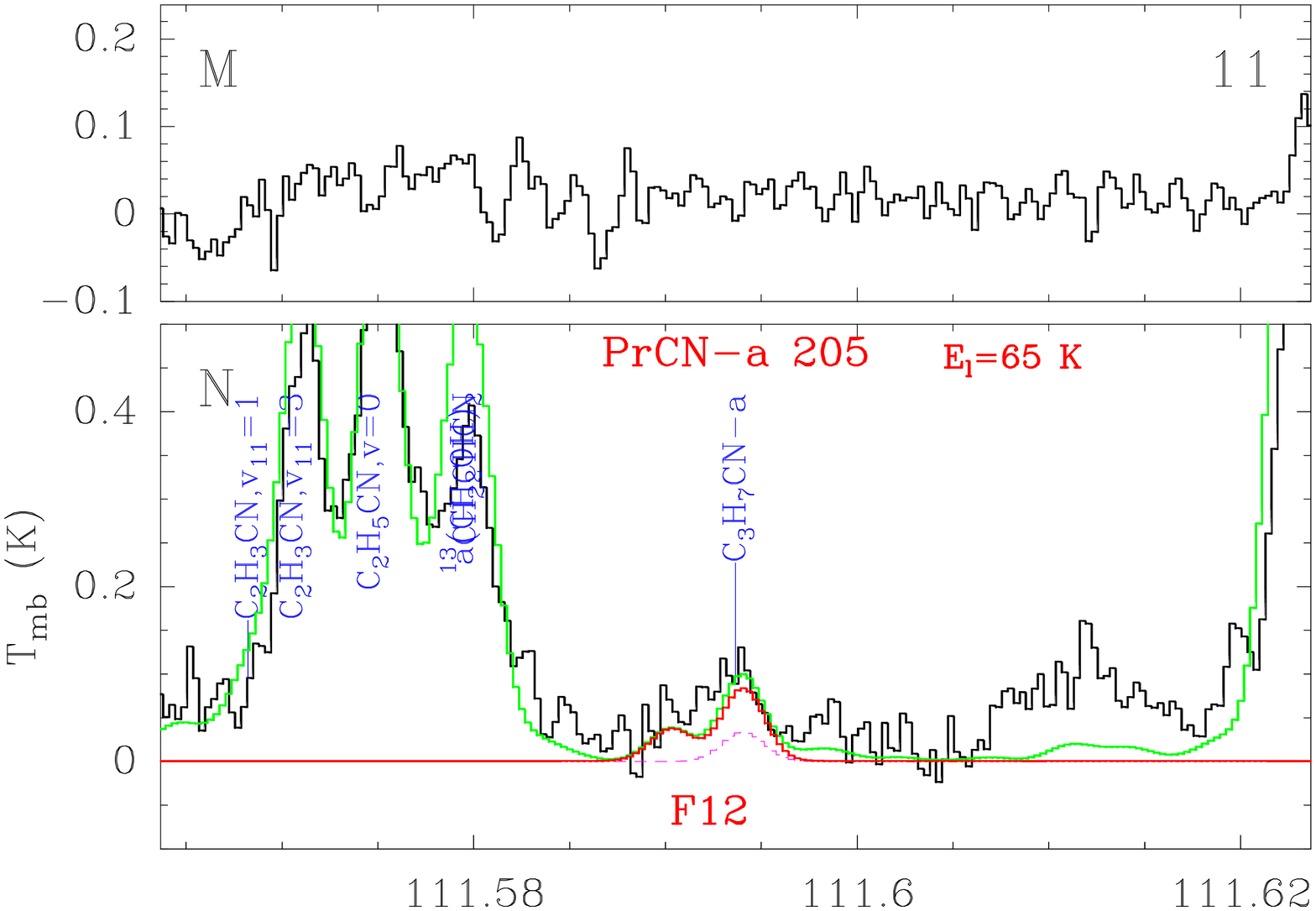}}}
\addtocounter{figure}{-1}
\caption{
(continued)
}
\label{f:detectprcn-a}
\end{figure*}

}


For the frequency range corresponding to each detected \textit{n}-propyl 
cyanide feature, we list in Table~\ref{t:detectprcn-a} the integrated 
intensities of the observed spectrum (Col.~10), of the best-fit model of 
\textit{n}-propyl cyanide (Col.~11), and of the best-fit model including all 
molecules (Col.~12). In these columns, 
the dash symbol indicates transitions belonging to the same feature. Columns 1 
to 7 of Table~\ref{t:detectprcn-a} are the same as in Table~\ref{t:prcn-a}. 
The $1\sigma$ uncertainty given for the integrated intensity in Col.~10 was 
computed using the estimated noise level of Col.~7. 

\begin{figure}
\centerline{\resizebox{1.0\hsize}{!}{\includegraphics[angle=270]{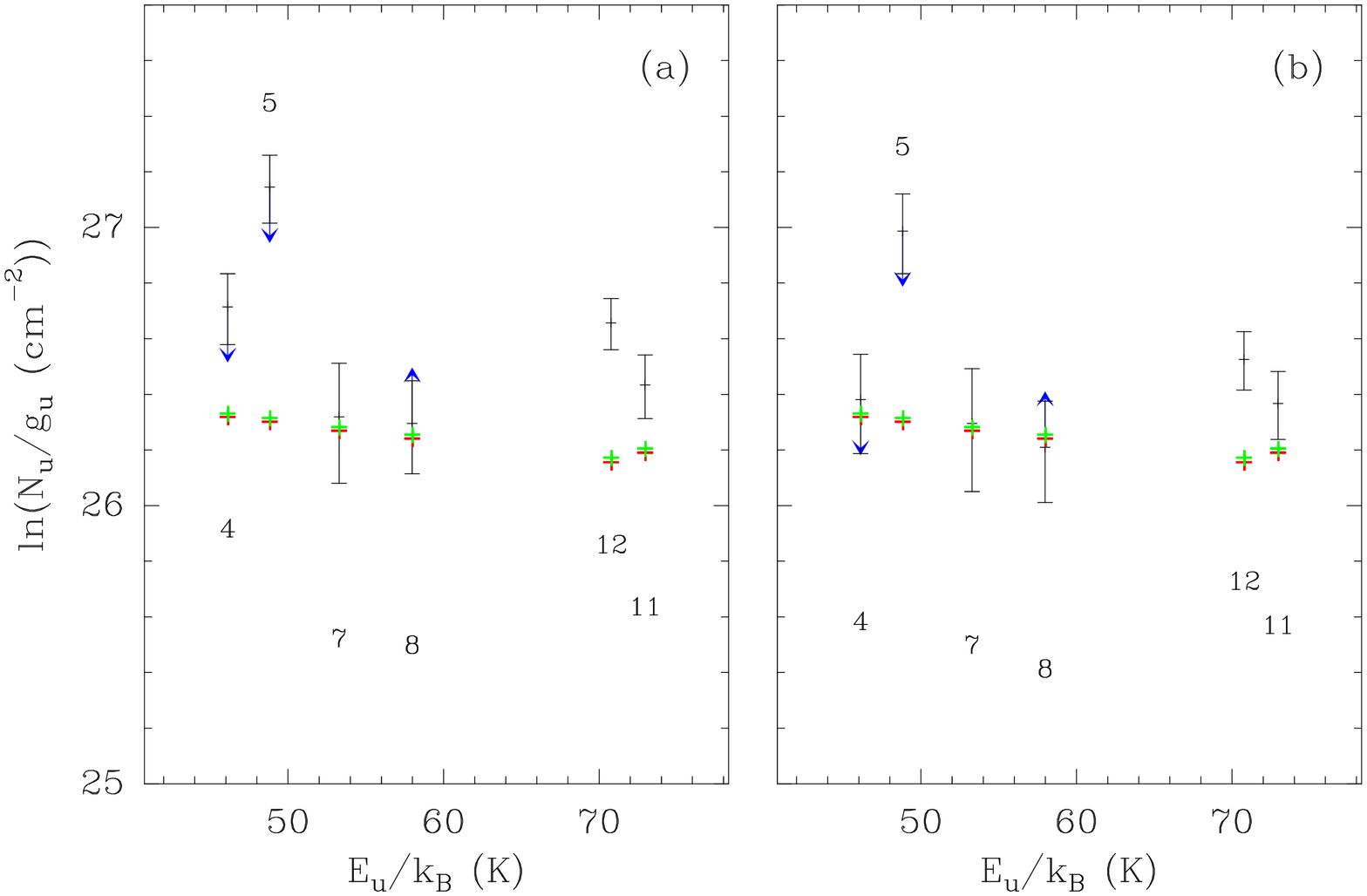}}}
\caption{Population diagram of the \textit{anti}-conformer of \textit{n}-propyl 
cyanide presented
in the same way as for ethyl formate in Fig.~\ref{f:popdiagetocho} (see the 
caption of that figure for details). Panel \textbf{a)} shows the population 
diagram derived from the measured integrated intensities while panel 
\textbf{b)} presents the population diagram after correction for the expected
contribution from contaminating molecules. Features
1, 2, 3, 6, 9, and 10 are blends of several transitions with different energy 
levels and were therefore omitted.}
\label{f:popdiagprcn}
\end{figure}

As we did for ethyl formate, we show in Fig.~\ref{f:popdiagprcn}a a population
diagram derived from the integrated intensities of the detected features of the 
\textit{anti}-conformer of \textit{n}-propyl cyanide. 
Figure~\ref{f:popdiagprcn}b 
displays the corresponding diagram after removing the expected contribution 
from contaminating molecules (see Sect.~\ref{ss:detetocho} for details).
This figure is less helpful than in the case of ethyl formate because all 
features containing several transitions (6 out of 12) have transitions with 
different energy levels and cannot be shown in a population diagram. 
Therefore, this diagram  does not help much for
the determination of the temperature. Feature 3, which is a blend of 
transitions with upper energy levels from 61 to 147~K, is however reasonably
well fitted by our 150~K model (see panel 2 of Fig.~\ref{f:detectprcn-a}) and
gives us some confidence in this high temperature. This is further confirmed
by the high temperatures measured in our survey for chemically related 
molecules (see Sect.~\ref{ss:compprcn} below). 

Our model for the \textit{anti}-conformer of \textit{n}-propyl cyanide 
consists of two components 
with different velocities. The need for a second component mainly comes from
the shape of features 2, 9, and 12. Its velocity is consistent with the 
velocity of the second component we find for many other, more abundant 
molecules in our survey toward Sgr~B2(N). It was shown interferometrically 
that this second velocity component is a physically distinct source located 
$\sim 5\arcsec$ to the North of the main hot core in Sgr~B2(N)
\citep[see, e.g., Sect.~3.4 of][]{Belloche08a}. Our data are consistent with
a second component about half as strong in \textit{n}-propyl cyanide as the 
first component (Table~\ref{t:prcnmodel}). This is also the ratio we found for 
the two components of ethyl cyanide (C$_2$H$_5$CN) with the IRAM Plateau de 
Bure interferometer and the 30~m telescope 
\citep[see Table~5 of][]{Belloche08b}. Finally, since all detected transitions 
are optically thin and the two regions emitting in \textit{n}-propyl 
cyanide are most likely compact given their high temperature, column density 
and source size are degenerated. We fixed the source size to 3$\arcsec$. This 
is approximately the size of the region emitting in the chemically
related molecule ethyl cyanide that we measured with the IRAM
Plateau de Bure interferometer \citep[see Table~5 of][]{Belloche08b}.

\begin{table}
 {\centering
 \caption{
 Parameters of our best-fit LTE model of \textit{n}-propyl cyanide with two velocity components.
}
 \label{t:prcnmodel}
 \vspace*{0.0ex}
 \begin{tabular}{ccccc}
 \hline\hline
 \multicolumn{1}{c}{Size$^{a}$} & \multicolumn{1}{c}{$T_{\mathrm{rot}}$$^{b}$} & \multicolumn{1}{c}{$N^{c}$} & \multicolumn{1}{c}{$\Delta V^{d}$} & \multicolumn{1}{c}{$V_{\mathrm{off}}$$^{e}$} \\ 
 \multicolumn{1}{c}{\scriptsize ($''$)} & \multicolumn{1}{c}{\scriptsize (K)} & \multicolumn{1}{c}{\scriptsize (cm$^{-2}$)} & \multicolumn{1}{c}{\scriptsize (km~s$^{-1}$)} & \multicolumn{1}{c}{\scriptsize (km~s$^{-1}$)} \\ 
 \multicolumn{1}{c}{(1)} & \multicolumn{1}{c}{(2)} & \multicolumn{1}{c}{(3)} & \multicolumn{1}{c}{(4)} & \multicolumn{1}{c}{(5)} \\ 
 \hline
 3.0 &  150 & $ 1.50 \times 10^{16}$ & 7.0 & -1.0 \\  3.0 &  150 & $ 6.60 \times 10^{15}$ & 7.0 & 9.0 \\  \hline
 \end{tabular}
 }\\[1ex] 
 Notes:
 $^a$ Source diameter (\textit{FWHM}).
 $^b$ Temperature.
 $^c$ Column density.
 $^d$ Linewidth (\textit{FWHM}).
 $^e$ Velocity offset with respect to the systemic velocity of Sgr~B2(N) V$_{\mathrm{lsr}} = 64$ km~s$^{-1}$.
 \end{table}

From this analysis, we conclude that our best-fit model for the 
\textit{anti}-conformer of \textit{n}-propyl cyanide is fully consistent with 
our 30~m data of Sgr~B2(N). This is, to our knowledge, the first clear 
detection of this molecule in space\footnote{\citet{Jones07} tentatively 
identified two lines detected with the Australia Telescope Compact Array at 
$\sim$86.9556 and $\sim$90.0560~GHz as transitions of the 
\textit{gauche}-conformer of \textit{n}-propyl cyanide. However, our model 
predicts a peak temperature of the \textit{n}-propyl cyanide transition at 
86.955466~GHz 15 times smaller than the peak temperature (0.13~K) of the line 
detected with the 30~m telescope at this frequency. The tentative 
identification of \citet{Jones07} at this frequency is therefore not confirmed. 
The origin of this line in our survey is still unknown. As far as the other 
transition is concerned, our model of \textit{n}-propyl cyanide predicts a peak 
intensity equal to only one quarter of the peak intensity (0.07~K) of the line 
detected with the 30~m telescope at $\sim$90.0560~GHz. Since this line is 
blended with a transition of $^{13}$CH$_3$CH$_2$CN that has a stronger 
contribution according to our modeling, the tentative identification of 
\citet{Jones07} should be viewed with caution too.}.

No feature of the \textit{gauche}-conformer of \textit{n}-propyl cyanide is 
clearly 
detected in our spectral survey of Sgr~B2(N). Only one feature at 211.4~GHz is 
possibly detected but the baseline in this frequency range is very uncertain 
and this feature is blended with a transition of acetone. If we consider this 
feature as a detection, it implies a column density a factor 2 smaller than 
for the model of the \textit{anti}-conformer, which may suggest a non-thermal 
distribution of the molecules in the two conformers. However, we first have to 
evaluate the uncertainty on the ratio of the \textit{anti}- and 
\textit{gauche}-conformer populations coming from the uncertainty on their 
energy difference ($\Delta E = 92 \pm 25$~cm$^{-1}$, see 
Sect.~\ref{ss:freqprcn}). For $\Delta E = 92$~cm$^{-1}$, the \textit{anti} to 
\textit{gauche} ratio is $0.51/0.49$ at 150~K, 
and increases to $0.57/0.43$ for $\Delta E = 117$~cm$^{-1}$, i.e. a variation
of $\sim 30 \%$. This is not enough to explain the factor 2 mentioned 
above, but it can have a significant contribution. Above all, the 
uncertainty on the baseline level at 211.4~GHz is quite large and the data are 
still consistent with a thermal distribution of the \textit{gauche}- and 
\textit{anti}-conformers.

\subsection{Upper limit in Sgr~B2(M)}
\label{ss:b2mprcn}

We do not detect \textit{n}-propyl cyanide in our spectral survey toward 
Sgr~B2(M). Using the same source size, linewidth, and temperature as for 
Srg~B2(N) (see Table~\ref{t:prcnmodel}), we find a $\sim 3 \sigma$ column 
density upper limit of $6 \times 10^{15}$~cm$^{-2}$ in the LTE approximation 
for both conformers. The column density of \textit{n}-propyl cyanide is thus 
at least a factor $\sim 2$ lower toward Sgr~B2(M) than toward Sgr~B2(N), 
which is again consistent with the results of, e.g., \citet{Nummelin00} for 
other molecules.

\subsection{Comparison to related species}
\label{ss:compprcn}

We easily detect the already known molecules methyl cyanide (CH$_3$CN) and 
ethyl cyanide (C$_2$H$_5$CN) in our survey toward Sgr~B2(N) 
\citep[see also, e.g.,][]{Miao95,Liu99,Nummelin00}. The parameters of 
our current best fit models of these two molecules are listed in 
Table~\ref{t:xcnmodel}. Our models use
also constraints from the weaker isotopologues containing $^{13}$C 
\citep[see, e.g.,][]{Mueller08}. The source
size is constrained by the optically thick transitions, once the temperature
has been fitted. For ethyl cyanide, we used in addition the constraints on the
source size derived from our high angular resolution observations with the IRAM
Plateau de Bure interferometer \citep[see Table~5 of][]{Belloche08b}. The first 
two velocity components detected in methyl cyanide and ethyl cyanide 
correspond to the two hot cores embedded in Sgr~B2(N) 
\citep[see, e.g.,][]{Hollis03,Belloche08a}. They are also seen in 
\textit{n}-propyl cyanide. In addition, methyl cyanide and ethyl cyanide show 
a third component that may arise from the blueshifted lobe of an outflow 
\citep[see the cyanoacetylene $\varv_7 = 1$ emission in Fig.5k 
to m of][]{Belloche08a}. The redshifted counterpart is blended with the
northern component in the single-dish beam \citep[see Fig.~3 of][]{Hollis03}. 
The third velocity component is too faint to be detected in \textit{n}-propyl 
cyanide. 

The model parameters for the compact sources listed for methyl cyanide in 
Table~\ref{t:xcnmodel} are mostly based on the $^{13}$C isotopologues with a 
$^{12}$C/$^{13}$C isotopic ratio of 20 because the transitions of the $^{12}$C 
main isotopologue are very optically thick and most likely dominated by large 
scale emission \citep[see maps of, e.g.,][]{deVicente97,Jones08}. 
\citet{deVicente97} analysed their maps of methyl cyanide emission in the
Large Velocity Gradient approximation. They found that the emission consists
of several components (hot core, warm envelope, diffuse and hot envelope), and 
mentioned that their modeling toward Sgr~B2(N) is uncertain because of the 
large opacities. However, their figure 5 suggests that the temperature and 
column density of methyl cyanide are strongly centrally peaked toward 
Sgr~B2(N). Therefore, the emission of the optically thin $^{13}$C 
isotopologues should be dominated by the compact hot cores which gives us some 
confidence (within a factor of 2) in the column densities listed in 
Table~\ref{t:xcnmodel}. \citet{Friedel04} measured similar intensities for 
CH$_3$$^{13}$CN with the NRAO 12~m telescope and the BIMA interferometer 
toward Sgr~B2(N), an additional evidence that the compact hot cores dominate 
the emission of the $^{13}$C isotopologues we detected with the 30~m telescope. 
For a source size of 2.7$\arcsec$, \citet{Nummelin00} found column densities of 
$0.7-1.1\times 10^{17}$~cm$^{-2}$ for the $^{13}$C isotopologues, which 
translates into a column density of $1.4-2.2 \times 10^{18}$~cm$^{-2}$ for
the main isotopologue assuming a $^{12}$C/$^{13}$C isotopic ratio of 20. This
is in very good agreement with our result (see Table~\ref{t:xcnmodel}).

\begin{table}
 {\centering
 \caption{
 Parameters of our best-fit LTE models of methyl cyanide, ethyl cyanide, vinyl cyanide, and aminoacetonitrile, and column density upper limit for cyanomethylidyne.
}
 \label{t:xcnmodel}
 \vspace*{0.0ex}
 \begin{tabular}{lccccc}
 \hline\hline
 \multicolumn{1}{c}{Molecule$^{a}$} & \multicolumn{1}{c}{Size$^{b}$} & \multicolumn{1}{c}{$T_{\mathrm{rot}}$$^{c}$} & \multicolumn{1}{c}{$N^{d}$} & \multicolumn{1}{c}{$\Delta V^{e}$} & \multicolumn{1}{c}{$V_{\mathrm{off}}$$^{f}$} \\ 
  & \multicolumn{1}{c}{\scriptsize ($''$)} & \multicolumn{1}{c}{\scriptsize (K)} & \multicolumn{1}{c}{\scriptsize (cm$^{-2}$)} & \multicolumn{1}{c}{\scriptsize (km~s$^{-1}$)} & \multicolumn{1}{c}{\scriptsize (km~s$^{-1}$)} \\ 
 \multicolumn{1}{c}{(1)} & \multicolumn{1}{c}{(2)} & \multicolumn{1}{c}{(3)} & \multicolumn{1}{c}{(4)} & \multicolumn{1}{c}{(5)} & \multicolumn{1}{c}{(6)} \\ 
 \hline
\noalign{\smallskip} CH$_3$CN & 2.7 &  200 & $ 2.00 \times 10^{18}$ & 6.5 & -1.0 \\  & 2.7 &  200 & $ 8.00 \times 10^{17}$ & 6.5 & 9.0 \\  & 2.7 &  200 & $ 1.00 \times 10^{17}$ & 8.0 & -11.0 \\ \noalign{\smallskip} C$_2$H$_5$CN & 3.0 &  170 & $ 1.20 \times 10^{18}$ & 6.5 & -1.0 \\  & 2.3 &  170 & $ 1.40 \times 10^{18}$ & 6.5 & 9.0 \\  & 1.7 &  170 & $ 9.00 \times 10^{17}$ & 8.0 & -11.0 \\ \noalign{\smallskip} C$_2$H$_3$CN & 2.3 &  170 & $ 8.00 \times 10^{17}$ & 7.0 & -1.0 \\  & 2.3 &  170 & $ 2.40 \times 10^{17}$ & 7.0 & 9.0 \\  & 2.3 &  170 & $ 1.00 \times 10^{17}$ & 10.0 & -9.0 \\ \noalign{\smallskip} NH$_2$CH$_2$CN & 2.0 &  100 & $ 2.80 \times 10^{16}$ & 7.0 & 0.0 \\ \noalign{\smallskip}\hline\noalign{\smallskip} CCN$^g$ & 3.0 &  200 & $ < 1.20 \times 10^{17}$ & 6.5 & -1.0 \\  & 3.0 &  200 & $ < 1.20 \times 10^{17}$ & 6.5 & 9.0 \\  \hline
 \end{tabular}
 }\\[1ex] 
 Notes:
 $^a$ We used the JPL entry for CH$_3$CN (version 3), and the CDMS entries for C$_2$H$_5$CN (ver. 1), C$_2$H$_3$CN (ver. 1), NH$_2$CH$_2$CN (ver. 1), and CCN (ver. 1). See references to the laboratory data therein.
 $^b$ Source diameter (\textit{FWHM}).
 $^c$ Temperature.
 $^d$ Column density.
 $^e$ Linewidth (\textit{FWHM}).
 $^f$ Velocity offset with respect to the systemic velocity of Sgr~B2(N) V$_{\mathrm{lsr}} = 64$ km~s$^{-1}$.
 $^g$ The column density upper limit is $\sim 3\sigma$. The other parameters were fixed.
 \end{table}

Assuming a temperature of 200~K and optically thin emission, \citet{Liu01} 
obtained a beam-averaged column density of 
$4.63 \pm 0.14 \times 10^{16}$~cm$^{-2}$ for ethyl cyanide with BIMA at 89.6~GHz 
($HPBW = 14\arcsec \times 4\arcsec$). For a source size of $3\arcsec$, this 
translates into a column density of $2.9 \times 10^{17}$~cm$^{-2}$, which is
a factor 4 smaller than the column density we derive for the main velocity
component. However, our model predicts peak line opacities of 4--6 for these 
transitions, which is supported by our simultaneous modeling of the $^{13}$C 
isotopologues of ethyl cyanide \citep[see][]{Mueller08}. As a result, 
\citet{Liu01} most likely underestimated the column densities of ethyl cyanide 
by a factor of a few, which reconciles the single-dish and interferometric
measurements and confirms that the source of ethyl cyanide emission is compact.
This is also confirmed by the reasonable agreement between the 30~m and 
Plateau de Bure Interferometer fluxes published by \citet{Belloche08b} at 
81.7~GHz (see their Table 5). The compactness of the source of ethyl cyanide
emission most likely explains the discrepancy with the column density found 
by \citet{Nummelin00} with SEST in the 1.3~mm wavelength range 
($HPBW \sim 23\arcsec$). These authors derived temperatures of 
$175^{+25}_{-20}$~K and $210^{+30}_{-30}$~K and beam-averaged column densities of 
$1.6^{+0.2}_{-0.1} \times 10^{15}$~cm$^{-2}$ and 
$1.5^{+0.4}_{-0.3} \times 10^{16}$~cm$^{-2}$ for the ethyl cyanide $a$- and
$b$-type lines, respectively. While they find an order of magnitude difference 
between the column densities of the $a$- and $b$-type lines, we successfully 
reproduce the ethyl cyanide emission in our 3~mm survey with a single model 
for the two types of lines, the former being optically thick while the latter 
are optically thin. Our model with a small source size predicts 
line opacities on the order of 
10--30 for the $a$-type lines in the 1.3~mm range. Hence, we believe 
that the column 
density derived by  \citet{Nummelin00} for these lines at 1.3~mm is 
underestimated by a large factor because they assumed a beam filling factor of 
1, yielding opacities for these lines that were too low. On the 
other hand, since our model predicts opacities $\la 1$ for 
the $b$-type lines at 1.3~mm, we would expect the column density 
derived by 
these authors to match ours. For a source size of $3\arcsec$, their column 
density of the $b$-type lines translates into a column density of 
$9.0 \times 10^{17}$~cm$^{-2}$, which is about a factor 2 smaller than the 
sum of the column densities of the two main velocity components in 
Table~\ref{t:xcnmodel} (after rescaling the second one to a source size of 
$3\arcsec$). As in Sect.~\ref{ss:competocho}, we think that the discrepancy 
arises from the uncertain baseline level and the partial dust 
absorption in the 
1.3~mm wavelength range. Our current model, which suffers from the same 
problems, also over-predicts intensities for the lines detected in our 
partial 1.3~mm survey.

After rescaling to the same size of 3$\arcsec$, the relative column densities
of the three related molecules \hbox{CH$_3$CN / C$_2$H$_5$CN / C$_3$H$_7$CN} 
are \hbox{108 / 80 / 1} for the first velocity component and 
\hbox{98 / 125 / 1} for the second velocity 
component. We discuss these ratios and the implications for the interstellar 
chemistry in Sect.~\ref{s:chemistry}.

In addition, we list in Table~\ref{t:xcnmodel} the best-fit parameters we found
for vinyl cyanide \citep[][]{Mueller08} and aminoacetonitrile 
\citep[][]{Belloche08a}, as well as an upper limit for the column 
density of cyanomethylidyne (CCN) for which the other parameters were fixed.

\section{Chemical modeling and discussion}
\label{s:chemistry}

To better understand the observational results, we model the chemistry of 
Sgr~B2(N) using a coupled gas-phase and grain-surface chemical code. 
\citet{Garrod08a} constructed a reaction network to account for the 
grain-surface formation of many complex molecules observed in hot cores. 
Surface formation was assumed to occur primarily by the addition of 
functional-group radicals derived from molecular ices or from other molecules 
formed in this way. Such reactions are viable when larger radicals 
become mobile at 
intermediate grain temperatures ($T_{d}\gtrsim 20$~K), achieved during the 
warm-up to typical hot-core temperatures ($>100$~K). The network also includes 
destruction mechanisms for all complex species, consisting of neutral--neutral 
reactions on the grain surfaces, ion--molecule reactions with simple ions in 
the gas phase, and cosmic ray-induced photodissociation both in the gas phase 
and on the grains. To this network we have added appropriate formation and 
destruction mechanisms for ethyl formate, ethyl and \textit{n}-propyl cyanide, 
and also the recently identified aminoacetonitrile 
\citep[NH$_2$CH$_2$CN,][]{Belloche08a,Belloche08b}, whose surface formation 
routes may be 
similar to the other cyanides. In addition, surface hydrogenation routes have 
been added to allow for the full hydrogenation of the carbon chains 
C$_3$ and 
C$_4$, which was not previously considered, as well as the associated 
hydrogenated species and their destruction channels. The techniques used to 
construct the new reaction set are presented in detail by \citet{Garrod08a}; 
the current model may be regarded as a consistent extension to that network.

We employ the single-point physical model used by \citet{Garrod06}, in 
which the isothermal collapse of a diffuse medium, to a density 
$n_{H}=10^{7}$~cm$^{-3}$, is followed by a warm-up from 10 to 
200~K. Their $T_{2}$ warm-up profile is assumed, in which the hot-core 
temperature has a $t^2$ dependence on the time, $t$, elapsed in the warm-up 
phase. Dust and gas temperatures are assumed to be well coupled, hence 
we let $T=T_{K}=T_{dust}$. The warm-up timescale is representative of the time 
required for a parcel of gas to achieve a temperature of 200~K, as the hot core 
forms; it therefore does not relate directly to the {\em current} infall 
timescale.

This model traces the evolution of the chemistry up to a temperature of 
200~K, associated with the central hot-core region. However, these 
time-dependent results may also be considered to represent differing 
spatial extents from the hot-core center, with the innermost regions being 
the most evolved and achieving the highest temperatures. As such, the 
time-dependent abundance profiles presented below also indicate a snapshot 
of the chemistry through the hot core.

Since we are interested mainly in specific features of the model, we choose 
not to fix the ice composition prior to the warm-up phase, but use the 
unadulterated composition computed in the collapse-phase.

Other details of the model may be found in \citet{Garrod08a}. One important 
difference is the removal, in keeping with prior chemical networks, of the 
activation energy barrier for the surface reaction 
OH + H$_2$CO $\rightarrow$ HCO + H$_2$O. Garrod et al. 
employed an activation energy merely for consistency with other 
hydrogen-abstraction reactions of OH. The available evidence, however, suggests 
there is no barrier\footnote{See the chemical kinetics database of the 
National Institute of Standards and Technology (NIST), 
\textit{http://kinetics.nist.gov/kinetics}.}. This change makes 
HCO radicals somewhat more abundant on the grains, tending to increase the 
final abundances of species such as methyl formate, which is 
consistent with our observational results.

\subsection{Surface Chemistry}
\label{ss:chemistry}

Surface chemical routes for the formation of methyl cyanide, CH$_3$CN, were 
already present in the \citet{Garrod08a} network, including direct addition 
of methyl and nitrile groups, and repetitive surface hydrogenation of gas 
phase-produced C$_2$N. Formation of ethyl cyanide, C$_2$H$_5$CN, 
was limited to repetitive surface hydrogenation of cyanoacetylene HC$_3$N and 
vinyl cyanide, C$_2$H$_3$CN, both of which may be formed in the gas phase. 
\textit{n}-Propyl cyanide and aminoacetonitrile were not present at all.

Table~\ref{t:chem-tab1} shows the full set of surface reactions employed in the 
current model to form methyl cyanide, ethyl cyanide, \textit{n}-propyl cyanide, 
aminoacetonitrile, and ethylformate, as well as a selection of significant 
cosmic ray-induced photodissociation processes that may occur on grain surfaces.
(The same CR-induced processes are assumed also to occur in the gas phase, at 
the same rates). A cosmic-ray ionization rate of 
$\zeta_{0} = 1.3 \times 10^{-17}$ s$^{-1}$ is assumed.

\begin{table}
\caption[]{Surface reactions and cosmic-ray induced surface
photodissociation processes related to the 
formation of cyanides, and ethyl formate.}
\label{t:chem-tab1}
\begin{center}
\begin{tabular}{lccc}
\hline
\hline
\noalign{\smallskip}
Reaction  & \hspace*{-5ex} Garrod et al.  & \hspace*{-2.5ex} \textit{Basic} & \hspace*{-2.5ex} \textit{Select}  \\
          & \hspace*{-5ex} (2008)         & \hspace*{-2.5ex} model & \hspace*{-2.5ex} model   \\
\noalign{\smallskip}
\hline
\noalign{\smallskip}
C$_2$N       +  H      $\rightarrow$  HCCN         & \hspace*{-5ex} $\bullet$ & \hspace*{-2.5ex} $\bullet$ & \hspace*{-2.5ex} $\bullet$  \\
\noalign{\smallskip}
HCCN         +  H      $\rightarrow$  CH$_2$CN     & \hspace*{-5ex} $\bullet$ & \hspace*{-2.5ex} $\bullet$ & \hspace*{-2.5ex} $\bullet$  \\
\noalign{\smallskip}
CH$_2$CN     +  H      $\rightarrow$  CH$_3$CN     & \hspace*{-5ex} $\bullet$ & \hspace*{-2.5ex} $\bullet$ & \hspace*{-2.5ex} $\bullet$  \\
\noalign{\smallskip}
\noalign{\smallskip}
\noalign{\smallskip}
HC$_3$N      +  H      $\rightarrow$  C$_2$H$_2$CN (1210~K) & \hspace*{-5ex} $\bullet$ & \hspace*{-2.5ex} $\bullet$ &    \\
\noalign{\smallskip}
C$_2$H$_2$CN +  H      $\rightarrow$  C$_2$H$_3$CN & \hspace*{-5ex} $\bullet$ & \hspace*{-2.5ex} $\bullet$ &    \\
\noalign{\smallskip}
C$_2$H$_3$CN +  H      $\rightarrow$  C$_2$H$_4$CN (750~K)  & \hspace*{-5ex} $\bullet$ & \hspace*{-2.5ex} $\bullet$ &    \\
\noalign{\smallskip}
C$_2$H$_4$CN +  H      $\rightarrow$  C$_2$H$_5$CN & \hspace*{-5ex} $\bullet$ & \hspace*{-2.5ex} $\bullet$ & \hspace*{-2.5ex} $\bullet$  \\
\noalign{\smallskip}
\noalign{\smallskip}
\noalign{\smallskip}
C$_3$H$_6$CN +  H      $\rightarrow$  C$_3$H$_7$CN &            & \hspace*{-2.5ex} $\bullet$ & \hspace*{-2.5ex} $\bullet$  \\
\noalign{\smallskip}
\noalign{\smallskip}
\noalign{\smallskip}
CH$_2$       +  CN             $\rightarrow$  CH$_2$CN     & \hspace*{-5ex} $\bullet$ & \hspace*{-2.5ex} $\bullet$ & \hspace*{-2.5ex} $\bullet$  \\
\noalign{\smallskip}
CH$_3$       +  CN             $\rightarrow$  CH$_3$CN     & \hspace*{-5ex} $\bullet$ & \hspace*{-2.5ex} $\bullet$ & \hspace*{-2.5ex} $\bullet$  \\
\noalign{\smallskip}
CH$_2$       +  CH$_2$CN       $\rightarrow$  C$_2$H$_4$CN &           & \hspace*{-2.5ex} $\bullet$ & \hspace*{-2.5ex} $\bullet$  \\
\noalign{\smallskip}
CH$_3$       +  CH$_2$CN       $\rightarrow$  C$_2$H$_5$CN &           & \hspace*{-2.5ex} $\bullet$ & \hspace*{-2.5ex} $\bullet$  \\
\noalign{\smallskip}
CH$_2$       +   C$_2$H$_4$CN  $\rightarrow$  C$_3$H$_6$CN &           & \hspace*{-2.5ex} $\bullet$ & \hspace*{-2.5ex} $\bullet$  \\
\noalign{\smallskip}
CH$_3$       +   C$_2$H$_4$CN  $\rightarrow$  C$_3$H$_7$CN &           & \hspace*{-2.5ex} $\bullet$ & \hspace*{-2.5ex} $\bullet$  \\
\noalign{\smallskip}
\noalign{\smallskip}
\noalign{\smallskip}
C$_2$H$_5$       +  CN         $\rightarrow$  C$_2$H$_5$CN &           & \hspace*{-2.5ex} $\bullet$ &    \\
\noalign{\smallskip}
C$_3$H$_7$       +  CN         $\rightarrow$  C$_3$H$_7$CN &           & \hspace*{-2.5ex} $\bullet$ &    \\
\noalign{\smallskip}
\noalign{\smallskip}
\noalign{\smallskip}
NH      +  CH$_2$CN   $\rightarrow$  NHCH$_2$CN     &       & \hspace*{-2.5ex} $\bullet$ & \hspace*{-2.5ex} $\bullet$  \\
\noalign{\smallskip}
NH$_2$  +  CH$_2$CN   $\rightarrow$  NH$_2$CH$_2$CN &       & \hspace*{-2.5ex} $\bullet$ & \hspace*{-2.5ex} $\bullet$  \\
\noalign{\smallskip}
H       +  NHCH$_2$CN $\rightarrow$  NH$_2$CH$_2$CN &       & \hspace*{-2.5ex} $\bullet$ & \hspace*{-2.5ex} $\bullet$  \\
\noalign{\smallskip}
\noalign{\smallskip}
\noalign{\smallskip}
CH$_2$NH      +  CN   $\rightarrow$  NHCH$_2$CN     &       & \hspace*{-2.5ex} $\bullet$ &   \\
\noalign{\smallskip}
CH$_2$NH$_2$  +  CN   $\rightarrow$  NH$_2$CH$_2$CN &       & \hspace*{-2.5ex} $\bullet$ &   \\
\noalign{\smallskip}
\noalign{\smallskip}
\noalign{\smallskip}
CH$_3$CN     +  $h \nu$  $\rightarrow$ CH$_2$CN     + H   &           & \hspace*{-2.5ex} $\bullet$ & \hspace*{-2.5ex} $\bullet$  \\
\noalign{\smallskip}
CH$_3$CN     +  $h \nu$  $\rightarrow$ CH$_3$       + CN  & \hspace*{-5ex} $\bullet$ & \hspace*{-2.5ex} $\bullet$ & \hspace*{-2.5ex} $\bullet$  \\
\noalign{\smallskip}
\noalign{\smallskip}
\noalign{\smallskip}
C$_2$H$_5$CN +  $h \nu$  $\rightarrow$ C$_2$H$_4$CN + H   &           & \hspace*{-2.5ex} $\bullet$ & \hspace*{-2.5ex} $\bullet$  \\
\noalign{\smallskip}
C$_2$H$_5$CN +  $h \nu$  $\rightarrow$ CH$_3$ + CH$_2$CN  &           & \hspace*{-2.5ex} $\bullet$ & \hspace*{-2.5ex} $\bullet$  \\
\noalign{\smallskip}
C$_2$H$_5$CN +  $h \nu$  $\rightarrow$ C$_2$H$_5$   + CN  & \hspace*{-5ex} $\bullet$ & \hspace*{-2.5ex} $\bullet$ & \hspace*{-2.5ex} $\bullet$  \\
\noalign{\smallskip}
\noalign{\smallskip}
\noalign{\smallskip}
C$_3$H$_7$CN +  $h \nu$  $\rightarrow$ CH$_3$ + C$_2$H$_4$CN &           & \hspace*{-2.5ex} $\bullet$ & \hspace*{-2.5ex} $\bullet$  \\
\noalign{\smallskip}
C$_3$H$_7$CN +  $h \nu$  $\rightarrow$ C$_2$H$_5$ + CH$_2$CN &           & \hspace*{-2.5ex} $\bullet$ & \hspace*{-2.5ex} $\bullet$  \\
\noalign{\smallskip}
C$_3$H$_7$CN +  $h \nu$  $\rightarrow$ C$_3$H$_7$ + CN       &           & \hspace*{-2.5ex} $\bullet$ & \hspace*{-2.5ex} $\bullet$  \\
\noalign{\smallskip}
\noalign{\smallskip}
\noalign{\smallskip}
NH$_2$CH$_2$CN + $h \nu$ $\rightarrow$ NH$_2$ + CH$_2$CN     &           & \hspace*{-2.5ex} $\bullet$ & \hspace*{-2.5ex} $\bullet$  \\
\noalign{\smallskip}
NH$_2$CH$_2$CN + $h \nu$ $\rightarrow$ NH$_2$CH$_2$ + CN     &           & \hspace*{-2.5ex} $\bullet$ & \hspace*{-2.5ex} $\bullet$  \\
\noalign{\smallskip}
\noalign{\smallskip}
\hline
\noalign{\smallskip}
\noalign{\smallskip}
CH$_3$  +  CH$_2$OCHO  $\rightarrow$  C$_2$H$_5$OCHO &           & \hspace*{-2.5ex} $\bullet$ & \hspace*{-2.5ex} $\bullet$ \\
\noalign{\smallskip}
H       +  CH$_2$OCHO  $\rightarrow$  CH$_3$OCHO &           & \hspace*{-2.5ex} $\bullet$ & \hspace*{-2.5ex} $\bullet$ \\
\noalign{\smallskip}
HCO     +  C$_2$H$_5$O $\rightarrow$  C$_2$H$_5$OCHO &           & \hspace*{-2.5ex} $\bullet$ & \hspace*{-2.5ex} $\bullet$ \\
\noalign{\smallskip}
H       +  C$_2$H$_5$O $\rightarrow$  C$_2$H$_5$OH &           & \hspace*{-2.5ex} $\bullet$ &  \hspace*{-2.5ex}$\bullet$ \\
\noalign{\smallskip}
\noalign{\smallskip}
\noalign{\smallskip}
CH$_3$OCHO   + $h \nu$ $\rightarrow$  CH$_2$OCHO + H &           & \hspace*{-2.5ex} $\bullet$ & \hspace*{-2.5ex} $\bullet$\\
\noalign{\smallskip}
C$_2$H$_5$OH + $h \nu$ $\rightarrow$  C$_2$H$_5$O + H &           & \hspace*{-2.5ex} $\bullet$ & \hspace*{-2.5ex} $\bullet$\\
\noalign{\smallskip}
\noalign{\smallskip}
\noalign{\smallskip}
C$_2$H$_5$OCHO + $h \nu$ $\rightarrow$  CH$_2$OCHO + CH$_3$ &           & \hspace*{-2.5ex} $\bullet$ & \hspace*{-2.5ex} $\bullet$\\
\noalign{\smallskip}
C$_2$H$_5$OCHO + $h \nu$ $\rightarrow$  C$_2$H$_5$O + HCO &           & \hspace*{-2.5ex} $\bullet$ & \hspace*{-2.5ex} $\bullet$\\
\noalign{\smallskip}
\noalign{\smallskip}
\hline
\end{tabular}
\end{center}
Notes: Reactions that were present in the hot core model of \citet{Garrod08a} 
are indicated. Activation energies required for reaction are shown in 
brackets, where applicable.
\end{table}

The new reactions allow each cyanide to be constructed by sequential formation 
of its carbon backbone by the addition of CH$_2$, CH$_3$, or yet larger 
hydrocarbon radicals; however, photodissociation also allows the break-down of 
these structures. The resultant radicals may further react with another 
functional-group radical, to extend the backbone, or with a hydrogen atom, to 
terminate this sequence. Similarly, aminoacetonitrile may be formed by the 
addition of NH or NH$_2$ groups, or by direct addition of CN to CH$_2$NH$_2$, 
or CH$_2$NH (followed by hydrogenation). Different routes will dominate 
according to the relative mobilities of competing radicals, and their 
availabilities. Hence, the net direction of inter-conversion between cyanides 
may change with temperature, or as the abundances of molecular precursors vary.

Ethyl formate may be formed on grain surfaces 
by the addition of a CH$_3$ or HCO radical to a 
CH$_2$OCHO or C$_2$H$_5$O radical, respectively. These latter species are 
formed directly by cosmic ray-induced photodissociation of methyl formate or 
ethanol on the grains; hence, methyl formate need not be the only precursor 
for ethyl 
formate, nor the most important one. We do not consider other routes to the 
formation of CH$_2$OCHO and C$_2$H$_5$O; radical addition to formaldehyde, 
H$_2$CO, would almost certainly be mediated by a substantial activation energy 
barrier. Alternatively, addition of an oxygen atom to C$_2$H$_5$ is unlikely to 
be important, due to the relative scarcity of atomic oxygen, which is mainly 
bound in the ice mantles as H$_2$O; however, this route cannot be entirely 
ruled out.

When the grain surface-produced molecules evaporate, they are subject to 
gas-phase destruction mechanisms. Whilst cosmic ray-induced photodissociation 
in the gas phase is also included for consistency, the gas-phase destruction 
of these molecules is dominated by reaction with the ions C$^+$, He$^+$, 
H$_3$$^+$, H$_3$O$^+$ and HCO$^+$ (followed by dissociative recombination, if 
a protonated molecule results). Ion-molecule and dissociative recombination 
reaction rates are of a similar order for all new species; 
see \citet{Garrod08a}.

\subsection{Results}
\label{ss:chem-res}

We analyse the model results for ethyl formate and the cyanides in the context 
of a selection of complex molecules to which they are chemically 
or observationally related. We consider first the results of the basic model 
described above (called hereafter \textit{Basic} model), using an intermediate 
warm-up timescale of $2 \times 10^{5}$~yr. This timescale was found by 
\citet{Garrod08a} to be most appropriate to match the abundances of Sgr~B2(N).

\subsubsection{Ethyl formate and related species}
\label{sss:etocho}

Table~\ref{t:chem-tab2} presents peak fractional abundances, and the 
temperatures at which they are achieved, derived from the chemical model. 
Model abundances are converted to values per mean particle with a mean 
molecular weight, $\mu$, of 2.33, for comparison to the observations. Also 
listed are the observed rotational temperatures and abundances (Cols.~7 and 8, 
respectively). The latter were 
derived from the column densities given in Tables~\ref{t:etochomodel}, 
\ref{t:xochomodel}, \ref{t:prcnmodel}, and \ref{t:xcnmodel}, assuming an H$_2$ 
column density of $1.8 \times 10^{25}$~cm$^{-2}$ for a source size of 2$\arcsec$ 
\citep[see][]{Belloche08b}, and an H$_2$ column density profile proportional to 
$r^{-0.5}$ that corresponds to an H$_2$ density profile proportional to 
$r^{-1.5}$ in spherical symmetry\footnote{\citet{Osorio99} expect a density
profile proportional to $r^{-p}$ with $p = 1.5$ for the central region of a 
hot core. On larger scales in Sgr~B2 ($20-200\arcsec$), \citet{Lis89} derived 
a density profile $p \sim 2-2.5$ while \citet{deVicente97} found $p \sim 0.9$.}.
Given that the dust properties are uncertain by a factor $\sim 2$ at least 
and that the contribution of the vibrationally or torsionally excited states 
of some molecules studied here \citep*[e.g. ethanol, see][]{Pearson08} to their 
partition function was not included, we estimate these observed abundances to 
be accurate within a factor $\sim 3$.

\begin{figure*}
\centering
\includegraphics[width=9cm]{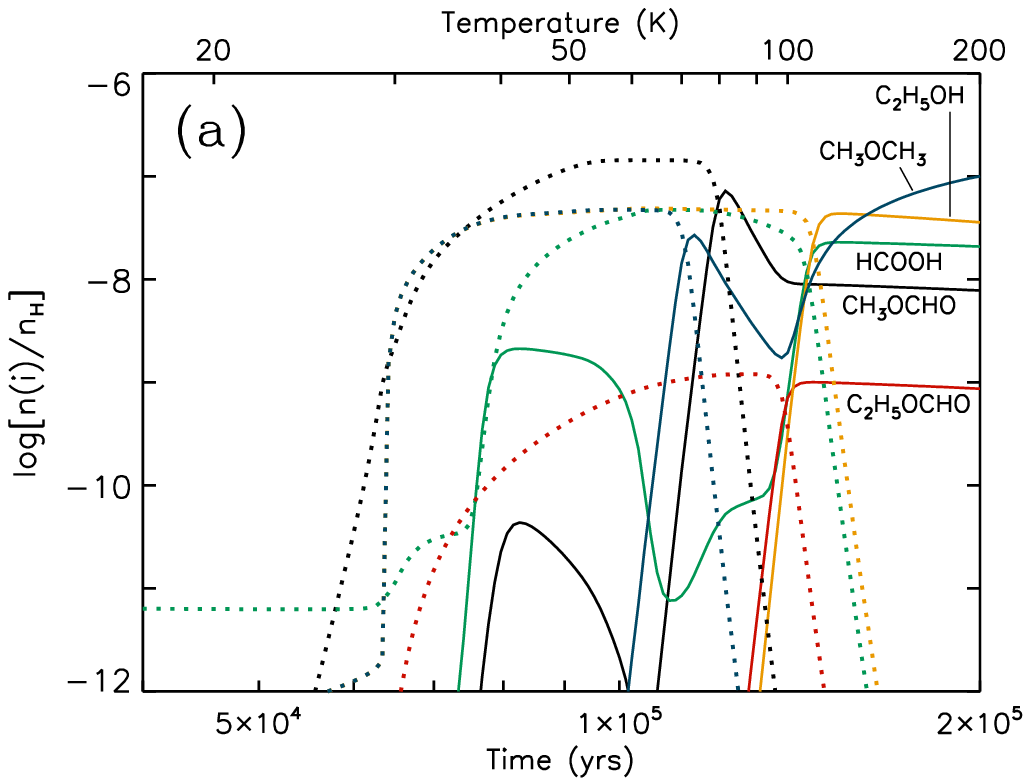}
\includegraphics[width=9cm]{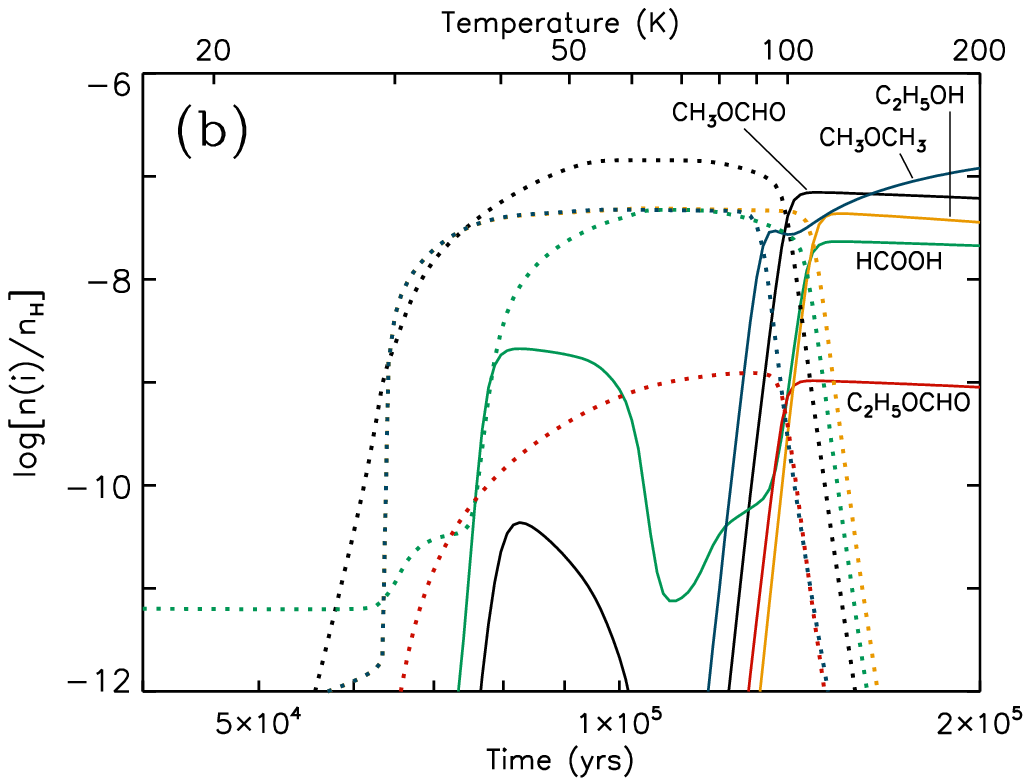}
\caption{\textbf{a)} \textit{Basic} model, showing methyl formate, ethyl 
formate, formic acid, and related species. \textbf{b)} The same species, 
following augmentation of methyl formate and dimethyl ether binding energies. 
Solid lines indicate gas-phase species; dotted lines of the same color 
indicate the same species on the grain surfaces.}
\label{f:chem1}
\end{figure*}

Ethyl formate is clearly formed most significantly at late times (see 
Fig.~\ref{f:chem1}a), and its grain-surface abundance (\textit{dotted} red 
lines) scales well with that of methyl formate. Grain-surface methyl formate 
is, in fact, the 
primary source of precursor radicals (via photodissociation) for the formation 
of ethyl formate. When methyl formate evaporates, 
and ethanol is left as the dominant source of precursor radicals, ethyl 
formate production becomes dependent on the addition of HCO to C$_2$H$_5$O. The 
post-evaporation gas-phase abundance of ethyl formate relative to methyl 
formate and formic 
acid appears to match observational abundances and rotational temperatures 
reasonably well.

The gas-phase methyl formate peak abundance is also relatively close to the 
observed abundance (within a factor 5), and the model temperature 
at this peak is in very good agreement with the observed rotational temperature 
(see Table~\ref{t:xochomodel}). However, the abundance quickly falls, 
and the ratio of gas phase CH$_3$OCHO to HCOOH, C$_2$H$_5$OH and CH$_3$OCH$_3$ 
at the higher temperatures most appropriate to the densest regions of the hot 
core is low compared to the observed values.

The \textit{Basic} model uses the same binding energies for methyl formate and 
dimethyl ether as were employed by \citet{Garrod08a}, appropriate to binding on 
amorphous water ice. These values cause 
relatively early evaporation of those species, resulting in significant 
destruction in the gas-phase, and low fractional abundances in comparison to 
observed values in the case of methyl formate. The binding energies of those 
molecules were obtained by simple interpolation of measured values obtained 
for other species. Laboratory data for methyl formate and dimethyl ether 
evaporation from appropriate ice surfaces are not currently available.

\begin{figure*}
\centering
\includegraphics[width=9cm]{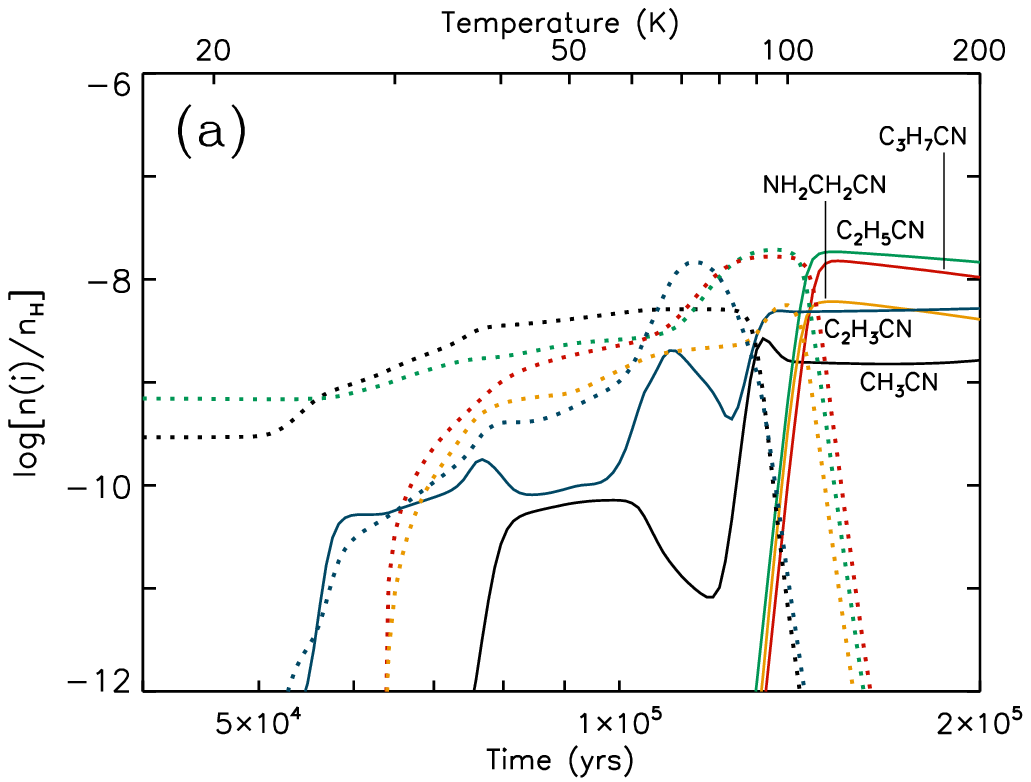}
\includegraphics[width=9cm]{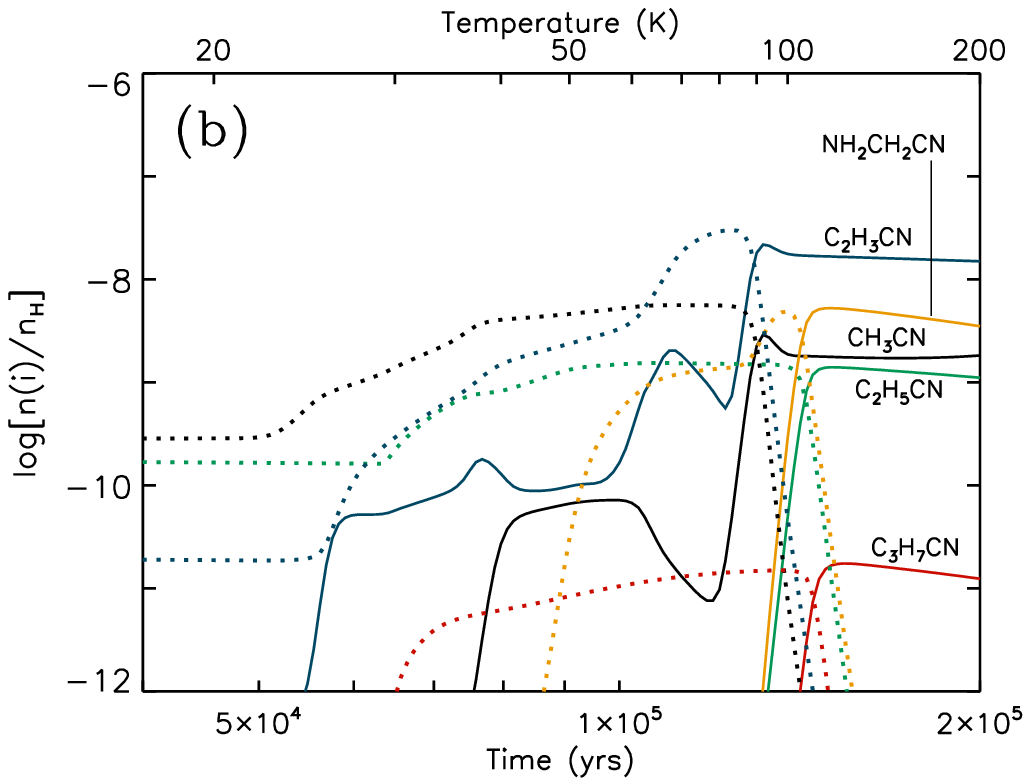}
\caption{\textbf{a)} \textit{Basic} model, showing cyanides. \textbf{b)} The 
same species, using the \textit{Select} model, in which selected 
grain-surface reactions are de-activated (see Table~\ref{t:chem-tab1}). Solid 
lines indicate gas-phase species; dotted lines of the same color indicate the 
same species on the grain surfaces.}
\label{f:chem2}
\end{figure*}

\begin{table*}
\caption[]{Peak gas-phase abundances from each model, with corresponding 
model temperatures, as 
well as source sizes, rotation temperatures, and 
gas-phase abundances derived from the observations of the main source in 
Sgr~B2(N).}
\label{t:chem-tab2}
\begin{center}
\begin{tabular}{lccccccccccc}
\hline
\hline
\noalign{\smallskip}
Species & \multicolumn{2}{c}{\textit{Basic} model} & & \multicolumn{2}{c}{\textit{Select} model} & & \multicolumn{3}{c}{Observations} & & Abundance ratio\\
\noalign{\smallskip}
\cline{2-3} \cline{5-6} \cline{8-10}
\noalign{\smallskip}
     & $n[i]/n_{{\mathrm{H}}_2}$ & $T^a$ & & $n[i]/n_{{\mathrm{H}}_2}$ & $T^a$ & & Size & $T_\mathrm{rot}$ & $n[i]/n_{{\mathrm{H}}_2}$ & & \textit{Select} model\\
     &                         & (K) & &                         & (K) & & ($\arcsec$) & (K)      &                         & & over observation \\
(1)  & (2)                     & (3) & & (4)                     & (5) & & (6)         & (7)      & (8)                     & & (9) \\
\hline
\noalign{\smallskip}
HCOOH (hot)    & 5.4e-08 & 120 & &  5.4e-08 & 120 & & 5.0 &  70 & 1.3e-09 & & 42 \\
\noalign{\smallskip}
HCOOH (cold)   & 4.9e-09 &  42 & &  4.9e-09 &  42 & & 5.0 &  70 & 1.3e-09 & &  3.8 \\
\noalign{\smallskip}
CH$_3$OCHO     & 1.7e-07 &  81 & &  1.6e-07 & 112 & & 4.0 &  80 & 3.5e-08 & &  4.6 \\
\noalign{\smallskip}
C$_2$H$_5$OCHO & 2.3e-09 & 110 & &  2.3e-09 & 110 & & 3.0 & 100 & 3.6e-09 & &  0.6 \\
\noalign{\smallskip}
C$_2$H$_5$OH   & 1.0e-07 & 120 & &  1.0e-07 & 120 & & 3.0 & 100 & 5.7e-08 & &  1.8 \\
\noalign{\smallskip}
CH$_3$OCH$_3$  & 2.3e-07 & 200 & &  2.8e-07 & 200 & & 2.5 & 130 & 1.4e-07 & &  2.0 \\
\noalign{\smallskip}
CH$_3$CN       & 6.3e-09 &  92 & &  6.8e-09 &  92 & & 2.7 & 200 & 1.3e-07 & &  0.05 \\
\noalign{\smallskip}
C$_2$H$_5$CN   & 4.4e-08 & 117 & &  3.3e-09 & 117 & & 3.0 & 170 & 8.1e-08 & &  0.04 \\
\noalign{\smallskip}
C$_3$H$_7$CN   & 3.5e-08 & 120 & &  4.0e-11 & 123 & & 3.0 & 150 & 1.0e-09 & &  0.04 \\
\noalign{\smallskip}
NH$_2$CH$_2$CN & 1.4e-08 & 117 & &  1.2e-08 & 117 & & 2.0 & 100 & 1.5e-09 & &  8.0 \\
\noalign{\smallskip}
C$_2$H$_3$CN   & 1.2e-08 & 200 & &  5.1e-08 &  92 & & 2.3 & 170 & 4.7e-08 & &  1.1 \\
\noalign{\smallskip}
\hline
\end{tabular}
\end{center}
Notes: $^a$ The model temperatures are the temperatures at which the peak 
gas-phase abundances are achieved.
\end{table*}

For species comprising at least one -OH functional group, binding-energy 
estimates take account of hydrogen-bonding interactions with the ice surface. 
Such species may act as both hydrogen-bond donors and acceptors, raising their 
binding strengths. However, both methyl formate and dimethyl ether have at 
least one unbonded electron pair attached to a strongly electro-negative atom 
(oxygen), allowing them to be hydrogen-bond acceptors. This may give them a 
somewhat stronger bond to the predominantly water-ice surface than has been 
assumed.

Here, the binding energy of methyl formate is raised 
beyond that of the {\em Basic} model, such that it falls 
approximately half way between its old value and that of ethanol, its most 
closely-matched counterpart with a single, fully hydrogen-bonding functional 
group. This augmentation constitutes an increase of approximately 1000~K, 
giving $E_{D}=5200$~K. The binding energy of dimethyl ether is similarly 
raised by 1000~K.

Augmentation of methyl formate binding energy allows it to remain on grains 
for longer, reducing the time available for gas-phase destruction, before the 
majority of other species evaporate, damping the effect of ion-molecule 
destruction pathways (see Fig.~\ref{f:chem1}b). This allows gas-phase 
methyl formate 
fractional abundances to remain high for longer, although the resulting 
peak-abundance temperature is somewhat greater, at 112~K. 

Dimethyl ether does have a viable gas-phase formation mechanism, and is largely 
produced in the gas phase, due to the large abundance of methanol 
($\sim$$10^{-5} n_{H}$); hence, the peak abundance is not strongly affected
by the augmentation of its binding energy. 
Its gas-phase abundance in the model is consistent with the observed value 
(within a factor 2, see Table~\ref{t:chem-tab2}). The peak-abundance 
temperature 
of the model is somewhat higher than that derived observationally. A slightly 
lower grain-surface methanol abundance would remedy this, as post-evaporation 
gas-phase methanol abundances should diminish more rapidly, reducing the rate 
of dimethyl ether formation. A slower warm-up subsequent to methanol 
evaporation would also produce a similar effect. Nevertheless, the observed 
rotational temperature of dimethyl ether seems consistent with gas-phase 
formation.

Surface formation rates of ethyl formate, methyl formate and ethanol are not 
strictly dependent on methanol abundance in the ices, but rather on the rate of
formation of its photodissociation products, CH$_3$O, CH$_2$OH, and CH$_3$. 
These rates are not well constrained; however, they seem appropriate for this 
model. A lower grain-surface methanol abundance, as suggested above, would 
therefore necessitate slightly greater methanol photodissociation rates, in 
order to achieve appropriate abundances for methyl formate and other 
surface-formed species. Gas-phase and grain-surface ethyl formate abundances 
are largely unaffected by the changes in methyl formate binding energy.
Both the gas-phase and grain-surface abundances of formic acid are strongly 
dependent on gas-phase processes 
\citep[see][]{Garrod08a}. As a result, there appears to be no simple 
correlation with ethyl or methyl formate abundances. However, the low 
rotational temperature reported in Sect.~\ref{ss:competocho} is qualitatively 
consistent with the low-temperature gas-phase formic acid peak at 40 -- 60~K, 
a point noted by \citet{Garrod08a} in comparison to other hot-core 
observations. This ``cold'' peak presents a fractional abundance very close to 
the observed value (within a factor 4, see Table~\ref{t:chem-tab2}). In 
Sect~\ref{ss:competocho}, we modeled the spectrum of formic acid using a 
single temperature component; however, a two-component model with 
rotational temperatures (and inferred spatial extents) appropriate to the 
chemical models is not noticeably worse than the single-component fit. As 
discussed in Sect.~\ref{ss:competocho}, the existence of both hot, compact and 
cold, extended components would be consistent with the lower flux measured 
with the BIMA interferometer by \citet{Liu01} compared to our lower-resolution
single-dish measurement.

\subsubsection{Cyanides}
\label{sss:xcn}

The \textit{Basic} model is capable of producing cyanide species in 
appropriate absolute 
quantities (see Fig.~\ref{f:chem2}a), however, their relative abundances are 
not well matched to the observationally determined values. In order to 
understand the behavior of the cyanide network, the different grain-surface
formation mechanisms, and combinations, were isolated by artificially 
de-activating 
particular reaction routes. In fact, all combinations that include either the 
hydrogenation of the cyanopolyyne HC$_3$N and of vinyl cyanide, C$_2$H$_3$CN, 
or the addition of large, pre-formed hydrocarbons directly to the CN radical, 
produce wildly inaccurate ratios. In some such cases, \textit{n}-propyl 
cyanide is 
the most abundant of all, often with methyl cyanide abundances deeply depressed.
The only combination in which the correct proportion is reproduced is that 
in which only the sequential addition of grain-surface CH$_2$ and CH$_3$ 
functional groups 
is allowed (see Fig.~\ref{f:chem2}b). We label this model, combined with the 
augmented binding energies of methyl formate and dimethyl ether, as the 
\textit{Select} model. In this scheme, formation of the larger cyanides begins 
with cosmic ray-induced photodissociation of a smaller grain-surface
alkyl cyanide molecule 
(resulting in the ejection of a hydrogen atom), or with the accretion of 
CH$_2$CN (which may be formed in the gas-phase following the evaporation of 
HCN). A methyl-group radical is then added to produce a larger alkyl cyanide 
molecule. 

Methyl cyanide itself is mainly formed on the grains by addition of CH$_3$ and 
CN radicals, but it may also be formed by gas-phase processes fuelled by the 
evaporation of HCN. Methyl cyanide evaporates from the dust grains around 90~K, 
producing its greatest gas-phase abundance; however, the subsequent 
evaporation of all molecular material from the grains promotes rapid gas-phase 
formation, maintaining methyl cyanide abundances for longer, and providing 
qualitative agreement with the large rotational temperature derived from the 
observational data.

The abundance obtained for aminoacetonitrile is in reasonable agreement with 
that obtained observationally (within a factor 8), suggesting that the 
addition of NH or NH$_2$ to CH$_2$CN on grain surfaces, similar to the 
suggested mechanism for ethyl cyanide, is a plausible 
route to its formation. There may therefore be some degree of correlation 
between these two species, which should be investigated in future. The removal 
of the other formation routes for aminoacetonitrile, comprising the addition of
grain-surface CN to either CH$_2$NH or CH$_2$NH$_2$, makes little difference 
to the results, mainly due to limited availability of the latter two radicals.

Vinyl cyanide, C$_2$H$_3$CN, a potential precursor of ethyl cyanide and 
\textit{n}-propyl cyanide, is formed predominantly in the gas-phase in both 
the \textit{Basic} and \textit{Select} models. This occurs through the 
reaction of CN with ethylene (C$_2$H$_4$), 
which has been shown experimentally to be rapid over a range of temperatures 
\citep{Carty01}. The resultant gas-phase vinyl cyanide then accretes onto the 
grains until greater temperatures are achieved. Following evaporation of the 
ice mantles at $T>100$~K, vinyl cyanide is again formed rapidly in the 
gas-phase by the same mechanism, allowing it, like methyl cyanide, to retain 
large fractional abundances longer than the other cyanides. This effect is 
also in qualititative agreement with its relatively high rotational 
temperature. Both models show good agreement with the 
observational abundance of this molecule, but the \textit{Select} model 
produces an excellent match (see Table~\ref{t:chem-tab2}).

For the \textit{Basic} model, ratios of peak abundance values are 
\hbox{HCOOH / CH$_3$OCHO / C$_2$H$_5$OCHO} = \hbox{23 / 72 / 1} and 
\hbox{CH$_3$CN / C$_2$H$_5$CN / C$_3$H$_7$CN}  = \hbox{0.18 / 1.3 / 1}. For 
the \textit{Select} model, these ratios are 
\hbox{HCOOH / CH$_3$OCHO / C$_2$H$_5$OCHO} = \hbox{23 / 70 / 1} and 
\hbox{CH$_3$CN / C$_2$H$_5$CN / C$_3$H$_7$CN} = \hbox{171 / 82 / 1}. These 
seem a fair match to the observed values of Sects.~\ref{ss:competocho} and 
\ref{ss:compprcn} (\hbox{0.8 / 15 / 1} and \hbox{108 / 80 / 1}, respectively). 
Consideration of only the low temperature formic acid peak 
in the models further improves its ratio with ethyl formate abundances.

The warm-up timescale of $t_{max}=2 \times 10^{5}$~yr appears to 
yield the most 
appropriate reproduction of observed cyanide ratios, although longer 
timescales are also plausible; the \textit{Select} model, with 
$t_{max}=10^{6}$~yr, produces peak abundance ratios of 
\hbox{HCOOH / CH$_3$OCHO / C$_2$H$_5$OCHO} = \hbox{4.2 / 3.3 / 1} 
and \hbox{CH$_3$CN / C$_2$H$_5$CN / C$_3$H$_7$CN} = \hbox{258 / 106 / 1}.

\subsection{Discussion}
\label{ss:discussion}

Based on the abundance ratios of the model, the dominant formation mechanism 
for alkyl cyanides is probably the sequential addition of CH$_2$ or CH$_3$ 
radicals to CN, CH$_2$CN and C$_2$H$_4$CN on the grain surfaces. 
Both the alternative routes -- the grain-surface hydrogenation of gas 
phase-formed HC$_3$N and C$_2$H$_3$CN, or the direct grain-surface addition of 
pre-formed large hydrocarbon radicals like C$_2$H$_5$ or 
C$_3$H$_7$ to a CN radical -- appear to be very much too fast, resulting in 
excessive quantities of the two largest alkyl cyanides.

To achieve the appropriate ratios, those two formation routes must be 
artificially disabled within the model. Why should these mechanisms be less 
efficient in reality than they would appear from the model? Firstly, gas-phase 
HC$_3$N and C$_2$H$_3$CN may be less abundant than the model suggests. 
The evaporation, and subsequent reaction, of HCN from the grains is a primary 
cause of gas-phase formation for each of these molecules. Variation in the 
evaporation characteristics or the composition/structure of the ices may weaken 
such mechanisms. However, the agreement between observed and modeled 
abundances of vinyl cyanide is very good. Indeed, the \textit{Select} model 
shows excellent agreement, providing further justification for the omission of 
its hydrogenation reactions. 

Alternatively, surface hydrogenation of HC$_3$N and C$_2$H$_3$CN, once they 
have accreted onto the grains, may be less efficient than has been assumed 
here. Importantly, activation energies are required for hydrogenation of both 
these species, whose values are poorly constrained. The fact that it is 
these very reactions that must be disabled suggests strongly that their 
activation energies should be significantly higher than has been assumed here. 
Additionally, our use of a ``deterministic'' gas-grain model may also produce 
somewhat more efficient hydrogenation than is really the case (although a 
test-run using the rate-modification method of \citet{Garrod08b} shows no 
great difference in this respect).

In the case of the addition of large hydrocarbon radicals to CN, the 
over-dominance of these channels is probably due to the incompleteness of the 
hydrocarbon chemistry as a whole, particularly on the grains. Whilst up to 10 
carbon atoms in a chain are considered in this model, the hydrogenation states 
of the larger chains are typically limited to 4 hydrogen atoms. Crucially, 
hydrogenation is the only type of reaction included in the network for most 
hydrocarbons, aside from the newly-added CN addition reactions. The 
hydrocarbon reaction set was largely devised with cold dark clouds in mind, 
where hydrogenation dominates. By including only a single new reaction 
(addition to CN) for any particular hydrocarbon, that reaction can easily 
become the dominant channel. The completion of the hydrocarbon network to 
include reactions with all major reactants would be beneficial, although this 
is not a trivial task.

The small hydrocarbons CH$_2$ and CH$_3$, on the other hand, as well as CN 
itself, have a much more comprehensive reaction network, making sequential 
addition and its apparent degree of efficiency more credible.

Ethyl formate and aminoacetonitrile also seem to be well reproduced with a 
similar addition scheme to that of the alkyl cyanides. Ethyl formate abundance 
may be dependent on ethanol as well as methyl formate, depending on the 
specific conditions. 

The \textit{Select} model reproduces well the abundance ratios for alkyl 
cyanides, but their absolute abundances are an order of magnitude lower than 
observational values. This also results in a poor match to abundance ratios 
relative to methyl formate and other methanol-related species. In fact, the
chemistries of the cyanides and the methanol-related species do not strongly
influence one another in the model. The overall abundances of each category of
molecule are mainly influenced by different, independent parameters: the 
formation rate of the products of methanol photodissociation (i.e. the product 
of the photodissociation rate and absolute grain-surface abundance of 
methanol), and the quantity of HCN or related nitrile-group species in the ice 
mantles, respectively. Similarly, the modeled abundance of aminoacetonitrile
relative to the alkyl cyanides is very high. The formation rate of this 
molecule is strongly dependent on the product of the abundance of NH$_3$ in the 
ices, and its rate of photodissociation. This indicates that one or both of 
these values may be too large, by at least an order of magnitude. A parameter 
search should yield the optimal values for all such quantities, but such is 
not the focus of this paper.

The augmentation of methyl formate binding energy allows its abundance to 
remain high at temperatures appropriate to the densest parts of the hot core. 
However, the low observed rotational temperatures suggest that methyl formate 
should still have a binding energy less than that of H$_2$O, which is indeed 
the case here, even with the highest value we use. A value somewhat lower than 
our maximum would also achieve quite acceptable results. Clearly, an 
experimental value for binding to astrophysically appropriate surfaces would 
be highly valuable for the chemical modeling of hot cores.

While certain crucial steps in the formation of these complex molecules 
occur only in the gas-phase or on the grain surfaces, processes in each 
phase are inter-dependent and cannot be understood in isolation.

\section{Conclusions}
\label{s:conclusions}

We used the complete 3\,mm and partial 2 and 1.3~mm line surveys obtained 
with the IRAM 30~m telescope toward the hot cores Sgr~B2(N) and (M) 
to search for emission from the organic molecules ethyl formate and 
\textit{n}-propyl cyanide. We report the detection of both molecules toward 
the hot core Sgr~B2(N), which are the first detections of these molecules in 
the interstellar medium. Our main results and conclusions are the following:

 \begin{enumerate}
   \item New entries for the CDMS catalog have been created for 
   \textit{n}-propyl cyanide and ethyl formate.
   \item 46 of the 711 significant transitions of the \textit{anti}-conformer 
   of ethyl formate covered by our 30~m line survey are relatively free of 
   contamination from other molecules and are detected in the form of 24
   observed features toward Sgr~B2(N). The emission of the
   \textit{gauche}-conformer is too weak to be clearly detected in our survey.
   \item 50 of the 636 significant transitions of the \textit{anti}-conformer 
   of \textit{n}-propyl cyanide covered by our 30~m line survey are relatively 
   free of contamination from other molecules and are detected in the form of 
   12 observed features toward Sgr~B2(N) with two velocity components. The 
   emission of the \textit{gauche}-conformer is too weak to be clearly 
   detected in our survey.
   \item With a source size of 3$\arcsec$, we derive an ethyl formate column 
   density of $5.4 \times 10^{16}$~cm$^{-2}$ for a temperature of 100~K and a 
   linewidth of 7~km~s$^{-1}$ in the LTE approximation. The abundance of 
   ethyl formate relative to H$_2$ is estimated to be $3.6 \times 10^{-9}$.
   \item The two velocity components detected in \textit{n}-propyl cyanide 
   have LTE column densities of $1.5 \times 10^{16}$ and 
   $6.6 \times 10^{15}$~cm$^{-2}$, respectively, with a temperature of 150~K,
   a linewidth of 7~km~s$^{-1}$, and a source size of 3$\arcsec$. The 
   fractional abundance of \textit{n}-propyl cyanide in the main source is 
   estimated to be $1.0 \times 10^{-9}$.
   \item We detected neither ethyl formate nor \textit{n}-propyl cyanide 
   toward the more evolved source Sgr~B2(M) and derived column density upper 
   limits of $2 \times 10^{16}$ and $6 \times 10^{15}$~cm$^{-2}$, respectively,
   for a source size of 3$\arcsec$.
   \item We modeled the emission of chemically related species also detected in 
   our survey of Sgr~B2(N) and derived column density ratios of 
   \hbox{0.8 / 15 / 1} for \hbox{$t$-HCOOH / CH$_3$OCHO / C$_2$H$_5$OCHO} and 
   \hbox{108 / 80 / 1} for \hbox{CH$_3$CN / C$_2$H$_5$CN / C$_3$H$_7$CN} in 
   the main hot core of Sgr~B2(N).
   \item The chemical models suggest that the sequential, piecewise 
   construction of ethyl and \textit{n}-propyl cyanide from their constituent 
   functional groups on the grain surfaces is their most likely formation 
   route. Aminoacetonitrile formation proceeds similarly, suggesting a possible 
   correlation with ethyl cyanide abundance. Vinyl cyanide is 
   formed predominantly in the gas-phase.
   \item Comparison of the observational and model results suggests 
   that the
   production of alkyl cyanides by the hydrogenation of less saturated
   species is much less efficient than functional-group addition.
   \item Ethyl formate can be formed on the grains by addition of HCO or CH$_3$ 
   to functional-group radicals derived from methyl formate and ethanol; 
   however, methyl formate appears to be the dominant precursor.
   \item Understanding of the complex interactions between gas-phase and 
   grain-surface processes may be necessary to fully explain the observational 
   features displayed by many complex molecules, including formic 
   acid and 
   methyl formate.
   \item The detection in Sgr~B2(N) of the next stage of complexity in two 
   classes of complex molecule, esters and alkyl cyanides, suggests that 
   greater complexity also may be present in other classes of molecule in the 
   interstellar medium.
 \end{enumerate}

Our results have demonstrated the power of the "complete spectrum 
fitting" approach used by us as a technique that is mandatory today for 
the identification of  new complex molecules by their generally weak 
signals. Ideally, one would want to verify identifications with 
interferometric observations as done for the case of aminoacetonitrile 
\citep[][]{Belloche08a,Belloche08b}. However, given the limited collecting 
area, 
bandwidth and spatial resolution of today's interferometer arrays, this 
would be very time consuming or even prohibitive. It will, however, be a 
trivial exercise for the Atacama Large Millimeter Array (ALMA) once it 
is fully operational.

\begin{acknowledgements}
We thank the anonymous referee and the editor for their careful 
reading of the manuscript and for their suggestions that helped improve the 
clarity of this article.
H.S.P.M. thanks Dr. J\"urgen Aschenbach from the library of the University of 
Kiel for providing the supplementary material to \citet{Vormann88}. 
We are grateful to Eric Herbst for providing the ethyl formate spectroscopic 
line list as well as a preprint of the manuscript prior to publication. 
H.S.P.M. thanks the Deutsche Forschungsgemeinschaft (DFG) for initial support 
through the collaborative research grant SFB~494. He is grateful to the 
Bundesministerium f\"ur Bildung und Forschung (BMBF) for recent support which 
was administered through Deutsches Zentrum f\"ur Luft- und Raumfahrt (DLR).
R.T.G thanks the Alexander von Humboldt Foundation for a Research Fellowship.
\end{acknowledgements}

\Online
\begin{appendix}
\section{\textit{a}-Type and \textit{b}-type lines of methyl formate}
\label{a:meocho}

\begin{figure*}
\centerline{\resizebox{0.9\hsize}{!}{\includegraphics[angle=270]{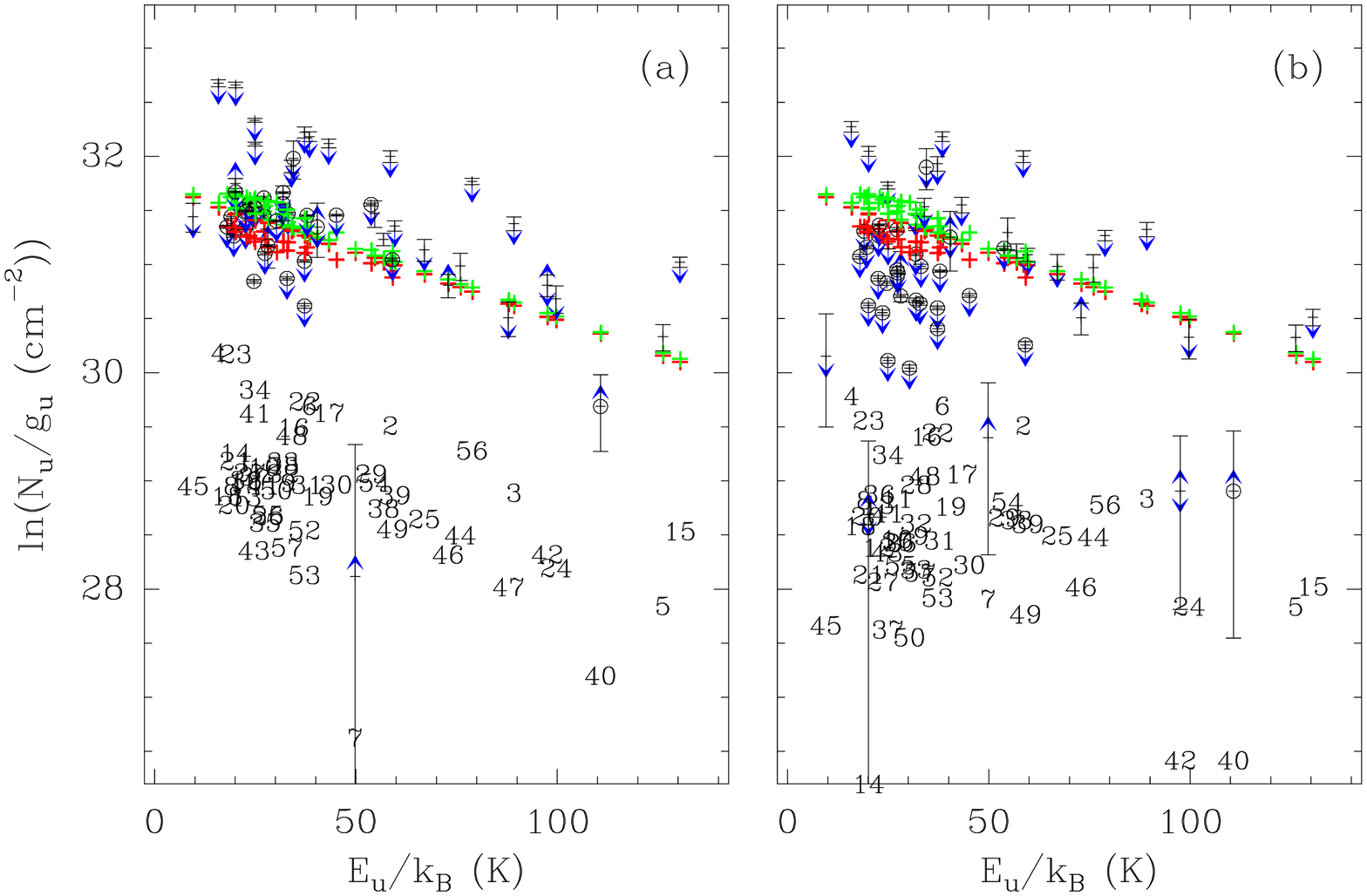}}}
\vspace*{2ex}
\centerline{\resizebox{0.9\hsize}{!}{\includegraphics[angle=270]{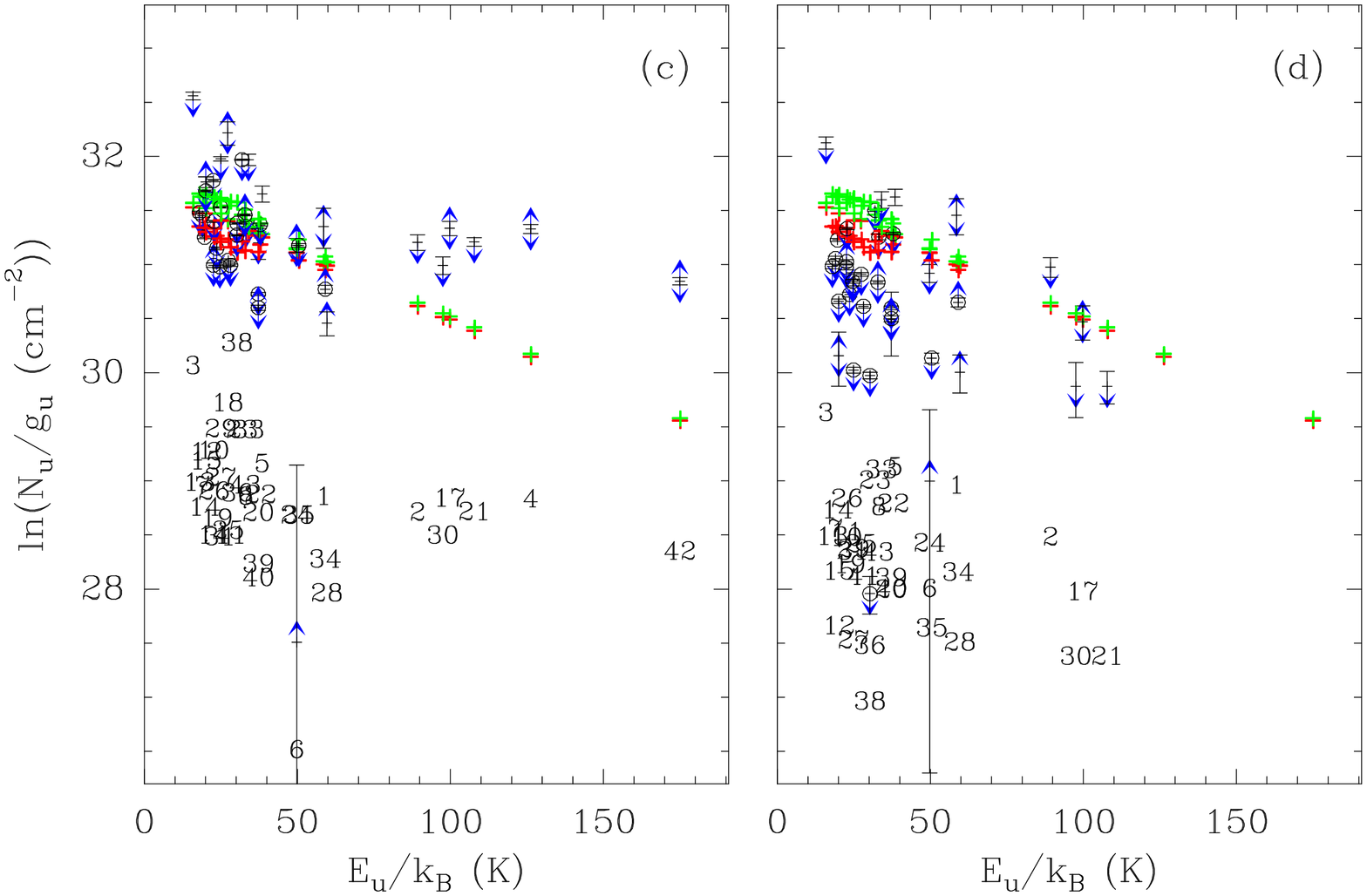}}}
\caption{Population diagrams of the \textit{A} and 
\textit{E} symmetry species of
methyl formate presented in the same way as for ethyl formate in 
Fig.~\ref{f:popdiagetocho} (see the caption of that figure for details). The 
\textit{a}-type lines are marked with a circle. Panel 
\textbf{a} and \textbf{c} show the population diagrams derived from the 
measured integrated intensities for the \textit{A} and \textit{E} species, 
respectively, while panels \textbf{b} and \textbf{d} present the respective 
population diagrams after removing the expected contribution from 
contaminating molecules. Features 4 and 42 with $E_u/k_B > 120$~K (see panel 
\textbf{c}) are missing in panel \textbf{d}
because the removal of the contaminating lines yields negative residuals. This 
is due to the uncertain level of the baseline that looks overestimated 
for both features in the observed spectrum.}
\label{f:popdiagmeocho}
\end{figure*}

Both \textit{A} and \textit{E} symmetry species of methyl formate 
(CH$_3$OCHO) are easily detected in our spectral survey of Sgr~B2(N) at 3~mm. 
Sixty four lines of the \textit{A} species are detected in the form of 57 
features in our 3~mm survey and 48 lines of the \textit{E} species in the 
form of 43 features. We followed the same procedure as described in 
Sect.~\ref{ss:detetocho} for ethyl formate to compute the population diagrams 
shown in Fig.~\ref{f:popdiagmeocho}. In these diagrams, the \textit{a}-type 
lines of methyl formate (with $\Delta K_a$= 0 [2] and $\Delta K_c$ = 1 [2]) 
are marked with an 
additional circle. As mentioned in Sect.~\ref{ss:competocho}, both 
\textit{a}- and \textit{b}-type lines are well fitted with the same physical 
model (see Table~\ref{t:xochomodel}). Although many \textit{a}-type 
transitions with $E_u/k_B < 50$~K look systematically too low in the 
population diagrams after removal of the contribution of contaminating lines 
(Fig.~\ref{f:popdiagmeocho}b and d), this can be explained by the limitations 
of our radiative transfer modeling: these \textit{a}-type  transitions (of 
the \textit{A} or \textit{E} species) have optical depths on the order of 
unity, as indicated by the significant shift between the red and green crosses 
in the lower energy range, and overlap with \textit{a}-type lines of the other 
symmetry species (\textit{E} or \textit{A}, respectively) that have significant 
optical depths too. Since our current complete model treats the two symmetry 
species as independent and our radiative transfer program computes the 
contributions of overlapping transitions of different species independently, 
the sum of the overlapping \textit{A} and \textit{E} transitions with 
significant optical depths is systematically overestimated. For a transition 
of, e.g., the \textit{A} species, the ``contamination'' by the 
\textit{E} species is overestimated and its removal in 
Fig.~\ref{f:popdiagmeocho}b yields an underestimated residual flux. Our model 
could be improved by treating both symmetry species as a single molecule but 
this would not significantly change the physical parameters found for methyl 
formate and is beyond the scope of this article focused on ethyl formate and 
\textit{n}-propyl cyanide.

\end{appendix}

\onltab{\value{aprcn}}{\longtab{\value{aprcn}}{\begin{longtable}{rclr@{}lrr}
\caption{\label{t:lines-anti-PrCN}
Transitions of \textit{anti}-\textit{n}-propyl cyanide, employed in 
the present fits, their frequencies (MHz), uncertainties
Unc. (kHz), and residuals O$-$C (kHz) between frequencies measured in the 
laboratory and those calculated from the final spectroscopic parameters. 
Unresolved asymmetry splitting (two transitions having the 
same $K_a$ and the same transition frequency) has been treated as 
intensity-weighted average of the two lines.}\\
\hline \hline
$J'$, $K_a'$, $K_c'$ & $-$ & $J''$, $K_a''$, $K_c''$ & \multicolumn{2}{c}{Frequency} & Unc. & O$-$C\\
\hline
\endfirsthead
\caption{continued.}\\
\hline\hline
$J'$, $K_a'$, $K_c'$ & $-$ & $J''$, $K_a''$, $K_c''$ & \multicolumn{2}{c}{Frequency} & Unc. & O$-$C\\
\hline
\endhead
\hline
\endfoot
  2,  1,  2 & $-$ &  1,  1,  1 &   8727&.068 & 10 &  $-$3 \\
  2,  0,  2 & $-$ &  1,  0,  1 &   8841&.749 & 10 &     4 \\
  2,  1,  1 & $-$ &  1,  1,  0 &   8957&.437 & 10 &     3 \\
  4,  1,  4 & $-$ &  3,  1,  3 &  17453&.162 & 10 &    12 \\
  3,  0,  3 & $-$ &  2,  0,  2 &  13261&.436 &  5 &     1 \\
  3,  1,  3 & $-$ &  2,  1,  2 &  13090&.304 &  5 &     8 \\
  3,  1,  2 & $-$ &  2,  1,  1 &  13435&.836 &  5 &     3 \\
  3,  2,  2 & $-$ &  2,  2,  1 &  13263&.547 &  5 &  $-$2 \\
  3,  2,  1 & $-$ &  2,  2,  0 &  13265&.404 &  5 &  $-$0 \\
  4,  2,  3 & $-$ &  3,  2,  2 &  17684&.317 &  5 & $-$10 \\
  4,  3,  2 & $-$ &  3,  3,  1 &  17686&.050 &  5 &     2 \\
  4,  3,  1 & $-$ &  3,  3,  0 &  17686&.050 &  5 &     2 \\
  4,  2,  2 & $-$ &  3,  2,  1 &  17688&.965 &  5 &     2 \\
  5,  3,  3 & $-$ &  4,  3,  2 &  22107&.924 &  5 &     1 \\
  5,  3,  2 & $-$ &  4,  3,  1 &  22107&.924 &  5 &     1 \\
  5,  4,  2 & $-$ &  4,  4,  1 &  22108&.119 &  5 &     1 \\
  5,  4,  1 & $-$ &  4,  4,  0 &  22108&.119 &  5 &     1 \\
  9,  1,  8 & $-$ &  9,  0,  9 &  24166&.242 &  5 &     7 \\
 10,  1,  9 & $-$ & 10,  0, 10 &  24799&.350 &  5 &     3 \\
 11,  1, 10 & $-$ & 11,  0, 11 &  25508&.602 &  5 &     0 \\
 12,  1, 11 & $-$ & 12,  0, 12 &  26297&.469 &  5 &  $-$2 \\
 11,  2, 10 & $-$ & 12,  1, 11 &   6897&.938 &  5 &  $-$4 \\
 16,  1, 15 & $-$ & 15,  2, 14 &  13992&.090 &  5 &  $-$1 \\
 17,  1, 16 & $-$ & 16,  2, 15 &  19339&.460 &  5 &     1 \\
 19,  1, 19 & $-$ & 18,  2, 16 &   6197&.075 &  5 &  $-$1 \\
 17,  3, 15 & $-$ & 18,  2, 16 &  26095&.072 &  5 &  $-$3 \\
 18,  3, 16 & $-$ & 19,  2, 17 &  21310&.053 &  5 &     1 \\
 19,  3, 16 & $-$ & 20,  2, 19 &  19831&.330 &  5 &     5 \\
 20,  3, 18 & $-$ & 21,  2, 19 &  11569&.979 &  5 &  $-$1 \\
 21,  3, 19 & $-$ & 22,  2, 20 &   6607&.331 &  5 &  $-$0 \\
 22,  3, 19 & $-$ & 23,  2, 22 &   7318&.534 &  5 &     2 \\
 26,  2, 25 & $-$ & 25,  3, 22 &   4813&.248 &  5 &  $-$1 \\
 26,  2, 24 & $-$ & 25,  3, 23 &  13932&.516 &  5 &  $-$7 \\
 28,  2, 26 & $-$ & 27,  3, 25 &  24644&.757 &  5 &  $-$0 \\
 28,  2, 27 & $-$ & 27,  3, 24 &  12629&.848 &  5 &     4 \\
 29,  2, 28 & $-$ & 28,  3, 25 &  16441&.572 &  5 &  $-$1 \\
 30,  2, 29 & $-$ & 29,  3, 26 &  20181&.553 &  5 &  $-$0 \\
 31,  2, 30 & $-$ & 30,  3, 27 &  23843&.940 &  5 &     0 \\
 27,  4, 24 & $-$ & 28,  3, 25 &  25160&.533 &  5 &     3 \\
 27,  4, 23 & $-$ & 28,  3, 26 &  26014&.363 &  5 &  $-$8 \\
 28,  4, 25 & $-$ & 29,  3, 26 &  20523&.876 &  5 &     5 \\
 28,  4, 24 & $-$ & 29,  3, 27 &  21572&.582 &  5 & $-$10 \\
 29,  4, 26 & $-$ & 30,  3, 27 &  15856&.700 &  5 &     6 \\
 29,  4, 25 & $-$ & 30,  3, 28 &  17135&.390 &  5 &  $-$4 \\
 30,  4, 27 & $-$ & 31,  3, 28 &  11155&.783 &  5 &     2 \\
 30,  4, 26 & $-$ & 31,  3, 29 &  12704&.192 &  5 &  $-$2 \\
 31,  4, 28 & $-$ & 32,  3, 29 &   6417&.726 &  5 &     3 \\
 31,  4, 27 & $-$ & 32,  3, 30 &   8280&.581 &  5 &     2 \\
 36,  3, 33 & $-$ & 35,  4, 32 &  12979&.624 &  5 &     4 \\
 36,  3, 34 & $-$ & 35,  4, 31 &   9299&.790 &  5 &  $-$2 \\
 37,  3, 35 & $-$ & 36,  4, 32 &  13655&.067 &  5 &  $-$4 \\
 38,  3, 35 & $-$ & 37,  4, 34 &  22999&.972 &  5 &     6 \\
 39,  3, 37 & $-$ & 38,  4, 34 &  22299&.091 &  5 &     3 \\
 37,  5, 32 & $-$ & 38,  4, 35 &  24129&.041 &  5 &     4 \\
 38,  5, 34 & $-$ & 39,  4, 35 &  19274&.638 &  5 &  $-$2 \\
 38,  5, 33 & $-$ & 39,  4, 36 &  19621&.858 &  5 &     1 \\
 39,  5, 34 & $-$ & 40,  4, 37 &  15110&.833 &  5 &     0 \\
 35,  0, 35 & $-$ & 34,  0, 34 & 152420&.07  & 50 &    10 \\
 35,  2, 33 & $-$ & 34,  2, 32 & 156362&.10  & 50 & $-$19 \\
 35,  6, 29 & $-$ & 34,  6, 28 & 154794&.60  & 50 &     5 \\
 35,  6, 30 & $-$ & 34,  6, 29 & 154794&.60  & 50 &     5 \\
 35,  7, 28 & $-$ & 34,  7, 27 & 154778&.85  & 50 & $-$24 \\
 35,  7, 29 & $-$ & 34,  7, 28 & 154778&.85  & 50 & $-$24 \\
 35,  8, 27 & $-$ & 34,  8, 26 & 154773&.84  & 50 & $-$40 \\
 35,  8, 28 & $-$ & 34,  8, 27 & 154773&.84  & 50 & $-$40 \\
 35,  9, 26 & $-$ & 34,  9, 25 & 154775&.62  & 50 & $-$38 \\
 35,  9, 27 & $-$ & 34,  9, 26 & 154775&.62  & 50 & $-$38 \\
 35, 10, 26 & $-$ & 34, 10, 25 & 154782&.18  & 50 &  $-$3 \\
 35, 10, 25 & $-$ & 34, 10, 24 & 154782&.18  & 50 &  $-$3 \\
 35, 11, 25 & $-$ & 34, 11, 24 & 154792&.35  & 50 &    22 \\
 35, 11, 24 & $-$ & 34, 11, 23 & 154792&.35  & 50 &    22 \\
 35, 12, 24 & $-$ & 34, 12, 23 & 154805&.40  & 50 & $-$28 \\
 35, 12, 23 & $-$ & 34, 12, 22 & 154805&.40  & 50 & $-$28 \\
 36,  0, 36 & $-$ & 35,  0, 35 & 156709&.22  & 50 &  $-$7 \\
 36,  1, 36 & $-$ & 35,  1, 35 & 156376&.78  & 50 &  $-$9 \\
 36,  2, 34 & $-$ & 35,  2, 33 & 160851&.40  & 50 & $-$55 \\
 36,  2, 35 & $-$ & 35,  2, 34 & 158511&.09  & 50 &    15 \\
 36,  3, 33 & $-$ & 35,  3, 32 & 159816&.09  & 50 & $-$55 \\
 36,  3, 34 & $-$ & 35,  3, 33 & 159295&.56  & 50 &  $-$7 \\
 36,  4, 32 & $-$ & 35,  4, 31 & 159360&.35  & 50 &  $-$6 \\
 36,  4, 33 & $-$ & 35,  4, 32 & 159324&.49  & 50 &  $-$4 \\
 36,  5, 31 & $-$ & 35,  5, 30 & 159258&.14  & 50 & $-$38 \\
 36,  5, 32 & $-$ & 35,  5, 31 & 159256&.97  & 50 &    14 \\
 36,  6, 31 & $-$ & 35,  6, 30 & 159218&.77  & 50 & $-$44 \\
 36,  6, 30 & $-$ & 35,  6, 29 & 159218&.77  & 50 & $-$44 \\
 36,  7, 30 & $-$ & 35,  7, 29 & 159201&.12  & 50 &     4 \\
 36,  7, 29 & $-$ & 35,  7, 28 & 159201&.12  & 50 &     4 \\
 36,  8, 29 & $-$ & 35,  8, 28 & 159194&.99  & 50 & $-$15 \\
 36,  8, 28 & $-$ & 35,  8, 27 & 159194&.99  & 50 & $-$15 \\
 36, 13, 24 & $-$ & 35, 13, 23 & 159241&.59  & 50 &     7 \\
 36, 13, 23 & $-$ & 35, 13, 22 & 159241&.59  & 50 &     7 \\
 36, 14, 23 & $-$ & 35, 14, 22 & 159259&.85  & 50 &     6 \\
 36, 14, 22 & $-$ & 35, 14, 21 & 159259&.85  & 50 &     6 \\
 36, 15, 22 & $-$ & 35, 15, 21 & 159280&.32  & 50 &    26 \\
 36, 15, 21 & $-$ & 35, 15, 20 & 159280&.32  & 50 &    26 \\
 36, 16, 21 & $-$ & 35, 16, 20 & 159302&.83  & 50 &    28 \\
 36, 16, 20 & $-$ & 35, 16, 19 & 159302&.83  & 50 &    28 \\
 36, 17, 20 & $-$ & 35, 17, 19 & 159327&.32  & 50 &    42 \\
 36, 17, 19 & $-$ & 35, 17, 18 & 159327&.32  & 50 &    42 \\
 36, 18, 19 & $-$ & 35, 18, 18 & 159353&.76  & 50 &   100 \\
 36, 18, 18 & $-$ & 35, 18, 17 & 159353&.76  & 50 &   100 \\
 37,  0, 37 & $-$ & 36,  0, 36 & 160998&.16  & 50 &    12 \\
 37,  1, 37 & $-$ & 36,  1, 36 & 160691&.64  & 50 &    25 \\
 48,  1, 47 & $-$ & 47,  1, 46 & 211856&.19  & 50 &    11 \\
 48,  2, 47 & $-$ & 47,  2, 46 & 210741&.48  & 50 &$-$115 \\
 48,  2, 46 & $-$ & 47,  2, 45 & 214343&.68  & 50 &    15 \\
 48,  3, 46 & $-$ & 47,  3, 45 & 212219&.54  & 50 &    25 \\
 48,  4, 45 & $-$ & 47,  4, 44 & 212517&.79  & 50 &     5 \\
 48,  4, 44 & $-$ & 47,  4, 43 & 212766&.64  & 50 & $-$12 \\
 48,  5, 44 & $-$ & 47,  5, 43 & 212415&.06  & 50 & $-$32 \\
 48,  5, 43 & $-$ & 47,  5, 42 & 212430&.89  & 50 &    12 \\
 48,  6, 43 & $-$ & 47,  6, 42 & 212320&.28  & 50 &    40 \\
 48,  6, 42 & $-$ & 47,  6, 41 & 212320&.77  & 50 & $-$59 \\
 48,  7, 42 & $-$ & 47,  7, 41 & 212267&.79  & 50 & $-$29 \\
 48,  7, 41 & $-$ & 47,  7, 40 & 212267&.79  & 50 & $-$29 \\
 48,  8, 41 & $-$ & 47,  8, 40 & 212241&.20  & 50 &     2 \\
 48,  8, 40 & $-$ & 47,  8, 39 & 212241&.20  & 50 &     2 \\
 48,  9, 40 & $-$ & 47,  9, 39 & 212230&.19  & 50 & $-$42 \\
 48,  9, 39 & $-$ & 47,  9, 38 & 212230&.19  & 50 & $-$42 \\
 48, 10, 39 & $-$ & 47, 10, 38 & 212229&.67  & 50 &    28 \\
 48, 10, 38 & $-$ & 47, 10, 37 & 212229&.67  & 50 &    28 \\
 48, 12, 37 & $-$ & 47, 12, 36 & 212249&.06  & 50 & $-$46 \\
 48, 12, 36 & $-$ & 47, 12, 35 & 212249&.06  & 50 & $-$46 \\
 48, 13, 36 & $-$ & 47, 13, 35 & 212266&.32  & 50 & $-$39 \\
 48, 13, 35 & $-$ & 47, 13, 34 & 212266&.32  & 50 & $-$39 \\
 48, 14, 35 & $-$ & 47, 14, 34 & 212287&.59  & 50 &    14 \\
 48, 14, 34 & $-$ & 47, 14, 33 & 212287&.59  & 50 &    14 \\
 48, 15, 34 & $-$ & 47, 15, 33 & 212312&.29  & 50 &  $-$1 \\
 48, 15, 33 & $-$ & 47, 15, 32 & 212312&.29  & 50 &  $-$1 \\
 48, 16, 33 & $-$ & 47, 16, 32 & 212340&.19  & 50 &     3 \\
 48, 16, 32 & $-$ & 47, 16, 31 & 212340&.19  & 50 &     3 \\
 48, 17, 32 & $-$ & 47, 17, 31 & 212371&.06  & 50 &    18 \\
 48, 17, 31 & $-$ & 47, 17, 30 & 212371&.06  & 50 &    18 \\
 48, 18, 31 & $-$ & 47, 18, 30 & 212404&.72  & 50 &    21 \\
 48, 18, 30 & $-$ & 47, 18, 29 & 212404&.72  & 50 &    21 \\
 48, 19, 30 & $-$ & 47, 19, 29 & 212441&.11  & 50 &    64 \\
 48, 19, 29 & $-$ & 47, 19, 28 & 212441&.11  & 50 &    64 \\
 48, 20, 29 & $-$ & 47, 20, 28 & 212480&.03  & 50 &    26 \\
 48, 20, 28 & $-$ & 47, 20, 27 & 212480&.03  & 50 &    26 \\
 48, 21, 28 & $-$ & 47, 21, 27 & 212521&.49  & 50 & $-$24 \\
 48, 21, 27 & $-$ & 47, 21, 26 & 212521&.49  & 50 & $-$24 \\
 48, 22, 27 & $-$ & 47, 22, 26 & 212565&.54  & 50 &     2 \\
 48, 22, 26 & $-$ & 47, 22, 25 & 212565&.54  & 50 &     2 \\
 48, 23, 26 & $-$ & 47, 23, 25 & 212611&.99  & 50 & $-$58 \\
 48, 23, 25 & $-$ & 47, 23, 24 & 212611&.99  & 50 & $-$58 \\
 49,  0, 49 & $-$ & 48,  0, 48 & 212476&.62  & 50 &   100 \\
 49,  1, 49 & $-$ & 48,  1, 48 & 212378&.48  & 50 &    62 \\
 49,  2, 48 & $-$ & 48,  2, 47 & 215077&.17  & 50 &    14 \\
 49,  4, 46 & $-$ & 48,  4, 45 & 216949&.09  & 50 & $-$14 \\
 49, 16, 34 & $-$ & 48, 16, 33 & 216757&.71  & 50 &     9 \\
 49, 16, 33 & $-$ & 48, 16, 32 & 216757&.71  & 50 &     9 \\
 49, 17, 33 & $-$ & 48, 17, 32 & 216789&.07  & 50 &    45 \\
 49, 17, 32 & $-$ & 48, 17, 31 & 216789&.07  & 50 &    45 \\
 49, 20, 30 & $-$ & 48, 20, 29 & 216899&.87  & 50 &     3 \\
 49, 20, 29 & $-$ & 48, 20, 28 & 216899&.87  & 50 &     3 \\
 50,  0, 50 & $-$ & 49,  0, 49 & 216767&.70  & 50 & $-$21 \\
 50,  3, 48 & $-$ & 49,  3, 47 & 221009&.11  & 50 &     7 \\
 50,  7, 44 & $-$ & 49,  7, 43 & 221112&.20  & 50 & $-$26 \\
 50,  7, 43 & $-$ & 49,  7, 42 & 221112&.20  & 50 & $-$26 \\
 50,  8, 43 & $-$ & 49,  8, 42 & 221080&.73  & 50 & $-$13 \\
 50,  8, 42 & $-$ & 49,  8, 41 & 221080&.73  & 50 & $-$13 \\
 50,  9, 42 & $-$ & 49,  9, 41 & 221066&.76  & 50 & $-$28 \\
 50,  9, 41 & $-$ & 49,  9, 40 & 221066&.76  & 50 & $-$28 \\
 50, 10, 41 & $-$ & 49, 10, 40 & 221064&.35  & 50 & $-$25 \\
 50, 10, 40 & $-$ & 49, 10, 39 & 221064&.35  & 50 & $-$25 \\
 50, 11, 40 & $-$ & 49, 11, 39 & 221070&.19  & 50 & $-$11 \\
 50, 11, 39 & $-$ & 49, 11, 38 & 221070&.19  & 50 & $-$11 \\
 50, 12, 39 & $-$ & 49, 12, 38 & 221082&.34  & 50 &    28 \\
 50, 12, 38 & $-$ & 49, 12, 37 & 221082&.34  & 50 &    28 \\
 50, 13, 38 & $-$ & 49, 13, 37 & 221099&.49  & 50 &  $-$5 \\
 50, 13, 37 & $-$ & 49, 13, 36 & 221099&.49  & 50 &  $-$5 \\
 50, 14, 37 & $-$ & 49, 14, 36 & 221121&.05  & 50 &    85 \\
 50, 14, 36 & $-$ & 49, 14, 35 & 221121&.05  & 50 &    85 \\
 50, 20, 31 & $-$ & 49, 20, 30 & 221319&.17  & 50 &$-$141 \\
 50, 20, 30 & $-$ & 49, 20, 29 & 221319&.17  & 50 &$-$141 \\
 51,  0, 51 & $-$ & 50,  0, 50 & 221059&.01  & 50 &  $-$5 \\
 51,  1, 51 & $-$ & 50,  1, 50 & 220979&.61  & 50 &    32 \\
 64,  6, 59 & $-$ & 63,  6, 58 & 283158&.51  & 50 & $-$21 \\
 64,  6, 58 & $-$ & 63,  6, 57 & 283171&.68  & 50 &    33 \\
 64,  7, 58 & $-$ & 63,  7, 57 & 283022&.98  & 50 &    14 \\
 64,  7, 57 & $-$ & 63,  7, 56 & 283022&.98  & 50 &    14 \\
 64,  8, 57 & $-$ & 63,  8, 56 & 282943&.15  & 50 &    24 \\
 64,  8, 56 & $-$ & 63,  8, 55 & 282943&.15  & 50 &    24 \\
 64,  9, 56 & $-$ & 63,  9, 55 & 282898&.81  & 50 &     1 \\
 64,  9, 55 & $-$ & 63,  9, 54 & 282898&.81  & 50 &     1 \\
 64, 10, 55 & $-$ & 63, 10, 54 & 282877&.00  & 50 & $-$26 \\
 64, 10, 54 & $-$ & 63, 10, 53 & 282877&.00  & 50 & $-$26 \\
 64, 11, 54 & $-$ & 63, 11, 53 & 282870&.71  & 50 & $-$11 \\
 64, 11, 53 & $-$ & 63, 11, 52 & 282870&.71  & 50 & $-$11 \\
 64, 12, 53 & $-$ & 63, 12, 52 & 282875&.72  & 50 & $-$41 \\
 64, 12, 52 & $-$ & 63, 12, 51 & 282875&.72  & 50 & $-$41 \\
 64, 13, 52 & $-$ & 63, 13, 51 & 282889&.55  & 50 & $-$38 \\
 64, 13, 51 & $-$ & 63, 13, 50 & 282889&.55  & 50 & $-$38 \\
 64, 14, 51 & $-$ & 63, 14, 50 & 282910&.51  & 50 & $-$38 \\
 64, 14, 50 & $-$ & 63, 14, 49 & 282910&.51  & 50 & $-$38 \\
 64, 15, 50 & $-$ & 63, 15, 49 & 282937&.54  & 50 &     6 \\
 64, 15, 49 & $-$ & 63, 15, 48 & 282937&.54  & 50 &     6 \\
 64, 16, 49 & $-$ & 63, 16, 48 & 282969&.81  & 50 &    25 \\
 64, 16, 48 & $-$ & 63, 16, 47 & 282969&.81  & 50 &    25 \\
 64, 17, 48 & $-$ & 63, 17, 47 & 283006&.76  & 50 &  $-$3 \\
 64, 17, 47 & $-$ & 63, 17, 46 & 283006&.76  & 50 &  $-$3 \\
 64, 18, 47 & $-$ & 63, 18, 46 & 283048&.15  & 50 &    64 \\
 64, 18, 46 & $-$ & 63, 18, 45 & 283048&.15  & 50 &    64 \\
 64, 19, 46 & $-$ & 63, 19, 45 & 283093&.46  & 50 & $-$14 \\
 64, 19, 45 & $-$ & 63, 19, 44 & 283093&.46  & 50 & $-$14 \\
 64, 20, 45 & $-$ & 63, 20, 44 & 283142&.77  & 50 &    47 \\
 64, 20, 44 & $-$ & 63, 20, 43 & 283142&.77  & 50 &    47 \\
 64, 21, 44 & $-$ & 63, 21, 43 & 283195&.66  & 50 & $-$22 \\
 64, 21, 43 & $-$ & 63, 21, 42 & 283195&.66  & 50 & $-$22 \\
\end{longtable}
}}
\onltab{\value{gprcn}}{\longtab{\value{gprcn}}{\begin{longtable}{rclr@{}lrr}
\caption{\label{t:lines-gauche-PrCN}
Transitions of \textit{gauche}-\textit{n}-propyl cyanide, employed in 
the present fits, their frequencies (MHz), uncertainties
Unc. (kHz), and residuals O$-$C (kHz) between frequencies measured in the 
laboratory and those calculated from the final spectroscopic parameters. 
Unresolved asymmetry splitting (two transitions having the same 
$K_a$ and the same transition frequency) has been treated as 
intensity-weighted average of the two lines.}\\
\hline \hline
$J'$, $K_a'$, $K_c'$ & $-$ & $J''$, $K_a''$, $K_c''$ & \multicolumn{2}{c}{Frequency} & Unc. & O$-$C\\
\hline
\endfirsthead
\caption{continued.}\\
\hline\hline
$J'$, $K_a'$, $K_c'$ & $-$ & $J''$, $K_a''$, $K_c''$ & \multicolumn{2}{c}{Frequency} & Unc. & O$-$C\\
\hline
\endhead
\hline
\endfoot
  2,  0,  2 & $-$ &  1,  0,  1 &  11912&.654 & 10 & $-$18 \\
  2,  1,  2 & $-$ &  1,  1,  1 &  11384&.045 & 10 &  $-$0 \\
  2,  1,  1 & $-$ &  1,  1,  0 &  12508&.374 & 10 & $-$10 \\
  3,  0,  3 & $-$ &  2,  0,  2 &  17785&.949 & 10 &     6 \\
  3,  1,  3 & $-$ &  2,  1,  2 &  17055&.610 &  5 &     1 \\
  2,  1,  2 & $-$ &  1,  0,  1 &  18176&.781 &  5 &     1 \\
  4,  1,  3 & $-$ &  4,  0,  4 &  10286&.601 &  5 &     0 \\
  5,  1,  4 & $-$ &  5,  0,  5 &  12167&.868 &  5 &  $-$3 \\
  6,  1,  5 & $-$ &  6,  0,  6 &  14609&.404 &  5 &     1 \\
  7,  1,  6 & $-$ &  7,  0,  7 &  17634&.822 &  5 &     1 \\
  2,  2,  1 & $-$ &  2,  1,  2 &  22064&.261 &  5 &     0 \\
 24,  3, 21 & $-$ & 25,  2, 24 &   5073&.952 &  5 &     0 \\
 10,  3,  7 & $-$ &  9,  4,  6 &  12670&.827 &  5 &  $-$2 \\
 11,  3,  9 & $-$ & 10,  4,  6 &  16546&.819 &  5 &  $-$3 \\
 28,  3, 26 & $-$ & 27,  4, 23 &  19891&.929 &  5 &     1 \\
 36,  4, 32 & $-$ & 37,  3, 35 &  19547&.899 &  5 &     1 \\
 13,  4, 10 & $-$ & 12,  5,  7 &  15362&.409 &  5 &     0 \\
 38,  4, 35 & $-$ & 37,  5, 32 &   9991&.617 &  5 &     0 \\
 39,  4, 36 & $-$ & 38,  5, 33 &   5311&.217 &  5 &     1 \\
 15,  5, 10 & $-$ & 14,  6,  9 &  13597&.769 &  5 &     1 \\
 12,  7,  6 & $-$ & 13,  6,  7 &  13549&.701 &  5 &  $-$1 \\
 12,  7,  5 & $-$ & 13,  6,  8 &  13551&.095 &  5 &     5 \\
 13,  7,  7 & $-$ & 14,  6,  8 &   7398&.401 &  5 &  $-$2 \\
 13,  7,  6 & $-$ & 14,  6,  9 &   7401&.855 &  5 &     4 \\
 15,  8,  8 & $-$ & 16,  7,  9 &   9498&.238 &  5 &  $-$4 \\
 15,  8,  7 & $-$ & 16,  7, 10 &   9498&.794 &  5 &  $-$5 \\
 19,  7, 12 & $-$ & 18,  8, 11 &   9123&.186 &  5 &  $-$1 \\
 19,  7, 13 & $-$ & 18,  8, 10 &   9116&.781 &  5 &  $-$1 \\
 20,  7, 14 & $-$ & 19,  8, 11 &  15398&.422 &  5 &     0 \\
 22,  8, 14 & $-$ & 21,  9, 13 &  13221&.090 &  5 &  $-$0 \\
 22,  8, 15 & $-$ & 21,  9, 12 &  13218&.792 &  5 &  $-$2 \\
 23,  8, 15 & $-$ & 22,  9, 14 &  19511&.211 &  5 &     2 \\
 23,  8, 16 & $-$ & 22,  9, 13 &  19506&.528 &  5 &     0 \\
 19, 10, 10 & $-$ & 20,  9, 11 &  13681&.725 &  5 &  $-$1 \\
 19, 10,  9 & $-$ & 20,  9, 12 &  13681&.725 &  5 &  $-$1 \\
 25,  9, 17 & $-$ & 24, 10, 14 &  17310&.401 &  5 &  $-$6 \\
 25,  9, 16 & $-$ & 24, 10, 15 &  17311&.214 &  5 &     2 \\
 26,  9, 18 & $-$ & 25, 10, 15 &  23599&.302 &  5 &     2 \\
 26,  9, 17 & $-$ & 25, 10, 16 &  23600&.945 &  5 &    11 \\
 28, 10, 19 & $-$ & 27, 11, 16 &  21388&.390 &  5 &  $-$9 \\
 28, 10, 18 & $-$ & 27, 11, 17 &  21388&.670 &  5 &  $-$6 \\
 26, 13, 14 & $-$ & 27, 12, 15 &  13813&.369 &  5 &  $-$3 \\
 26, 13, 13 & $-$ & 27, 12, 16 &  13813&.369 &  5 &  $-$3 \\
 22,  4, 18 & $-$ & 22,  4, 19 &  25543&.715 &  5 &  $-$1 \\
 25,  5, 20 & $-$ & 25,  5, 21 &  16254&.957 &  5 &     2 \\
 26,  5, 21 & $-$ & 26,  5, 22 &  21090&.090 &  5 &  $-$1 \\
 30,  6, 24 & $-$ & 30,  6, 25 &  16246&.388 &  5 &     1 \\
 31,  6, 25 & $-$ & 31,  6, 26 &  21145&.902 &  5 &     1 \\
 34,  7, 27 & $-$ & 34,  7, 28 &  11727&.780 &  5 &  $-$2 \\
 36,  7, 29 & $-$ & 36,  7, 30 &  20572&.880 &  5 &  $-$2 \\
 37,  7, 30 & $-$ & 37,  7, 31 &  26237&.933 &  5 &     4 \\
 39,  8, 31 & $-$ & 39,  8, 32 &  11013&.037 &  5 &  $-$2 \\
 25,  5, 21 & $-$ & 24,  4, 20 & 182990&.23  & 50 &    19 \\
 26,  6, 21 & $-$ & 25,  6, 20 & 157628&.25  & 50 &    16 \\
 26,  7, 19 & $-$ & 25,  7, 18 & 157477&.90  & 50 &     8 \\
 26, 10, 17 & $-$ & 25, 10, 16 & 156296&.53  & 50 & $-$59 \\
 26, 10, 16 & $-$ & 25, 10, 15 & 156296&.53  & 50 & $-$59 \\
 26, 11, 16 & $-$ & 25, 11, 15 & 156141&.79  & 50 &    65 \\
 26, 11, 15 & $-$ & 25, 11, 14 & 156141&.79  & 50 &    65 \\
 26, 12, 15 & $-$ & 25, 12, 14 & 156032&.20  & 50 & $-$22 \\
 26, 12, 14 & $-$ & 25, 12, 13 & 156032&.20  & 50 & $-$22 \\
 26, 13, 14 & $-$ & 25, 13, 13 & 155954&.49  & 50 & $-$56 \\
 26, 13, 13 & $-$ & 25, 13, 12 & 155954&.49  & 50 & $-$56 \\
 26, 14, 13 & $-$ & 25, 14, 12 & 155900&.13  & 50 &  $-$0 \\
 26, 14, 12 & $-$ & 25, 14, 11 & 155900&.13  & 50 &  $-$0 \\
 26, 15, 12 & $-$ & 25, 15, 11 & 155863&.29  & 50 & $-$10 \\
 26, 15, 11 & $-$ & 25, 15, 10 & 155863&.29  & 50 & $-$10 \\
 26, 17, 10 & $-$ & 25, 17,  9 & 155827&.91  & 50 & $-$33 \\
 26, 17,  9 & $-$ & 25, 17,  8 & 155827&.91  & 50 & $-$33 \\
 26, 18,  9 & $-$ & 25, 18,  8 & 155824&.64  & 50 & $-$20 \\
 26, 18,  8 & $-$ & 25, 18,  7 & 155824&.64  & 50 & $-$20 \\
 26, 19,  8 & $-$ & 25, 19,  7 & 155828&.81  & 50 & $-$17 \\
 26, 19,  7 & $-$ & 25, 19,  6 & 155828&.81  & 50 & $-$17 \\
 28,  1, 27 & $-$ & 27,  1, 26 & 157495&.25  & 50 & $-$33 \\
 28,  2, 27 & $-$ & 27,  1, 26 & 157503&.23  & 50 & $-$51 \\
 28,  1, 27 & $-$ & 27,  2, 26 & 157482&.64  & 50 &    46 \\
 28,  2, 27 & $-$ & 27,  2, 26 & 157490&.60  & 50 &     8 \\
 29,  5, 25 & $-$ & 28,  4, 24 & 189805&.28  & 50 &    48 \\
 30,  4, 27 & $-$ & 29,  3, 26 & 178167&.69  & 50 &    31 \\
 30,  4, 26 & $-$ & 29,  4, 25 & 182703&.13  & 50 & $-$23 \\
 30,  5, 26 & $-$ & 29,  4, 25 & 192210&.43  & 50 &    32 \\
 30,  7, 23 & $-$ & 29,  7, 22 & 182961&.27  & 50 & $-$29 \\
 31,  2, 29 & $-$ & 30,  2, 28 & 177803&.90  & 50 & $-$13 \\
 31,  3, 29 & $-$ & 30,  2, 28 & 177857&.94  & 50 &  $-$1 \\
 31,  2, 29 & $-$ & 30,  3, 28 & 177721&.22  & 50 &     6 \\
 31,  3, 29 & $-$ & 30,  3, 28 & 177775&.23  & 50 & $-$12 \\
 31,  3, 28 & $-$ & 30,  3, 27 & 182248&.22  & 50 & $-$22 \\
 31,  4, 28 & $-$ & 30,  3, 27 & 183077&.55  & 50 &    26 \\
 31,  3, 28 & $-$ & 30,  4, 27 & 181050&.65  & 50 &    36 \\
 31,  4, 27 & $-$ & 30,  5, 26 & 178318&.43  & 50 &   156 \\
 31,  5, 26 & $-$ & 30,  5, 25 & 192803&.61  & 50 &     6 \\
 31,  7, 25 & $-$ & 30,  7, 24 & 188156&.51  & 50 & $-$20 \\
 31,  7, 24 & $-$ & 30,  7, 23 & 189515&.12  & 50 & $-$74 \\
 31,  8, 23 & $-$ & 30,  8, 22 & 187858&.13  & 50 &     9 \\
 32,  2, 30 & $-$ & 31,  2, 29 & 183183&.36  & 50 &    18 \\
 32,  3, 30 & $-$ & 31,  2, 29 & 183218&.50  & 50 &    28 \\
 32,  2, 30 & $-$ & 31,  3, 29 & 183129&.35  & 50 &    37 \\
 32,  3, 30 & $-$ & 31,  3, 29 & 183164&.45  & 50 &     6 \\
 32,  4, 29 & $-$ & 31,  3, 28 & 188119&.54  & 50 &  $-$5 \\
 32,  4, 28 & $-$ & 31,  4, 27 & 192927&.96  & 50 & $-$13 \\
 32,  4, 28 & $-$ & 31,  5, 27 & 185714&.09  & 50 &     6 \\
 32, 10, 23 & $-$ & 31, 10, 22 & 192852&.63  & 50 &  $-$4 \\
 32, 10, 22 & $-$ & 31, 10, 21 & 192854&.42  & 50 &  $-$4 \\
 32, 12, 21 & $-$ & 31, 12, 20 & 192320&.23  & 50 &     5 \\
 32, 12, 20 & $-$ & 31, 12, 19 & 192320&.23  & 50 &     5 \\
 32, 13, 20 & $-$ & 31, 13, 19 & 192158&.76  & 50 & $-$19 \\
 32, 13, 19 & $-$ & 31, 13, 18 & 192158&.76  & 50 & $-$19 \\
 32, 14, 19 & $-$ & 31, 14, 18 & 192040&.62  & 50 &    14 \\
 32, 14, 18 & $-$ & 31, 14, 17 & 192040&.62  & 50 &    14 \\
 33,  2, 31 & $-$ & 32,  2, 30 & 188564&.34  & 50 & $-$33 \\
 33,  3, 31 & $-$ & 32,  2, 30 & 188587&.13  & 50 &    13 \\
 33,  2, 31 & $-$ & 32,  3, 30 & 188529&.24  & 50 &  $-$2 \\
 33,  3, 31 & $-$ & 32,  3, 30 & 188551&.94  & 50 & $-$47 \\
 33,  3, 30 & $-$ & 32,  3, 29 & 192871&.17  & 50 &  $-$3 \\
 33,  4, 30 & $-$ & 32,  3, 29 & 193259&.00  & 50 &    18 \\
 33,  3, 30 & $-$ & 32,  4, 29 & 192301&.88  & 50 & $-$16 \\
 33,  4, 30 & $-$ & 32,  4, 29 & 192689&.70  & 50 &  $-$5 \\
 33,  4, 29 & $-$ & 32,  5, 28 & 192663&.51  & 50 &    12 \\
 34,  2, 32 & $-$ & 33,  2, 31 & 193946&.21  & 50 & $-$59 \\
 34,  3, 32 & $-$ & 33,  2, 31 & 193960&.95  & 50 &    14 \\
 34,  2, 32 & $-$ & 33,  3, 31 & 193923&.55  & 50 &    26 \\
 34,  3, 32 & $-$ & 33,  3, 31 & 193938&.16  & 50 & $-$32 \\
 34,  6, 28 & $-$ & 33,  6, 27 & 212120&.06  & 50 &     3 \\
 35,  6, 29 & $-$ & 34,  6, 28 & 218287&.17  & 50 & $-$26 \\
 35,  7, 29 & $-$ & 34,  7, 28 & 212483&.75  & 50 &    16 \\
 35,  8, 28 & $-$ & 34,  8, 27 & 212458&.29  & 50 & $-$13 \\
 35,  9, 27 & $-$ & 34,  9, 26 & 211828&.06  & 50 &   142 \\
 35,  9, 26 & $-$ & 34,  9, 25 & 211938&.01  & 50 &     7 \\
 36,  4, 32 & $-$ & 35,  5, 31 & 211527&.01  & 50 &    14 \\
 36, 10, 27 & $-$ & 35, 10, 26 & 217402&.38  & 50 & $-$15 \\
 36, 10, 26 & $-$ & 35, 10, 25 & 217418&.12  & 50 &  $-$2 \\
 36, 11, 26 & $-$ & 35, 11, 25 & 216940&.59  & 50 & $-$30 \\
 36, 11, 25 & $-$ & 35, 11, 24 & 216941&.74  & 50 &    33 \\
 36, 12, 25 & $-$ & 35, 12, 24 & 216609&.17  & 50 & $-$33 \\
 36, 12, 24 & $-$ & 35, 12, 23 & 216609&.17  & 50 & $-$33 \\
 36, 13, 24 & $-$ & 35, 13, 23 & 216367&.23  & 50 &  $-$1 \\
 36, 13, 23 & $-$ & 35, 13, 22 & 216367&.23  & 50 &  $-$1 \\
 36, 14, 23 & $-$ & 35, 14, 22 & 216188&.17  & 50 &    23 \\
 36, 14, 22 & $-$ & 35, 14, 21 & 216188&.17  & 50 &    23 \\
 36, 16, 21 & $-$ & 35, 16, 20 & 215956&.57  & 50 & $-$26 \\
 36, 16, 20 & $-$ & 35, 16, 19 & 215956&.57  & 50 & $-$26 \\
 36, 17, 20 & $-$ & 35, 17, 19 & 215885&.18  & 50 & $-$12 \\
 36, 17, 19 & $-$ & 35, 17, 18 & 215885&.18  & 50 & $-$12 \\
 36, 18, 19 & $-$ & 35, 18, 18 & 215835&.30  & 50 &     9 \\
 36, 18, 18 & $-$ & 35, 18, 17 & 215835&.30  & 50 &     9 \\
 36, 20, 17 & $-$ & 35, 20, 16 & 215784&.95  & 50 &    23 \\
 36, 20, 16 & $-$ & 35, 20, 15 & 215784&.95  & 50 &    23 \\
 36, 21, 16 & $-$ & 35, 21, 15 & 215779&.22  & 50 &    45 \\
 36, 21, 15 & $-$ & 35, 21, 14 & 215779&.22  & 50 &    45 \\
 36, 22, 15 & $-$ & 35, 22, 14 & 215783&.85  & 50 &     5 \\
 36, 22, 14 & $-$ & 35, 22, 13 & 215783&.85  & 50 &     5 \\
 36, 23, 14 & $-$ & 35, 23, 13 & 215797&.53  & 50 & $-$19 \\
 36, 23, 13 & $-$ & 35, 23, 12 & 215797&.53  & 50 & $-$19 \\
 36, 24, 13 & $-$ & 35, 24, 12 & 215819&.22  & 50 &    36 \\
 36, 24, 12 & $-$ & 35, 24, 11 & 215819&.22  & 50 &    36 \\
 36, 25, 12 & $-$ & 35, 25, 11 & 215847&.81  & 50 & $-$52 \\
 36, 25, 11 & $-$ & 35, 25, 10 & 215847&.81  & 50 & $-$52 \\
 36, 26, 11 & $-$ & 35, 26, 10 & 215882&.85  & 50 & $-$12 \\
 36, 26, 10 & $-$ & 35, 26,  9 & 215882&.85  & 50 & $-$12 \\
 36, 27, 10 & $-$ & 35, 27,  9 & 215923&.60  & 50 &    10 \\
 36, 27,  9 & $-$ & 35, 27,  8 & 215923&.60  & 50 &    10 \\
 36, 28,  9 & $-$ & 35, 28,  8 & 215969&.54  & 50 & $-$15 \\
 36, 28,  8 & $-$ & 35, 28,  7 & 215969&.54  & 50 & $-$15 \\
 36, 32,  5 & $-$ & 35, 32,  4 & 216198&.45  & 50 &     4 \\
 36, 32,  4 & $-$ & 35, 32,  3 & 216198&.45  & 50 &     4 \\
 37,  3, 34 & $-$ & 36,  3, 33 & 214265&.30  & 50 &  $-$8 \\
 37,  3, 34 & $-$ & 36,  4, 33 & 214147&.29  & 50 & $-$26 \\
 37,  4, 34 & $-$ & 36,  4, 33 & 214225&.81  & 50 & $-$13 \\
 38,  1, 37 & $-$ & 37,  2, 36 & 211401&.39  & 50 &    59 \\
 38,  3, 36 & $-$ & 37,  2, 35 & 215476&.23  & 50 &    36 \\
 38,  2, 36 & $-$ & 37,  3, 35 & 215469&.85  & 50 & $-$35 \\
 38,  3, 35 & $-$ & 37,  3, 34 & 219628&.21  & 50 &    19 \\
 38,  3, 35 & $-$ & 37,  4, 34 & 219549&.67  & 50 & $-$13 \\
 38,  4, 35 & $-$ & 37,  4, 34 & 219601&.64  & 50 & $-$40 \\
 39,  0, 39 & $-$ & 38,  1, 38 & 212743&.02  & 50 &     6 \\
 39,  2, 37 & $-$ & 38,  2, 36 & 220854&.26  & 50 &  $-$8 \\
 39,  3, 37 & $-$ & 38,  2, 36 & 220855&.81  & 50 & $-$18 \\
 39,  2, 37 & $-$ & 38,  3, 36 & 220851&.79  & 50 & $-$24 \\
 39,  3, 37 & $-$ & 38,  3, 36 & 220853&.34  & 50 & $-$33 \\
 40,  0, 40 & $-$ & 39,  1, 39 & 218129&.30  & 50 &    18 \\
 47, 14, 34 & $-$ & 46, 14, 33 & 282874&.18  & 50 &    42 \\
 47, 17, 31 & $-$ & 46, 17, 30 & 282070&.21  & 50 &     4 \\
 47, 17, 30 & $-$ & 46, 17, 29 & 282070&.21  & 50 &     4 \\
 47, 19, 29 & $-$ & 46, 19, 28 & 281801&.94  & 50 &    20 \\
 47, 19, 28 & $-$ & 46, 19, 27 & 281801&.94  & 50 &    20 \\
 47, 20, 28 & $-$ & 46, 20, 27 & 281716&.83  & 50 &    33 \\
 47, 20, 27 & $-$ & 46, 20, 26 & 281716&.83  & 50 &    33 \\
 47, 21, 27 & $-$ & 46, 21, 26 & 281656&.67  & 50 &    10 \\
 47, 21, 26 & $-$ & 46, 21, 25 & 281656&.67  & 50 &    10 \\
 47, 22, 26 & $-$ & 46, 22, 25 & 281617&.41  & 50 & $-$44 \\
 47, 22, 25 & $-$ & 46, 22, 24 & 281617&.41  & 50 & $-$44 \\
 47, 23, 25 & $-$ & 46, 23, 24 & 281596&.02  & 50 & $-$16 \\
 47, 23, 24 & $-$ & 46, 23, 23 & 281596&.02  & 50 & $-$16 \\
 47, 24, 24 & $-$ & 46, 24, 23 & 281589&.94  & 50 &     8 \\
 47, 24, 23 & $-$ & 46, 24, 22 & 281589&.94  & 50 &     8 \\
 47, 25, 23 & $-$ & 46, 25, 22 & 281597&.15  & 50 & $-$18 \\
 47, 25, 22 & $-$ & 46, 25, 21 & 281597&.15  & 50 & $-$18 \\
 47, 26, 22 & $-$ & 46, 26, 21 & 281616&.14  & 50 & $-$10 \\
 47, 26, 21 & $-$ & 46, 26, 20 & 281616&.14  & 50 & $-$10 \\
 47, 27, 21 & $-$ & 46, 27, 20 & 281645&.57  & 50 &  $-$9 \\
 47, 27, 20 & $-$ & 46, 27, 19 & 281645&.57  & 50 &  $-$9 \\
 47, 28, 20 & $-$ & 46, 28, 19 & 281684&.43  & 50 &    48 \\
 47, 28, 19 & $-$ & 46, 28, 18 & 281684&.43  & 50 &    48 \\
 47, 29, 19 & $-$ & 46, 29, 18 & 281731&.64  & 50 & $-$30 \\
 47, 29, 18 & $-$ & 46, 29, 17 & 281731&.64  & 50 & $-$30 \\
 47, 30, 18 & $-$ & 46, 30, 17 & 281786&.66  & 50 & $-$36 \\
 47, 30, 17 & $-$ & 46, 30, 16 & 281786&.66  & 50 & $-$36 \\
 47, 31, 17 & $-$ & 46, 31, 16 & 281848&.86  & 50 &    31 \\
 47, 31, 16 & $-$ & 46, 31, 15 & 281848&.86  & 50 &    31 \\
 47, 33, 15 & $-$ & 46, 33, 14 & 281992&.38  & 50 &    34 \\
 47, 33, 14 & $-$ & 46, 33, 13 & 281992&.38  & 50 &    34 \\
 47, 34, 14 & $-$ & 46, 34, 13 & 282072&.85  & 50 & $-$22 \\
 47, 34, 13 & $-$ & 46, 34, 12 & 282072&.85  & 50 & $-$22 \\
 48,  5, 43 & $-$ & 47,  5, 42 & 281724&.06  & 50 &     0 \\
 48,  6, 43 & $-$ & 47,  6, 42 & 281635&.78  & 50 & $-$33 \\
\end{longtable}
}}
\end{document}